# Feasibility of Dark Matter in Neutron Stars: A Quantitative Analysis

**THESIS**

Submitted in partial fulfillment
of the requirements for the degree of
**DOCTOR OF PHILOSOPHY**

by

**Prashant Thakur**
**(Roll No. 2019PHXF0072G)**

Under the guidance of
**Prof. Tarun Kumar Jha (Supervisor)**

DEPARTMENT OF PHYSICS
BIRLA INSTITUTE OF TECHNOLOGY AND SCIENCE, PILANI
February - 2025

*This Thesis is Dedicated to Prof. Tarun Kumar Jha,*

*Dr. Tuhin Malik and*

*my family*

# BIRLA INSTITUTE OF TECHNOLOGY AND SCIENCE, PILANI (RAJASTHAN)

## CERTIFICATE

This is to certify that the thesis entitled **"Feasibility of Dark Matter in Neutron Stars: A Quantitative Analysis"** submitted by **Prashant Thakur**, ID No 2019PHXF0072G for award of **Ph.D.** of the Institute embodies original work done by him under our supervision.

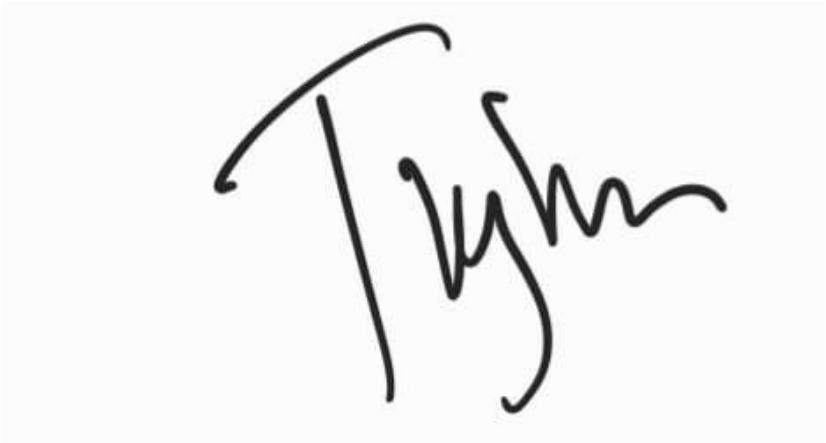

Signature of the Supervisor

Name: **Prof. Tarun Kumar Jha**

Designation: Associate professor

Affiliation: BITS Pilani, Goa, India

Date: February 12, 2025

## Abstract

This thesis explores the intricate relationship between dark matter and neutron star properties, focusing on the effects of dark matter on neutron star macroscopic characteristics such as mass, radius, and tidal deformability. Utilizing both two-fluid and single-fluid models, dark matter is incorporated into neutron star equations of state (EOS) to assess its influence. The study employs a Relativistic Mean Field (RMF) approach, generating a comprehensive set of EOS to simulate nuclear and dark matter interactions within neutron stars.

The research highlights the correlation between dark matter parameters and neutron star properties, demonstrating that an increase in dark matter fraction leads to a decrease in the maximum mass, radius, and tidal deformability of neutron stars. Bayesian inference, coupled with the latest astrophysical observations (PSR J0437-4715), including gravitational wave data from LIGO-Virgo and mass-radius measurements from NICER, is employed to refine the models and provide insights into the dark matter content within neutron stars.

With our robust and extensive dataset, we delve deeper and demonstrate that even in the presence of dark matter, the semi-universal C-Love relation remains intact. Advanced machine learning techniques are also applied to classify neutron stars based on their dark matter content, achieving promising results in identifying dark matter admixed neutron stars.

Furthermore, the thesis explores the inclusion of a $\sigma$-cut potential in the EOS, which simulates the effect of a quarkyonic phase or an exclusion volume, leading to a stiffening of the EOS at higher densities. This modification significantly impacts neutron star properties, favoring models that predict larger radii and smaller $f$-mode frequencies. The investigation of non-radial oscillations, specifically $f$ and $p$ modes, reveals a strong sensitivity to neutron star composition and EOS, with implications for future observational constraints.

In conclusion, this research advances our understanding of the interplay between dark matter and neutron star properties, offering new perspectives on the internal structure of neutron stars and contributing to the broader quest to unravel the mysteries of dark matter. The findings underscore the need for continued observational efforts and the development of more flexible theoretical models to accommodate emerging data and refine our understanding of these enigmatic astrophysical objects.

# Declaration of Academic Integrity

I, **Prashant Thakur**, S/O Mr. Har Krishan Singh, declare that this written submission represents my ideas in my own words and where others' ideas or words have been included, and I have adequately cited and referenced the original sources. I also declare that I have adhered to all principles of academic honesty and integrity and have not misrepresented, or fabricated, or falsified any idea/data/fact/source in my submission. I understand that any violation of the above will be cause for disciplinary action as per the rules and regulations of the Institute.

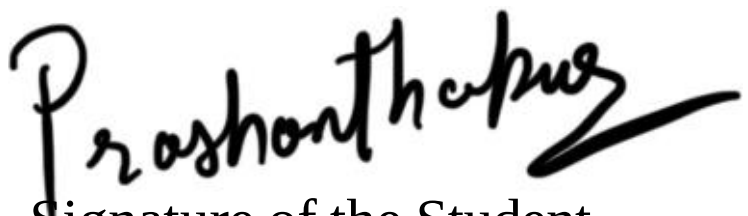

Signature of the Student

Name: Prashant Thakur

ID No.: 2019PHXF0072G

Date: February 12, 2025

# Acknowledgement

I would like to express my deepest gratitude to **my supervisor, Prof. Tarun Kumar Jha**. He was the first to believe in me and provided countless opportunities for learning and growth. His unwavering support, both in my academic journey and in my personal life, has been invaluable, and he has always been there when I needed it, especially during difficult times. Our discussions on the physics of neutron stars, astrophysics, and other critical topics have significantly deepened my understanding. **Tarun sir's** clear explanations and encouragement to think critically have been instrumental in my development as a researcher. His guidance has instilled in me a newfound confidence in my work, which I will carry forward into my future endeavors.

Beyond academics, **Tarun sir** has also taught me the importance of balancing work with personal life. I am immensely grateful for the warm hospitality that he and his wife **Anupma Ma'am** extended to my wife and me, as well as the thoughtful gifts they gave us. He made handling official tasks and paperwork much easier, taking a lot of stress off my shoulders. **Tarun sir** often took his Ph.D. students out for meals, making us feel valued and supported as part of a team. He frequently invited me to his home for lunch and dinner, where the warm and welcoming atmosphere made me feel like part of his extended family.

I owe much of my growth, both academically and personally, to **Tarun sir's** guidance. His influence will continue to shape my future work, and I wish him happiness and all the best in the future.

I would like to express my deepest gratitude to **my Ph.D. Mentor Dr. Tuhin Malik**, who has been the backbone of my PhD journey. Since we met in 2020, he has been much more than just a mentor: He has been a constant source of guidance, support, and encouragement, helping me become self-independent and shaping me into the person I am today. **Dr. Malik** not only taught me the essentials of computation and data analysis but also helped me grow by pointing out my mistakes and pushing me to improve. His scoldings were as valuable as his praises, teaching me the importance of hard work and dedication. He appreciated my efforts when I did well, which motivated me to strive even harder.

During my research trip to Portugal, **Dr. Malik** provided exceptional hospitality, ensuring I was well taken care of, both professionally and personally. He also played a significant role in supporting my mental health during challenging times, for which I am incredibly grateful. His wife's delicious meals added a comforting touch to my stay, making me feel at home even when I was far away.

Without **Dr. Malik's** support, my PhD journey would be incomplete. He has been instrumental not only in my academic work—helping me with my thesis and providing new ideas—but also in shaping my overall experience. I am truly fortunate to have had him by my side, and I wish him and his family a future filled with happiness and success.

I would like to express my deep gratitude to **Prof. Bharat Kishore Sharma**, a valued collaborator and mentor. He has consistently guided me through every aspect of my journey, from solving complex physics problems to making important life decisions. Whether it was discussing our research or helping me with personal challenges, **Bharat sir** was always there to offer advice and support. His guidance on identifying good opportunities and avoiding potential pitfalls has been especially valuable. I am truly thankful for his constant support, which has greatly influenced both my work and personal growth.

I am deeply grateful to **Prof. Constança Providência**, a core member of our research group and my esteemed collaborator. Her exceptional hospitality during my visit to Portugal made my stay both comfortable and productive. She played a crucial role in the editing, analysis, and discussion of my manuscripts, contributing significantly to their quality and depth. Our long discussions on the physics of neutron stars during my visit were particularly enlightening and had a profound impact on my work. Her guidance and support have been invaluable throughout this journey.

I would like to extend my heartfelt thanks to **Dr. Arpan Das**, our collaborator from BITS Pilani (Pilani Campus). His insightful analysis of every theoretical aspect of my work has been profoundly impactful, shaping the manuscripts in meaningful ways. I am especially appreciative of his efforts in ensuring our research reached a wider audience, which has greatly contributed to its visibility and impact. His guidance and collaboration have been essential to the success of this work.

I thank my Doctoral Advisory Committee members, **Prof. Deepak Pachattu** and **Prof. Sunil Kumar Vattezhath**, Department of Physics, for their insightful comments, useful suggestions, and encouragement that have helped me greatly to strengthen my research ideas from various perspectives. I would also like to thank **Prof. Radhika Vathsan** (Ex-Head of the Department) and **Prof. P. Nandakumar** (Head of the Department) and all the other faculty members of the department for their continuous encouragement and support.

I would like to express my sincere gratitude to **Prof. Chandradew Sharma** and **Prof. Ram Shanker Patel** for their invaluable support and resources during my research. I particularly appreciate their assistance in providing access to Turnitin, which was instrumental in ensuring the originality of my work.

I would also like to express my heartfelt thanks to my pre-Ph.D. coursework instructors, **Prof. Prasanta Kumar Das**, **Prof. Toby Joseph**, **Prof. Kinjal Banerjee**, **Prof. Tarun Kumar Jha**, **Prof. Raghunath Anand Ratabole**, and **Prof. Arun Venkatesh Kulkarni**, for their kind assistance and guidance during the courses.

I am also grateful to **Prof. Senthamarai Kannan Ethirajulu** (former Department Research Committee (DRC) member) and **Prof. Kinjal Banerjee** (current Department Research Committee (DRC) member) for their invaluable support and guidance. Their assistance ensured the smooth functioning of all official work throughout my five-year journey in the department.

I am also deeply grateful to my friend **Mrinmoy**, who was my companion in exploring Goa, making my time here even more memorable. I would also like to thank **Ram Singh** for his support and assistance during difficult times, which helped me navigate through challenging phases.

I would like to thank **Prof. Ramgopal Rao** (Vice Chancellor, BITS Pilani), **Prof. Suman Kundu** (Director, BITS Pilani, K K Birla Goa Campus), **Prof. M. B. Srinivas** (Dean, Academic Graduate Studies and Research Division, BITS Pilani Goa), and **Prof. Bharat Deshpande** (Associate Dean, Academic Graduate Studies and Research Division).

I thank our laboratory staff of the Department of Physics, **Mr. Krishanu Paul**, **Mr. Rupesh Walke**, and **Ms. Meenakshi Ambegar**.

I also thank **Mr. Pratap Behera** of the Academic Graduate Studies and Research Division Office for their kind help throughout the period of my Ph.D.

I would like to express my heartfelt gratitude to **my family** for their unwavering support and encouragement throughout my academic journey. They stood by me in every difficult situation, providing the strength I needed to overcome challenges and stay focused on my goals.

To **my in-laws**, thank you for your understanding and support during this demanding period. Your encouragement and patience have meant the world to me, and I am deeply grateful for your kindness and love.

A special thank you to **my wife Prachie Sharma** . Her love, patience, and constant support have been my anchor. She stood by me through every late night, every setback, and every success, providing comfort and motivation when I needed it most. Her sacrifices and understanding have allowed me to dedicate myself fully to my research, and I could not have reached this point without her by my side.

Prashant Thakur

Date: February 12, 2025

# Contents

# List of Tables

# List of Figures

# Nomenclature

## Physical constants

| Quantity | Symbol | Value |
|---|---|---|
| speed of light in vacuum | $c$ | $299,792,458\ m/s$ |
| Planck constant | $h$ | $6.62607 \times 10^{-34}\ Js$ |
| Planck constant, reduced | $\hbar$ | $1.0545 \times 10^{-34}\ Js$ |
| electron charge magnitude | $e$ | $1.6021 \times 10^{-19}\ C$ |
| conversion constant | $\hbar c$ | $197.326\ MeV\ \ fm$ |
| conversion constant | $(\hbar c)^2$ | $0.3893\ GeV^{2}\ mbarn$ |
| electron mass | $m_e$ | $0.510\ MeV/c^2$ |
| proton mass | $m_p$ | $938.272\ MeV/c^2$ |
| fine-structure constant | $\alpha = e^2/4\pi\epsilon_0\hbar c$ | $1/137.035$ |
| Fermi coupling constant | $G_F/(\hbar c)^3$ | $6.6740 \times 10^{-11}\ m^3 Kg^{-1} s^{-2}$ |

# Abbreviation

| | |
|---|---|
| ANM | Asymmetric Nuclear matter |
| BNS | Binary Neutron Star |
| BPS | Baym-Pethick-Sutherland |
| CMB | Cosmic Microwave Background |
| EFT | Effective Field Theory |
| EOS | Equation of State |
| GW | Gravitational-Wave |
| GEO | Gravitational-Wave Observatory |
| GTR | General Theory of Relativity |
| HIC | Heavy Ion Collisions |
| ISGMR | Iso-scalar Giant Monopole Resonance |
| IVGDR | Iso-vector Giant Dipole Resonance |
| LIGO | Laser Interferometer Gravitational-Wave Observatory |
| NICER | Neutron Star Interior Composition Explorer |
| CREX | Coherent Reactions Experiment |
| PREX | Parity Radius Experiment |
| NN | Nucleon-Nucleon |
| NS | Neutron Star |
| CCSN | Core Collapse Supernova |
| DM | Dark Matter |
| DANS | Dark Matter Admixed Neutron Star |
| PNS | Proto-Neutron Star |
| SGR | Soft Gamma Repeater |
| AXP | Anomalous X-ray Pulsar |
| NMP | Nuclear Matter Parameter |
| PNM | Pure Neutron Matter |
| BEM | $\beta-$euilibrium Matter |
| SNM | Symmetric Nuclear Matter |
| TOV | Tolman-Oppenheimer-Volkoff |
| QCD | Quantum Chromodynamics |
| pQCD | Perturbative Quantum Chromodynamics |
| QGP | Quark-Gluon Plasma |
| BA | Bayesian Analysis |
| PDF | probability density function |
| ML | Machine Learning |

# Keywords

# List of Publications

1. **”Impact of $\sigma$-cut potential on Antikaon condensation in neutron stars within relativistic mean field model”**

   **Prashant Thakur**, B. K. Sharma, Lakshana Sudarsan, Krishna Kunnampully, T. K. Jha

   To be communicate soon...

2. **”Feasibility of dark matter admixed neutron star based on recent observational constraints”**

   **Prashant Thakur**, Tuhin Malik, Arpan Das, T.K. Jha, B.K. Sharma and Constança Providência

   https://doi.org/10.48550/arXiv.2408.03780

   arXiv no. **arXiv:2408.03780** (2024) In communication to Astronomy and Astrophysics

3. **”Exploring robust correlations between fermionic dark matter model parameters and neutron star properties: A two-fluid perspective”**

   **Prashant Thakur**, Tuhin Malik, Arpan Das, T.K. Jha and Constança Providência

   DOI:10.1103/PhysRevD.106.043024

   Phys. Rev. D **109** (2024) 4, **043030**

4. **”Towards Uncovering Dark Matter Effects on Neutron Star Properties: A Machine Learning Approach”**

   **Prashant Thakur**, Tuhin Malik, T.K. Jha

   DOI:10.3390/particles7010005

   Particles **7** (2024) 80-95, **7010005**

5. **”Hyperon bulk viscosity and r-modes of neutron stars”**

   O P Jyothilakshmi, P E Sravan Krishnan, **Prashant Thakur**, V Sreekanth, T.K. Jha

   DOI:10.1093/mnras/stac2360

   Monthly Notices of the Royal Astronomical Society **516** (2022) 3, 3381-3388

6. **”Influence of the symmetry energy and $\sigma$-cut potential on the properties of pure nucleonic and hyperon-rich neutron star matter”**

**Prashant Thakur**, B.K. Sharma, A. Ashika, S.Srivishnu and T.K. Jha

## This thesis is based on following publications

1. **"Exploring robust correlations between fermionic dark matter model parameters and neutron star properties: A two-fluid perspective"**
   **Prashant Thakur**,Tuhin Malik, Arpan Das, T.K. Jha and Constança Providência
   DOI:10.1103/PhysRevD.106.043024
   Phys. Rev. D **109** (2024) 4, **043030**

2. **"Towards Uncovering Dark Matter Effects on Neutron Star Properties: A Machine Learning Approach"**
   **Prashant Thakur**, Tuhin Malik, T.K. Jha
   DOI:10.3390/particles7010005
   Particles **7** (2024) 80-95, **7010005**

3. **"Feasibility of dark matter admixed neutron star based on recent observational constraints"**
   **Prashant Thakur**, Tuhin Malik, Arpan Das, T.K. Jha, B.K. Sharma and Constança Providência
   https://doi.org/10.48550/arXiv.2408.03780
   arXiv no. **arXiv:2408.03780** (2024) Communicated to Astronomy and Astrophysics

# Chapter 1

# Introduction

## 1.1 Neutron Stars: Overview

### 1.1.1 Where it all begun

In the early 20th century, Ernest Rutherford described the atom as a system of nucleus at the center surrounded by electrons in orbit. The positively charged protons and the negatively charged electrons made a charge neutral atom and the protons being heavier accounted for most of the atom's mass. However, when mass of the nucleus was experimentally determined, the underlying discrepancy validated the presence of another particle 'neutrons'. In 1932, the neutron was discovered by James Chadwick. Neutrons are electrically neutral particles having mass very similar to that of the protons. Subsequently, lots of different configuration of the number of neutrons and protons in the nucleus led to identification of isotopes, isotones, neutron rich nuclei and proton rich nuclei as well.

The discovery of the neutron changed our understanding of nuclear physics entirely, particularly the aspects of nuclear force and nuclear binding energy, the energy required to hold the nucleons (collectively protons and neutrons) in the nucleus. This was crucial for explaining the stability of nuclei. Additionally, neutron's being charge neutral, seems to be an ideal particle for probing the nucleus and nuclear force as they remain unaffected to the presence of charge in matter, thereby allowing for more detailed studies of nuclear structures and reactions. In 1934, Enrico Fermi used neutrons to induce nuclear reactions. Reactions such as nuclear fission transformed the world outright leading to the development of nuclear reactors and atomic bombs.

The concealed power of the nucleus is now known to mankind. The mass of an atom is predominantly concentrated in its nucleus of radius of a few femtometers ($10^{-15}$ meters), compared to the entire atom's size of a few angstroms ($10^{-10}$ meters). An atom, represented as ${}^{A}_{Z}X$, has a mass number $A = N + Z$, where $N$ is the number of neutrons, and $Z$ is the number of protons, with an equal number of electrons orbiting the nucleus. The nuclear radius $r$ is empirically given by $r \approx r_0 A^{1/3}$, where $r_0 \approx 1.2$ fm. Assuming the nucleus to be a sphere, its volume $V$ can be expressed as

$$V = \frac{4}{3}\pi r^3 = \frac{4}{3}\pi (r_0 A^{1/3})^3 = \frac{4}{3}\pi r_0^3 A.$$

The average density of nucleons in the nucleus is

$$\rho_0 = \frac{A}{V} = \left(\frac{4}{3}\pi r_0^3\right)^{-1},$$

which remains constant irrespective of $A$. This density, approximately 0.16 fm$^{-3}$, is known as the saturation density of nuclear matter, conceptualized as a large system of neutrons and protons with no Coulomb interaction.

### 1.1.2 History of Neutron Stars

In 1931, Lev Landau, hypothesized the existence of extremely dense stars Landau [1932], deriving a formula for the maximum mass of white dwarfs, independently reaching conclusions similar to those of Chandrasekhar, though he did not specifically mentioned them as "neutron stars". The term "neutron stars" was first coined by Walter Baade and Fritz Zwicky Baade & Zwicky [1934], Baade & Zwicky [1934] in 1934, who, suggested that these dense remnants could form from the explosive deaths (supernovae) of massive stars. Interestingly, Richard Tolman along with J. Robert Oppenheimer and George Volkoff, independently derived the equations governing hydrostatic equilibrium for a spherically symmetric star. Their work, published in 1939 Oppenheimer & Volkoff [1939], Tolman [1939], led to the formulation of the Tolman-Oppenheimer-Volkoff (TOV) equations in line with Albert Einstein's theory of general relativity. Two decades later, in 1959, Alastair Cameron Cameron [1959] solved the Tolman-Oppenheimer-Volkoff equations for a star composed of a degenerate gas of relativistic neutrons with repulsive neutron-neutron interactions. He determined a maximum mass of about twice the mass of the Sun, a limit that remains approximately valid today.

In 1967, Dame Susan Jocelyn Bell Burnell, then a graduate student, was researching interplanetary scintillations in the radio wavelength. Accidentally she detected a highly regular signal with a periodicity indicative of a compact star, the findings led to the publication of the paper "Observation of a Rapidly Pulsating Radio Source" Hewish *et al.* [1968], Pilkington *et al.* [1968] the following year.

### 1.1.3 Formation and Structure of Neutron Stars

Neutron stars are a possible end product marking the death of massive main-sequence star. These explosions occur after the progenitor star exhausts its nuclear fuel, starting with hydrogen, followed by helium, and finally heavier elements like oxygen and magnesium. The core eventually accumulates isotopes of iron-group elements. The electron Fermi gas pressure is the only force preventing the iron-nickel core from collapsing under gravity. Once the core's mass exceeds the Chandrasekhar limit of 1.44 $M_{\odot}$, gravitational collapse ensues. This catastrophic event releases immense gravitational energy (greater than $10^{53}$ erg) and generates a shock wave that expels the star's outer layers at speeds up to 10% of the speed of light, while the inner core continues to collapse at a similar rate. The atomic nuclei merge into a single massive nucleus. If the core's mass exceeds the Oppenheimer–Volkoff limit, approximately (2-3) $M_{\odot}$ according to modern theoretical models, the pressure of degenerate neutrons cannot halt the compression, resulting in the formation of a black hole. This collapse may produce hypernovae, significantly brighter than supernovae, and could be the source of mysterious gamma-ray bursts from distant galaxies. If the mass remains below the Oppenheimer–Volkoff limit, the result is a neutron star, stabilized by the pressure of nuclear matter against gravitational compression. Not every star concludes its evolution as a supernova, let alone a hypernova. Only stars with masses greater than approximately 8 $M_{\odot}$ are destined for such an end. Stars with lower masses go through a giant phase at the end of their lifetimes, gradually shedding their outer layers. The core of these less massive stars contracts to form a white dwarf.

Neutron stars are incredibly dense compact objects, with observed masses typically ranging from around 1.2 to 2.5 solar masses, possibly confined within a radius of approximately (10-15) km. These remarkable objects play a crucial role in astrophysical measurements and provide a unique insight into the behavior of matter at extreme densities. The wide range of expected densities, from about $10^{11}$ to $10^{15}$ g/cm$^3$, in neutron stars necessitates modeling under vastly different physical conditions. Based on our current understanding, neutron stars are composed of an atmosphere and four primary internal regions as shown in pictorial representation fig.1.1: the outer crust, inner crust, outer core, and inner core. The neutron star's atmosphere is a thin plasma layer responsible for forming the spectrum of its thermal electromagnetic radiation. This layer, only a few millimeters thick, contributes insignificantly to the star's total mass.

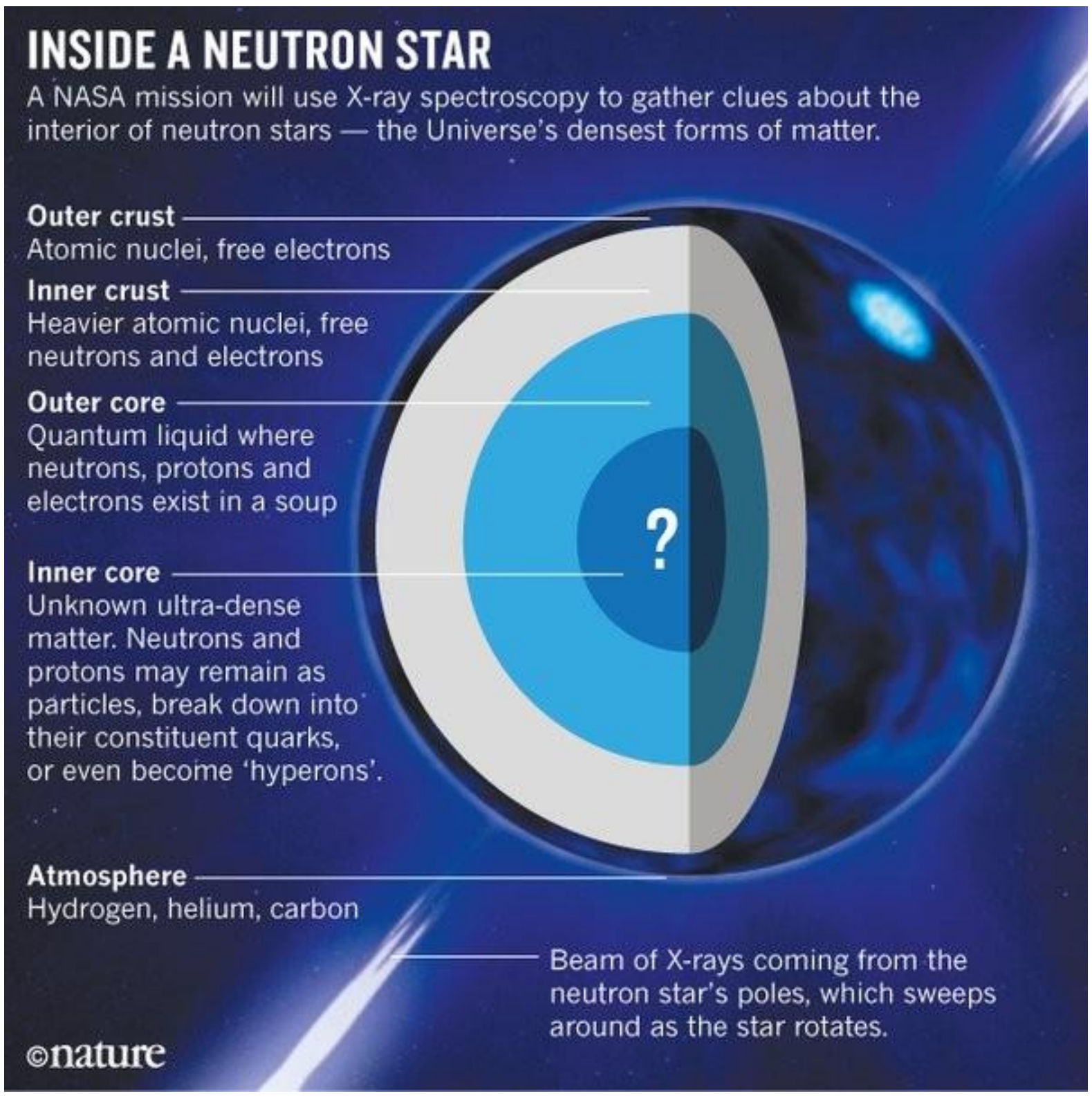

Figure 1.1: Pictorial representation of the structure of a neutron star, published on Nature Gibney [2017] on the occasion of the start of Nasa's NICER mission (2017)

- **Outer crust:** The outer crust is a region with a thickness ranging from 0.3 to 0.5 km, consisting of matter with a density less than $\rho_{\text{drip}} \approx 4 \times 10^{11}$ g/cm$^3$. The ”drip point” is the density at which nuclei can no longer retain additional neutrons Piekarewicz [2022]. At this density, the matter consists of a Coulomb lattice of nuclei immersed in a gas of electrons.

- **Inner crust:** The inner crust spans approximately 1 to 2 km Haensel *et al.* [2007], Piekarewicz [2022] and features a density ranging from $\rho_{\text{drip}}$ to about half of the saturation density $\rho_0 = 2.8 \times 10^{14}$ g/cm$^3$ (or $\rho_0 = 0.15\,\text{fm}^{-3}$). In this region, highly neutron-rich nuclei are immersed in a superfluid neutron gas. Due to the limited understanding of the precise nucleon-nucleon (NN) interaction, theoretical approaches based on many-body theory are required.

- **Outer core:** The outer core spans a density range of $0.5\rho_0 \leq \rho \leq 2\rho_0$ and is several kilometers thick. It consists primarily of neutrons, with a certain fraction of protons and electrons to maintain charge neutrality. As one moves deeper into the star, muons ($\mu$) appear at a specific threshold density, which varies depending on the model. Muons become energetically favorable in high-density matter because the energy associated with their fermionic nature increases rapidly with density, making slow muons more stable than electrons.

- **Inner core:** The inner core is the central region of a neutron star, and its properties and extent remain poorly understood. Typically, baryonic densities in this part of the neutron star range from approximately $2\rho_0$ to very high values $(10-15\rho_0)$ Haensel *et al.* [2007]. Several hypotheses have been proposed to describe matter under these extreme conditions, including:

  – The **hyperonization hypothesis**, which suggests the appearance of hyperons, primarily $\Lambda$ and $\Sigma^-$ particles.

  – A **phase transition to quark matter**, consisting of deconfined u, d, and s quarks.

  The presence of hyperons at high densities is attributed to the fermionic nature of nucleons. According to the Fermi model, the energy of protons and neutrons increases rapidly with density. At a certain threshold, it becomes energetically favorable for neutrons near the Fermi surface to decay into $\Lambda$ particles. Incorporating hyperons into the nuclear matter model significantly reduces the maximum mass that such a system can support, leading to results that strongly contradict astrophysical observations of neutron stars. This discrepancy is commonly referred to in the literature as the "hyperon puzzle" Bombaci [2017].

The possible composition of the inner core is discussed in section1.4.

## 1.2 Nuclear Matter

Nuclear matter is a theoretical and idealized concept, referring to a state consisting of an infinitely large number of nucleons, with equal numbers of neutrons and protons (N = Z), known as symmetric nuclear matter (SNM) in the absence of Coulomb interaction. This matter is also described as "saturated," meaning that even as more nucleons are added, the density of the core remains unchanged. This phenomenon arises from the strong repulsive force between nucleons at short distances combined with the effects of the Pauli exclusion principle Deshalit & Feshbach [1974], Gomes *et al.* [1958], Thaler [1962]. Such nuclear matter is believed to characterize various extreme environments, including compact stellar matter, the early universe, and quark-gluon plasma (QGP). Saturated nuclear matter is characterized by the following properties:

- **Binding Energy:**

  The Bethe-Weizsäcker semi-empirical mass formula calculates the binding energy per nucleon based on the number of protons and neutrons. In the case of saturated nuclear matter, only the volume term of the formula is relevant, leading to a binding energy per nucleon that reflects the interactions within the bulk of the matter, unaffected by surface, Coulomb, or asymmetry effects.

$$\frac{E_B}{A} = -16.3\,\text{MeV} \tag{1.1}$$

where the total number of nucleons $A = N + Z$ counts up to infinity.

- **Saturation Density:**

Saturation density Bethe [1971] refers to the density at which nuclear matter is in its most stable configuration, corresponding to a balance between attractive and repulsive forces within the nucleus.

As the radius parameter $r_0 = 1.16$ fm for symmetric matter, the saturation density can be readily calculated as

$$\rho_0 = \left(\frac{4\pi}{3} r_0^3\right)^{-1} \sim 0.16\,\text{fm}^{-3} \tag{1.2}$$

The Fermi momentum in general is related to density as

$$\rho = \frac{\gamma}{6\pi^2} k_F^3 \tag{1.3}$$

where $\gamma$ is the spin degeneracy factor, related to the isospin $J$ as $\gamma = (2J + 1)$. Therefore for symmetric nuclear matter (SNM), the Fermi momentum is obtained as

$$k_F = \left(\frac{3\pi^2}{2} \rho_0\right)^{1/3} \sim 1.33\,\text{fm}^{-1} \tag{1.4}$$

In terms of energy density $\varepsilon(\rho)$, the binding energy is written as

$$\frac{E_B}{A} = \frac{\varepsilon}{\rho} - m \tag{1.5}$$

where $m = 939$ MeV is the mass of nucleons.

- **Effective Nucleon Mass:**

The effective nucleon mass $m^*$ is a modified version of the nucleon mass that results from interactions within a medium, primarily due to the Dirac field. This quantity is strongly dependent on the density, and thus on the collective Fermi momentum of all energy states of the nucleons. It plays a crucial role in determining the equation of state (EoS) of dense matter. The EOS describes the relationship between pressure, energy density, and other thermodynamic properties in a neutron star. It encodes the interactions among the microscopic constituents, such as nucleons, hyperons, or quarks, and determines the macroscopic properties of the star. To solve the Tolman-Oppenheimer-Volkoff (TOV) equations, one needs information about the internal composition of the star, which is provided by the EOS. The EOS plays a fundamental role in predicting observables like mass, radius, and tidal deformability, making it essential for interpreting astrophysical measurements, including gravitational wave and X-ray observations.

A lower value of $m^*$ contributes to a stiffer EoS, which, in turn, supports the existence of more massive neutron star configurations. However, there are currently no direct measurements of the effective mass. Non-relativistic theoretical analyses, such as those based on neutron scattering from $^{208}$Pb nuclei, provide estimates of the isoscalar component of the mean field effective mass Glendenning [1992], Jaminon & Mahaux [1989], Johnson *et al.* [1987], Mahaux & Sartor [1989].

$$m^*_S/m \approx 0.74 - 0.82 \tag{1.6}$$

This mass is equivalent to the Landau effective mass $m^*_L$ which is related to the Dirac effective mass as

$$m^*_S \equiv m^*_L = \sqrt{m^{*2} + k_F^2} \tag{1.7}$$

The range of Dirac effective mass is therefore prescribed to be $m^*/m \approx (0.7 - 0.8)$ at saturation density.

- **Nuclear Incompressibility:**

  Nuclear incompressibility, which reflects the curvature of the energy density curve at saturation density, can be expressed mathematically as:

$$K_0 = 9\rho^2 \frac{\partial^2}{\partial \rho^2}\left(\frac{\epsilon}{\rho}\right)_{\rho_0} \tag{1.8}$$

  This quantity plays a significant role in understanding high-density nuclear equations of state (EoS), particularly in relation to neutron stars (NSs). A higher incompressibility value corresponds to a "stiffer" equation of state, implying greater pressure at high densities and potentially leading to the formation of more massive neutron stars. Experimentally, nuclear incompressibility has been measured through collective nuclear vibrations known as giant resonances, which are induced by changes in nuclear density. Specifically, isoscalar giant monopole resonances (GMR)Pearson *et al.* [2010] and isovector giant dipole resonances (GDR) Lipparini & Stringari [1989] have been instrumental in setting the bounds on nuclear incompressibility. Theoretical approaches, including models like Hartree-Fock with random-phase approximation (RPA), estimate the value of incompressibility to be around $K = 210 \pm 30$ MeV Blaizot [1980]. More recent studies have narrowed the expected range to between 220 and 260 MeV, with some research suggesting a range from 230 $\pm$ 40 MeV, while others propose $K$ could be between 250 and 315 MeV Stone *et al.* [2014].

## 1.3 Neutron Star Matter

The internal composition of a neutron star represents a delicate equilibrium between the relentless pull of gravity, driving matter inward, and the outward pressure from neutron degeneracy. To understand the inner structure of a neutron star, we must examine the behavior of matter under extreme conditions. Central to this investigation is the equation of state (EOS) for infinite nuclear matter, which becomes crucial at the incredibly high densities found in neutron stars. In an environment dominated by neutrons, the chemical potential of neutrons significantly exceeds that of protons, creating favorable conditions for phenomena like $\beta$- decay. This decay process transforms some neutrons into protons and electrons, as illustrated by the reactions:

$$n \rightarrow p + e^- + \overline{\nu} \tag{1.9}$$

$$n + \nu \rightarrow p + e^- \tag{1.10}$$

The charge neutrality condition is expressed as:

$$\rho_{\mathrm{p}} = \rho_{\mathrm{e}} + \rho_{\mu}, \tag{1.11}$$

where $\rho_{\mathrm{n}}$, $\rho_{\mathrm{p}}$, and $\rho_{\mu}$ represent the number densities of neutrons, protons, and muons, respectively. The condition for $\beta$-equilibrium is given by:

$$\mu_{\mathrm{n}} = \mu_{\mathrm{p}} + \mu_{\mathrm{e}} \quad \text{and} \quad \mu_{\mathrm{e}} = \mu_{\mu}, \tag{1.12}$$

where $\mu_{\mathrm{n}}$, $\mu_{\mathrm{p}}$, $\mu_{\mathrm{e}}$, and $\mu_{\mu}$ are the chemical potentials of neutrons, protons, electrons, and muons. Unlike terrestrial experiments that produce and study high-density matter through heavy-ion collisions, the study of neutron stars presents unique challenges due to their significant asymmetry. Matter in neutron stars is characterized by a high degree of isospin asymmetry. To analyze this asymmetry in purely nucleonic matter, we introduce the following concepts:

- **Isoscalar density ($\rho_s$)**: Defined as $\rho_s = \rho_n + \rho_p$, where $\rho_n$ and $\rho_p$ are the neutron and proton densities, respectively. The isoscalar density remains invariant under the exchange of neutrons and protons.
- **Isovector density ($\rho_v$)**: Defined as $\rho_v = \rho_n - \rho_p$. The isovector density changes sign when neutrons and protons are exchanged.

The isospin asymmetry parameter denoted $\delta$ is defined as the ratio between the isovector and the isoscalar density.

$$\delta = \frac{\rho_n - \rho_p}{\rho} \tag{1.13}$$

Pure neutron matter corresponds to $\delta = 1$, and symmetric matter corresponds to $\delta = 0$. The equation of state for catalyzed matter within a given nuclear model is derived using $\delta$ values appropriate for $\beta$-equilibrated matter. However, this nuclear model can also be used to determine results for pure neutron matter ($\delta = 1$) and symmetric matter ($\delta = 0$). The energy per baryon can be approximated using a parabolic expansion around the isospin asymmetry parameter.

$$E(\rho, \delta) = E(\rho, \delta = 0) + \delta^2 E_{sym}(\rho) + O(\delta^4) \tag{1.14}$$

The energy per baryon for symmetric matter can be expanded around this quantity as:

$$E(\rho, \delta = 0) = E_{sat} + K_{sat}\frac{u(\rho)^2}{2!} + Q_{sat}\frac{u(\rho)^3}{3!} + \ldots \tag{1.15}$$

where

$$u(\rho) = \frac{\rho - \rho_{sat}}{3\rho_{sat}} \tag{1.16}$$

with:

- $E_{sat}$ the energy per baryon at saturation density for symmetric matter,
- $K_{sat}$ the isoscalar incompressibility modulus,
- $Q_{sat}$ the isoscalar skewness,
- and etc. for higher order parameters.

$$E_{\text{sym}}(\rho) = J + Lu(\rho) + K_{\text{sym}}\frac{u(\rho)^2}{2!} + Q_{\text{sym}}\frac{u(\rho)^3}{3!} + \ldots \quad (1.17)$$

with:

- $J$ the symmetry energy at saturation density,
- $L$ the slope of the symmetry energy at saturation density,
- $K_{\text{sym}}$ the isovector incompressibility,
- $Q_{\text{sym}}$ the isovector skewness,
- and etc. for higher order parameters.

The parameters established in the vicinity of saturation density can be defined using laboratory experiments, particularly through the isoscalar parameters since symmetric matter is more straightforward to investigate. Comprehensive discussions regarding the constraints on the symmetry energy can be found in the works of Tsang *et al.* [2012] and Oertel *et al.* [2017]

## 1.4 Gaps in Existing Research

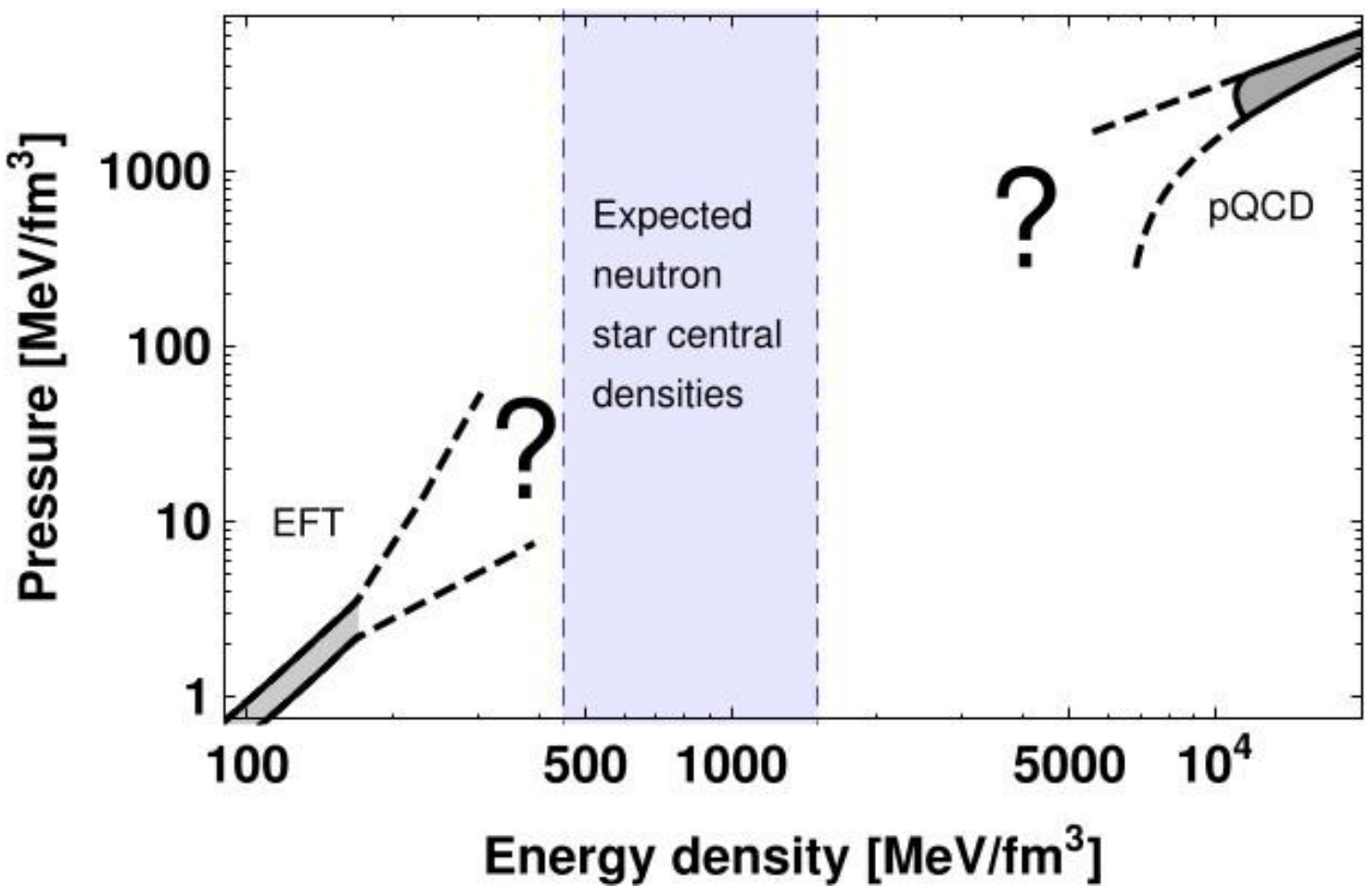

Figure 1.2: Predictions for the equation of state (EOS) at low and high densities compared to the estimated range of central densities in neutron stars.Ja¨rvinen [2022]

Interpreting high-energy astrophysical phenomena, such as neutron stars and their collisions, requires a profound understanding of matter at supranuclear densities. However, our knowledge of the equation of state (EOS) of dense matter, particularly within the cores of neutron stars, remains limited. Dense matter is studied not only through astrophysical observations but also in terrestrial heavy-ion collision experiments. The nuclear EOS of neutron stars is pivotal in determining their internal structure and observable characteristics, including mass and radius. At these extreme densities, particle interactions are incredibly strong. While Quantum Chromodynamics (QCD) theoretically governs these interactions, we currently lack a systematic computational method to accurately calculate the properties of high-density matter in the inner cores of massive neutron stars. Consequently, at densities well above nuclear saturation density $\rho_0$=0.16 $fm^{-3}$, where neither experimental nor theoretical information is robust, the EOS remains highly uncertain. Many questions remain open, such as whether a possible phase transition to an exotic phase of matter occurs in nature.

Despite this, we have two successful theories at the extreme ends of the density spectrum as shown in fig.1.2. For densities up to 1-2 $\rho_0$, chiral effective field theory (ChEFT) is effective, while at densities exceeding 50-100 $\rho_0$, perturbative quantum chromodynamics (pQCD) provides a reliable framework. We will briefly discuss these two theories below.

- **Chiral Effective Field Theory (ChEFT)**: ChEFT is based on the symmetries of Quantum Chromodynamics (QCD) Kaplan *et al.* [1996], Weinberg [1990, 1991]. At momentum scales relevant to nuclear physics, around $p \sim m_\pi$ (where $m_\pi$ is the pion mass), QCD becomes strongly coupled, leading to nonperturbative effects such as the spontaneous breaking of chiral symmetry and the confinement of quarks and gluons into hadrons. ChEFT provides an effective description of low-energy interactions, including those between two nucleons and among multiple nucleons. While direct applications of QCD to hadronic physics at finite density are extremely challenging due to the sign problem in lattice QCD, making such approaches impractical for the foreseeable future, ChEFT offers a systematic framework for studying nuclear physics by focusing on the relevant low-energy degrees of freedom—nucleons, pions, and delta isobars Epelbaum *et al.* [2009], Hammer *et al.* [2013], Machleidt & Entem [2011]. These interactions are essential for understanding the behavior of nucleons in dense nuclear matter, such as that found in neutron stars. It permits and organizes all possible operators that adhere to chiral symmetry by using a power counting scheme Kaplan *et al.* [1996], Weinberg [1990]. This approach results in a systematic expansion based on nucleon momenta, which not only facilitates the calculation of the EOS but also allows for error estimation through order-by-order comparisons.

- **Perturbative Quantum Chromodynamics (pQCD)**: As discussed earlier, QCD is the theory that describes the strong interactions of hadrons. It stands as one of the most precise and complex microscopic theories in physics, being a self-consistent relativistic quantum field theory. In QCD, the interactions between quarks and gluons are governed by a property known as ”color charge,” and these interactions are mediated by the exchange of gluons. The strength of these interactions varies with the energy scale or the distance between quarks and gluons. At very high energies, or at very short distances, the interaction strength decreases, allowing for a perturbative approach—a phenomenon known as ”asymptotic freedom,” which is a fundamental aspect of QCD. Perturbative Quantum Chromodynamics (pQCD) is the framework within QCD that facilitates the study of strong interactions at these high energies, where the strong coupling constant becomes sufficiently small to enable perturbative methods. In pQCD, calculations are expanded into a series of terms, with each term corresponding to a specific power of the strong coupling constant, which quantifies the strength of the strong interaction.

At densities above approximately 50 to $100\rho_0$, the asymptotic freedom of QCD suggests that perturbative treatments become valid and useful.

### 1.4.1 The Missing Link

While ChEFT and pQCD effectively describe densities at the two extremes, neither provides direct insight into the equation of state (EOS) at the densities found in the cores of neutron stars. Although pQCD calculations are intriguing, they are applicable only at densities above 50 to $100\rho_0$, whereas stable neutron stars exhibit densities of up to just $5 - 7\rho_0$. The significant difference between these density regimes means that the relevance of pQCD calculations in the study of neutron star matter is not straightforward to assess. In Refs. Komoltsev & Kurkela [2022], Kurkela *et al.* [2014], various authors have demonstrated that understanding pQCD constraints is crucial for determining the EOS at neutron star densities. These constraints arise from the requirement that the EOS at low densities must smoothly transition to the high-density regime while maintaining mechanical stability, causality, and thermodynamic consistency across all densities $\rho$. However, there is still room for further refinement, as the current uncertainties in pQCD predictions are too large to establish meaningful bounds.

In recent pioneering work Kurkela *et al.* [2014], authors have proposed a novel approach to connect both density regimes. This method allows for ’integrating backwards,’ enabling the propagation of pQCD constraints to lower densities in a completely general, analytical, and model-independent manner. They concluded that, without considering neutron star observations, pQCD calculations rule out approximately 65% of the region in the pressure-energy density plane at $\rho = 5\rho_0$.

Based on the above discussion, it is evident that we still don't fully understand the internal composition of a neutron star's core—it remains a mystery. Additionally, there is no comprehensive theory that can fully explain the density regime from $2\rho_0$ to $7\rho_0$.

**This leaves us with two open questions.**

1. **What is the true internal composition of a neutron star's core, and how can we uncover this mystery?**

2. **Current Status and Challenges of Equation of State Models in Neutron Stars**

### 1.4.2 Theoretical Speculations and Models of Neutron Star Core Composition

The nature of matter at extremely high densities ($\rho > 2.8 \times 10^{14}$ g/cm$^3$), where there is a large imbalance between the number of protons and neutrons, and temperatures are below $10^{10}$ K, remains one of the most significant unsolved problems in modern physics. This is due to the substantial challenges involved in both experimental and theoretical studies. Despite these difficulties, various well-supported theoretical models have been proposed to describe matter under these conditions. These models suggest a range of possibilities, from normal nucleonic matter to more exotic states, such as hyperons, deconfined quarks, color superconducting phases, and Bose-Einstein condensates .

- **Hyperon Matter**:

  Hyperons are baryons that are heavier than nucleons and possess a strange quantum number. Unlike nucleons, hyperons contain one or more strange quarks in their quark composition as shown in Table 1.1

Table 1.1: Tabulated below are the different hyperons with their mass ($M$), charge ($Q$), spin with parity ($J^P$), isospin ($I$), and third component of isospin ($I_3$) states with respective strangeness ($S$).

| Baryon | $M$ (MeV) | Q | $J^P$ | I | $I_3$ | S | Quark Structure |
|---|---|---|---|---|---|---|---|
| $N$ | 939 | 0 | $1/2^+$ | $1/2$ | $-1/2$ | 0 | udd |
| | | 1 | $1/2^+$ | $1/2$ | $1/2$ | 0 | uud |
| $\Lambda$ | 1116 | 0 | $1/2^+$ | 0 | 0 | -1 | uds |
| $\Sigma$ | 1193 | -1 | $1/2^+$ | 1 | -1 | -1 | dds |
| | | 0 | $1/2^+$ | 0 | 0 | -1 | uds |
| | | +1 | $1/2^+$ | 1 | +1 | -1 | uus |
| $\Xi$ | 1318 | -1 | $1/2^+$ | $1/2$ | $-1/2$ | -2 | dss |
| | | 0 | $1/2^+$ | $1/2$ | $1/2$ | -2 | uss |

Hyperons are unstable in terrestrial conditions and decay into nucleons through the weak interaction. Contrary to this, the equilibrium conditions in neutron stars can make the inverse process happen. Hyperons may appear in the inner core of neutron stars at densities of about $2-3\rho_0$ Bednarek *et al.* [2012], Oertel *et al.* [2017], Schaffner & Mishustin [1996], Vidaña [2016], Weissenborn *et al.* [2012]. At extremely high densities within neutron star matter (NSM), the neutron chemical potential and Fermi momentum can reach or even exceed the rest masses of heavier baryons, such as the $\Lambda$ hyperon (1116 MeV), $\Sigma^-$, $\Sigma^0$, $\Sigma^+$ hyperons (1193 MeV), and $\Xi^-$, $\Xi^0$ hyperons (1318 MeV). Under these conditions, these hyperons are formed in the NSM, governed by the principles of charge neutrality and chemical potential equilibrium. As a result, they are treated on an equal footing with nucleons, becoming integral components of the matter within neutron stars.

The behavior of hyperons in a vacuum is very different from what occurs in the high-density matter of a neutron star's core. Simple thermodynamical considerations suggest that hyperons are produced in neutron star matter when the baryonic density becomes larger than a certain threshold, which depends on the features of the interaction between particles. If the baryonic density, namely the sum of the neutron and proton numerical densities, exceeds this threshold value, the system starts to convert some nucleons into hyperons. Additionally, the immense density of the system and the Pauli-blocking mechanism prevent hyperon decay. In contrast, in a vacuum, hyperons are unstable and decay on a typical timescale set by the weak interaction. Typical processes that lead to the formation of $\Lambda$ and $\Sigma^-$ hyperons include: $p + e^- \rightarrow \Lambda + \nu_e$ and $n + n \rightarrow p + \Sigma^-$. Although there is no experimental evidence supporting the presence of hyperons in neutron stars, their presence in NSM is theoretically speculated and was first suggested by V. A. Ambartsumyan and G. S. Saakyan in 1960 **?**

The presence of hyperons can significantly impact the equation of state (EOS) for neutron stars, leading to a softer EOS Banik *et al.* [2014], Chatterjee & Vidaña [2016], Glendenning [1982], Schaffner-Bielich [2008], which in turn influences the mass-radius relationship and other key astrophysical properties of neutron stars.

- **$\Delta$ Resonances:**

  The formation of $\Delta$ resonances $\Delta^-$, $\Delta^0$, $\Delta^+$, $\Delta^{++}$ in NSM as metastable states occurs when the nucleon chemical potential reaches or surpasses the mass of the $\Delta$ particles. Similar to hyperons, the onset of $\Delta$ isobars in the matter is highly dependent on the strength of their coupling with mesons. This coupling influences the conditions under which these $\Delta$ particles become energetically favorable and can thus appear within the dense environment of a neutron star. These particles are of particular interest due to their quark structures, which are similar to those of nucleons, particularly in terms of strangeness. Different types of $\Delta$ particles are shown in table 1.2. This similarity allows these $\Delta$ particles to be treated on an equal footing with nucleons in various theoretical and computational models. These $\Delta$ particles are formed as resonance states and therefore exhibit a Breit-Wigner mass distribution. The centroid of this distribution is located at approximately 1232 MeV, with a width of around 120 MeV Cai *et al.* [2015], Li *et al.* [2018]. This distribution reflects the inherent instability of $\Delta$ resonances, as they have a finite lifetime and decay into nucleons and pions. In the absence of any conclusive experimental data for the potential depth of $\Delta$s in normal nuclear matter, the $\Delta$-meson couplings are poorly known Griegel & Cohen [1990], Riek *et al.* [2009]. Refs. Kolomeitsev *et al.* [2017], Li *et al.* [2018, 2020], Raduta [2021], Ribes *et al.* [2019] have suggested various ways to solve the delta puzzle.

Table 1.2: Different types of $\Delta$ particles and their properties.

| **Particle** | $M$ **(MeV)** | **Q** | $J^P$ | **I** | $I_3$ | **S** | **Quark Structure** |
|---|---|---|---|---|---|---|---|
| $\Delta$ | $1232 \pm 120$ | -1 | $3/2^+$ | $3/2$ | $-3/2$ | 0 | ddd |
| | | 0 | $3/2^+$ | $3/2$ | $-1/2$ | 0 | udd |
| | | 1 | $3/2^+$ | $3/2$ | $1/2$ | 0 | uud |
| | | 2 | $3/2^+$ | $3/2$ | $3/2$ | 0 | uuu |

- **Boson Condensation:**

  In the dense cores of neutron stars, various particle interactions can significantly affect the energy levels of pions ($\pi$) and kaons ($K$). When these particles' energies become very low, the interactions may lead to the formation of condensates. This occurs when the particles undergo Bose-Einstein condensation, where they all occupy the same ground state Kaplan & Nelson [1988], Pethick *et al.* [2015], Sawyer [1972], Scalapino [1972], forming a coherent state in the boson fields. The strength of this condensation depends on the repulsive forces between the bosons.

  For kaons, particularly, experimental data shows that they experience a strong attractive potential within dense nuclear matter Friedman *et al.* [1994], Migdal [1978]. This potential can make kaons more likely to form condensates. In contrast, pions, especially the negatively charged ones ($\pi^-$), can condense at much lower densities, closer to the density of normal nuclear matter, but only in the absence of interactions. When interactions are present, the condensation of $\pi^-$ is hindered due to weak and repulsive s-wave interactions Friedman *et al.* [1994], Migdal [1978]. As the density increases inside a neutron star, the excitation energy of antikaons (particles with strangeness = -1) decreases, making their condensation more likely at sufficiently high densities. Theoretical studies, especially those focusing on $K^-$ condensation Glendenning & Schaffner-Bielich [1998], Gupta & Arumugam [2012], Maruyama *et al.* [2005], Pons *et al.* [2000], suggest that this type of condensation is dominant because it occurs at relatively lower densities. With the onset of $K^-$ condensation, the process $n \rightarrow p + K^-$ becomes favored, leading to a significant increase in the proton fraction, potentially surpassing the neutron fraction at higher densities. However, when $\bar{K}^0$ condensation occurs, it alters this scenario. In this case, there is competition between the processes $N \rightarrow N + \bar{K}^0$ and $n \rightarrow p + K^-$, potentially resulting in a perfectly symmetric matter of nucleons and antikaons inside neutron stars Banik & Bandyopadhyay [2003], Glendenning & Schaffner-Bielich [1999], Koch [1994], Kolomeitsev *et al.* [1995], Ramos *et al.* [2001], Waas & Weise [1997].

- **Quark Matter and Quark Stars:**

Quarks are fundamental particles and the building blocks of matter, combining to form hadrons, such as protons and neutrons. Due to the phenomenon of color confinement, quarks cannot exist in isolation and are only found within hadrons. Uniquely, quarks are the only particles with electric charges that are fractional multiples of the elementary charge. As spin-$\frac{1}{2}$ particles, quarks follow Pauli's exclusion principle. There are six known types (flavors) of quarks: up (u), down (d), strange (s), charm (c), bottom (b), and top (t) as depicted in Table 1.3. Quarks have intrinsic properties, including electric charge, mass, color charge, and spin.

Table 1.3: Different types of quarks and their properties. Quantum numbers for each flavor like third component of isospin ($I_3$), baryon number ($B$), charm ($C$), strangeness ($S$), topness ($T$), and bottomness ($B^j$) are given.

| **Quark** | $M$ **(MeV)** | $Q$ **(e)** | $J^P$ | **B** | $I_3$ | **C** | **S** | **T** | $B^j$ |
|---|---|---|---|---|---|---|---|---|---|
| u | ≈ 5.0 | 2/3 | $1/2^+$ | +1/3 | 1/2 | 0 | 0 | 0 | 0 |
| d | ≈ 7.0 | -1/3 | $1/2^+$ | +1/3 | -1/2 | 0 | 0 | 0 | 0 |
| c | ≈ 1275.0 | 2/3 | $1/2^+$ | +1/3 | 0 | 1 | 0 | 0 | 0 |
| s | ≈ 100.0 | -1/3 | $1/2^+$ | +1/3 | 0 | 0 | -1 | 0 | 0 |
| t | ≈ 173210.0 | 2/3 | $1/2^+$ | +1/3 | 0 | 0 | 0 | 1 | 0 |
| b | ≈ 4180.0 | -1/3 | $1/2^+$ | +1/3 | 0 | 0 | 0 | 0 | -1 |

Within hadrons, quarks exhibit a property known as ”asymptotic freedom,” meaning that under extreme conditions of density or temperature, the hadrons lose their distinct identities, allowing quarks to explore a broader, ’colorless’ region. Under these conditions, a phase transition occurs, leading to the deconfinement of hadronic matter into quark matter (QM). Fig.1.3 illustrates the well-known schematic QCD Phase Diagram, depicting the deconfinement of hadronic matter into quark-gluon plasma (QGP) at extremely high temperatures or densities.

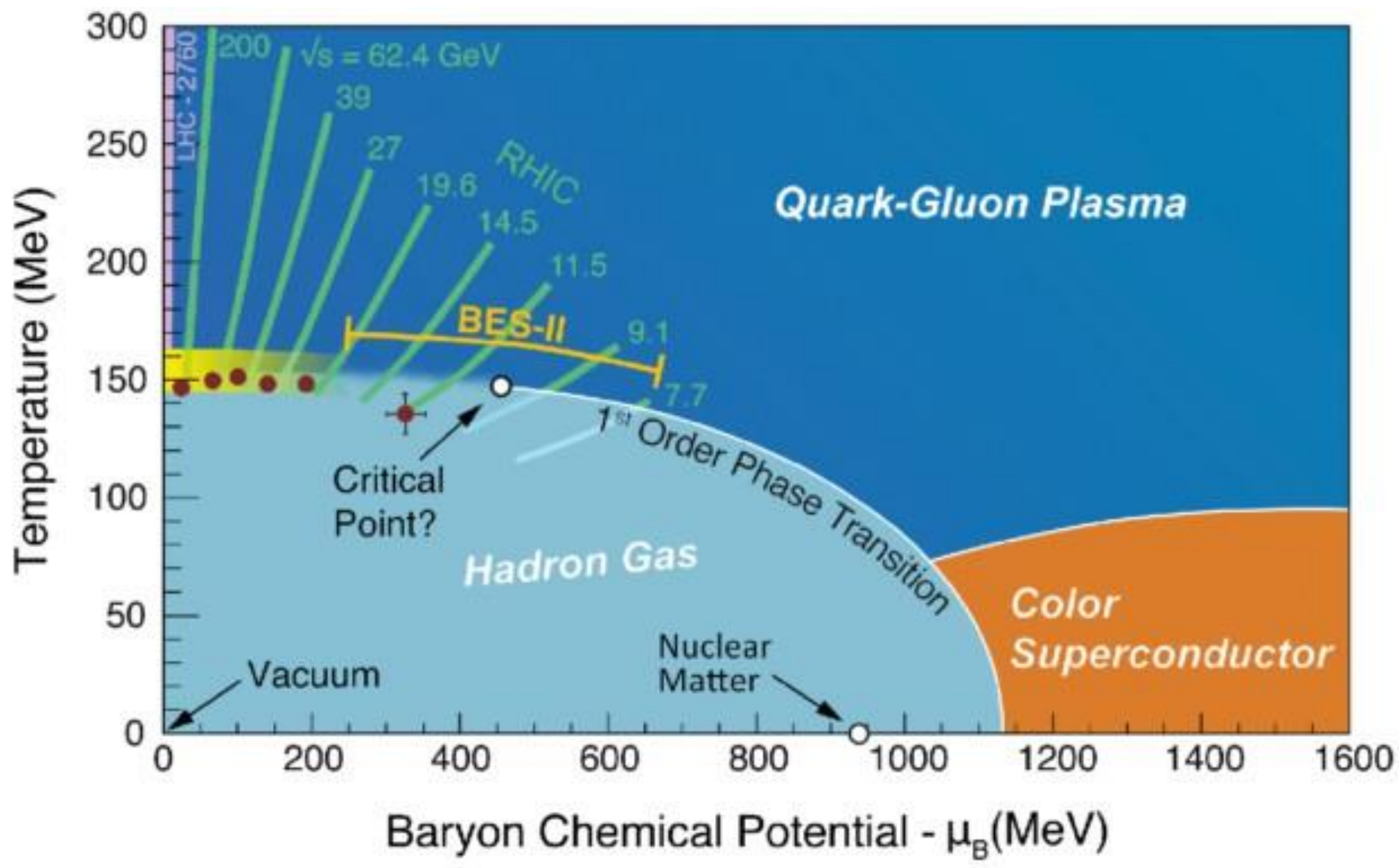

Figure 1.3: The QCD Phase Diagram. Credit: Fukushima & Hatsuda [2011]

In NS cores, the conditions might be so extreme that the dense hadronic matter, composed of particles like nucleons, hyperons, and $\Delta$ baryons, could undergo a phase transition, leading to the formation of a deconfined quark phase. This implies that within such stars, hadrons could break down into their fundamental components—quarks. Since nucleons, hyperons, and $\Delta$ baryons are made up of up ($u$), down ($d$), and strange ($s$) quarks, the resulting quark phase inside neutron stars or hybrid stars (HSs) is expected to contain only these three types of quarks. Heavier quarks, such as charm ($c$), top ($t$), and bottom ($b$), are unlikely to be present because the energy required to produce them would be too high under the conditions found in NS cores. Therefore, quark or hybrid stars would primarily consist of $u$, $d$, and $s$ quarks.

Quark stars (QSs) could form when hadronic matter undergoes a transition to strange QM through various processes, such as the clustering of $\Lambda$ hyperons, kaon condensation, the burning of hadrons into strange QM, or even seeding from external sources through accretion. QSs are theorized to have a core made of unpaired or superconducting QM composed of up (u), down (d), and strange (s) quarks. This dense core is thought to be surrounded by a thin outer layer made of degenerate electrons. The idea that a quark star could have a smaller gravitational mass than a hadronic star of the same baryon mass makes this transition energetically favorable. If this occurs, the complete conversion of hadronic matter into strange quark matter, resulting in a quark star, becomes a compelling possibility. However, despite the theoretical interest, there is still no strong experimental or observational evidence confirming the existence of quark stars or stable QGP. Nevertheless, this area remains an exciting and active field of research, particularly at facilities like the RHIC (Relativistic Heavy Ion Collider) and LHC (Large Hadron Collider).

- **Superconducting Quark Matter:**

  Superconducting quark matter refers to a theoretical state of quark matter in which quarks, pair up in a manner similar to the Cooper pairs in conventional superconductors. This pairing occurs at extremely high densities and low temperatures, conditions that might be present in the cores of NSs or in the early universe Alford & Reddy [2003], Alford *et al.* [2001], Sharma *et al.* [2007], Shifman & Ioffe [2001]. In the regime of asymptotically high densities, QCD predicts that QM can undergo a phase transition into a color superconducting state. This occurs because quarks, which are fermions, can pair up near the Fermi surface to form Cooper pairs, similar to how electrons pair up in conventional superconductivity. The pairing leads to the formation of a color-superconducting phase, and the associated energy gap in the quark spectrum is referred to as the BCS gap, denoted by $\Delta$. This gap $\Delta$ is crucial because it dictates the strength of the interaction between quarks in the superconducting phase. QCD calculations suggests that the quark phase may also be color-flavour locked (CFL) Alford & Reddy [2003], Alford *et al.* [2001], Shifman & Ioffe [2001]. The thermodynamic potential related to this CFL phase is controlled by the contributions from quark, lepton and the Goldstone bosons due to chiral symmetry breaking Alford & Reddy [2003], Alford *et al.* [2001], Sharma *et al.* [2007], Shifman & Ioffe [2001]. It has been proposed that at sufficiently high asymptotic densities, boson condensation may occur. However, such effects are relevant when $\Delta^2\mu^2 \ll m_s^4$. Therefore, by selecting an appropriate value for the gap parameter, the impact of boson condensation can be disregarded Alford *et al.* [2001]. Typically, a gap parameter value of $\Delta = 100$ MeV is chosen, as suggested in references Alford *et al.* [2001], Sharma *et al.* [2007].

### 1.4.3 Current Status and Challenges in Equation of State

As discussed in section 1.4.1, there is no comprehensive theory that fully explains the density regime from approximately $2\rho_0$ to $7\rho_0$. Although continuous efforts have been made by the aforementioned facilities to enhance our understanding and establish stringent constraints on the composition and EOS of NSs, significant challenges remain. Numerous studies have aimed to determine the EOS, using phenomenological models that fall into two broad categories: Non-Relativistic and Relativistic, each differing in their descriptive approach. Non-Relativistic models have successfully described finite nuclei, but when dealing with infinite dense nuclear matter, relativistic effects must be considered, ensuring that the speed of sound remains below the speed of light. In the context of Relativistic Mean Field (RMF) theory, two distinct approaches have been developed. One involves incorporating non-linear meson terms into the Lagrangian density to accurately capture the density dependence of the equation of state (EOS) and symmetry energy, as described by Boguta & Bodmer [1977], Mueller & Serot [1996a], Steiner *et al.* [2005a], Todd-Rutel & Piekarewicz [2005]. The other approach describes the non-linearities by introducing density-dependent coupling parameters, thereby avoiding the inclusion of non-linear mesonic terms, as proposed by Lalazissis *et al.* [2005], Typel & Wolter [1999], Typel *et al.* [2010]. To cover the entire phase space from low to high-density constraints, various interpolation schemes have been employed, such as piecewise polytropic interpolation Annala *et al.* [2021], Kurkela *et al.* [2014], spectral interpolation Lindblom [2010], speed-of-sound interpolation Altiparmak *et al.* [2022], Annala *et al.* [2020], Somasundaram *et al.* [2022], meta-models based on Taylor expansions Ferreira & Providência [2021], Ferreira *et al.* [2020], Margueron *et al.* [2018], Xie & Li [2019], non-parametric inference of the EOS Essick *et al.* [2020], Gorda *et al.* [2022], Landry & Essick [2019], Zhou *et al.* [2023b] using Gaussian Processes (GPs) Essick *et al.* [2021a], or Machine Learning (ML) Han *et al.* [2021] techniques. However, these EOS models face significant limitations, as they do not assume any specific composition of matter in the intermediate density regime.

At these densities, our knowledge of dense neutron-rich matter is largely derived from observations of NSs. Recent multimessenger observations, encompassing radio Antoniadis *et al.* [2013], Cromartie *et al.* [2019], Demorest *et al.* [2010], Fonseca *et al.* [2021], X-ray Miller *et al.* [2019, 2021], Riley *et al.* [2019, 2021], gravitational waves (GW Abbott *et al.* [2017a], Abbott *et al.* [2018], De *et al.* [2018], and their electromagnetic counterparts Abbott *et al.* [2017c, e], De *et al.* [2018], have offered significant new insights into the EOS of dense matter. We will explore the constraints from astrophysical observations in the next section.

## 1.5 Available Constraints

Data from terrestrial experiments, such as those involving finite nuclei and Heavy Ion Collisions (HIC), along with insights from Perturbative Quantum Chromodynamics (pQCD) and astrophysical observations of neutron stars, provide crucial constraints on the equation of state of nuclear matter and nucleon-nucleon interactions at both low and high densities. The empirical bounds identified through these methods will be summarized in the following sections.

### 1.5.1 Finite nuclei

The parameters for symmetric nuclear matter (SNM) at saturation density, such as the energy per nucleon $e_0$ and nuclear incompressibility $K_0$, along with coefficients like $J_0$, $L_0$, and $K_{\text{sym},0}$ that dictate the density dependence of the symmetry energy, are detailed in Table 1.5.1. These parameters are essential for describing infinite nuclear matter but also influence the properties of finite nuclear systems significantly. By correlating these parameters with various observables at saturation density in finite nuclei, researchers can derive precise insights. Such observables include nuclear masses, isoscalar giant monopole resonances (ISGMR), and isovector giant dipole resonances (IVGDR) energies, all of which are closely linked to these coefficients related to SNM and symmetry energy.

The average values of $e_0$ and $\rho_0$ are $e_0 = -15.88 \pm 0.24$ MeV and $\rho_0 = 0.163 \pm 0.005$ fm$^{-3}$, respectively, obtained from a selection of the most fitting nuclear equations of state according to nuclear data Dutra *et al.* [2012, 2014]. Research indicates that $K_0$, the measure of nuclear incompressibility, is around $K_0 \approx 240 \pm 20$ MeV, aligning well with the centroids of ISGMR measured in several nuclei including $^{208}$Pb, $^{90}$Zr, and $^{144}$Sm Agrawal *et al.* [2005], Avogadro & Bertulani [2013], Garg & Colò [2018], Niksic *et al.* [2008], Todd-Rutel & Piekarewicz [2005]. Additionally, a comprehensive examination of the experimental differences in symmetry energies relative to mass numbers in finite nuclei points to $J_0$ being about $32.10 \pm 0.31$ MeV, as reported in Jiang *et al.* [2012]

Recent microscopic calculations involving the neutron skin of heavy nuclei have been utilized to refine the determination of $L_0$, resulting in a value of $59 \pm 13$ MeV Agrawal *et al.* [2012, 2013]. Additionally, studies of the isovector giant dipole and quadrupole resonances in the $^{208}$Pb nucleus suggest that $L_0$ is approximately $43 \pm 26$ and $37 \pm 18$ MeV, respectively Roca-Maza *et al.* [2013a, b]. The slope parameter for symmetry energy has also been determined using various methods applied to different nuclei. For instance, neutron-skin thickness measurements of the $^{48}$Ca nucleus by the Coherent Reactions Experiment (CREX) collaboration Adhikari *et al.* [2022] and of $^{208}$Pb by the Parity Radius

Table 1.4: The current nuclear parameters, including those related to symmetric nuclear matter such as $e_0$ and $K_0$, as well as the coefficients governing from the density-dependent of symmetry energy ($J_0$, $L_0$, and $K_{\mathrm{sym},0}$), at saturation density $\rho_0$ are listed.

| NMP | empirical value (MeV) | References |
|---|---|---|
| $e_0$ | $-15.88 \pm 0.24$ | Dutra *et al.* [2012, 2014] |
| $K_0$ | $240 \pm 20$ | Garg & Colò [2018] |
| $J_0$ | $32.10 \pm 0.31$ | Jiang *et al.* [2012] |
| $L_0$ | $53 \pm 15$ | Essick *et al.* [2021b] |
| $K_{\mathrm{sym},0}$ | $-111.8 \pm 71.8$ | Mondal *et al.* [2017] |

Experiment (PREX-II) collaboration Adhikari *et al.* [2021] have provided valuable data.

Recently, comprehensive data analysis from the PREX-II experiment by Reed & Horowitz [2020], Reed *et al.* [2021] reported $L_0 = 106 \pm 37$ MeV. Combining astronomical observations with PREX-II data, Essick *et al.* [2021b] estimated $L_0$ to be $53^{+14}_{-15}$ MeV. Another analysis of the PREX-II data by Reinhard *et al.* [2021] suggested a slightly lower value of $54 \pm 8$ MeV. Moreover, the CREX experiment's data indicates a possible range for $L_0$ from 0 to 51 MeV Tagami *et al.* [2022].

When examining the density-dependent behavior of symmetry energy at densities significantly higher than the saturation density ($\rho \gg \rho_0$), the parameter $K_{\mathrm{sym},0}$ becomes crucial. However, $K_{\mathrm{sym},0}$ remains poorly constrained due to the lack of direct probes. Various nuclear models suggest a wide range of values for $K_{\mathrm{sym},0}$, spanning from $-700$ MeV to 400 MeV Dutra *et al.* [2012, 2014], Ferreira & Providência [2021]. Recent research Mondal *et al.* [2017] highlights a significant correlation between $K_{\mathrm{sym},0}$ and the pair $3J_0 - L_0$, constraining $K_{\mathrm{sym},0}$ to approximately $-111.8 \pm 71.3$ MeV.

### 1.5.2 Heavy Ion Collisions and *ab-initio* calculations

The current empirical constraints on the equation of state, excluding those associated with the saturation density $\rho_0$, are summarized in Table 1.5. The first two rows pertain to the pressure of symmetric nuclear matter (SNM), derived from analyses of directed and elliptic flow Danielewicz *et al.* [2002] and kaon production in heavy ion collisions Fuchs [2006]. The following five rows focus on the energy density and pressure of pure neutron matter (PNM). 'Best-fit' Skyrme EDFs provide its energy density ($E(\rho)$) at a density of $\rho = 0.1$ $\text{fm}^{-3}$ Brown [2013]. Analytical calculations involving effective degrees of freedom at low density, such as chiral effective theory, indicate negligible uncertainty. Precise next-to-next-to-next-to-leading-order ($N^3LO$) calculations, typically fitted to nucleon–deuteron scattering cross sections or few-body observables and even saturation properties of heavier nuclei, are referenced in Drischler *et al.* [2021]. The EOS for pure neutron matter at low density (0.04-0.16 $\text{fm}^{-3}$) was established using the $N^3LO$ framework within chiral effective field theory, as detailed in Hebeler *et al.* [2013]. This EOS has been extended up to twice the saturation density $\rho_0$ in Lattimer [2021]. The last three rows address the symmetry energy $E_{\text{sym}}(\rho)$ at various densities. The symmetry energy is derived from simulations of low-energy HIC in $^{112}Sn+^{112}Sn$ and $^{124}Sn+^{124}Sn$ Tsang *et al.* [2009], nuclear structure studies using Isobaric Analogue States (IAS) Danielewicz & Lee [2014], and Asy-EOS experiments at GSI Russotto *et al.* [2016]. Additionally, $E_{\rm sym}(\rho)$ at $\rho$ = 0.1 fm$^{-3}$ is obtained from a microscopic examination of IVGDR in $^{208}$Pb Trippa *et al.* [2008].

Table 1.5: The current empirical constraints on the equation of state which include pressure ($P(\rho)$), energy per particle ($E(\rho)$), and symmetry energy ($E_{\text{sym}}(\rho)$) corresponding to different nuclear matter configurations, such as symmetric nuclear matter (SNMX), pure neutron matter (PNMX), and the symmetry energy (SYMX). The limitations are presented in conjunction with the corresponding density ranges from which they are generated.

| | Quantity | Density region ($\text{fm}^{-3}$) | Band/Range (MeV) | References |
|---|---|---|---|---|
| SNM1 | $P(\rho)$ | 0.32 to 0.74 | HIC | Danielewicz *et al.* [2002] |
| SNM2 | $P(\rho)$ | 0.19 to 0.33 | Kaon exp. | Fuchs [2006] |
| | | | | |
| PNM1 | $E(\rho)$ | 0.1 | $10.9 \pm 0.5$ | Brown [2013] |
| PNM2 | $E(\rho)$ | 0.04 to 0.16 | $N^3LO$ | Hebeler *et al.* [2013] |
| PNM3 | $P(\rho)$ | 0.04 to 0.16 | $N^3LO$ | Hebeler *et al.* [2013] |
| PNM4 | $E(\rho)$ | 0.01 to 0.33 | $N^3LO$ | Lattimer [2021] |
| PNM5 | $P(\rho)$ | 0.01 to 0.33 | $N^3LO$ | Lattimer [2021] |
| | | | | |
| SYM1 | $E_{\text{sym}}(\rho)$ | 0.1 | $24.1 \pm 0.8$ | Trippa *et al.* [2008] |
| SYM2 | $E_{\text{sym}}(\rho)$ | 0.01 to 0.19 | IAS,HIC | Danielewicz & Lee [2014], Tsang *et al.* [2009] |
| SYM3 | $E_{\text{sym}}(\rho)$ | 0.01 to 0.31 | ASY-EOS | Russotto *et al.* [2016] |

### 1.5.3 Astrophysical Observations

The theoretical upper limits for the maximum achievable mass and radius of neutron stars are determined by the properties of the nuclear equation of state (EOS) across the entire density spectrum, from low to high densities. Precise measurements of both mass and radius can provide critical constraints on the EOS of nuclear matter. However, due to the immense distances involved, directly determining the radius of neutron stars is a formidable challenge. The observation of neutron stars with masses around $2M_\odot$ Antoniadis *et al.* [2013], Arzoumanian *et al.* [2018] has established a lower limit on the maximum mass that any EOS must predict.

The tidal deformability parameter of neutron stars, which encodes information about the EOS, was inferred for the first time from the gravitational wave event GW170817, observed by the advanced LIGO Aasi *et al.* [2015] and advanced Virgo Acernese *et al.* [2015] detectors during a binary neutron star (BNS) merger with a total system mass of $2.74^{+0.04}_{-0.01}M_\odot$ Abbott *et al.* [2019a, b]. Another subsequent event, GW190425, also likely originated from a BNS coalescence Abbott *et al.* [2020a]. Future observations of BNS signals are expected to become more frequent with the LIGO-Virgo-KAGRA and future detectors like the Einstein Telescope Punturo *et al.* [2010] and Cosmic Explorer Reitze *et al.* [2019]. The constraints on the EOS promised by gravitational wave astronomy have triggered numerous theoretical investigations into the properties of neutron stars Abbott *et al.* [2017d, 2020a], Biswas *et al.* [2021], De *et al.* [2018], Fattoyev *et al.* [2018], Forbes *et al.* [2019], Landry & Essick [2019], Malik *et al.* [2018], Piekarewicz & Fattoyev [2019], Thi *et al.* [2021].

Recently, two different groups using the Neutron Star Interior Composition Explorer (NICER) X-ray telescopes provided simultaneous mass and radius measurements for PSR J0030+0451, reporting $R = 13.02^{+1.24}_{-1.06}$ km for a mass of $1.44^{+0.15}_{-0.14}M_\odot$ Miller *et al.* [2019] and $R = 12.71^{+1.14}_{-1.19}$ km for a mass of $1.34^{+0.15}_{-0.16}M_\odot$ Riley *et al.* [2019], offering complementary constraints on the EOS. For the heavier pulsar PSR J0740+6620, they reported $R = 13.7^{+2.6}_{-1.5}$ km for a mass of $2.08 \pm 0.07M_\odot$ Miller *et al.* [2021] and $R = 12.39^{+1.30}_{-0.98}$ km for a mass of $2.072^{+0.067}_{-0.066}M_\odot$ Riley *et al.* [2021]. The current observational lower bound on the maximum neutron star mass is $M_{\rm max} = 2.35 \pm 0.17M_\odot$ for the black-widow pulsar PSR J0952-0607 Romani *et al.* [2022], which exceeds any previous measurements, including $M_{\rm max} = 2.27^{+0.17}_{-0.15}M_\odot$ for PSR J2215-5135 Linares *et al.* [2018]. If these observational bounds are reliable, stiffer EOSs are required to support neutron stars with masses exceeding $2M_{\odot}$.

## 1.6 The Dark Side

According to $\Lambda$-CDM model Deruelle & Uzan [2018] dark matter constitutes about 26.5% of the mass-energy density of the universe. The remaining 4.9% comprises all baryonic matter, while 68.6 % contributes to dark energy. Dark matter (DM), a non-luminous and elusive substance, does not emit, absorb, or reflect light, making it invisible across the electromagnetic spectrum. Its presence is inferred primarily through its gravitational interactions with visible matter, playing a critical role in the formation and dynamics of galaxies and galaxy clusters. Despite its significant influence on the universe, the precise nature of DM remains one of the most profound mysteries in modern physics. Researchers across multiple disciplines, including astrophysics, cosmology, and particle physics, are actively exploring its properties and composition. Several theoretical models have been developed to explain DM, complemented by a range of experimental efforts aimed at its detection. Direct detection experiments, such as those conducted by CDMS II, CRESST Angloher *et al.* [2014], and COGeNT Aalseth *et al.* [2013], focus on measuring the scattering cross-section of DM particles in terrestrial detectors. Additionally, particle accelerators Feng [2010] are employed to investigate potential dark matter candidates further.

Another indirect method for studying dark matter involves examining its influence on compact objects such as neutron stars. The dense cores of neutron stars can serve as gravitational traps for dark matter particles, potentially altering their behavior and properties.

This thesis aims to address the gaps identified in Section 1.4.1 by exploring potential solutions within the framework of Dark Matter Admixed Neutron Stars (DANS). The following sections will delve into the historical background of dark matter, its characteristics, potential candidates, ongoing research in the field of DANS, and the underlying motivation for this thesis.

## 1.7 Historical Background: Dark Matter

Over eight decades ago, the concept of dark matter was first hypothesized by Swiss-American astronomer Fritz Zwicky Zwicky [1937]. He noticed that galaxies within the Coma cluster were moving at such high speeds that, theoretically, they should have dispersed into space. Yet, they remained gravitationally anchored to the cluster, suggesting the presence of some invisible mass providing the necessary gravitational force to maintain their orbits.

Further investigations Freeman [1970], Rubin & Ford [1970] into galaxy dynamics have consistently revealed that the rotational velocities of stars in galaxies remain surprisingly constant—or ”flat”—as the distance from the galactic center increases (as shown in fig.1.4). This observation contradicts the expectations set by Newton’s law of gravity, which would predict a decrease in rotational velocity for stars located further from the center. This is analogous to the way inner planets in our Solar System orbit the Sun more quickly than the outer planets. If additional mass were added between the planets in our Solar System, the outer planets would orbit faster; similarly, the flat rotation curves of galaxies suggest the existence of significant amounts of unseen mass surrounding each galaxy.

This led to the widely accepted theory that dark matter forms a massive, roughly spherical halo around each galaxy, accounting for the gravitational effects observed but not explained by visible matter alone.

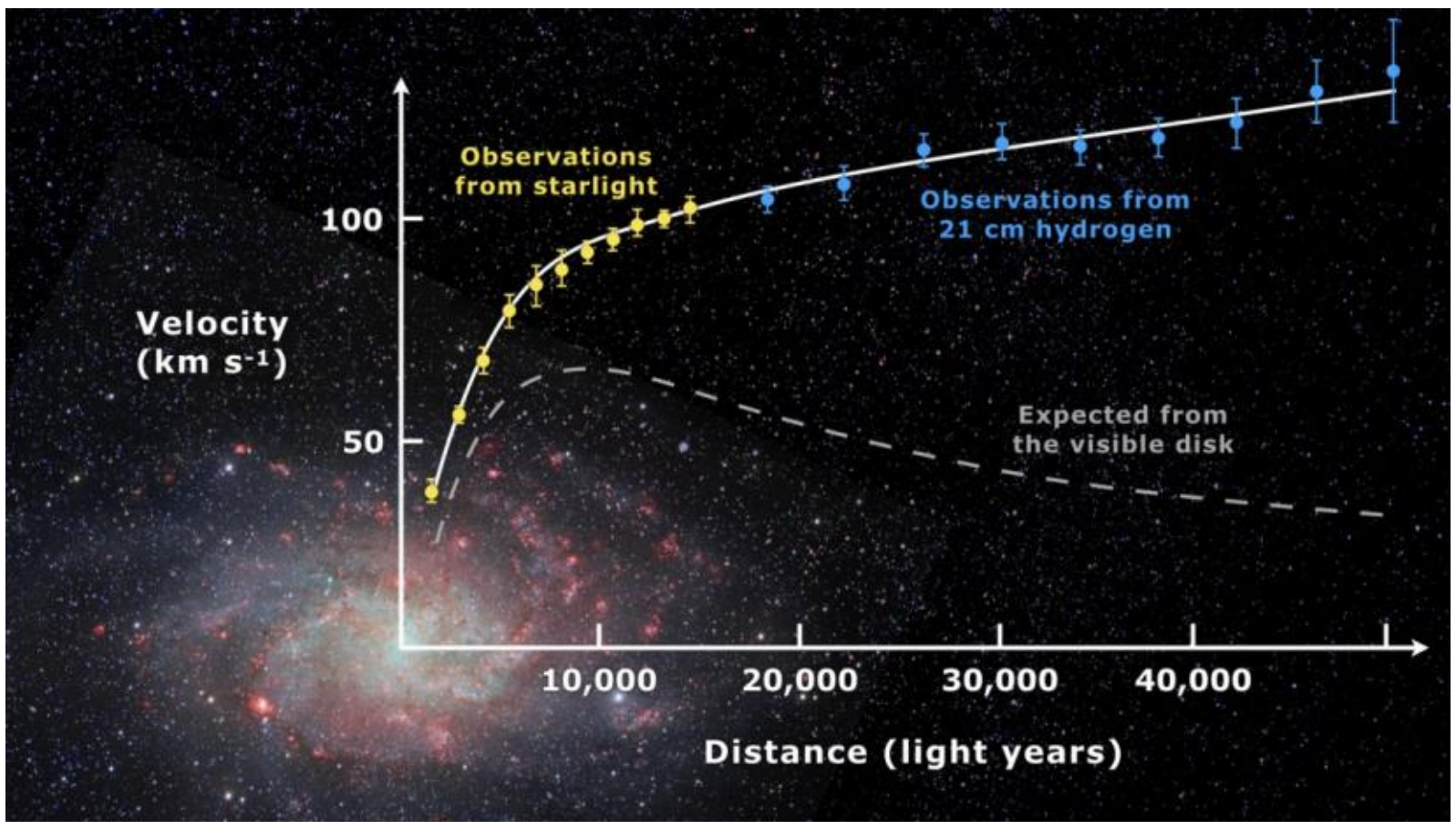

Figure 1.4: Rotation curve of spiral galaxy Messier 33 (yellow and blue points with error bars), and a predicted one from distribution of the visible matter (gray line). The discrepancy between the two curves can be accounted for by adding a dark matter halo surrounding the galaxy. Image sourced from Bergström [2000]

### Large and Cosmological Scale

Dark matter is crucial in understanding the large-scale structure of the universe, as evidenced by the distribution of galaxies and galaxy clusters across vast cosmic expanses. These structures align with the hypothesis of substantial quantities of dark matter, which acts as the gravitational framework necessary for the aggregation of visible matter into galaxies and clusters.

Cosmological simulations, which trace the evolution of these structures over time, consistently require the inclusion of dark matter to accurately mirror the observed spatial distribution of galaxies. This modeling confirms the integral role of dark matter in the formation and stability of cosmic structures.

Additionally, indirect evidence supporting the existence of dark matter emerges from the phenomena of gravitational lensing. This effect, where the gravitational field of massive objects distorts and bends the path of light coming from more distant galaxies, further points to dark matter's pervasive influence. By analyzing the lensing effects attributed to intervening dark matter halos, scientists can map the distribution of dark matter throughout the universe, enhancing our understanding of its profound impact on cosmic architecture.

**Nature of dark matter**

The first consideration regarding the nature of the missing matter involves identifying its composition. Based on its elusive nature, two possibilities arise: (1) dark matter is composed of ordinary matter, though highly dispersed, or (2) it consists of exotic matter that is widespread yet undetectable by our instruments.

The first category suggests that dark matter is made of protons and neutrons, the same components as ordinary matter, technically known as 'baryonic' matter. Conversely, the second category refers to 'non-baryonic' matter, indicating a fundamentally different composition.

Baryonic dark matter is a hypothetical form of dark matter composed of baryons. It is believed that only a small fraction of the universe's dark matter is baryonic. This type of dark matter may exist as non-luminous gas or as Massive Astrophysical Compact Halo Objects (MACHOs) Alcock *et al.* [2000]. Examples of these condensed objects include black holes, neutron stars, white dwarfs, very faint stars, and non-luminous bodies such as planets and brown dwarfs.

Non-baryonic dark matter candidates consist of exotic subatomic particles that do not interact electromagnetically; otherwise, their radiation would be detectable. Individually, these particles interact with matter only on extremely rare occasions. However, collectively, their gravitational effects become significant due to the enormous quantities of these particles that are believed to exist. These weakly interacting particles must be very stable, implying that non-baryonic candidates are likely relic particles from the early universe, where conditions were radically different.

Non-baryonic candidates have been further subdivided by cosmologists between low mass particles moving very fast – referred to as 'hot dark matter' Primack & Blumenthal [1984] – or high mass slow moving particles – 'cold dark matter. Hot dark matter (HDM) consists of particles that remain relativistic for significantly longer times. This requires their masses to be less than $\approx$ 100 eV,. Cold dark matter (CDM) Sommer-Larsen & Dolgov [2001] consists of weakly interacting massive particles (WIMPs) that become non-relativistic at temperatures well above $10^4$ K. In contrast to the CDM case in which there are no experimentally verified candidates, HDM has a definite candidate, the neutrino.

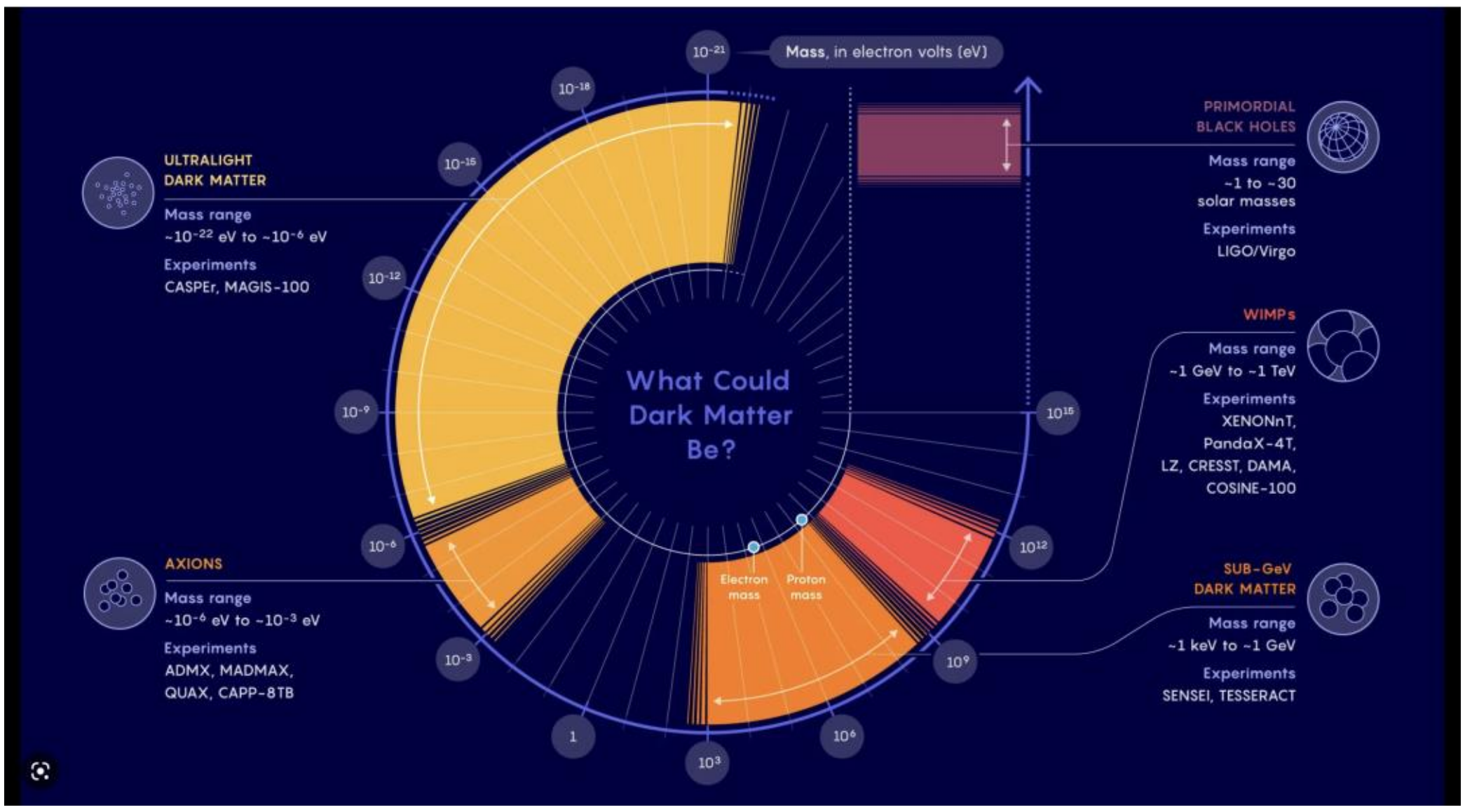

Figure 1.5: Candidates of Dark Matter. Source: Samuel Velasco/ Quanta Magazine

**Dark Matter Candidates**

Over the years, a wide range of dark matter models have been proposed following the initial suggestion of its presence in the universe. In fig.1.5 we show the different candidates of dark matter. The catalog of potential dark matter candidates is extensive, encompassing a variety of possibilities that are each promising in their own right. Although certain candidates have emerged as more favorable compared to others, it's important to note that no single candidate perfectly addresses all the complexities associated with dark matter. Here are some of the notable candidates that have been extensively investigated:

- **Weakly Interacting Massive Particles (WIMPs)**:

  The category of dark matter known as Weakly Interacting Massive Particles (WIMPs) Steigman *et al.* [2012] is characterized by its significant mass and minimal interaction with visible matter. This class of particles was initially proposed by Steigman and Turner in 1985. WIMPs are thought to possess a mass around 100 GeV and engage in weak-scale interactions, with coupling strengths estimated at about $10^{-2}$

- **Axions**:

  Axions stand out as one of the leading candidates for dark matter due to their dual role in addressing both the dark matter problem and the strong CP violation issue in particle physics Olive *et al.* [2014], Tanabashi *et al.* [2018]. Originating from the proposed solutions to the strong CP violation, axions are either non-interacting or interact very weakly with other particles. They are believed to be produced through processes related to symmetry breaking, highlighting their unique position in theoretical physics.

- **Dark Photons**:

  The light boson, with mass $m_V < 2m_e$, is a viable candidate for dark matter. Dark photons can be stable and may have been produced in the early universe through annihilation or scattering processes such as $e^+e^- \rightarrow V\gamma$ or $\gamma e^\pm \rightarrow Ve^\pm$, where $V$ denotes a dark photon An *et al.* [2015].

- **Sterile Neutrinos**:

  Sterile neutrinos are theoretical particles that are believed to interact only through gravitational forces, remaining inert to the forces described in the Standard Model. The label ”sterile” indicates that these neutrinos are distinct from the ”active” neutrinos included in the Standard Model, as they do not participate in electroweak interactions. These neutrinos have been suggested in various scenarios to solve a range of problems Feng *et al.* [2003]. As candidates for dark matter, it is suggested that sterile neutrinos could have been formed in the early universe through various mechanisms. The specific process of their creation can influence their detectability, as their impacts on the small-scale cosmic structures can be used to set constraints Abazajian [2006] .

## 1.8 Dark Matter Capture and Accumulation in Neutron Stars

The amount of DM in NS is expected to vary depending on the evolution history and position of the star. Thus, the environment from which it originates and its distance from the galactic centre (GC) define properties of the progenitor and DM accretion rate during the whole lifetime of the star, including (i) progenitor, (ii) main sequence (MS) star,(iii) supernova explosion with formation of a proto-NS and (iv) equilibrated NS phases. The key stages of DM accumulation are outlined as follows:

### 1.8.1 Progenitor Stage

The formation of massive stars, particularly in the innermost regions of galaxies, is influenced by DM through gravitational effects such as tidal energy, which enhances star formation. During this phase, DM and baryonic matter (BM) contract together, leading to the initial trapping of DM. Through scattering interactions with baryons, DM undergoes thermalization and energy loss, with its effect becoming more significant in later evolutionary stages.

### 1.8.2 Main Sequence (MS) Accretion

DM accretion during the MS phase has been extensively studied in the literature Lopes *et al.* [2019], Scott *et al.* [2009]. The gravitational potential of the star increases, facilitating further capture of DM. The amount of DM accreted depends on the star's distance from the GC, the accreted mass ranging from $10^{-9}M_\odot$ to $10^{-5}M_\odot$ in the central galactic region.

### 1.8.3 Supernova Collapse and Proto-NS Formation

Gravitational collapse, followed by a supernova explosion, marks the end of a massive star that has exhausted all its nuclear fuel. The proto-neutron star (NS) formed during the supernova is characterized by dynamical instabilities, inhomogeneities in its interior, and a high fraction of heavy elements, which enhance interactions with dark matter (DM) particles. The newly-born NS will be surrounded by a dense cloud of DM particles, with a temperature and radius corresponding to the last stage of main sequence (MS) star evolution, i.e., a star with a silicon core. Consequently, the accretion onto the proto-NS and the DM density in the surrounding medium may be influenced by the prior evolutionary stage Kouvaris & Tinyakov [2010a]. Furthermore, a significant amount of DM can be produced during the supernova explosion and largely remain gravitationally trapped inside the star Nelson *et al.* [2019].

### 1.8.4 Equilibrated Neutron Star Phase

Once the NS reaches equilibrium, it continues to capture DM through spherically symmetric accretion. The amount of accreted DM mass is given by:

$$M_{\rm acc} \approx 10^{-14} \; \frac{\rho_\chi}{0.3\ {\rm GeV/cm^3}} \; \frac{\sigma_{\chi n}}{10^{-45}{\rm cm^2}} \; \frac{t}{\rm Gyr} \; M_\odot, \tag{1.18}$$

where $t$ is the accretion time, $\rho_\chi$ is the local DM density, and $\sigma_{\chi n}$ represents the DM-nucleon interaction cross-section. For a typical cross-section of $10^{-45}$ cm$^2$, the mean free path of DM within the NS is a few kilometers, leading to rapid saturation of captured DM particles.

For the obtained dark matter (DM) densities around the two heaviest pulsars, calculated using the NFW [see Eq. (1.19)] and Einasto [see Eq. (1.19)] profiles, the amount of accreted matter is estimated to be in the range of $\sim 10^{-13} - 10^{-14} M_\odot$. Since the NFW profile exhibits a divergent inner density distribution, $\rho_\chi(d) \propto d^{-1}$ [see Eq. (1.19)], the Einasto profile is preferred for describing the central regions of the Galaxy.

In the most central Galactic region, the accreted DM mass can vary between $10^{-5}M_\odot$ and $10^{-8}M_\odot$, depending on the value of the parameter $\alpha$ Del Popolo *et al.* [2020].

Furthermore, in the innermost parts of the Galaxy, DM clumps with masses in the range of $10^{-6}M_\odot$ to $10^{2}M_\odot$ Profumo *et al.* [2006] could be accreted onto neutron stars. Such a mechanism may significantly enhance the DM fraction within compact stars.

$$\rho_\chi(d) = \rho_{-2} e^{-\frac{2}{\alpha}\left[\left(\frac{d}{r_{-2}}\right)^{\alpha}-1\right]} \tag{1.19}$$

with $\rho_{-2}$ being the DM density at the distance $r_{-2}$, where the logarithmic gradient $\frac{\mathrm{d}\ln\rho}{\mathrm{d}\ln r} = -2$.

The accretion and retention of DM inside neutron stars can have significant implications for their internal structure and observable properties. If a considerable fraction of DM accumulates within the star over time, it may affect the equation of state (EoS), mass-radius relation, and tidal deformability of NSs. The efficient capture of DM in neutron stars, particularly in regions with high ambient DM density, supports the feasibility of DM-admixed neutron stars as viable astrophysical objects.

## 1.9 Recent Advances in the Study of Dark Matter Admixed Neutron Stars

As previously discussed, a further indirect approach to understanding dark matter involves examining its influence on compact stars. NS, owing to their extreme density, function as natural laboratories for this purpose. The presence of dark matter in or around neutron stars can significantly influence the interpretation of observational data on these stars, including their mass, radius, and tidal deformability Das *et al.* [2019, 2020a], de Lavallaz & Fairbairn [2010], Ellis *et al.* [2018], Goldman & Nussinov [1989], Güver *et al.* [2014], Ivanytskyi *et al.* [2020], Kain [2021], Kouvaris [2008], Kouvaris & Tinyakov [2010b], Panotopoulos & Lopes [2017], Raj *et al.* [2018], Shahrbaf *et al.* [2022, 2024], Shakeri & Karkevandi [2022], Shirke *et al.* [2023], Tolos *et al.* [2015]. Neutron stars capture dark matter through scattering interactions with nucleons and leptons, leading to a loss of kinetic energy. There are multiple pathways for dark matter (DM) to be present within neutron stars (NSs), with the primary mode of DM accretion occurring throughout their evolutionary stages, from progenitor stars to fully equilibrated neutron stars. Events such as neutron star mergers and supernova explosions can also introduce dark matter into neutron stars. Additionally, a neutron star traveling through regions of high dark matter density in the galaxy can accumulate significant amounts of dark matter Del Popolo *et al.* [2020],

Sandin & Ciarcelluti [2009]. Several theoretical models suggest that dark matter, including Weakly Interacting Massive Particles (WIMPs), could be gravitationally attracted and concentrated within the cores of neutron stars.Goldman & Nussinov [1989]. Dark matter (DM) can either be asymmetric or symmetric, based on the presence of particle-antiparticle asymmetry. Symmetric DM particles have the potential to self-annihilate, which could enable their detection through X-ray, $\gamma$-ray, or neutrino telescopes Kouvaris [2008]. Studies show that self-annihilating DM in the inner regions of neutron stars (NSs) may significantly influence their kinematic properties. In addition, DM particle annihilation within the NS core may result in late-time heating that may be observed in the surface temperatures of older NS population . Asymmetric DM, on the other hand, will accumulate inside a star, altering its properties. From the perspective of particle physics, there is a plethora of candidates as discussed in section 1.7 for dark matter particles, including bosonic dark matter, axions, sterile neutrinos, and various forms of Weakly Interacting Massive Particles (WIMPs), as discussed in the literature Bauer & Plehn [2019], Calmet & Kuipers [2021], Silk *et al.* [2010]. Given the uncertain nature of particle dark matter, both bosonic and fermionic particles have been explored to understand their impacts on neutron star dynamics Diedrichs *et al.* [2023], Ellis *et al.* [2018], Leung *et al.* [2022], Narain *et al.* [2006], Panotopoulos & Lopes [2017].

Fermionic dark matter, composed of particles that adhere to Fermi-Dirac statistics, has been a significant subject of study due to its potential effects on neutron stars. Research suggests that such dark matter could form a core within a star, leading to observable phenomena including the heating of cold neutron stars Ángeles Pérez-Garc´ıa *et al.* [2022], Bell *et al.* [2018], Bertone & Fairbairn [2008], alterations in pulsar scintillation Á ngeles Pérez-Garc´ıa *et al.* [2013], and changes in rotational patterns Kouvaris & Perez-Garcia [2014]. These effects may be detectable with future observations. Integrating observations across electromagnetic bands—with instruments like the James Webb Space Telescope, gamma-ray telescopes such as e-ASTROGAM Tavani *et al.* [2018] and AMEGO, the upcoming Square Kilometer Array in radio astronomy, ATHENA in X-ray astronomy, and third-generation gravitational wave detectors like Advanced LIGO, Virgo, KAGRA, Cosmic Explorer Reitze *et al.* [2019], or the Einstein Telescope Branchesi *et al.* [2023]—offers a promising approach to detecting signals indicative of dark matter-induced phenomena Boddy *et al.* [2022]. In extreme scenarios, the accumulation of dark matter could even trigger a star's collapse into a black hole McDermott *et al.* [2012], Singh *et al.* [2023], Zurek [2014] or its transformation into a quark star Herrero *et al.* [2019], producing bursts of radiation Zenati *et al.* [2024].

In contrast, bosonic dark matter models, which include scalar fields, axions, and sexaquarks,

have attracted significant interest in astrophysics and have been extensively studied for their profound effects on compact objects Chavanis [2023], Karkevandi *et al.* [2024], Shakeri & Hajkarim [2023]. At the extreme densities within neutron stars, these particles, at zero temperature, could form a Bose-Einstein condensate (BEC), leading to the gravitational collapse of the bosonic dark matter into a black hole Kouvaris [2013]. Light dark matter particles such as axions, which are pseudo-Goldstone bosons associated with a spontaneously broken symmetry Kim [1987], Zhang *et al.* [2024], could contribute additional cooling channels in neutron stars Zhang *et al.* [2024], affecting the thermal evolution of the star. Depending on the distribution of dark matter, neutron stars can exhibit two distinct spatial configurations: a dark matter core within the star and a dark matter halo extending beyond its baryonic surface. These configurations significantly influence the star's mass-radius relationship and tidal deformability, attributes measurable through the analysis of gravitational waves from neutron star mergers. The mass and interaction strength of these bosonic particles are crucial in determining the star's structure and overall stability. Given that the exact nature of dark matter and its in-medium properties, such as self-interaction and coupling with standard model particles, remain unknown, there are two main approaches to studying dark matter-admixed neutron stars. One approach involves considering non-gravitational interactions between ordinary matter and dark matter, such as the Higgs portal mechanism Das *et al.* [2019, 2020b], Dutra *et al.* [2022], Flores *et al.* [2024], Hong & Ren [2024], Lenzi *et al.* [2023], Sen & Guha [2021], and self-interacting dark matter models Shirke *et al.* [2023]. In this scenario, the system can be effectively treated as a single fluid due to the non-gravitational interaction between normal matter and dark matter. The other approach involves ignoring the non-gravitational interaction between normal matter and dark matter, resulting in a two-fluid system where normal matter and dark matter interact solely through gravitational interactions. Two fluid approach has been extensively studied in Ref. Buras-Stubbs & Lopes [2024], Collier *et al.* [2022], Emma *et al.* [2022], Hong & Ren [2024], Ivanytskyi *et al.* [2020], Karkevandi *et al.* [2022], Liu *et al.* [2023], Miao *et al.* [2022], Rü ter *et al.* [2023], Rutherford *et al.* [2023]. Recent advancements in observational astronomy, spearheaded by instruments such as LIGO, Virgo, and KAGRA Abbott *et al.* [2021], are revolutionizing our understanding of neutron stars and their potential dark matter components. These gravitational wave observatories have transformed the field by offering unprecedented insights into neutron star mergers and the fundamental physics governing these extreme objects. The emergence of multimessenger astronomy, which integrates gravitational wave data with electromagnetic observations, has substantially enhanced our knowledge of neutron stars. This approach is crucial for testing dark matter theories, as gravitational waves from neutron star mergers serve as precise indicators of the equation of state and dark matter properties like interaction cross-section and mass. The groundbreaking detection of gravitational waves from the GW170817 event Abbott *et al.* [2019] was a pivotal moment in astrophysics, setting stringent constraints on tidal deformabilities and providing valuable data for probing hadronic models. Observatories like NICER Miller *et al.* [2019], Raaijmakers *et al.* [2020] have complemented these findings by offering crucial mass-radius measurements of neutron stars, enabling more accurate tests of theoretical models. NICER, an x-ray telescope aboard the International Space Station, has pioneered advanced mass and radius measurement techniques through Pulse Profile Modeling (PPM) Watts [2019]. Although current PPM methods do not yet account for potential dark matter components within neutron stars, NICER's measurements of stars like pulsar PSR J0030+0451 and PSR J0740+6620 have provided valuable information on the mass-radius relation of neutron stars. Recent accurate mass measurements have led to a determination of M = $2.08^{+0.07}_{-0.07}\ M_{\odot}$ for massive neutron stars. Future missions like STROBE-X (the Spectroscopic Time-Resolving Observatory for Broadband Energy x-rays Wilson-Hodge *et al.* [2017]) and the Chinese-European mission concept eXTP (the enhanced x-ray Timing and Polarimetry mission Zhang *et al.* [2016]) aim to extend PPM to a broader population of neutron stars, promising even deeper insights. The detection of gravitational waves has also opened new avenues for exploring particle interactions in celestial objects. Additionally, radio observations from SKA Weltman *et al.* [2020] can provide a comprehensive approach to studying dark matter in and around neutron stars. Upcoming projects like the Einstein Telescope Branchesi *et al.* [2023] and Cosmic Explorer Reitze *et al.* [2019]. will further revolutionize our understanding by offering greater sensitivity and resolution in gravitational wave detection. These advancements will deepen our understanding of neutron stars and open new pathways for exploring the enigmatic nature of dark matter, pushing the boundaries of contemporary astrophysical research. This integrated approach not only enriches our understanding of neutron stars but also propels us toward uncovering the mysteries of dark matter, driving the frontier of astrophysical exploration forward.

## 1.10 Motivation And Objectives:

Historically, much of the prior research in the literature has prematurely committed to constraining dark matter model parameters from observational properties of neutron stars without acknowledging the inherent uncertainties in the behaviour of baryonic matter. This oversight has often led to simplistic interpretations. One of the motivations for this thesis is to investigate possible links between the dark matter sector and the overall characteristics of neutron stars, considering the inherent uncertainties in the Equation of State (EOS) of the baryonic matter. It is crucial to consider the uncertainties associated with the EOS in the baryonic sector, as they can affect our understanding of the overall behavior of neutron stars. We use a two-fluid model to study the properties of neutron stars, where one fluid represents nuclear matter and the other represents dark matter. For the nuclear matter, we apply several realistic equations of state (EOS) derived from the relativistic mean field (RMF) model, each with different levels of stiffness and composition. On the dark matter side, we consider fermionic matter with repulsive interactions, described by a relativistic mean field Lagrangian, and explore a wide range of parameter values. The main goal of our study is to investigate the relationships between the parameters of the dark matter model and various neutron star properties, using a comprehensive set of EOSs.

Second, if we accurately know the sequence of NS properties, can we differentiate the admixed dark matter in the NS case from other scenarios? In the near future, advancements in measuring Neutron Star (NS) properties aim to achieve high precision through upcoming X-ray pulse profile observatories such as the enhanced X-ray Timing and Polarimetry mission (eXTP) Zhang *et al.* [2019] and the Spectroscopic Time-Resolving Observatory for Broadband Energy X-rays (STROBE-X) Ray *et al.* [2019], alongside third-generation gravitational-wave detectors like the Einstein Telescope Sathyaprakash *et al.* [2012] and Cosmic Explorer Abbott *et al.* [2020b], Evans *et al.* [2021], Hild *et al.* [2010]. Recent studies and simulations, including those in Bandopadhyay *et al.* [2024], Chatziioannou [2022], Finstad *et al.* [2023], Pacilio *et al.* [2022], Walker *et al.* [2024], indicate that these future instruments could determine the radius of a neutron star with 1.4 solar masses ($R_{1.4}$) with a precision exceeding 2.0%. instance, just the 75 loudest events among the anticipated more than $3 \times 10^5$ binary NS mergers within a year of operation by a network comprising one Cosmic Explorer and the Einstein Telescope could constrain NS radii in the mass range of 1 to 1.97 solar masses to an accuracy of $\Delta R \leq 0.2$ km at 90% credibility Walker *et al.* [2024]. Additionally, it has been noted that a single 40 km Cosmic Explorer detector could determine the NS radius with an accuracy of 10 meters within a decade, whereas the current generation of ground-based detectors, such as the Advanced LIGO-Virgo network, would require on the order of $10^5$ years to achieve similar precision Bandopadhyay *et al.* [2024]. This thesis additionally seeks to categorize the series of observed NS characteristics based on their core composition, specifically whether NS contains dark matter or not. We use machine learning classification algorithms that are trained on extensive datasets of neutron star properties, encompassing scenarios with and without dark matter, to investigate the feasibility of distinguishing neutron star characteristics based on the presence of dark matter. This machine learning endeavor has the potential to revolutionize our understanding of neutron stars. We also plan to determine the number of observations needed to conclusively classify the existence of dark matter within neutron stars. This could provide a new way to view and understand these enigmatic celestial bodies. Our research combines traditional astrophysical methods with cutting-edge machine learning techniques to venture into uncharted territories of neutron star and dark matter research.

Last but not least, a key objective of our thesis is to conduct a statistical evaluation of admixed dark matter scenarios in the context of the latest observational constraints on neutron stars. Studies have shown that the presence of dark matter within neutron stars generally leads to a reduction in both mass and radius compared to scenarios without dark matter. To explore this further, we calculate the Bayes factor for three distinct scenarios: one without dark matter, another with admixed dark matter, and a third where the neutron star properties are affected in the opposite manner—specifically, where the EOS stiffens at high densities, leading to increased mass and radius. By calculating the Bayesian evidence for each scenario, we aim to identify which model aligns most closely with recent observational data. This approach allows us to rank the models and gain deeper insights into the interior structure of neutron stars and the role of dark matter and modified nucleonic interactions.

**Objectives**

- Our first aim is to explore correlations between dark matter and neutron star properties, while carefully considering the uncertainties in the baryonic matter's equation of state.
- This thesis lays the foundation for future studies by using current data and machine learning techniques to explore the impact of dark matter on neutron star properties, providing insights that can guide upcoming observational studies.
- it also aims to to statistically evaluate admixed dark matter scenarios in neutron stars, comparing different scenarios to determine which best aligns with recent observational data and sheds light on the star's interior structure.

## 1.11 Organization of the Thesis

The present thesis is organized in the following manner: After a brief introduction in Chapter 1, we proceed to examine Relativistic Mean Field (RMF) models, along with its utilization in the context of neutron stars in Chapter 2 2. In this chapter, we provide a thorough explanation of the formalism related to Dark Matter(DM) Equation of State (EOS), two-fluid formalism, f and p modes oscillation. These frameworks subsequently serve as the foundation for exploring various quantities in the chapters that follow.

In Chapter 3-3, We investigate the potential existence of dark matter within neutron stars, acknowledging that current observations have not definitively ruled out this possibility. Our research aims to explore how neutron star properties like mass, radius, and tidal deformability might constrain dark matter models. Using a two-fluid model, we calculate neutron star properties with several realistic nuclear matter equations of state (EOS) derived from the relativistic mean field (RMF) model, alongside a dark matter EOS based on fermionic matter with repulsive interactions. We carefully sample a range of parameters to explore correlations between dark matter model parameters and neutron star properties. Our findings reveal a promising correlation between dark matter parameters and stellar properties when nuclear matter EOS uncertainties are ignored. However, this correlation weakens when these uncertainties are considered, making it challenging to conclusively constrain specific dark matter models using global properties alone. We also observe that dark-matter admixed stars tend to have higher central baryonic densities, potentially allowing for non-nucleonic degrees of freedom or direct Urca processes in lower-mass stars. Additionally, stars with identical masses but different surface temperatures could indicate the presence of dark matter. Despite this complexity, our study demonstrates that the semi-universal C-Love relation remains intact even in the presence of dark matter. .

In Chapter 4-4, We employ Random Forest classifiers to examine NS properties and determine whether these stars show indications of dark matter admixture. The effectiveness of these classifiers is meticulously evaluated using confusion matrices, which indicate that NS with dark matter admixture can be identified with a roughly 17% chance of being misclassified as nuclear matter NS. This work is the first to reveal that the radii of neutron stars at 1.4 and 2.07 solar masses, measured by NICER data from pulsars PSR J0030+0451 and PSR J0740+6620, strongly imply that the presence of dark matter in a neutron star is more probable than a purely hadronic composition.

In Chapter 5-5, We utilize a Bayesian Inference framework to evaluate various neutron star EOS models, including analyses of dark matter influences and high-density phenomena. Significantly, it is the first to compute Bayesian evidence for models incorporating both dark matter and sigma cut potential effects, using the latest NICER data from PSR J0437-4715. Our findings reveal notable tension with older datasets, highlighting the necessity for more flexible EOS models.

Our results demonstrate that EOS models informed by recent PREX-II data are less favored, indicating conflicts with other leading constraints, such as the pure neutron matter EOS at low density derived from chiral effective field theory. In contrast, the sigma cut model suggests a stiffer EOS at high densities, which aligns more closely with the current data.

In Chapter 6-6, Finally, a summary and future perspective are given in the last chapter (Chapter-6).

# Chapter 2

# Formalism

In this chapter, we delineate the formalism employed in the examination of neutron stars and dark matter.

The primary challenge in studying neutron stars lies in their exceedingly high densities. The model must not only accommodate the characteristics of matter under such extreme densities but also accurately represent properties observed under normal density conditions. Physicists commonly use three basic types of models to generate these equations of state: (i) Microscopic Models, (ii) Phenomenological Models, and (iii) Meta-models, which offer additional avenues for approaching this problem and understanding nuclear matter. Phenomenological models are further divided into relativistic and non-relativistic models. Therefore, a formalism capable of portraying relativistic, quantum mechanical many-body systems is necessary, and Relativistic Mean-Field models (RMF) fulfill this role. Neutron star matter is predominantly neutron-rich, yet these systems are not self-bound, meaning there are no direct experimental data available to determine appropriate values for RMF couplings in this regime. Consequently, RMF are typically calibrated by fitting the couplings to data extrapolated from experimental results for atomic nuclei, utilizing the Bethe-Weizsacker liquid-drop model. This calibration is based on properties of symmetric nuclear matter (with equal numbers of protons and neutrons) around nuclear saturation density ($\rho_0$), or the ground state properties of nearly symmetric finite nuclei, such as charge radii and binding energies of closed-shell nuclei. Subsequently, the RMF formalism is extrapolated to conditions prevalent in neutron stars, where matter is far from symmetric. To describe neutron stars accurately, RMF needs to be extrapolated from nuclear densities to neutron star densities of up to several times $\rho_0$.

## 2.1 Relativistic Mean Field Approach (RMF)

In 1974, J.D. Walecka Walecka [1974] introduced a model for nuclei and nuclear matter, which relied on the exchange of $\sigma$ and $\omega$ mesons. This model, known as quantum hadrodynamics (QHD), has seen refinement over the past few decades, particularly through the incorporation of non-linearities associated with the $\sigma$ meson. Various models have emerged, each distinguished by different parameter sets. These parameters may encompass additional meson fields, varied couplings among these fields, or recalibrated coupling constants. The $\sigma$ meson introduces a potent attractive central force and a spin-orbit force in nucleon-nucleon interactions. Conversely, the $\omega$ meson introduces a robust repulsive central force along with a spin-orbit force, which aligns with the direction of the scalar meson's spin-orbit force. An additional enhancement comes from including the charged (isovector) vector rho meson triplet ($\rho^0, \rho^\pm$) Boguta & Bodmer [1977], Boguta & Stoecker [1983], Serot & Walecka [1997], Walecka [1974]. Given the close similarity between protons and neutrons in terms of their isospin projections, incorporating rho mesons allows for a more nuanced distinction between these baryons. Furthermore, their inclusion contributes to a more accurate representation of symmetry energy within the system. In this theoretical framework, nuclear interactions are mathematically described using $\mathcal{L}$-denoted Lagrangian density functions. The Euler-Lagrange equation regulating a field $\varphi$ is expressed as follows within the covariant formalism.

$$\partial_\mu \frac{\partial \mathcal{L}}{\partial(\partial_\mu\varphi)} = \frac{\partial \mathcal{L}}{\partial\varphi} \tag{2.1}$$

and the energy-momentum tensor $T^{\mu\nu}$ is,

$$T^{\mu\nu} = \frac{\partial \mathcal{L}}{\partial(\partial_\mu\varphi)}\partial^\nu\varphi - g^{\mu\nu}\mathcal{L} \tag{2.2}$$

Here $g^{\mu\nu}$ are the components of metric tensor and is given by $g^{\mu\nu} = \mathrm{Diag}[1, -1, -1, -1]$. The energy $E$ and pressure $P$ of the system in the static case is then,

$$\begin{aligned} E &= < T_{00} > \\ P &= \tfrac{1}{3} < T_{ii} > \end{aligned} \tag{2.3}$$

In this work, we consider a more general, nonlinear finite-range Relativistic Mean-Field model, which is represented by the following Lagrangian density Dutra *et al.* [2014], The Lagrangian density is given by [Dutra *et al.* 2014, Fattoyev *et al.* 2010, Malik *et al.* 2023]

$$\mathcal{L} = \mathcal{L}_N + \mathcal{L}_M + \mathcal{L}_{NL} + \mathcal{L}_{leptons} \tag{2.4}$$

where

$$\mathcal{L}_N = \bar{\Psi}\left[\gamma^\mu\left(i\partial_\mu - g_\omega\omega_\mu - \frac{1}{2}g_\varrho \boldsymbol{t}\cdot\boldsymbol{\varrho}_\mu\right) - (m_N - g_\sigma\sigma)\right]\Psi$$

denotes the Dirac equation for the nucleon doublet (neutron and proton) with bare mass $m_N$ , $\Psi$ is a Dirac spinor, $\gamma^\mu$ are the Dirac matrices, and $\boldsymbol{t}$ is the isospin operator. $\mathcal{L}_M$ is the Lagrangian density for the mesons, given by

$$
\begin{aligned}
\mathcal{L}_M = & \frac{1}{2}\left(\partial_\mu\sigma\partial^\mu\sigma - m_\sigma^2\sigma^2\right) - \frac{1}{4}F^{(\omega)}_{\mu\nu}F^{(\omega)\mu\nu} + \frac{1}{2}m_\omega^2\omega_\mu\omega^\mu \\
& - \frac{1}{4}\boldsymbol{F}^{(\varrho)}_{\mu\nu}\cdot\boldsymbol{F}^{(\varrho)\mu\nu} + \frac{1}{2}m_\varrho^2\boldsymbol{\varrho}_\mu\cdot\boldsymbol{\varrho}^\mu
\end{aligned}
$$

where $F^{(\omega,\varrho)\mu\nu} = \partial^\mu A^{(\omega,\varrho)\nu} - \partial^\nu A^{(\omega,\varrho)\mu}$ are the vector meson tensors, and

$$
\begin{aligned}
\mathcal{L}_{NL} = & -\frac{1}{3}b\, m_N\, (g_\sigma\sigma)^3 - \frac{1}{4}c(g_\sigma\sigma)^4 + \frac{\xi}{4!}g_\omega^4(\omega_\mu\omega^\mu)^2 \\
& + \Lambda_\omega g_\varrho^2\boldsymbol{\varrho}_\mu\cdot\boldsymbol{\varrho}^\mu g_\omega^2\omega_\mu\omega^\mu
\end{aligned}
$$

contains the non-linear mesonic terms with parameters $b, c, \xi, \Lambda_\omega$ to take care of the high-density behavior of nuclear matter. The parameters $g_i$'s are the couplings of the nucleons to the meson fields $i = \sigma, \omega, \varrho$, with masses $m_i$. Finally, the Lagrangian density for the leptons is given as

$\mathcal{L}_{leptons} = \bar{\Psi}_l\, \gamma^\mu\left(i\partial_\mu - m_l\right)\Psi_l$ , where $\Psi_l$ ($l = e^-, \mu^-$) denotes the lepton spinor for electrons and muons; leptons are considered non-interacting.

The equations of motion for the meson fields are obtained from the Euler-Lagrange equations:

$$\sigma = \frac{g_\sigma}{m^2_{\sigma,\mathrm{eff}}}\sum_i \rho^s_i \tag{2.5}$$

$$\omega = \frac{g_\omega}{m^2_{\omega,\mathrm{eff}}}\sum_i \rho_i \tag{2.6}$$

$$\varrho = \frac{g_\varrho}{m^2_{\varrho,\mathrm{eff}}}\sum_i t_3\rho_i, \tag{2.7}$$

where $\rho^s_i$ and $\rho_i$ are, respectively, the scalar density and the number density of nucleon $i$, and the scalar and vector densities are given as,

$$\rho_s = \bar{\psi}\psi = \rho_{s_p} + \rho_{s_n}, \quad \rho_{s3} = \bar{\psi}\tau_3\psi = \rho_{s_p} - \rho_{s_n}, \tag{2.8}$$

$$\rho = \bar{\psi}\gamma^0\psi = \rho_p + \rho_n, \quad \rho_3 = \bar{\psi}\gamma^0\tau_3\psi = \rho_p - \rho_n = (2y-1)\rho, \tag{2.9}$$

where as the effective meson masses $m_{i,\mathrm{eff}}$ are defined as

$$m^2_{\sigma,\mathrm{eff}} = m^2_\sigma + b\, m_N\, g^3_\sigma\sigma + c g^4_\sigma\sigma^2 \tag{2.10}$$

$$m^2_{\omega,\mathrm{eff}} = m^2_\omega + \frac{\xi}{3!}g^4_\omega\omega^2 + 2\Lambda_\omega g^2_\varrho g^2_\omega\varrho^2 \tag{2.11}$$

$$m^2_{\varrho,\mathrm{eff}} = m^2_\varrho + 2\Lambda_\omega g^2_\omega g^2_\varrho\omega^2,. \tag{2.12}$$

The meson equations are solved using the relativistic mean-field approximation for a given $\rho_n$ and $\rho_p$. Note, that the effect of the nonlinear terms on the magnitude of the meson fields enters through the effective meson masses, $m_{i,\mathrm{eff}}$. The energy density of the baryons and leptons are given by the following expressions:

$$
\begin{aligned}
\epsilon = & \sum_{i=n,p,e,\mu} \frac{1}{\pi^2} \int_0^{k_{Fi}} \sqrt{k^2 + m_i^{*2}}\, k^2\, dk \\
& + \frac{1}{2} m_\sigma^2 \sigma^2 + \frac{1}{2} m_\omega^2 \omega^2 + \frac{1}{2} m_\varrho^2 \varrho^2 \\
& + \frac{b}{3} (g_\sigma \sigma)^3 + \frac{c}{4} (g_\sigma \sigma)^4 + \frac{\xi}{8} (g_\omega \omega)^4 + \Lambda_\omega (g_\varrho g_\omega \varrho \omega)^2,
\end{aligned}
\tag{2.13}
$$

where $m_i^* = m_i - g_\sigma \sigma$ for protons and neutrons and $m_i^* = m_i$ for electrons and muons, and $k_{Fi}$ is the Fermi moment of particle $i$.

Once we have the energy density for a given EoS model, we can compute the chemical potential of neutron ($\mu_n$) and proton ($\mu_p$). The chemical potential of electron ($\mu_e$) and muon ($\mu_\mu$) can be computed using the condition of $\beta$-equilibrium : $\mu_n - \mu_p = \mu_e$ and $\mu_e = \mu_\mu$ and the charge neutrality: $\rho_p = \rho_e + \rho_\mu$. Where $\rho_e$ and $\rho_\mu$ are the electron and muon number density. The pressure is then determined from the thermodynamic relation:

$$
p = \sum_i \mu_i \rho_i - \epsilon. \tag{2.14}
$$

therefore p can be expressed as :

$$
\begin{aligned}
p = & \sum_{i=n,p,e,\mu} \frac{1}{\pi^2} \int_0^{k_{Fi}} \sqrt{k^2 + m_i^{*2}}\, k^2\, dk \\
& - \frac{1}{2} m_\sigma^2 \sigma^2 - \frac{1}{2} m_\omega^2 \omega^2 - \frac{1}{2} m_\varrho^2 \varrho^2 \\
& - \frac{b}{3} (g_\sigma \sigma)^3 - \frac{c}{4} (g_\sigma \sigma)^4 + \frac{\xi}{8} (g_\omega \omega)^4 + \Lambda_\omega (g_\varrho g_\omega \varrho \omega)^2,
\end{aligned}
\tag{2.15}
$$

## 2.2 Non Linear-$\sigma$ Cut Model

A $\sigma$-cut scheme Maslov *et al.* [2015] was developed to stiffen the equation of state (EoS) at high densities without altering the properties of nuclear matter near saturation density $\rho_0$. This approach modifies the self-interaction term of the $\sigma$ meson at high densities, effectively mitigating the pronounced decrease in the effective mass of nucleons in this region. Consequently, the EoS becomes stiffer at high densities. The $U_{\mathrm{cut}}(\sigma)$ has a logarithmic form, as in Maslov *et al.* [2015], which is given by:

$$
U_{\mathrm{cut}}(\sigma) = \alpha \ln[1 + \exp\{\beta(g_\sigma \sigma / m_N - f_s)\}] \tag{2.16}
$$

where $\alpha = m_\pi^4$ and $\beta = 120$ Maslov *et al.* [2015], and the parameter $f_s$ is determined by Bayesian inference.

The Lagrangian density equation (1) for $\mathcal{L}_{NL,\sigma cut}$ with $U_{cut}(\sigma)$ is given by,

$$\mathcal{L}_{NL,\sigma cut} = \mathcal{L}_{NL} - U_{cut}(\sigma) \tag{2.17}$$

and the meson fields are determined from the equations

$$\sigma = \frac{g_\sigma}{m^2_{\sigma,\text{eff}}} \sum_i \rho_i^s \tag{2.18}$$

$$\omega = \frac{g_\omega}{m^2_{\omega,\text{eff}}} \sum_i \rho_i \tag{2.19}$$

$$\varrho = \frac{g_\varrho}{m^2_{\varrho,\text{eff}}} \sum_i I_3 \rho_i, \tag{2.20}$$

where $\rho_i^s$ and $\rho_i$ are, respectively, the scalar density and the number density of nucleon $i$, and

$$m^2_{\sigma,\text{eff}} = m^2_\sigma + b\, m_N g^3_\sigma \sigma + c g^4_\sigma \sigma^2 + \frac{U'_{cut}(\sigma)}{\sigma} \tag{2.21}$$

$$m^2_{\omega,\text{eff}} = m^2_\omega + \frac{\xi}{3!} g^4_\omega \omega^2 + 2\Lambda_\omega g^2_\varrho g^2_\omega \varrho^2 \tag{2.22}$$

$$m^2_{\varrho,\text{eff}} = m^2_\varrho + 2\Lambda_\omega g^2_\omega g^2_\varrho \omega^2 \tag{2.23}$$

where $U'_{cut}(\sigma)$ is the derivative of $U_{cut}(\sigma)$ with respect to $\sigma$. In these equations, the meson fields should be interpreted as their expectation values. The energy density and pressure for this case are Maslov *et al.* [2015]

$$\epsilon_{\sigma cut} = \epsilon + U_{cut}(\sigma) \tag{2.24}$$

$$P_{\sigma cut} = P - U_{cut}(\sigma). \tag{2.25}$$

## 2.3 Neutron Star Asteroseismology

Neutron stars can undergo various types of oscillations, which stem from different restoring forces acting on displaced mass elements. These forces include gravity, pressure gradients, the elasticity of the crust, magnetic fields, and the effects of centrifugal and Coriolis forces in rotating neutron stars, such as pulsars. When an NS is perturbed, it oscillates in radial or nonradial modes. Nonradial oscillations refer to oscillations in which a star changes shape, deviating from its spherical form. In contrast, radial oscillations involve the star expanding and contracting around its equilibrium shape while maintaining its spherical symmetry. Radial oscillations can be considered a specific form of the broader category of nonradial oscillations. The key oscillation modes in NS include the fundamental mode (f), the first and second pressure modes (p), and the first gravitational mode (g). These modes are significant because they are responsible for the majority of the gravitational waves (GWs) emitted by NS. To analyze these different oscillation modes, it's necessary to solve the perturbed fluid equations within the context of general relativity. Thorne and Campolattaro were the first to calculate the non-radial oscillations of NS using the framework of general relativity. To solve non-radial oscillations, one can also apply the Cowling approximation, which simplifies the problem by neglecting space perturbations. When using the Cowling approximation, the calculated frequencies can differ by 10-30% compared to those obtained from the full linearized equations of general relativity (GR). In this study, we use the Cowling approximation to calculate the non-radial oscillations of NS. In the Cowling approximation, the spacetime metric for a spherically symmetric background is given by

$$ds^2 = -e^{2\Phi(r)}dt^2 + e^{2\Lambda(r)}dr^2 + r^2 d\theta^2 + r^2 \sin^2\theta d\phi^2. \tag{2.26}$$

In order to find mode frequencies, one has to solve the following differential equations (Pradhan & Chatterjee [2021]):

$$\begin{aligned} \frac{dW(r)}{dr} &= \frac{d\epsilon}{dp}\omega^2 r^2 e^{\Lambda(r)-2\Phi(r)} V(r) + \frac{d\Phi(r)}{dr} W(r) \\ &\quad - l(l+1)e^{\Lambda(r)} V(r) \\ \frac{dV(r)}{dr} &= 2\frac{d\Phi(r)}{dr} V(r) - \frac{1}{r^2} e^{\Lambda(r)} W(r) \end{aligned} \tag{2.27}$$

where,

$$\frac{d\Phi(r)}{dr} = \frac{-1}{\epsilon(r) + p(r)} \frac{dp}{dr}.$$

The solution of Eq. (2.27) with the fixed background metric Eq. (2.26) near the origin will behave as follows:

$$W(r) = Ar^{l+1}, \ V(r) = -\frac{A}{l} r^l. \tag{2.28}$$

The vanishing perturbed Lagrangian pressure at the surface will provide another constraint to be included while solving Eq. 2.27, which is given by,

$$\omega^2 e^{\Lambda(R)-2\Phi(R)} V(R) + \frac{1}{R^2} \left.\frac{d\Phi(r)}{dr}\right|_{r=R} W(R) = 0. \tag{2.29}$$

Eqs. (2.27) are eigenvalue equations. Among the solutions, those that satisfy the boundary condition given by Eq. (2.29) are the eigenfrequencies of the star.

## 2.4 Dark matter admixed neutron stars

As discussed in Section 1.9, the in-medium properties of dark matter, such as its mass and self-interaction, are still not precisely known. Therefore, there are two main approaches to study the effects of dark matter on neutron stars. The first approach is the single-fluid formalism, which takes into account non-gravitational interactions between normal matter and dark matter. The second approach assumes there are no non-gravitational interactions between normal matter and dark matter, creating a two-fluid system where these two types of matter interact only through gravitational forces. In the following section, we will delve into these two formalisms.

### 2.4.1 Single fluid formalism

For single fluid formalism, in this work, we adopt a DM model based on a recent work by Motta *et al.* [2018], Shirke *et al.* [2023] which is motivated by the fact that the neutron decay anomaly can be explained via the decay of the neutron to the dark sector. The current interpretation of experimental data on neutron decay suggests the potential presence of phenomena beyond the standard model of physics Baym *et al.* [2018], Czarnecki *et al.* [2018], Fornal & Grinstein [2018]. Neutrons predominantly undergo $\beta$-decay:

$$n \to p + e^- + \bar{\nu}_e\,.$$

The two experiments that measure neutron lifetime, namely the beam experiment and the bottle experiment, yield two different neutron lifetimes. The bottle experiment yields $\tau_{\mathrm{bottle}} = 879.6 \pm 0.6$ s Arzumanov *et al.* [2015], Mampe *et al.* [1993], Pattie *et al.* [2018], Pichlmaier *et al.* [2010], Serebrov *et al.* [2005], Steyerl *et al.* [2012], and the beam experiment measurement gives $\tau_{\mathrm{beam}} = 888.0 \pm 2.0$ s Byrne & Dawber [1996], Yue *et al.* [2013]. These two neutron lifetime measurements differ by $4\sigma$, indicating a need to reconcile our understanding of fundamental interactions. In Ref. Fornal & Grinstein [2018], authors have come up with an intriguing suggestion where they propose that new decay channels of neutrons into dark matter particles could account for the anomaly in the neutron lifetime measurement. These new decay channels, where neutrons decay into dark matter particles, could be potentially interesting for neutron star physics. Several recent studies have investigated this possibility, suggesting that neutron stars can serve as powerful laboratories to test the proposed decay of neutrons into dark matter particlesBastero-Gil *et al.* [2024], Baym *et al.* [2018], Husain *et al.* [2022], Motta *et al.* [2018], Shirke *et al.* [2023, 2024]. In this work, we examine the effect of neutron decay on neutron star dynamics using the decay channel involving baryon-number- violating beyond the standard model (BSM) interaction,

$$n \to \chi + \phi\,, \tag{2.30}$$

where $\chi$ is a dark spin-1/2 fermion, and $\phi$ is a light dark boson. Other decay channels of neutrons are also possible, e.g., $n \to \chi + \gamma$, and $n \to \chi + e^+e^-$. However, phenomenologically all decay channels are not favored; e.g., laboratory experiment puts stringent constraints on the decay channel $n \to \chi + \gamma$ Tang *et al.* [2018]. The decay channel $n \to \chi + \phi$ is especially intriguing in the context of neutron star physics, as it can be argued that the light dark matter boson $\phi$ would quickly escape the neutron star, rendering it insignificant. Conversely, some of the neutrons within the neutron star will transform into fermionic dark matter $\chi$ due to the BSM interaction. Physically these dark matter particles will experience the gravitational potential of the neutron star and will reach thermal equilibrium with the surrounding neutron star matter. This sets the equilibrium condition

$$\mu_\chi = \mu_n\,. \tag{2.31}$$

Nuclear stability requires $937.993\,\text{MeV} < m_\chi + m_\phi < m_n = 939.565\,\text{MeV}$ Motta *et al.* [2018], Shirke *et al.* [2023]. For the dark particles to remain stable and avoid further beta decay, the condition $|m_\chi - m_\phi| < m_p + m_e = 938.783\,\text{MeV}$ must be met Fornal & Grinstein [2020].

To account for dark matter (DM) self-interactions, we introduce vector interactions between dark particles, described by:

$$\mathcal{L} \supset -g_V \bar{\chi}\gamma^\mu \chi V_\mu - \frac{1}{4} V_{\mu\nu}V^{\mu\nu} + \frac{1}{2} m_V^2 V_\mu V^\mu\,, \tag{2.32}$$

where $g_V$ is the coupling strength and $m_V$ is the mass of the vector boson. This introduces an additional interaction term in the energy density, beyond the free fermion part. The energy density of DM is given by:

$$\epsilon_{DM} = \frac{1}{\pi^2}\int_0^{k_{F_\chi}} k^2 \sqrt{k^2 + m_\chi^2}\, dk + \frac{1}{2} G_\chi n_\chi^2, \tag{2.33}$$

where,

$$G_\chi = \left(\frac{g_V}{m_V}\right)^2, \qquad n_\chi = \frac{k_{F_\chi}^3}{3\pi^2} \tag{2.34}$$

and

$$\mu_\chi = \sqrt{k^2 + m_\chi^2} + G_\chi n_\chi.$$

#### 2.4.1.1 Structure and Dynamics of Neutron Star

The determination of neutron star properties relies on the resolution of the Tolman-Oppenheimer-Volkoff (TOV) equations as outlined by Oppenheimer & Volkoff [1939] and Tolman [1939]. Within the framework of general relativity, the TOV equation characterizes the hydrostatic stability of a non-rotating, spherically symmetric stellar entity consisting of a perfect fluid. The metric for the static case can generally be written as:

$$ds^2 = c^2dt^2e^{2\Phi} - e^{2\lambda}dr^2 - r^2(d\theta^2 + \sin^2\theta d\phi^2) \tag{2.35}$$

The function $\phi(r)$ serves as the gravitational potential, particularly evident in the Newtonian limit. Meanwhile, the variable $\lambda(r)$ is closely connected to the enclosed mass m(r), through a specific relationship

$$e^{-\lambda} = \sqrt{1 - \frac{2Gm}{rc^2}}. \tag{2.36}$$

#### 2.4.1.2 Masses and Radii

The Tolman-Oppenheimer-Volkoff (TOV) equations can be summarized as follows

$$\frac{dP}{dr} = -\frac{G\epsilon(r)m(r)}{c^2r^2}\left[1 + \frac{P(r)}{\epsilon(r)}\right]\left[1 + \frac{4\pi r^3P(r)}{m(r)c^2}\right]\left[1 - \frac{2Gm(r)}{c^2r}\right]^{-1}, \tag{2.37}$$

$$\frac{dm}{dr} = 4\pi r^2\epsilon(r), \tag{2.38}$$

$$\frac{d\phi}{dr} = -\frac{1}{\rho c^2}\frac{dP}{dr}\left[1 + \frac{P}{\rho c^2}\right]^{-1}, \tag{2.39}$$

In this context, $G$ and $c$ denote the gravitational constant and the speed of light, respectively. The functions $P(r)$, $\epsilon(r)$, and $m(r)$ represent the pressure, energy density, and mass of the neutron star as a function of the radial distance $r$ from its center. The boundary conditions are given as $P(0) = P_c$ and $m(0) = 0$, where $P_c$ is the central pressure, and $m(0)$ is the mass at the neutron star's center. Equation 2.38 is integrated from $r = 0$ to the star's surface at $r = R$, with $R$ being the radius of the neutron star. This integration yields the total gravitational mass $M$ of the neutron star. Equation 2.39 illustrates the behavior of the metric function $\phi(r)$ within a relativistic framework. The core physics of how matter influences the equations governing stellar structure is encapsulated in the equation of state, which describes the relationship between pressure $P$ and energy density $\epsilon$. These quantities are determined using the models outlined in the preceding sections.

In this case, our primary focus is on solving equations (2.37) and (2.38), which provide profiles for $P(r)$, $\rho(r)$, and $m(r)$ for a given equation of state (EOS). We start by selecting a random value for the central mass density, $\rho_c$, in order to solve the hydrostatic equilibrium equations for a specific EOS. Next, we determine the central pressure, $P_c = P(\rho_{B,c})$, corresponding to $r = 0$, while imposing the boundary condition $m(r = 0) = 0$. Using the Runge-Kutta method, we integrate up to $r = R$, where $R$ is the radius at which $P(r = R) = 0$. The pressure must remain below $5 \times 10^{-4}$ MeV/fm$^3$, a value chosen to ensure consistent total mass results.

At each step where the pressure changes, we use a precomputed table to interpolate the mass density, $\rho$. Finally, we obtain the profiles $P(r)$, $\rho(r)$, and $m(r)$ for a given central mass density, $\rho_c$, allowing us to determine the neutron star's mass $M$ and its radius $R$. According to Haensel (2007) Haensel *et al.* [2007], the canonical neutron star mass is approximately $M = 1.4M_{\odot}$, corresponding to a radius of $R = 10$ to 14 km. It's important to note that the exact central density of a given neutron star remains uncertain, potentially ranging from approximately $4.6 \times 10^{14}$ to $4 \times 10^{15}$ g/cm$^3$. Therefore, determining the relationship between mass and radius is of particular interest. By applying the numerical method described earlier, we can determine this relationship by calculating $M$ and $R$ for this range of central mass densities.

#### 2.4.1.3 Tidal deformability

The tidal deformability characterizes the degree to which a celestial body deforms in response to tidal forces, which arise when two massive bodies orbit each other. In the context of orbiting neutron stars, the specific configuration of the gravitational wave signal is shaped by both their masses and their tidal deformabilities. This parameter, known as the neutron star tidal deformability, is crucial in determining the gravitational-wave signal before the neutron stars merge. The detection of gravitational waves (GW) resulting from the merger of two neutron stars, illustrated by the GW170817 event, stands out for the valuable knowledge it has imparted about the tidal deformability of neutron stars (NS). The Tidal deformability parameter $\lambda$ is defined as

$$Q_{ij} = -\lambda E_{ij}, \tag{2.40}$$

where $Q_{ij}$ is the induced quadrupole moment of a star in a binary due to the static external tidal field $E_{ij}$ of the companion star. The parameter $\lambda$ can be expressed in terms of the dimensionless quadrupole tidal Love number $k_2$ as

$$\lambda = \frac{2}{3} k_2 R^5, \tag{2.41}$$

where $R$ is the radius of the NS. The value of $k_2$ is typically in the range $\sim 0.05 - 0.15$ Hinderer [2008], Hinderer *et al.* [2010], Postnikov *et al.* [2010] for NSs and depends on the stellar structure. This quantity can be calculated using the following expression Hinderer [2008]:

$$k_2 = \frac{8C^5}{5}(1-2C)^2\left[2+2C(y_R-1)-y_R\right] \times \Big\{2C\left(6-3y_R+3C(5y_R-8)\right) + 4C^3\left[13-11y_R+C(3y_R-2)+2C^2(1+y_R)\right] + 3(1-2C)^2\left[2-y_R+2C(y_R-1)\right]\log(1-2C)\Big\}^{-1}, \tag{2.42}$$

where $C$ ($\equiv M/R$) is the compactness parameter of the star of mass $M$ with radius $R$. The quantity $y_R$ ($\equiv y(R)$) for a two fluid system can be obtained by solving the following differential equation.

$$r\frac{dy(r)}{dr} + y(r)^2 + y(r)F(r) + r^2Q(r) = 0, \tag{2.43}$$

with

$$F(r) = \frac{r - 4\pi r^3\left(\epsilon(r) - P(r)\right)}{r - 2m(r)}, \tag{2.44}$$

$$Q(r) = \frac{4\pi r\left[5\epsilon(r) + 9p(r) + \frac{\epsilon(r)+p(r)}{\partial p(r)/\partial\epsilon(r)} - \frac{6}{4\pi r^2}\right]}{r - 2m(r)} - 4\left[\frac{m(r) + 4\pi r^3 p(r)}{r^2\left(1-2m(r)/r\right)}\right]^2. \tag{2.45}$$

In the preceding expressions, $m(r)$ denotes the mass enclosed within the radius $r$, while $\epsilon(r)$ and $P(r)$ represent the energy density and pressure, respectively, as functions of the radial coordinate $r$ within a star. These parameters are computed using the selected nuclear matter model to characterize the stellar Equation of State (EoS). For a given EoS, Eq. (2.43) can be integrated alongside the Tolman-Oppenheimer-Volkoff equations Oppenheimer & Volkoff [1939], employing boundary conditions $y(0) = 2$, $P(0) = P_c$, and $m(0) = 0$, where $y(0)$, $P_c$, and $m(0)$ denote the dimensionless quantity, pressure, and mass at the center of the NS, respectively. The tidal deformabilities of the NSs present in the binary neutron star system can be combined to yield the weighted average as,

$$\tilde{\Lambda} = \frac{16}{13}\frac{(12q+1)\Lambda_1 + (12+q)q^4\Lambda_2}{(1+q)^5}, \tag{2.46}$$

where $\Lambda_1$ and $\Lambda_2$ are the individual tidal deformabilities corresponding to the two components in the NS binary with masses $m_1$ and $m_2$, respectively Favata [2014], Flanagan & Hinderer [2008a] with $q = m_2/m_1 < 1$.

During the binary's gradual inspiral, the signature of neutron star matter in gravitational waves, arises from tidal interactions, where the gravity gradient across the neutron star causes it to deform from sphericity Hinderer *et al.* [2016]. Since tidal deformations are sensitive to the mass, radius and internal structure of NSs, measuring the tidal deformability can be used to constrain the EoS. During a binary NS's inspiral the signature of tidal deformation in the GW signal mainly appears through a change in the phase of the waveform Flanagan & Hinderer [2008b], Takátsy *et al.* [2024]. The dominant effect results from the neutron star's adiabatic tide (static tidal deformability) where the distorted neutron star remains in hydrostatic equilibrium and tracks the companian's tidal force which varies periodically due to orbital motion Hinderer *et al.* [2016]. These adiabatic tides can be characterized by a single constant for each multipole moment, the tidal deformability Damour & Nagar [2009], Flanagan & Hinderer [2008b]. Adiabatic tides provide a good approximation in the regime where the eigenfrequencies of the tidal modes are much higher than the frequency of the acting force. However, finite frequency effects, also known as dynamical tides, can become relevant when approaching the late inspiral phase ( near merger) of the binary system Andersson & Pnigouras [2021], Hinderer *et al.* [2016], Schmidt & Hinderer [2019]. Dynamical tides arise when tidal forcing frequencies (orbital frequency) approaches a resonance with one of the internal modes of neutron star, resulting in an enhanced, more complex tidal response than adiabatic tides. The effects of dynamical tides caused by the most dominant, fundamental-mode (f-mode) describing the neutron star's quadrupole (l = 2) have been extensively studied Kokkotas & Schaefer [1995], Lai [1994], Shibata [1994]. During the early inspiral phase, the orbital separation of the system distance (d) is much larger than the size of single objects radius (R). At this stage the compact objects are modeled as pointlike massive particles and orbital dynamics is described using post-Newtonian theory (PN). The PN approximation consists in an expansion of the Einstein equations around the Newtonian limit. It is commonly expressed in terms of the dimensionless parameter $x = M^{2/3}$, where $M$ is the total mass of the system. It can also be written as $x = v^2$, where $v$ is the orbital velocity. The static tide enters the post-Newtonian expansion at order $v^5 = x^{5/2}$, representing a fifth-order contribution Andersson & Pnigouras [2021]. According to authors Hinderer *et al.* [2016], Schmidt & Hinderer [2019] noted that the effective Love number is not (overall) proportional to the static one. However, the difference appears at a higher post-Newtonian order. Specifically, the difference enters as we, formally, add an order $x^{11/2} = v^{11}$ term to the expansion. The fact that this is a very high-order contribution, which one would normally safely neglect. Neverthless, in late inspiral phase (near merger) when two neutron stars come closer, the approximation $d >> R$ breaks down, and Numerical Relativity simulations, which integrate the fully non-linear Einstein equations, are needed to describe the merger phase Pretorius [2005]. The observation of GW170817 allowed for the first measurement of the tidal deformability. Abbott *et al.* [2017b] constrain the effective tidal deformability for GW170817 to be $\tilde{\Lambda} \lesssim 800$ at the 90% confidence level. Current gravitational wave detectors are not sufficiently sensitive at frequencies above $\gtrsim$ 800 Hz, where dynamical tides become more prominent, and hence the measurement of the f-mode frequency is difficult. However, third-generation (3G) detectors such as the Einstein Telescope (ET) Punturo *et al.* [2010] and Cosmic Explorer Reitze *et al.* [2019] will have significantly improved sensitivity in the high-frequency regime. As a result, the complete binary neutron star (BNS) signal through merger will be detectable for many of the anticipated $10^4$ detections per year Baibhav *et al.* [2019], allowing us to also measure such higher-order tidal effects. Therefore next generation of ground-based GW detectors will have unprecedented sensitivities between $\sim$ 10 Hz and a few kHz. Therefore, there is no doubt that tidal effects are dominant during the merger of binary neutron stars, when the orbital separation is comparable to the size of the compact objects. However, tidal deformations play a fundamental role also in the inspiral phase, even if they are subdominant with respect to the point-particle contributions Abdelsalhin [2019]. In this thesis, we adopt the static tidal deformability approximation. The late inspiral phase, particularly near merger, is not considered in this work as it is beyond its scope. Therefore, the use of the static approximation in our analysis is justified

### 2.4.2 Two Fluid Formalism

In following sections, we will explore the frameworks and dynamics of dark matter admixed neutron stars using the two-fluid approach.

#### 2.4.2.1 Dark matter Equation of State

In the realm of theoretical physics, an extension of the nuclear mean field approach to the fermionic dark matter sector involves the construction of a Lagrangian that encapsulates the dynamics of dark matter. In this specific scenario, we focus on a minimalist dark matter Lagrangian featuring a single fermionic component denoted as $\chi_D$. Accompanying this fermion is a dark vector meson, represented by $V^\mu_D$, which interacts with the conserved dark matter current through the coupling term $g_{vd}\bar{\chi}_D\gamma_\mu\chi_D V^\mu_D$.

The Lagrangian that characterizes this dark matter model within the mean field approximation is expressed as Das *et al.* [2022], Xiang *et al.* [2014]:

$$\mathcal{L}_\chi = \bar{\chi}_D \left[\gamma_\mu(i\partial^\mu - g_{vd}V^\mu_D) - m_\chi\right]\chi_D - \frac{1}{4}V_{\mu\nu,D}V^{\mu\nu}_D + \frac{1}{2}m^2_{vd}V_{\mu,D}V^\mu_D. \tag{2.47}$$

Here, $\chi_D$ is the fermionic dark matter field, $V_{\mu\nu,D}$ is the field strength tensor for the dark vector meson, and $m_\chi$ stands for the bare mass of the fermionic dark matter. The interaction between the dark matter and the dark vector meson is dictated by the coupling constant $g_{vd}$, while $m_{vd}$ represents the mass of the dark vector meson.

Moving beyond the Lagrangian, we delve into the energy density ($\varepsilon_\chi$) and pressure ($P_\chi$) associated with the fermionic dark matter in the mean field approximation:

$$\varepsilon_\chi = \frac{1}{\pi^2}\int_0^{k_D} dk\, k^2 \sqrt{k^2 + m_\chi^2} + \frac{1}{2} c_\omega^2 \rho_D^2 , \tag{2.48}$$

$$P_\chi = \frac{1}{3\pi^2}\int_0^{k_D} dk \frac{k^4}{\sqrt{k^2 + m_\chi^2}} + \frac{1}{2} c_\omega^2 \rho_D^2, \tag{2.49}$$

Here, $c_\omega \equiv \frac{g_{vd}}{m_{vd}}$ stands as the ratio of the coupling constant $g_{vd}$ to the mass $m_{vd}$ of the dark vector meson. The parameters $m_\chi$ and $c_\omega$, coupled with the dark matter Fermi momenta denoted by $k_D$, collaboratively shape the equation of state (EOS) for the dark matter. This EOS, in essence, becomes a pivotal factor influencing the characteristics of dark matter-infused neutron stars.

#### 2.4.2.2 Structure and Dynamics of Neutron Star

Following the method outlined in Das *et al.* [2022], we employ a two-fluid Tolman-Oppenheimer-Volkoff (TOV) formalism to study neutron stars that include dark matter, referred to as Dark Matter Admixed Neutron Stars (DANSs).

We begin by considering the background Einstein equation, which accounts for the presence of two distinct fluids, each with its own conserved energy-momentum tensor $\bar{T}^{(1)}_{\mu\nu}$ and $\bar{T}^{(2)}_{\mu\nu}$ respectively,

$$\begin{aligned}\bar{G}_{\mu\nu} &= 8\pi \bar{T}_{\mu\nu} = 8\pi(\bar{T}^{(1)}_{\mu\nu} + \bar{T}^{(2)}_{\mu\nu}),\\ &= \mathrm{diag}(e^{\alpha}\varepsilon, e^{\beta}P, r^2P, r^2 \sin^2\theta P),\end{aligned} \tag{2.50}$$

We have identified the total energy density ($\varepsilon$) and total pressure ($P$) of the two-fluid system. The TOV equations for this system, with individual energy densities and pressures, denoted as $\epsilon_1, P_1$ and $\epsilon_2, P_2$, are derived as follows.

$$\varepsilon(r) = \varepsilon_1(r) + \varepsilon_2(r), \tag{2.51}$$

$$P(r) = P_1(r) + P_2(r), \tag{2.52}$$

respectively. The background spherically symmetric static metric and its inverse in the spherical polar coordinate system $x^{\mu} \equiv (t, r, \theta, \phi)$, can be given as,

$$g^{(0)}_{\mu\nu} = \text{diag}\left(-e^{\alpha(r)}, e^{\beta(r)}, r^2, r^2 \sin^2\theta\right), \tag{2.53}$$

$$g^{(0)\mu\nu} = \text{diag}\left(-e^{-\alpha(r)}, e^{-\beta(r)}, 1/r^2, 1/r^2 \sin^2\theta\right). \tag{2.54}$$

The "$tt$" component of the background Einstein equation gives us,

$$\begin{aligned} \bar{G}_{tt} &= 8\pi\bar{T}_{tt} \\ \Longrightarrow r\beta' &= 1 - e^{\beta} + 8\pi r^2(\varepsilon_1 + \varepsilon_2)e^{\beta}. \end{aligned} \tag{2.55}$$

Similarly the "$rr$" component of the background Einstein equation gives us,

$$\begin{aligned} \bar{G}_{rr} &= 8\pi\bar{T}_{rr} \\ \Longrightarrow r\alpha' &= 8\pi r^2(P_1 + P_2)e^{\beta} + e^{\beta} - 1. \end{aligned} \tag{2.56}$$

Eq.(2.55) can be recasted as,

$$\begin{aligned} r\beta' &= 1 - e^{\beta} + 8\pi r^2(\varepsilon_1 + \varepsilon_2)e^{\beta} \\ \Longrightarrow -\frac{d}{dr}\left(r(e^{-\beta} - 1)\right) &= 8\pi(\varepsilon_1 + \varepsilon_2)r^2 \\ \Longrightarrow e^{-\beta} &= 1 - \frac{2m(r)}{r}, \end{aligned} \tag{2.57}$$

where the mass function $m(r)$ for the two fluid system is defined as,

$$m(r) \equiv 4\pi \int_0^r (\varepsilon_1(r') + \varepsilon_2(r'))r'^2 dr'. \tag{2.58}$$

In a similar way using Eq.(2.57), Eq.(2.56) can be written as,

$$\begin{aligned} r\alpha' &= 8\pi r^2(P_1 + P_2)e^{\beta} + e^{\beta} - 1 \\ \Longrightarrow \frac{d\alpha}{dr} &= \frac{8\pi r^3(P_1 + P_2) + 2m(r)}{r(r - 2m(r))}. \end{aligned} \tag{2.59}$$

Furthermore, the conservation of the energy-momentum tensors for these two fluids implies that:

$$\nabla_{\mu} T^{\mu\nu}_{(1)} = 0, \tag{2.60}$$

$$\nabla_{\mu} T^{\mu\nu}_{(2)} = 0. \tag{2.61}$$

Similar to the single fluid case, in a two fluid scenario conservation of energy-momentum tensor as given by Eqs.(2.60) and (2.61) boil down to,

$$\frac{dP_1}{dr} = -\frac{1}{2}(P_1 + \varepsilon_1)\frac{d\alpha}{dr}, \tag{2.62}$$

$$\frac{dP_2}{dr} = -\frac{1}{2}(P_2 + \varepsilon_2)\frac{d\alpha}{dr}. \tag{2.63}$$

Using Eq.(2.59) back into Eqs.(2.62) and Eq.(2.63) we obtain,

$$\frac{dP_1}{dr} = -(P_1 + \varepsilon_1)\frac{4\pi r^3(P_1 + P_2) + m(r)}{r(r - 2m(r))}, \tag{2.64}$$

$$\frac{dP_2}{dr} = -(P_2 + \varepsilon_2)\frac{4\pi r^3(P_1 + P_2) + m(r)}{r(r - 2m(r))}, \tag{2.65}$$

with,

$$\frac{dm(r)}{dr} = 4\pi(\varepsilon_1(r) + \varepsilon_2(r))r^2. \tag{2.66}$$

Equations (2.64), (2.65), and (2.66) together form the TOV equations for a two-fluid system. The standard TOV equation for a single-fluid system can be recovered by taking the appropriate limit of these two-fluid TOV equations. For an alternative derivation of the two-fluid TOV equations using Lagrange multiplier techniques, refer to Ref. Xiang *et al.* [2014].

#### 2.4.2.3 Tidal deformability:

Retaining only the terms up to first order in the metric perturbation, the perturbed metric and its inverse can be expressed as shown in Hinderer [2008],

$$g_{\mu\nu} = g^{(0)}_{\mu\nu} + h_{\mu\nu},$$
$$g^{\mu\nu} = g^{(0)\mu\nu} - h^{\mu\nu}, \tag{2.67}$$

respectively. Here $g^{(0)}_{\mu\nu}$ is the background spherically static metric and $h_{\mu\nu}$ is the associated metric perturbation. Note that $h_{\mu\nu}$ has correct tensorial properties in the background metric. Otherway stated space-time indices of $h_{\mu\nu}$ can be raised by $g^{(0)\mu\nu}$. Hence,

$$h^{\mu\nu} = g^{(0)\mu\lambda}g^{(0)\nu\sigma}h_{\lambda\sigma}. \tag{2.68}$$

The spherically symmetric static background metric and it's inverse can be expressed as,

$$g^{(0)}_{\mu\nu} = \mathrm{diag}\left(-e^{\alpha(r)}, e^{\beta(r)}, r^2, r^2\sin^2\theta\right), \tag{2.69}$$

$$g^{(0)\mu\nu} = \mathrm{diag}\left(-e^{-\alpha(r)}, e^{-\beta(r)}, 1/r^2, 1/r^2\sin^2\theta\right), \tag{2.70}$$

respectively. For $l = 2$, $m = 0$, static, even-parity metric perturbations ($h_{\mu\nu}$ and $h^{\mu\nu}$) in the Regge-Wheeler gauge can be expressed as (for a detailed discussion see Hinderer [2008]),

$$h_{\mu\nu} = \mathrm{diag}\left(-e^{\alpha(r)}H(r)Y_{20}(\theta,\phi), -e^{\beta(r)}H(r)Y_{20}(\theta,\phi), r^2K(r)Y_{20}(\theta,\phi), r^2\sin^2\theta K(r)Y_{20}(\theta,\phi)\right), \tag{2.71}$$

$$h^{\mu\nu} = \text{diag}\left[-e^{-\alpha(r)}H(r)Y_{20}(\theta,\phi),\ -e^{-\beta(r)}H(r)Y_{20}(\theta,\phi),\ \frac{1}{r^2}K(r)Y_{20}(\theta,\phi),\ \frac{1}{r^2\sin^2\theta}K(r)Y_{20}(\theta,\phi)\right]. \tag{2.72}$$

Using the perturbed metric and energy-momentum tensor, the first order Einstein equations can be written as,

$$\delta G^t{}_t = 8\pi\delta T^t{}_t = -8\pi\delta\varepsilon, \tag{2.73}$$

$$\delta G^r{}_r = 8\pi\delta T^r{}_r = 8\pi\delta P, \tag{2.74}$$

$$\delta G^\theta{}_\theta = 8\pi\delta T^\theta{}_\theta = 8\pi\delta P, \tag{2.75}$$

$$\delta G^\phi{}_\phi = 8\pi\delta T^\phi{}_\phi = 8\pi\delta P, \tag{2.76}$$

$$\delta G^r{}_\theta = 0. \tag{2.77}$$

Here the components of the perturbed Einstein tensor are,

$$\delta G^t{}_t = \frac{e^{-\beta}}{2r^2}Y_{20}(\theta,\phi)\left[r\left(2H' + K'(6-r\beta') + 2rK''\right) + H\left(2+6e^\beta - 2r\beta'\right) - 4Ke^\beta\right], \tag{2.78}$$

$$\delta G^r{}_r = \frac{Y_{20}(\theta,\phi)}{2r^2}e^{-\beta}\left[-4Ke^\beta + H(2+2r\alpha' - 6e^\beta) + r\left(2H' + K'(2+r\alpha')\right)\right], \tag{2.79}$$

$$\delta G^\theta{}_\theta = \frac{Y_{20}(\theta,\phi)}{4r}e^{-\beta}\Big[2H(\alpha'-\beta') + Hr\alpha'(\alpha'-\beta') + H'(4+3r\alpha'-r\beta')$$
$$+ K'(4+r\alpha'-r\beta') + 2rH'' + 2rK'' + 2rH\alpha''\Big], \tag{2.80}$$

$$\delta G^r{}_\theta = -\frac{1}{2}e^{-\beta}\frac{\partial Y_{20}(\theta,\phi)}{\partial\theta}\left[\alpha' H(r) + H' + K'\right]. \tag{2.81}$$

Using Eq.(2.75), Eq.(2.76) and Eq.(2.80) perturbation in the total pressure of the two fluid systems can be expressed as,

$$\delta P = \frac{\delta G^\theta{}_\theta + \delta G^\phi{}_\phi}{16\pi} = \frac{\delta G^\theta{}_\theta}{8\pi}$$
$$= \frac{Y_{20}(\theta,\phi)}{32\pi r}e^{-\beta}\Big[2H(\alpha'-\beta') + Hr\alpha'(\alpha'-\beta') + H'(4+3r\alpha'-r\beta')$$
$$+ K'(4+r\alpha'-r\beta') + 2rH'' + 2rK'' + 2rH\alpha''\Big]. \tag{2.82}$$

The $(r,\theta)$ component of the perturbed Einstein equation allows us to write,

$$K'' = -H'' - \alpha' H' - \alpha'' H. \tag{2.83}$$

Using the "$tt$" and "$rr$" component of the perturbed Einstein tensor as given in Eq.(2.73) and Eq.(2.74) we obtain,

$$\delta G^t{}_t - \delta G^r{}_r = 8\pi(\delta T^0{}_0 - \delta T^r{}_r),$$

$$\Longrightarrow \delta G^{t}_{\ t} - \delta G^{r}_{\ r} + 8\pi\left(1+\frac{d\varepsilon}{dP}\right)\delta P = 0. \tag{2.84}$$

Here we have expressed $\delta\varepsilon = (d\varepsilon/dP)\delta P$. Further using Eq.(2.78), Eq.(2.79), Eq.(2.82) and Eq.(2.83) allows us to recast Eq.(2.84) as,

$$H'' + C_1 H' + C_0 H = 0, \tag{2.85}$$

where the coefficients of $H'$ and $H$ are,

$$C_1 = \frac{2}{r} + \frac{\alpha' - \beta'}{2}, \tag{2.86}$$

and,

$$C_0 = -\frac{6}{r^2}e^{\beta} + \left(\alpha'' - \frac{\alpha'\beta'}{2} - \frac{1}{2}\alpha'^2\right) + \frac{7}{2}\frac{\alpha'}{r} + \frac{1}{2r}\frac{d\varepsilon}{dP}\alpha' + \frac{3}{2}\frac{\beta'}{r} + \frac{1}{2r}\frac{d\varepsilon}{dP}\beta', \tag{2.87}$$

respectively. Till this point, the derivation of the evolution equation of $H(r)$ is general and does not assume any specific form of matter energy momentum tensor. Recall for a two fluid system,

$$r\alpha' = 8\pi(P_1 + P_2)r^2 e^{\beta} + e^{\beta} - 1, \tag{2.88}$$

$$r\beta' = 1 - e^{\beta} + 8\pi(\varepsilon_1 + \varepsilon_2)r^2 e^{\beta}. \tag{2.89}$$

Therefore $C_1$ as given in Eq.(2.86) can be expressed as,

$$\begin{aligned} C_1 &= \frac{2}{r} + 4\pi r e^{\beta}\left((P_1 + P_2) - (\varepsilon_1 + \varepsilon_2)\right) + \frac{e^{\beta} - 1}{r} \\ &= \frac{2}{r} + e^{\beta}\left(\frac{2m}{r^2} + 4\pi r\left((P_1 + P_2) - (\varepsilon_1 + \varepsilon_2)\right)\right). \end{aligned} \tag{2.90}$$

Recall,

$$C_0 = -\frac{6}{r^2}e^{\beta} + \left(\alpha'' - \frac{\alpha'\beta'}{2} - \frac{\alpha'^2}{2}\right) + \frac{7}{2}\frac{\alpha'}{r} + \frac{1}{2r}\frac{d\varepsilon}{dP}\alpha' + \frac{3}{2}\frac{\beta'}{r} + \frac{1}{2r}\frac{d\varepsilon}{dP}\beta'. \tag{2.91}$$

Further using the background Einstein equation for $(\theta\theta)$ component allows us to write,

$$\begin{aligned} \bar{G}_{\theta\theta} &= 8\pi(P_1 + P_2)r^2 \\ \Longrightarrow \alpha'' - \frac{\alpha'\beta'}{2} &= 16\pi(P_1 + P_2)e^{\beta} - \frac{\alpha'^2}{2} + \frac{\beta'}{r} - \frac{\alpha'}{r}. \end{aligned} \tag{2.92}$$

Using Eq.(2.92), $C_0$ as given in Eq.(2.91) can be simplified to,

$$C_0 = -\frac{6}{r^2}e^{\beta} + 4\pi e^{\beta}\left[(P_1 + P_2) + (\varepsilon_1 + \varepsilon_2)\right]\frac{d\varepsilon}{dP} + 4\pi e^{\beta}\left[5(\varepsilon_1 + \varepsilon_2) + 9(P_1 + P_2)\right] - \alpha'^2. \tag{2.93}$$

For two fluid system assuming both fluids have a barotropic equation of state we get,

$$\frac{d\varepsilon}{dP} = \frac{d\varepsilon_1}{dP_1}\frac{\delta P_1}{\delta P_1 + \delta P_2} + \frac{d\varepsilon_2}{dP_2}\frac{\delta P_2}{\delta P_1 + \delta P_2}. \tag{2.94}$$

Further using two fluid TOV equations (Eqs.(2.64) and (2.65)) we can write,

$$\frac{\delta P_1}{\delta P_1 + \delta P_2} = \frac{P_1 + \varepsilon_1}{(P_1 + P_2) + (\varepsilon_1 + \varepsilon_2)}, \tag{2.95}$$

and

$$\frac{\delta P_2}{\delta P_1 + \delta P_2} = \frac{P_2 + \epsilon_2}{(P_1 + P_2) + (\varepsilon_1 + \varepsilon_2)}. \tag{2.96}$$

Therefore for two fluid case,

$$C_0 = -6\frac{e^{\beta}}{r^2} + 4\pi e^{\beta}\left[5\varepsilon_1 + 9P_1 + (P_1 + \varepsilon_1)\frac{d\varepsilon_1}{dP_1} + 5\varepsilon_2 + 9P_2 + (P_2 + \varepsilon_2)\frac{d\varepsilon_2}{dP_2}\right] - \alpha'^2. \tag{2.97}$$

Now let us introduce the dimensionless variable $y \equiv \frac{rH'}{H}$. In terms of $y(r)$ the evolution equation of $H(r)$ as given in Eq.(2.85) can be expressed as,

$$r\frac{dy(r)}{dr} + y(r)^2 + y(r)F(r) + r^2Q(r) = 0, \tag{2.98}$$

with,

$$F(r) = \frac{r - 4\pi r^3\left((\varepsilon_1(r) + \varepsilon_2(r)) - (P_1(r) + P_2(r))\right)}{r - 2m(r)},$$

and

$$Q(r) = \frac{4\pi r\left[5(\varepsilon_1(r) + \varepsilon_2(r)) + 9(P_1(r) + P_2(r)) + \frac{\varepsilon_1(r)+P_1(r)}{\partial P_1(r)/\partial \varepsilon_1(r)} + \frac{\varepsilon_2(r)+P_2(r)}{\partial P_2(r)/\partial \varepsilon_2(r)} - \frac{6}{4\pi r^2}\right]}{r - 2m(r)} - 4\left[\frac{m(r) + 4\pi r^3(P_1(r) + P_2(r))}{r^2(1 - 2m(r)/r)}\right]^2,$$

For a two-fluid system, the Einstein equation is modified only within the matter, while the evolution equation for $y$ outside the matter remains unchanged (see Ref. Hinderer [2008] for further details). The boundary conditions for $y(r)$ can be determined by ensuring regularity in $r = 0$. By considering only the leading-order contribution to $y(r)$ at $r = 0$, one can easily derive the boundary condition, which is identical to that of a single-fluid system (refer to Ref. Hinderer [2008]). This is because, at $r = 0$, the leading-order contribution from the matter does not affect $y(r)$. Therefore, even in the case of multiple fluids, $y(r = 0) = 2$.

## 2.5 Bayesian Approach

In this section, we will give a brief overview of the Bayesian inference technique. This thesis employs a Bayesian Inference framework in chapter 5 to assess various neutron star EOS models, including the evaluations of dark matter effects and high-density phenomena.

Bayesian inference Malik & Providência [2022a], Malik *et al.* [2022b], Zhou *et al.* [2023a] is a statistical approach that relies on Bayes' theorem to calculate the probability of a particular outcome, known as the posterior probability, based on prior knowledge and new evidence. The theorem provides a formula to update our beliefs about an event $A$ given the occurrence of another event $B$:

$$P(A|B) = \frac{P(B|A)P(A)}{P(B)}, \tag{2.99}$$

In this equation, $P(A)$ represents our initial belief or prior probability about $A$, $P(B)$ is the prior probability of $B$, and $P(B|A)$ is the likelihood of observing $B$ given that $A$ is true. The essence of Bayesian inference is that it allows us to combine prior information with new data to refine our understanding of uncertain events.

When applying this to a set of data $B$ and parameters $\mathbf{X}$ representing $A$, we can express Bayes' theorem in a more specific form:

$$p(\mathbf{X}|\text{data}) = \frac{L(\text{data}|\mathbf{X})p(\mathbf{X})}{p(\text{data})}. \tag{2.100}$$

Here, $p(\mathbf{X}|\text{data})$ is the updated probability distribution of the parameters given the data, often referred to as the posterior distribution. The denominator $p(\text{data})$ ensures that the total probability is normalized, meaning it sums to 1:

$$p(\text{data}) = L(\text{data}|\mathbf{X})p(\mathbf{X})d\mathbf{X}. \tag{2.101}$$

In many cases, the relationship is simplified to:

$$p(\mathbf{X}|\text{data}) \propto L(\text{data}|\mathbf{X})p(\mathbf{X}), \tag{2.102}$$

which indicates that the posterior distribution is proportional to the product of the likelihood and the prior. From this posterior distribution, we can extract marginal distributions for individual or paired parameters:

$$p(X_j|\text{data}) = \left(\prod_{i\neq j} \int dX_i\right) p(\mathbf{X}|\text{data}) \tag{2.103}$$

$$p(X_j, X_k|\text{data}) = \left(\prod_{i\neq j,k} \int dX_i\right) p(\mathbf{X}|\text{data}) \tag{2.104}$$

The prior distribution $p(\mathbf{X})$ represents what is known about the parameters before considering the current data, while the likelihood function $L(\text{data}|\mathbf{X})$ reflects how well the data supports different parameter values.

In summary, Bayesian inference is a powerful method for updating our knowledge about a model or parameters as new data becomes available. The choice of prior distribution is crucial because it influences the outcome, making it important to carefully consider the prior to ensure accurate and meaningful results.

*Data*:- Within our inference analysis, we have taken into account various constraints ranging from nuclear physics experiments to astrophysical observations. The data set is presented in Table 5.1. The constraints are established based on empirical data derived from experimental observations of finite nuclei properties, such as nuclear masses, neutron skin thickness in $^{208}$Pb, dipole polarizability, and isobaric analog states, alongside heavy ion collision (HIC) data spanning densities from 0.03 to 0.32 fm$^{-3}$. Moreover, our analysis includes astrophysical data like the mass-radius relationships observed in pulsars PSR J0030+0451 and PSR J0740+6620, as well as the tidal deformability of binary neutron star systems as demonstrated by the GW170817 event. We apply these constraints within a Bayesian framework to refine the EOSs. The terrestrial and astrophysical data sources used in our analysis are listed in Table 5.1 of chapter 5

*Likelihood*:- The Likelihood of the given data is described as follows:

- **Experimental data:** For experimental data, $D_{\text{expt}} \pm \sigma$ having a symmetric Gaussian distribution, the likelihood is given as,

$$L(D_{\text{expt}}|\theta) = \frac{1}{\sqrt{2\pi\sigma^2}} \exp\left(-\frac{(D(\theta) - D_{\text{expt}})^2}{2\sigma^2}\right) = L_{\text{expt}}.$$

  Here, $D(\theta)$ is the model value for a given model parameter set $\theta$.

- **GW observation**: For GW observations, information about EOS parameters comes from the masses $m_1, m_2$ of the two binary components and the corresponding tidal deformabilities $\Lambda_1, \Lambda_2$.

In this case,

$$P(d_{\rm GW}|{\rm EOS}) = \int_{m_l}^{M_u} dm_1 \int_{M_l}^{m_1} dm_2 P(m_1, m_2|{\rm EOS}) \times P(d_{\rm GW}|m_1, m_2, \Lambda_1(m_1, {\rm EOS}), \Lambda_2(m_2, {\rm EOS})) = L^{\rm GW} \quad (2.105)$$

where P(m|EOS) [Agathos *et al.* 2015, Biswas *et al.* 2021, Landry *et al.* 2020, Wysocki *et al.* 2020] can be written as,

$$P(m|{\rm EOS}) = \begin{cases} \frac{1}{M_u - M_l} & \text{iff} \quad M_l \le m \le M_u, \\ 0 & \text{else,} \end{cases} \quad (2.106)$$

In our calculation, we set $M_l = 1\ M_\odot$ and $M_u$ as the maximum mass for a given EOS.

- **X-ray observation(NICER)**: X-ray observations give the mass and radius measurements of NS. Therefore, the corresponding evidence takes the following form,

$$P(d_{\rm X\text{-}ray}|{\rm EOS}) = \int_{M_l}^{M_u} dm P(m|{\rm EOS}) \times P(d_{\rm X\text{-}ray}|m, R(m, {\rm EOS})) = L^{\rm NICER}. \quad (2.107)$$

Here again, $M_l$ represents a mass of 1 $M_\odot$, and $M_u$ denotes the maximum mass of a neutron star according to the respective EOS.

The final likelihood for the three scenarios:

$$L = L^{\rm EXPT} L^{\rm GW} L^{\rm NICERI} L^{\rm NICERII}. \quad (2.108)$$

NICER I and NICER II refer to the mass-radius measurements of the pulsars PSR J0030+0451 and PSR J0740+6620, respectively.

# Chapter 3

# Investigating Correlations between Fermionic Dark Matter Parameters and Neutron Star Properties: A Two-Fluid Approach

## 3.1 Motivation

In this chapter, we employ a two-fluid approach, as discussed in Section 2.4.2.2, to calculate the properties of neutron stars. For the nuclear matter equation of state (EOS), we use several realistic EOS derived from the relativistic mean field (RMF) model, each one with different levels of stiffness and composition. Concurrently, we explore the dark matter EOS, focusing on fermionic matter with repulsive interactions described by a relativistic mean field Lagrangian. The motivation behind this study is to investigate potential correlations between the dark matter sector and the global properties of neutron stars, while accounting for the inherent uncertainties in the baryonic sector's equation of state (EOS). These uncertainties are critical, as they influence our overall understanding of neutron star behavior.

By employing the well-known Kendall rank correlation studies, we examine the relationship between the dark matter sector and neutron star properties. Our goal is to identify any connections or influences that may exist, offering valuable insights into both dark matter properties and the behavior of these enigmatic cosmic objects. Additionally, we aim to explore new ways to constrain the dark matter sector by investigating its impact on neutron star cooling, a key observable, and by assessing the viability of direct Urca processes.

## 3.2 Sampling

To compute the structure of neutron stars (NS), we have utilized four distinct nucleonic equation-of-states (EOSs) and a diverse range of dark matter EOSs constructed through the Relativistic Mean Field (RMF) formalism, as detailed in sections 2.1 and 2.4.2.1 respectively. For the nuclear matter, we utilize four realistic EOS, namely EOS1, EOS2, EOS3, and EOS4. We have sampled a total of 50,000 dark matter parameter combinations, namely $c_\omega$, $m_\chi$, and $F_\chi$ from uniform distributions within the ranges specified in Table 3.3. These large combinations of nuclear matter and dark matter EOS give rise to 200K mass-radius curves by solving the two-fluid TOV equations. All of these mass-radius relations are not practically relevant as we use the filter that NS must have a mass greater than 1.9 $M_\odot$, set by the pulsars PSR J0348+0432 and PSR J0740+6620 within $\sim 3\sigma$. We also consider that the dark matter admixed neutron stars only produce non-halo configurations, which means the dark matter admixed radius is smaller than the luminous radius. Following the application of this filter, the remaining values of samples for EOS1, EOS2, EOS3, and EOS4 are 25K, 14K, 10K, and 20K, respectively. The median values, along with the lower and upper bounds of the 90% confidence intervals (CI), for various neutron star properties for these sets are listed in Table 3.4. These properties include the maximum mass of neutron stars ($M_{max}$), the total radius ($R_{t,x}$) for values of $x$ in the range of [1.2, 1.4, 1.6, 1.8, 2.0], the dimensionless tidal deformability ($\Lambda_x$) for $x$ in the range of [1.4, 1.6, 1.8], the fraction of dark matter energy density over nuclear matter ($f_{d,x}$), and the nuclear matter baryon density ($\rho_{B,x}$) for $x$ in the range of [1.4, 1.6, 2.0].

## 3.3 Equation of State

Figure 3.1 displays the four EOS as discussed in section 3.1 and 3.2, along with their corresponding neutron stars (NS) properties. The graph depicts the pressure, denoted as $P$, as a function of baryon density $\rho_B$ in the left plot. The middle plot showcases the NS mass, denoted as $M$, as a function of radius $R$. Additionally, the right plot illustrates the relation between the NS mass ($M$) and the square of the speed of sound ($c_s^2$) for the four nuclear matter EOSs. They have been constrained to several nuclear matter properties, in particular, the saturation density, binding energy, incompressibility, symmetry energy at saturation, and the pure neutron matter pressure calculated with chiral effective field theory. It was also imposed that the pure neutron matter pressure must be an increasing function of the baryonic density and stars with at least 2.0 $M_\odot$ must be described. The parameters of all these four models are presented in Table 3.1. The EOS1 is the stiffest and EOS3 is the softest one. The nuclear saturation properties along with star properties can be accessed from Table 3.2. It can be seen from the table that the NS maximum mass of these four EOSs ranges from 2.10 to 2.74 $M_\odot$. The radius and tidal deformability for a 1.4 $M_\odot$ NS are in the range of 12.55–13.78 km and 462–844, respectively.

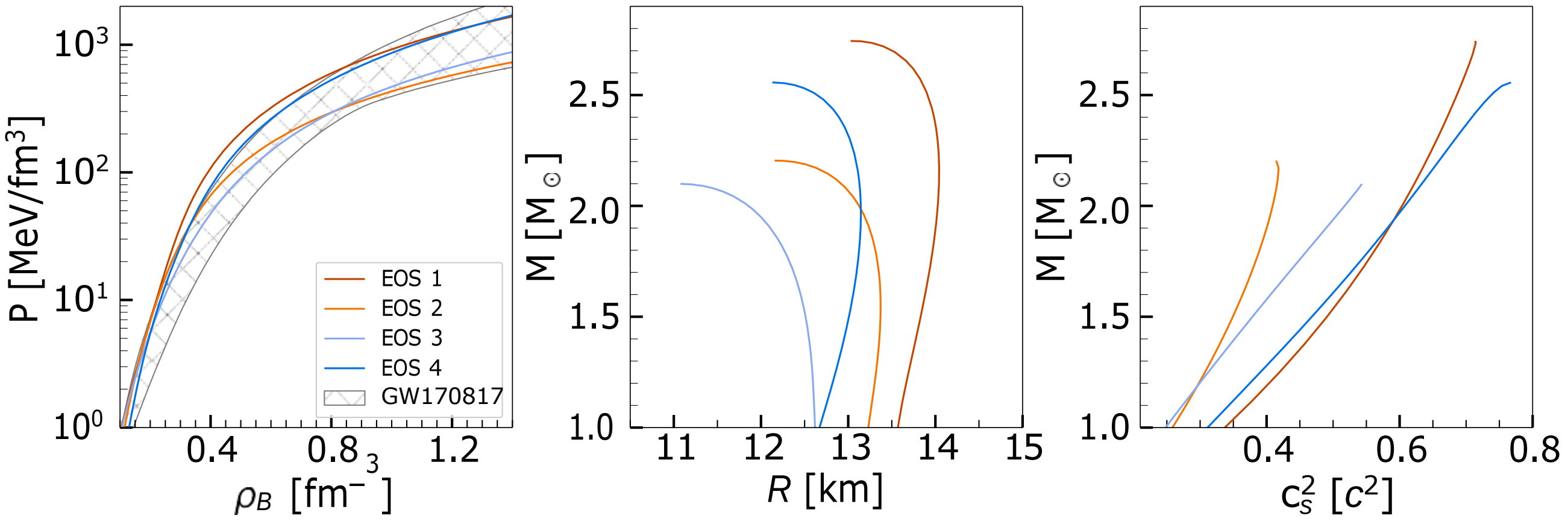

Figure 3.1: (left plot) The pressure $P$ as a function of baryon density $\rho_B$, (middle plot) the NS mass $M$ as a function of radius $R$, and (right plot) NS mass $M$ as a function of the square of the speed of sound $c_s^2$ for nuclear matter EOS: EOS1, EOS2, EOS3, and EOS4, respectively.

Table 3.1: Parameters of the employed nuclear matter EOS: EOS1, EOS2, EOS3 and EOS4. The B and C are $b \times 10^3$, and $c \times 10^3$ respectively Malik *et al.* [2023].

| EOS | $g_\sigma$ | $g_\omega$ | $g_\rho$ | $B$ | $C$ | $\xi$ | $\Lambda_\omega$ |
|---|---|---|---|---|---|---|---|
| EOS1 | 10.411847 | 13.219028 | 11.180337 | 2.541001 | -3.586261 | 0.000845 | 0.027999 |
| EOS2 | 11.150279 | 14.420375 | 13.806001 | 2.036239 | -1.635468 | 0.018019 | 0.037600 |
| EOS3 | 8.695491 | 10.431351 | 9.821776 | 3.975509 | -2.615425 | 0.006394 | 0.039323 |
| EOS4 | 9.608190 | 11.957725 | 12.191950 | 3.117923 | -4.098400 | 0.000255 | 0.058744 |

## 3.4 Dark Matter Equation Of State

In Fig. 3.2 we represent EOS for normal matter and dark matter components. There were 50K dark matter EOSs solved individually for each nucleonic EOS, resulting in a total of 200,000 mass-radius (M-R) calculations.

To determine the dark matter equation of state (EOS), we consider dark matter within the mass range 0.5 GeV $\leq m_\chi \leq$ 4.5 GeV Calmet & Kuipers [2021]. This mass range is chosen based on both theoretical considerations and observational constraints. The lower bound of 0.5 GeV is motivated by the fact that for $m_\chi < 500$ MeV, dark matter tends to form extended halos rather than a concentrated core within neutron stars. Since our analysis focuses solely on dark matter cores and does not account for halo configurations, we impose this lower limit to align with the scope of our study. The upper bound of 4.5 GeV is informed by constraints from Ref. Essig *et al.* [2012], which provides the first direct detection limits on dark matter in the MeV–GeV mass range using XENON10 data. The strongest constraint is reported at $m_\chi = 100$ MeV, where the dark matter-electron scattering cross-section is limited to $\sigma_e < 3 \times 10^{-38}$ cm$^2$ at 90% confidence level (C.L.). Furthermore, dark matter masses in the range 20 MeV to 1 GeV are constrained by $\sigma_e < 10^{-37}$ cm$^2$ at 90% C.L. These results provide an empirical justification for our chosen upper limit. Additionally, recent improvements in the S2-only DarkSide-50 analysis Agnes *et al.* [2023] have resulted in the leading spin-independent (SI) sensitivity in the 1.2 to 10 GeV/$c^2$ mass range, further supporting our selection of mass constraints.

Table 3.2: Nuclear Saturation Properties - (i) For Symmetric Nuclear Matter - energy per nucleon $\varepsilon_0$, incompressibility coefficient $K_0$, and skewness $Q_0$; and (ii) For Symmetry Energy - symmetry energy at saturation $J_{\text{sym},0}$, its slope $L_{\text{sym},0}$; and Neutron Star Properties - maximum mass $M_{\text{max}}$, radius $R_{\text{max}}$, radius $R_{1.4}$ for 1.4 $M_\odot$ and $R_{2.08}$ for 2.08 $M_\odot$ neutron stars, tidal deformability $\Lambda_{1.4}$ for 1.4 solar mass neutron stars, the square of speed-of-sound $c_s^2$ at the center of maximum mass neutron stars, the neutron star mass at which the direct Urca process occurs $M_{\text{dUrca}}$, and the direct Urca density $\rho_{\text{B,x}}$ for $x \in [1.4, 1.6, 1.8]$ solar mass neutron stars.

| | NMP | | | | | | NS | | | | | | | | | |
|---|---|---|---|---|---|---|---|---|---|---|---|---|---|---|---|---|
| EOS | $\rho_0$ | $\varepsilon_0$ | $K_0$ | $Q_0$ | $J_{\text{sym},0}$ | $L_{\text{sym},0}$ | $M_{\text{max}}$ | $R_{\text{max}}$ | $R_{1.4}$ | $R_{2.08}$ | $\Lambda_{1.4}$ | $c_s^2$ | $M_{\text{dUrca}}$ | $\rho_{\text{dUrca}}$ | $\rho_{B,1.4}$ | $\rho_{B,1.6}$ | $\rho_{B,1.8}$ |
| | [fm$^{-3}$] | | | [MeV] | | | [$M_\odot$] | | [km] | | [...] | [$c^2$] | [$M_\odot$] | | [fm$^{-3}$] | | |
| EOS1 | 0.155 | -16.08 | 177 | -74 | 33 | 64 | 2.74 | 13.03 | 13.78 | 14.04 | 844 | 0.713 | 2.06 | 0.366 | 0.298 | 0.316 | 0.336 |
| EOS2 | 0.154 | -15.72 | 190 | 614 | 32 | 60 | 2.20 | 12.16 | 13.36 | 13.00 | 709 | 0.414 | 1.83 | 0.443 | 0.344 | 0.382 | 0.432 |
| EOS3 | 0.157 | -16.24 | 260 | -400 | 32 | 57 | 2.10 | 11.08 | 12.55 | 11.53 | 462 | 0.543 | 2.07 | 0.829 | 0.432 | 0.491 | 0.570 |
| EOS4 | 0.156 | -16.12 | 216 | -339 | 29 | 42 | 2.56 | 12.13 | 12.95 | 13.14 | 638 | 0.767 | 2.55 | 0.747 | 0.345 | 0.370 | 0.399 |

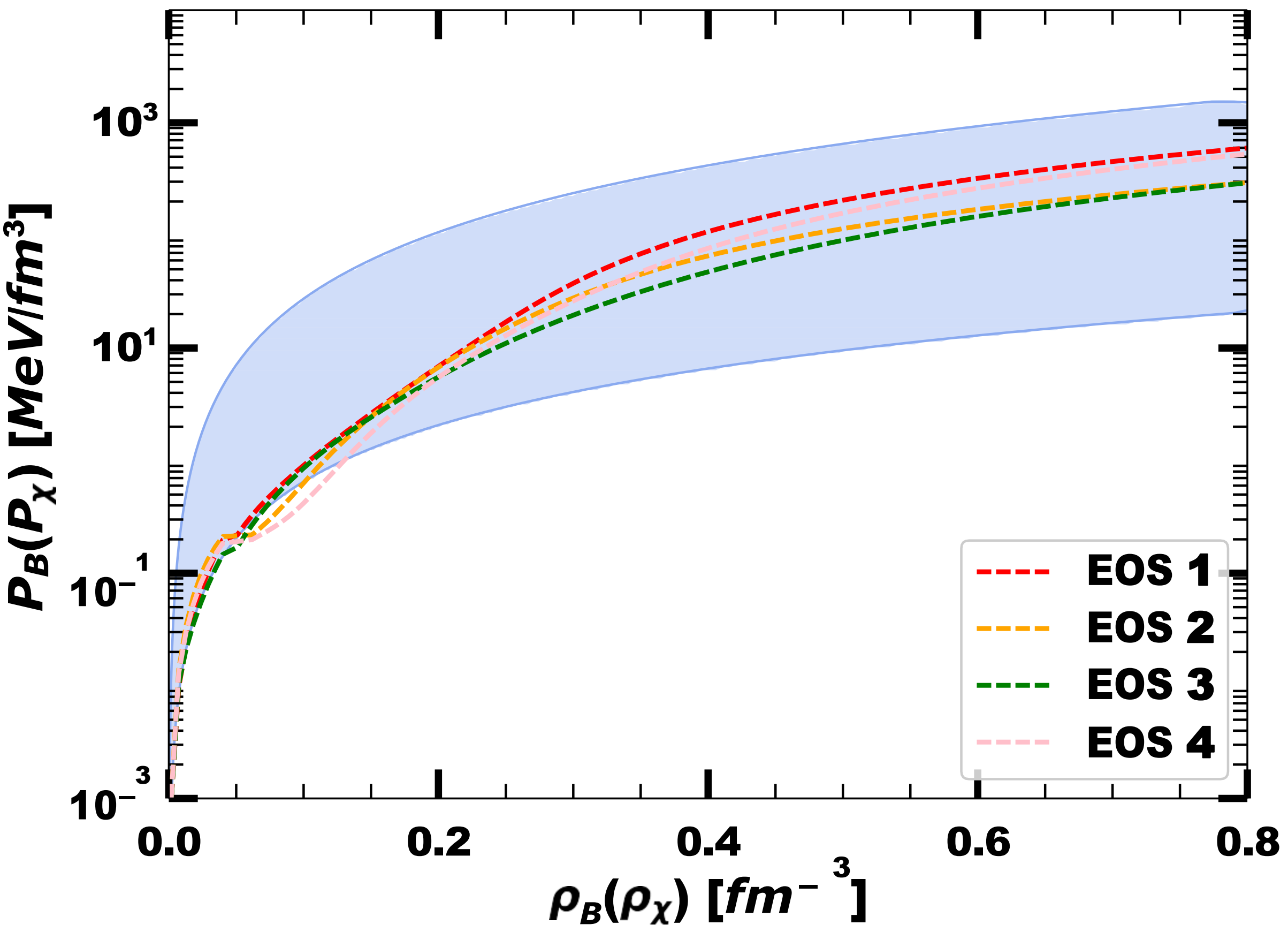

Figure 3.2: The shaded blue domain represents the sampled dark matter EOS, i.e., the pressure ($P_\chi$) as a function of density ($\rho_\chi$). The colored dashed lines depict the nuclear matter EOS, i.e., the variation of baryonic matter pressure ($P_B$) with baryon number density ($\rho_B$).

Table 3.3: The prior set for dark matter model parameters.

| $m_\chi$ | | $c_\omega$ | | $F_\chi$ | |
|---|---|---|---|---|---|
| GeV | | fm | | % | |
| min | max | min | max | min | max |
| 0.5 | 4.5 | 0.1 | 5 | 0 | 25 |

Dark matter self-interaction is considered within the range 0.1 fm $\leq c_\omega \leq$ 5 fm. This parameter is not strictly constrained within the dark matter model adopted in this study. In principle, larger values of $c_\omega$ can be chosen; however, the vector interaction primarily governs the repulsive nature of the interaction, leading to a stiffening of the equation of state (EOS). If the repulsive interaction becomes too strong, it can result in the formation of a dark matter halo. In this analysis, we specifically avoid such halo configurations and focus solely on dark matter cores within neutron stars. Additionally, previous studies, such as Ref. Das *et al.* [2022], have considered both scalar ($C_\sigma$) and vector interactions in the dark matter sector. When both interactions are present, larger values of $C_\omega$ can be accommodated because the scalar interaction softens the EOS while the vector interaction stiffens it. This interplay effectively balances the overall impact on the EOS. For a more detailed discussion on these interactions and their implications, readers may refer to Refs. Das *et al.* [2022], Xiang *et al.* [2014]

The fraction of dark matter ($f_\chi$) within neutron stars remains an open question, as its typical values are yet to be precisely determined. Consequently, selecting a specific range for $f_\chi$ in the two-fluid neutron star formalism is inherently model-dependent, guided by both observational constraints and theoretical considerations. For instance, Koehn *et al.* [2024] acknowledge the uncertainty in the dark matter fraction within neutron stars and adopt an arbitrary upper limit of 1%. Similarly, Ivanytskyi *et al.* [2020] estimate the local dark matter fraction in the Milky Way by evaluating the ratio of dark matter mass density to the total mass density (including baryonic matter). Their analysis suggests values of $f_\chi^* = 1.6 \pm 0.4\%$ near PSR J0348+0432 and $f_\chi^* = 1.35 \pm 0.35\%$ near PSR J0740+6620. Additional constraints arise from studies on sub-GeV bosonic dark matter, where Karkevandi *et al.* [2022] conclude that existing observational data favor low dark matter fractions below 5%. Meanwhile, Rutherford *et al.* [2023] propose an upper bound of $f_\chi \leq 20\%$. Furthermore, Ciarcelluti & Sandin [2011] demonstrate that dark matter can contribute to the formation of highly compact neutron stars and explain the existence of massive pulsars, such as PSR J1614-2230 with a mass of $1.97 \pm 0.04 M_\odot$ and PSR J0348+0432 with a mass of $2.01 \pm 0.04 M_\odot$. According to Ciarcelluti & Sandin [2011], a neutron star can reach $2M_\odot$ for a dark matter fraction of 15%, while a mass of $1.8M_\odot$ can be obtained with a dark matter fraction of 70%. Additionally, Goldman *et al.* [2013] suggest that a neutron star can achieve $2M_\odot$ with a dark matter fraction of 50%. In this study, we consider a dark matter mass fraction in the range $0 \leq F_\chi \leq 25\%$ Ciancarella *et al.* [2021], motivated by previous studies on dark matter-admixed neutron stars. As summarized in Table 3.2, we employ four different equations of state (EOS) to investigate the role of dark matter in neutron star structure. Among these, EOS1 and EOS4 are relatively stiff, yielding maximum neutron star masses of 2.74 $M_\odot$ and 2.56 $M_\odot$, respectively, whereas EOS2 and EOS3 are softer, with maximum masses of 2.20 $M_\odot$ and 2.10 $M_\odot$, respectively. Given that observations of massive pulsars, such as PSR J0348+0432 and PSR J0740+6620, set a lower mass constraint of approximately 1.9 $M_\odot$, we impose this threshold as a filtering criterion in our analysis. Our objective is to determine the maximum fraction of dark matter that neutron stars can sustain while satisfying this mass constraint. By applying this filter, we aim to assess the impact of dark matter on the mass-radius relation and other fundamental neutron star properties across different EOS models. Also Note that the dark matter mass fraction crucially depends on the dark matter capture rate inside neutron stars. Depending upon the generic modeling of the nucleon-dark matter interaction inside the high-density region of neutron stars, the dark matter capture rate can be of the order of $10^{25}$ GeV/sec for dark matter mass near 1 GeV Bell *et al.* [2020]. Such an estimate of the capture of dark matter particles in the interior of neutron stars over its lifetime ($\sim 10^{17}$ seconds) indicates that it may be difficult for a neutron star to accumulate a significant fraction of dark matter. Instead, some other mechanism, such as the production of dark matter in the star during supernovae (SN) DeRocco *et al.* [2019], may be necessary for a neutron star to have a large dark matter fraction. Conversion of the neutrons into dark matter particles might also allow a significant fraction of the nuclear matter to be converted into dark matter particles inside neutron stars. Non-standard conversion of neutrons into scalar dark matter particles has been explored in the context of dark matter admixed neutron stars in Ref. Ellis *et al.* [2018] and references therein. The dark matter capture rate can crucially depend on the Pauli blocking factor in the degenerate nuclear matter, multiple scattering among dark matter particles and nucleons, neutron star internal structure, momentum dependent form factors of hadrons, etc Anzuini *et al.* [2021], Bell *et al.* [2021]. Note that apart from the degenerate neutrons, protons, electrons, and muons are also present in the beta-equilibrated nuclear matter. The interaction between the leptonic sector and the dark matter sector can also account for the dark matter capture within neutron stars Anzuini *et al.* [2021]. Furthermore, the capture of dark matter particles by neutron stars can be enhanced in a close binary system. This amplification stems from the energy loss of dark matter particles resulting from their gravitational interaction with moving companions (gravitational slingshot) Brayeur & Tinyakov [2012]. This effect is maximum when the velocities of the companions are comparable to the asymptotic velocity of dark matter particles Brayeur & Tinyakov [2012].

## 3.5 Corelation studies

To study the relation between dark matter parameters and various neutron star properties, we employ the Kendall rank correlation coefficient analysis. The Kendall rank correlation can be considered a non-parametric test that allows us to quantify the strength of dependence between two variables. Unlike Pearson correlation, it accounts for non-linearity in the correlation, making it suitable for analyzing relationships that may not follow a linear pattern Ghosh *et al.* [2022]. Our strategy involves first selecting a fixed nuclear EOS and then assessing the correlation coefficients using the chosen set of dark matter EOS. In this way, we can explore the uncertainty only in the dark matter sector. Lastly, we explore the combined effects of all four combinations of nuclear equations and the entire set of dark matter equation-of-states. This approach helps us to comprehend the impact of uncertainty in the nuclear EOS on these correlations. In Fig. 3.3 we plot the Kendall rank correlation coefficients, where we consider the nuclear EOS1 and the entire set of filtered dark matter equation-of-state sets, i.e., with NS having a maximum mass above 1.9 $M_{\odot}$ and NS having to be non-halo (as mentioned earlier). As portrayed in Fig. 3.3, Fig. 3.4 includes the complete data set with all four nuclear EOSs employed. Please note that in this context, the subscript '$t$' denotes the total radius of the neutron star, while the subscript '$dm$' represents the radius specifically associated with the admixed dark matter.

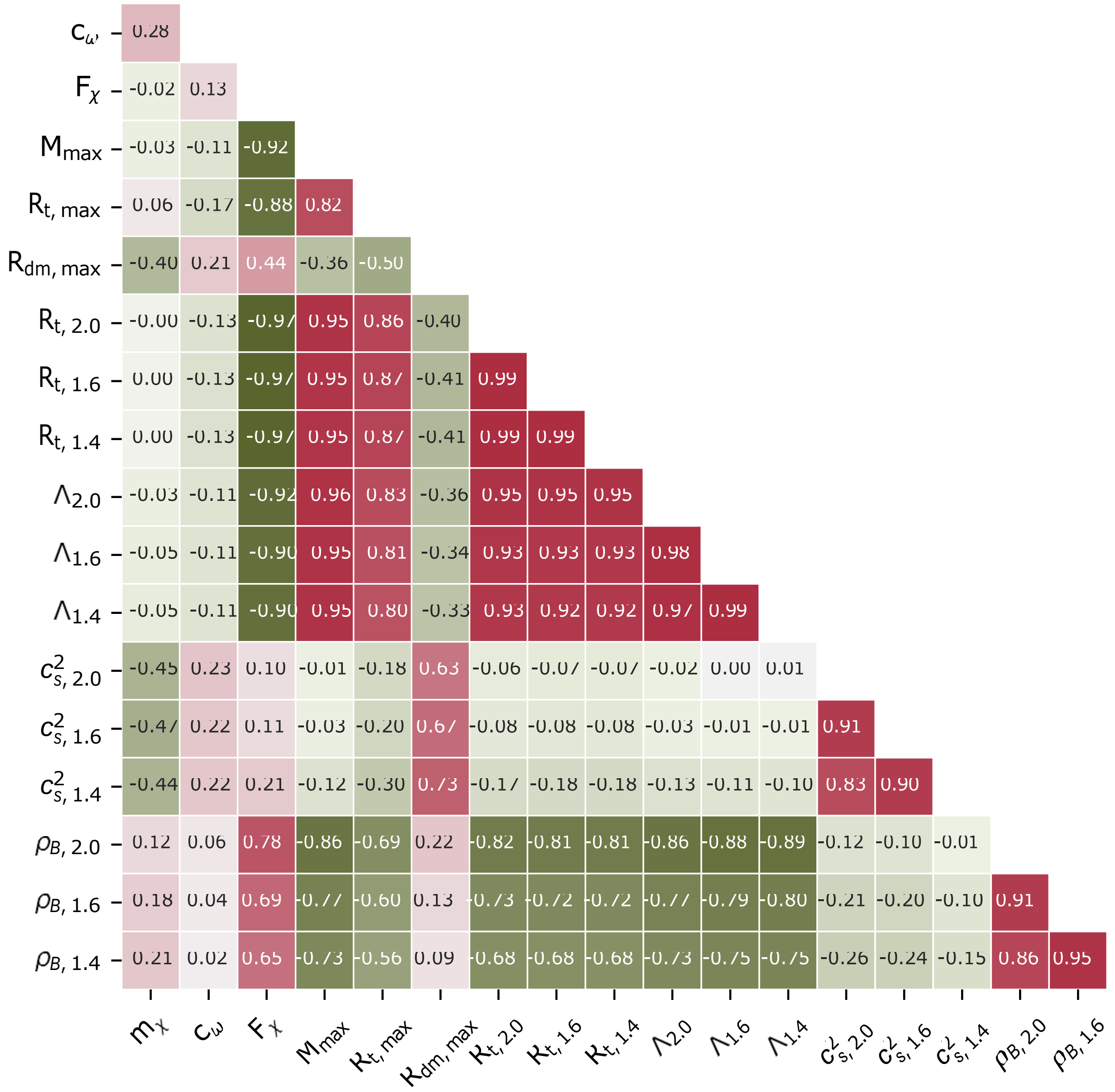

Figure 3.3: With only one nuclear EOS, namely EOS1, we compute the Kendall rank correlation coefficients linking various dark matter parameters to neutron star properties, and with the entire set of dark matter EOS sets after applying the filter (see text).

The following comments are in order:

- We have observed in Fig. 3.3 a strong negative correlation $\sim$ 0.9 between the dark matter mass fraction, $F_\chi$, and the maximum gravitational mass of a neutron star (NS). Additionally, there is a notable correlation $\sim$ 0.9 between $F_\chi$ and the radius as well as the tidal deformability of NS at masses of 1.4, 1.6, and 2.0 $M_\odot$ respectively. However in Fig.3.4, when incorporating uncertainty in the nuclear sector and considering all four nuclear matter equation-of-states (EOSs) alongside the sampled dark matter EOSs, the previously mentioned correlation disappears.

- Furthermore, our findings have revealed a notable positive correlation between $F_\chi$ and the central baryonic density $\rho_B$ (at different NS masses 1.4, 1.6, and 2.0 $M_\odot$) for EOS1. However, once again when we account for the uncertainties associated with nuclear matter, as can be seen in Fig.3.4, those correlations disappear.

- Nonetheless, the correlations between radii, tidal deformability, mass, and central baryon density persist even when considering uncertainties in the nuclear sector, i.e., including the entire dark matter admixed set for all nuclear EOS. It is worth noting that there is a robust and strong correlation between the central baryon density and the star radius even when dark matter is not considered Jiang *et al.* [2022], Malik *et al.* [2023]. However, this correlation is found to break in the case of modified gravity Nobleson *et al.* [2023]. This interestingly allows us to distinguish between the effects of modified gravity and dark matter in neutron stars (NS).

## 3.6 Neutron star properties

Figure 3.5 displays the corner plot featuring the dark matter parameters $c_\omega$, $m_\chi$, and $F_\chi$ corresponding to EOS1, EOS2, EOS3, and EOS4 after applying the filtration process. A corner plot is a visualization used to explore the multivariate distribution of parameters in a dataset. It allows us to observe the relationships and correlations between different variables simultaneously, providing valuable insights into the underlying data structure. When considering a fixed nuclear EOS and only varying the dark matter EOS within the two-fluid formalism, there exists a direct correlation between the fraction of dark matter $F_\chi$ and $M_{\max}$, as depicted in Fig. 3.3. Consequently, the maximum mass constraints of 1.9

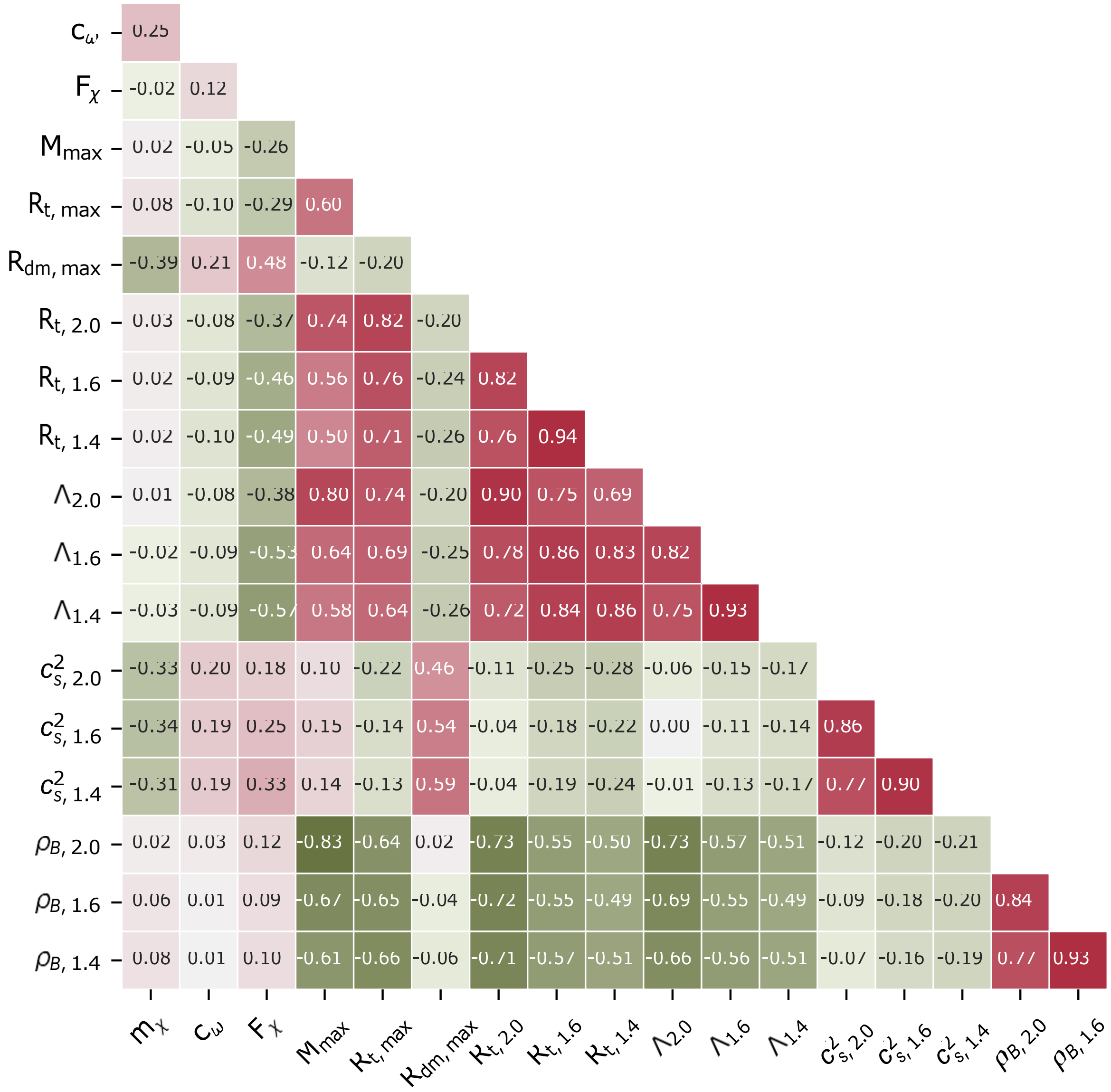

Figure 3.4: With only one nuclear EOS, namely EOS1, we compute the Kendall rank correlation coefficients linking various dark matter parameters to neutron star properties, and with the entire set of dark matter EOS sets after applying the filter (see text).

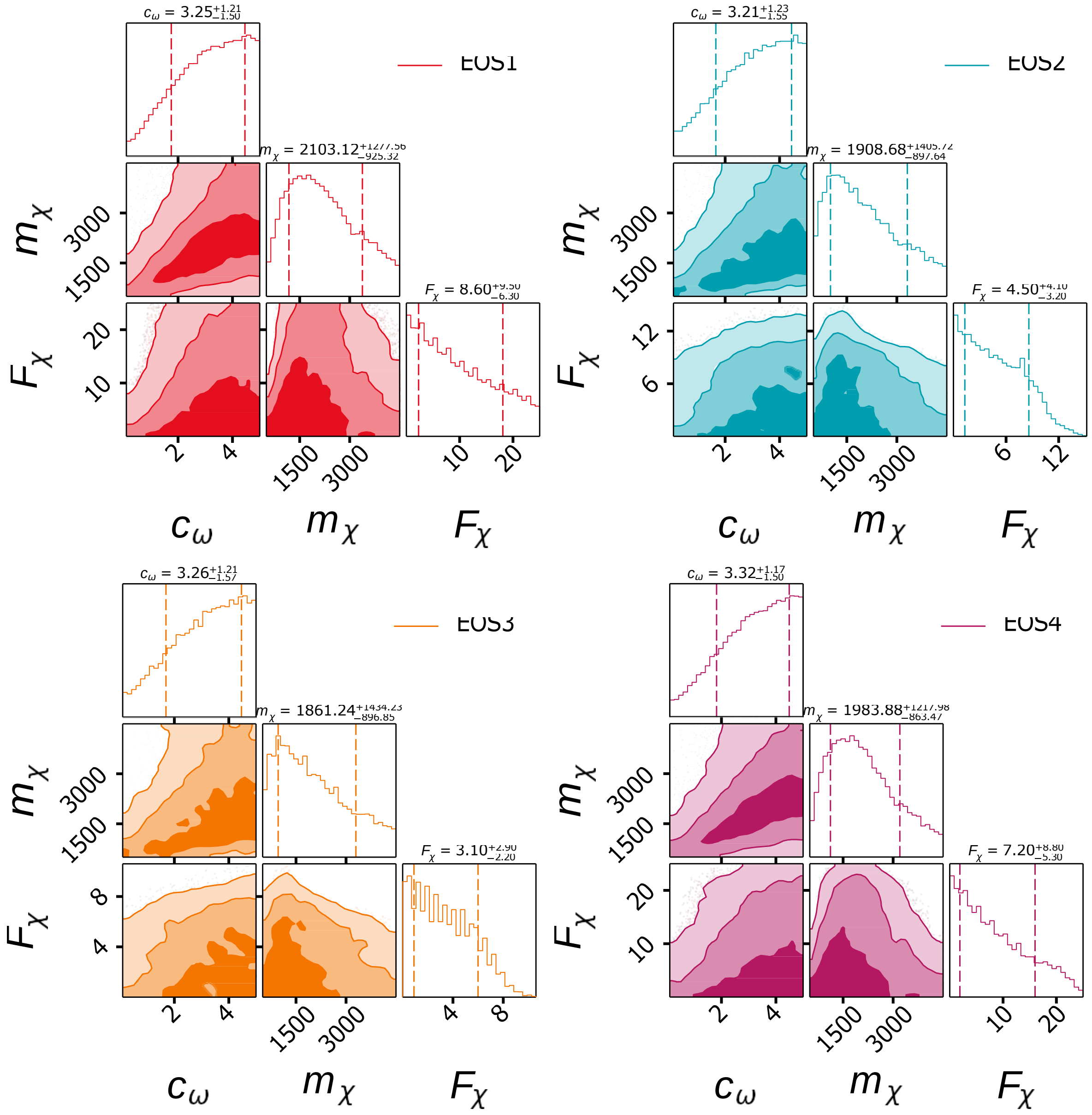

Figure 3.5: The distribution of dark matter parameters, i.e., $c_\omega$, $m_\chi$ (in MeV units), and $F_\chi$ for the prior set mentioned in Table 3.3 after applying the filter that NS must have a mass greater than 1.9 $M_\odot$.

$M_\odot$ depend on the softness or stiffness of the EOS for each individual nuclear matter EOS. Each nuclear matter EOS is capable of sustaining a different percentage of dark matter $F_\chi$. For example, the softest EOS3 can sustain only up to $\approx$ 10% of dark matter. whereas EOS1, EOS2 and EOS4 can sustain $\approx$ 24 %, 14% and 22% of dark matter respectively. Therefore, depending on the stiffness of the employed EOS, we can observe variations in the percentage of dark matter fraction $F_\chi$ranging from 0% to 25%.

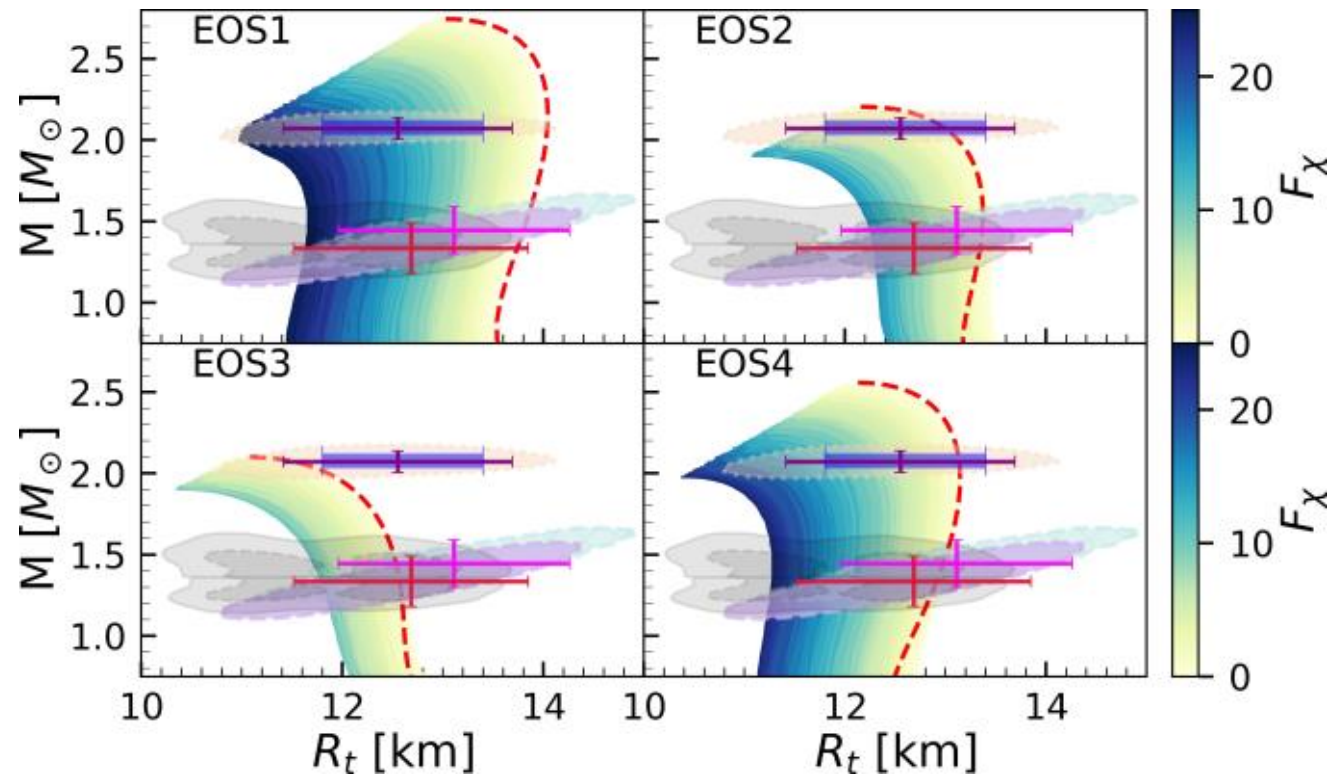

Figure 3.6: The domain of neutron star (NS) mass-radius using a two-fluid scenario, considering our entire set of dark matter EOSs in conjunction with different nuclear EOSs. The vertical color bar with the panel depicts the dark matter mass fraction $F_\chi$ from 0 to 25 %. The dashed lines on each plot correspond to NS properties computed using only nuclear EOS in single fluid TOV, without dark matter. We compare the M-R domains with current observational constraints. The gray region depicts the constraints from the binary components of GW170817 Abbott *et al.* [2019], along with their 90% and 50% credible intervals(CI). The 1$\sigma$(68%) CI for the 2D posterior distribution in the mass-radii domain for millisecond pulsar PSRJ0030+0451 (cyan and yellow) Miller *et al.* [2019], Riley *et al.* [2019] as well as PSRJ0740+6620 (violate) Miller *et al.* [2021], Riley *et al.* [2021] from the NICER x-ray data are also shown.

In Figure 3.6, we demonstrate the outcomes of mass-radius calculations obtained through a collection of EOSs ensemble. The upper panels illustrate the results for EOS1 and EOS2, whereas the lower panels correspond to EOS3 and EOS4 respectively. The vertical color bar, which ranges from 0% to 25%, visually illustrates the spread of dark matter mass fractions. It provides insight into the resulting variations in the mass-radius curve influenced by the parameter $F_\chi$. Interestingly, when the percentage of $F_\chi$ increases, it noticeably leads to a decrease in the maximum mass of neutron stars. The dashed lines present in each plot correspond to the properties of neutron stars computed exclusively based on the nuclear EOS, without taking into account the influence of dark matter. To assess the validity of our findings, we compare them with recent observational constraints, represented by skin lines. These constraints encompass the binary components of GW170817 Abbott *et al.* [2019], along with their corresponding 90% and 50% credible intervals (CI). Furthermore, we illustrate the 1$\sigma$ (68%) CI for the two-dimensional posterior distribution in the mass radii domain obtained from NICER x-ray data for the millisecond pulsars PSRJ0030+0451 (cyan and yellow) and PSRJ0740+6620 (violet). The horizontal (radius) and vertical (mass) error bars reflect the $1\sigma$ credible interval derived from the 1-dimensional marginalized posterior distribution of the same NICER data. From this figure, one may observe that as the percentage of the dark matter component increases, both the mass and radius decrease for different neutron star mass sequences. It is worth noting that the current observational constraints on mass and radius, whether from NICER or GW observations, are not able to precisely determine the dark matter fraction $F_\chi$. Therefore, using the robust investigation presented in this Fig.3.6, we suggest that the dark matter fraction can be as high as 25% when 1.9 $M_\odot$ NS maximum mass constraint is imposed.

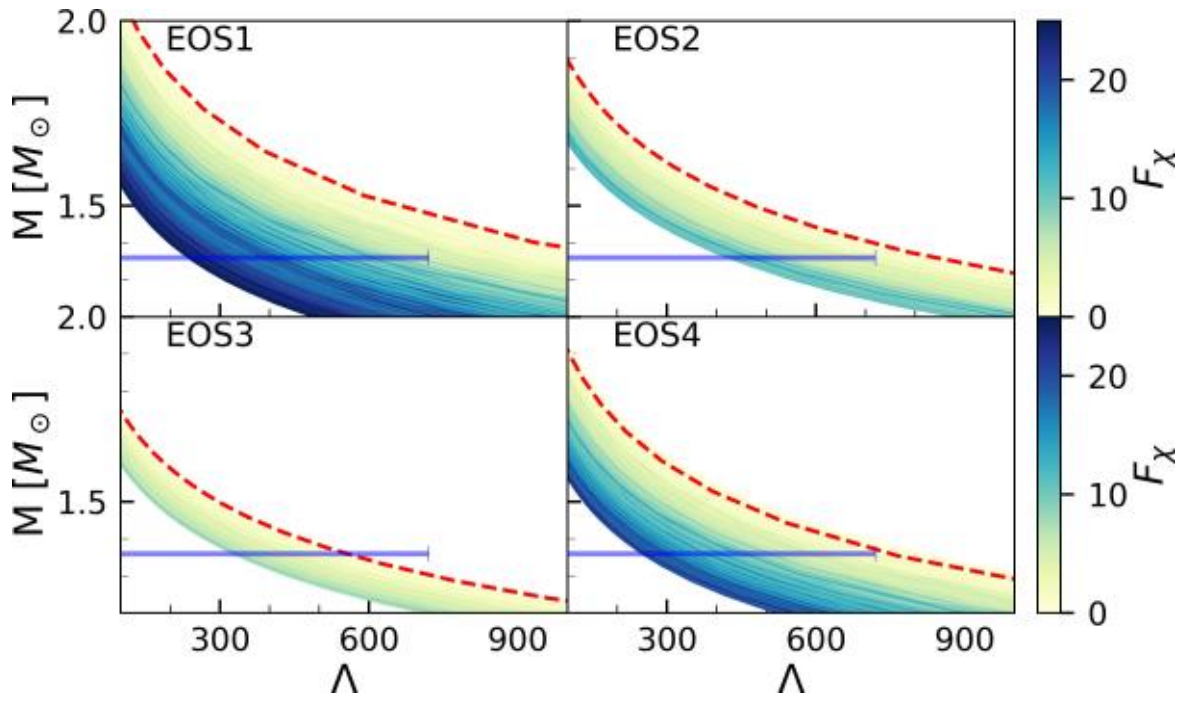

Figure 3.7: The graphical representation illustrates the relationship between tidal deformability (Λ) and the mass ($M_\odot$) of neutron stars (NS). Meanwhile, the dashed lines depicted in each plot correspond to the computed properties of neutron stars solely based on the nuclear EOS in single fluid TOV, incorporating the observational constraint (blue bars) depict the tidal deformability at 1.36 $M_\odot$ Abbott *et al.* [2019]. Additionally, the color gradient on the side denotes the proportional dark matter mass fraction $F_\chi$, encompassing values from 0% to 25%.

Figure 3.7 illustrates the relation between the dimensionless tidal deformability (Λ) and the mass of neutron stars (NS) for different nuclear equation-of-states (EOSs) in separate panels. The dashed lines represent the properties of NS computed solely using the nuclear EOS in a single fluid TOV calculation, excluding the presence of dark matter. The color bar on the side indicates the dark matter mass fraction ($F_\chi$), with the color tone varying from yellow to blue, representing DM mass fractions ranging from 0% to 25%. From the figure, it is clear that the inclusion of dark matter leads to a decrease in tidal deformability for all masses, the same as obtained in the previous figure for the radius. As can be seen from the figure, dimensionless tidal deformability of different NS masses was negatively correlated with dark matter mass fraction $F_\chi$. The inclusion of observational constraints from GW170817 is represented by the blue bars, depicting the tidal deformability at 1.36 $M_\odot$, $\tilde{\Lambda}_{1.36} < 720$, Abbott *et al.* [2019]. The inclusion of dark matter could potentially lead to a reduction in the higher tidal deformability attributed to the stiff nuclear EOS. This similarity holds even for bosonic dark matter when employing a two-fluid approach, as demonstrated in previous studies Refs. Ellis *et al.* [2018], Ivanytskyi *et al.* [2020], Karkevandi *et al.* [2022], Rutherford *et al.* [2023]. The same behavior was obtained with a fermionic dark matter model based on RMF description incorporating short-range correlations within a single fluid approach Lourenço *et al.* [2022], and with the linear sigma-omega fermionic dark matter model together with a two-fluid approach Das *et al.* [2022].

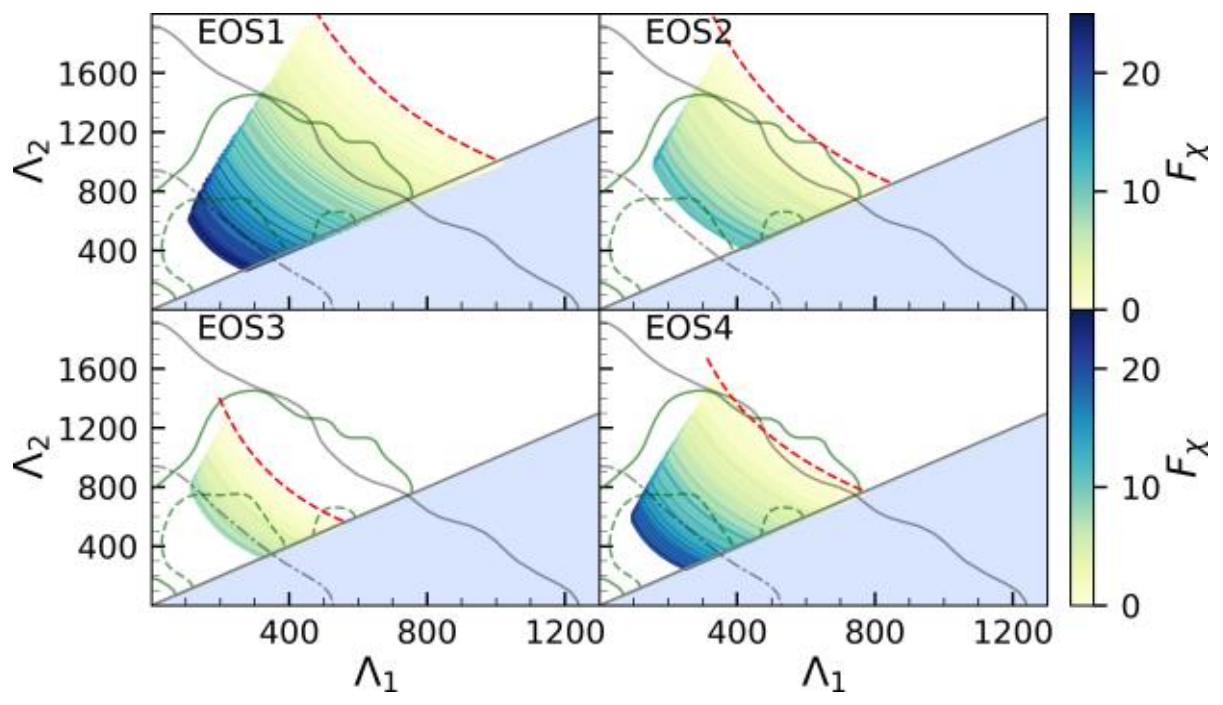

Figure 3.8: The graphical representation of $\Lambda_1$, and $\Lambda_2$ with a fraction of dark matter $F_\chi$ from 0 to 25%, where $\Lambda_1$, and $\Lambda_2$ are the dimensionless tidal deformability parameters of the binary neutron star merger from the GW170817 event, using the observed chirp mass of $M_{\text{chirp}} = 1.186\ M_\odot$. The green and gray solid (dashed) lines represent the 90% (50%) CI from the marginalized posterior for the tidal deformabilities of the two binary components of GW170817 using a parametrized EOS, with (green) and without (gray) a maximum mass of 1.97 $M_\odot$ requirement.

Fig.3.8 illustrates the representation of $\Lambda_1$ and $\Lambda_2$, the dimensionless tidal deformability parameters obtained with nuclear matter EOS: EOS1, EOS2, EOS3, and EOS4 for the binary neutron star merger event GW170817. For this calculation, we have fixed the chirp mass ($M_{\text{chirp}}$) 1.186 $M_\odot$ which is observed in the GW170817 event. In the plot, for the comparison we have included the constraints in the gray solid (dashed) line corresponding to the 90% (50%) confidence interval (CI) obtained from the marginalized posterior, which represents the tidal deformability of the two binary components of neutron star merger event GW170817. Furthermore, the green solid (dashed) lines depict the 90% (50%) CI derived from the marginalized posterior, indicating the tidal deformability of the two binary components of GW170817 based on an equation-of-state which is parameterized with a requirement that the maximum mass of at least 1.97 $M_\odot$. Here it can be seen that, for EOS1 in the absence of any dark matter component lies outside the boundary of observational constraints, but in the presence of dark matter as $F_\chi$ increases this comes inside the boundaries which results that the stiff nuclear EOS with admixed dark matter, comes inside the boundaries defined by the constraints on tidal deformability.

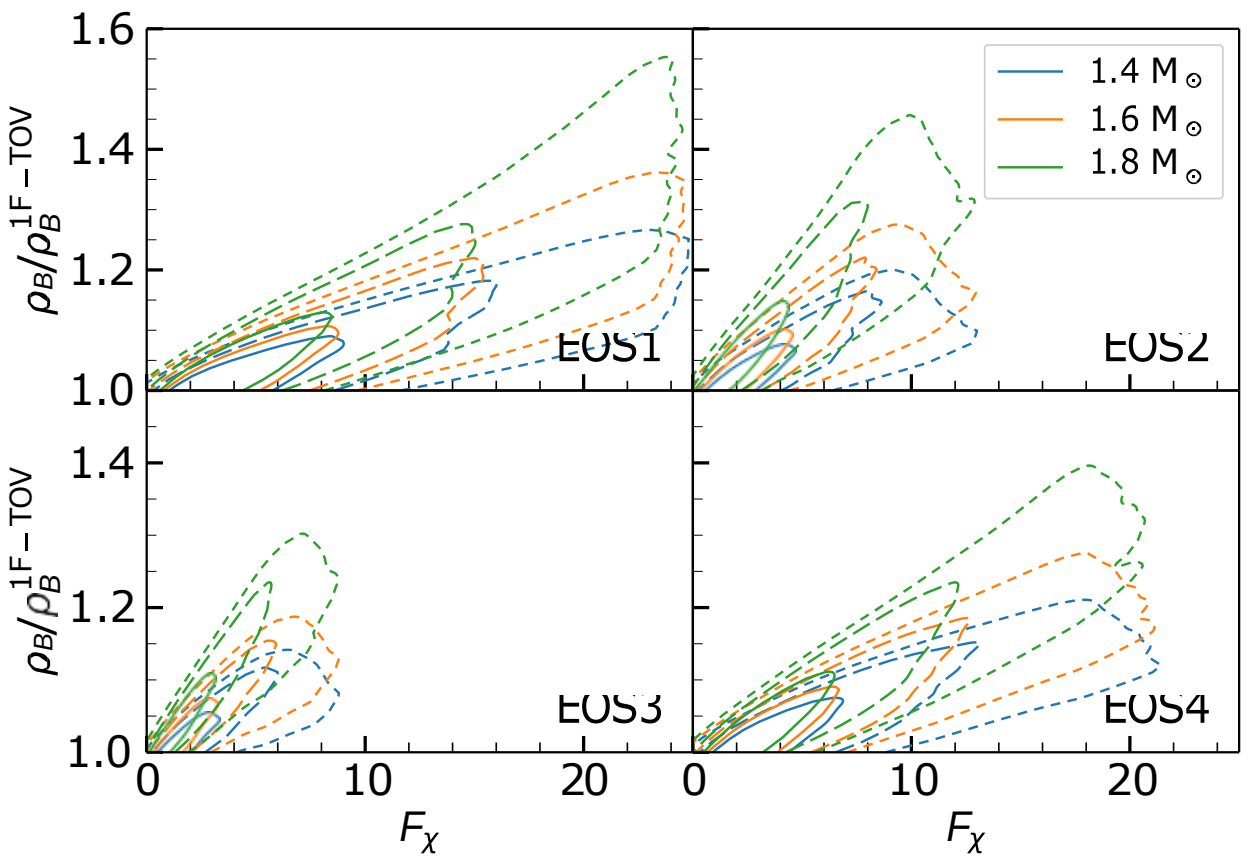

Figure 3.9: The central baryonic density $\rho_B/\rho_B^{1F-TOV}$ as a function of dark matter mass fraction $F_\chi$ for NS masses equal to 1.4 $M_\odot$(blue), 1.6 $M_\odot$(orange), 1.8 $M_\odot$(green). The top panels are for EOS1 and EOS2, whereas the bottom panels are for EOS3 and EOS4. The solid line in each panel represents 68% confidence interval whereas dashed and dotted lines represent 95% and 99%, respectively.

As it is discussed in Fig. 3.5 EOS2 and EOS3 can only sustain upto $\approx$ 14% and 10% of dark matter respectively, if the 1.9 $M_\odot$ constraint is imposed. As a consequence, the acceptable $\Lambda_1$-$\Lambda_2$ domain is quite small, whereas for EOS1 and EOS4 the domain is wider because these EOS can sustain, respectively, 24% and 22% of dark matter.

In Figure 3.9, we investigate the effect of DM on the NS central density. The main effect is the compression of matter inside a star which results in a decrease of the NS radius as the fraction of DM increases and the gravitation mass is kept constant. We consider all four equation-of-states (EOSs) to examine how dark matter affects this compression. The plot shows the scaled central density $\rho_B/\rho_B^{1F-TOV}$ of stars where $\rho_B^{1F-TOV}$ represents the central baryon density in the absence of dark matter (single fluid TOV) for masses 1.4, 1.6, and 1.8 $M_\odot$ as a function of the percentage of dark matter $F_\chi$. The solid line in each panel represents 68% confidence interval whereas dashed and dotted lines represent 95% and 99% CI, respectively. From the figure, it is evident that, for each EOS, as the percentage of dark matter ($F_\chi$) increases, the central density for masses ranging from 1.4 to 1.8 $M_\odot$ increases in all cases.

In Fig. 3.10, we plot the density profile of 1.4$M_\odot$ NS with different fractions of DM. The presence of DM increases the gravitational interaction at the star center. As a consequence, mass is pushed to the center, the central baryonic density increases, and the radius of the star decreases. This is an interesting result because it indicates that due to the presence of DM, processes that are otherwise not favorable are now allowed, e.g., the onset of hyperons or of the nucleonic direct Urca, etc. may open in smaller mass stars with the DM presence. The increase in central baryon density due to the compression of matter may also give rise to quark hadron phase transition inside the core of dark matter admixed neutron stars. Model calculations indicate that onset densities for hadron-quark pasta phases and pure quark matter phase can be of the order of 0.4-0.7 fm$^{-3}$ Ju *et al.* [2021]. These results also imply that the accumulation of dark matter inside neutron stars can trigger the QCD phase transition. This is a novel but model-dependent result and it needs further detailed studies.

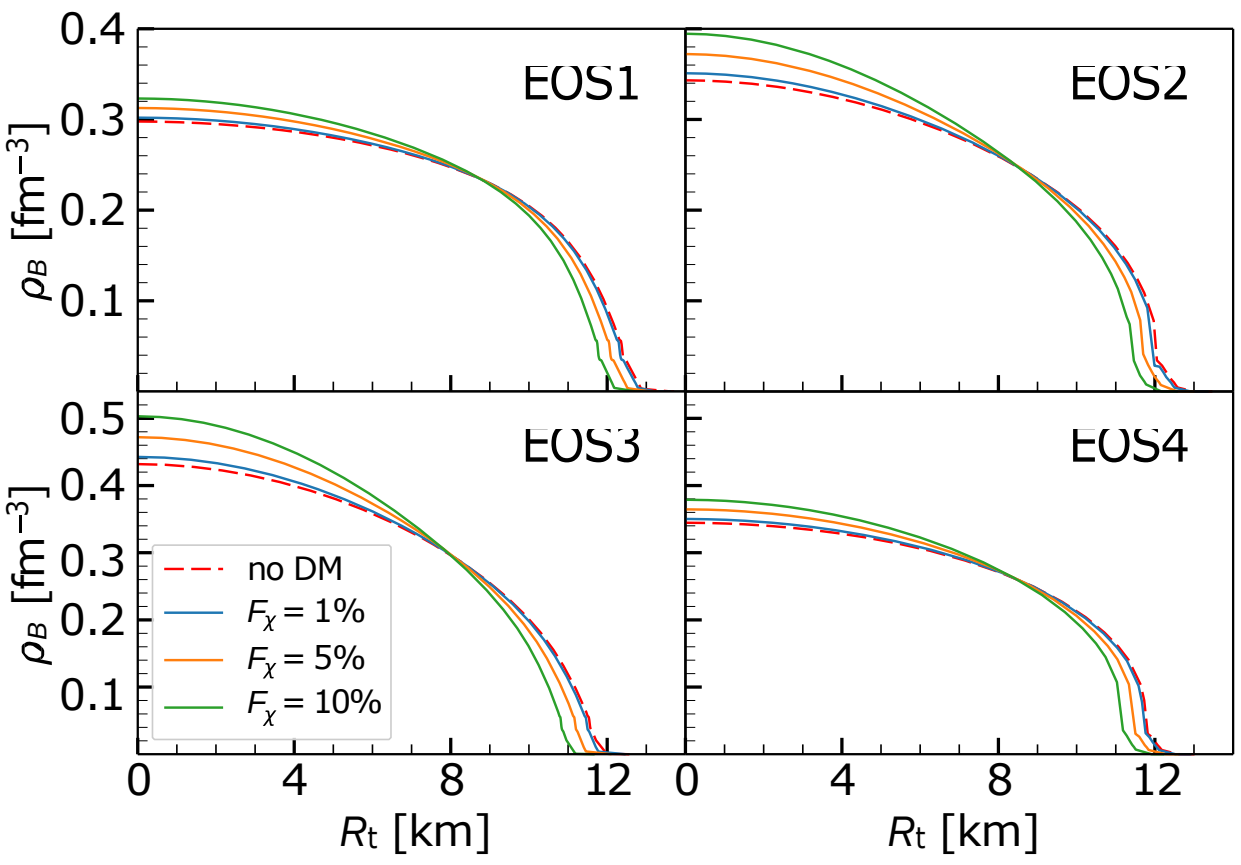

Figure 3.10: The baryon density $\rho_B$ plotted against the NS radius $R_t$ for a 1.4 $M_\odot$ NS. The plot includes four panels, each representing a different nuclear EOS, along with a generic dark matter EOS.

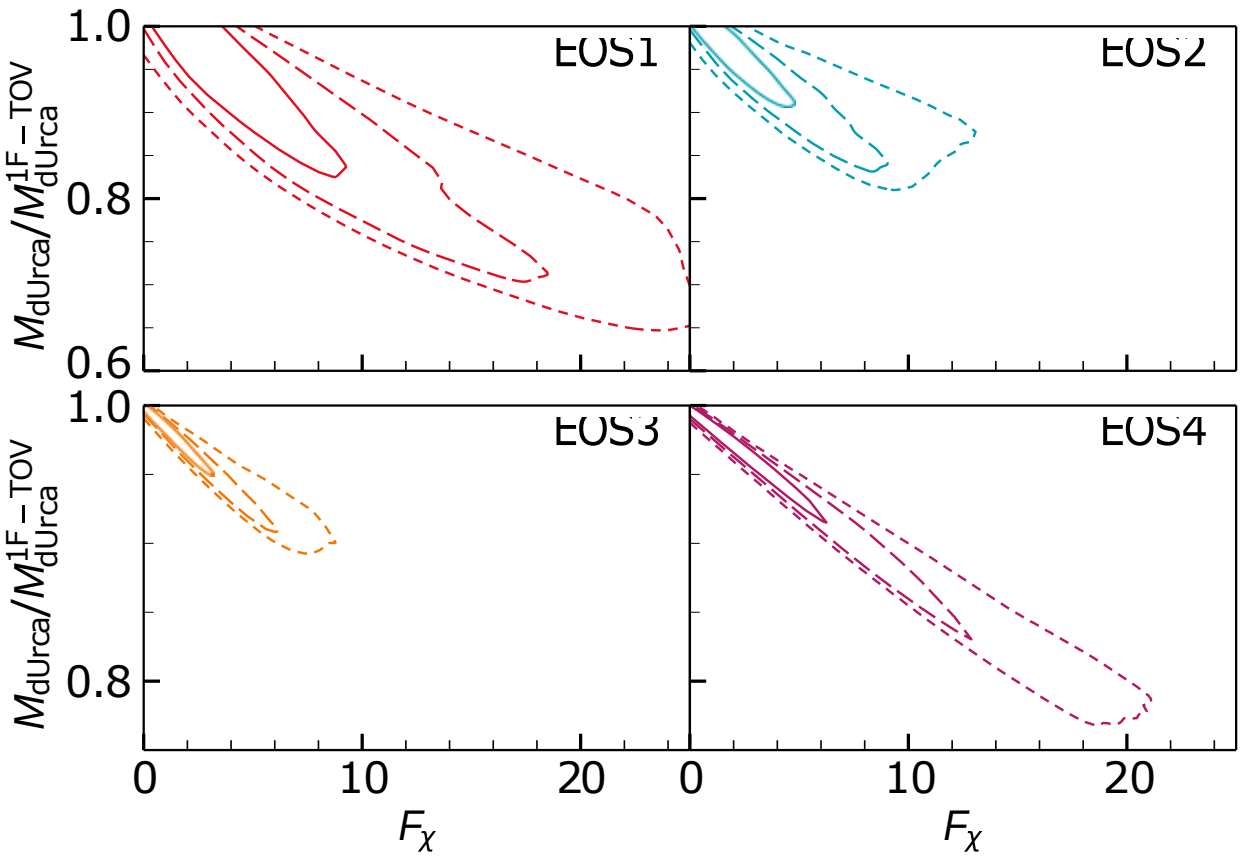

Figure 3.11: The $M_{\rm dUrca}/M^{\rm 1F-TOV}_{\rm dUrca}$ as a function of dark matter mass fraction $F_\chi$. The top panels are for EOS1 and EOS2, whereas the bottom panel represents results for EOS3 and EOS4. The solid line in each panel represents 68% confidence interval whereas dashed and dotted lines represent 95% and 99% CI, respectively.

In the following, we discuss how the presence of dark matter leads to a decrease in the mass of the star where nucleonic direct Urca processes start to occur, which we designate as $M_{\rm dUrca}$. In Figure 3.11, the scaled Urca mass $M_{\rm dUrca}/M^{\rm 1F-TOV}_{\rm dUrca}$ is plotted as a function of the fraction of dark matter, where $M^{\rm 1F-TOV}_{\rm dUrca}$ represents the Urca mass with single fluid TOV, obtained for the four nuclear matter EOSs. A strong correlation between the dark matter fraction and the Urca mass for each individual case is observed. Notice that the decrease in the $M_{\mathrm{dUrca}}$ can be large for stiff EOSs. This is because stiff EOSs can allow a large dark matter fraction inside the neutron stars.

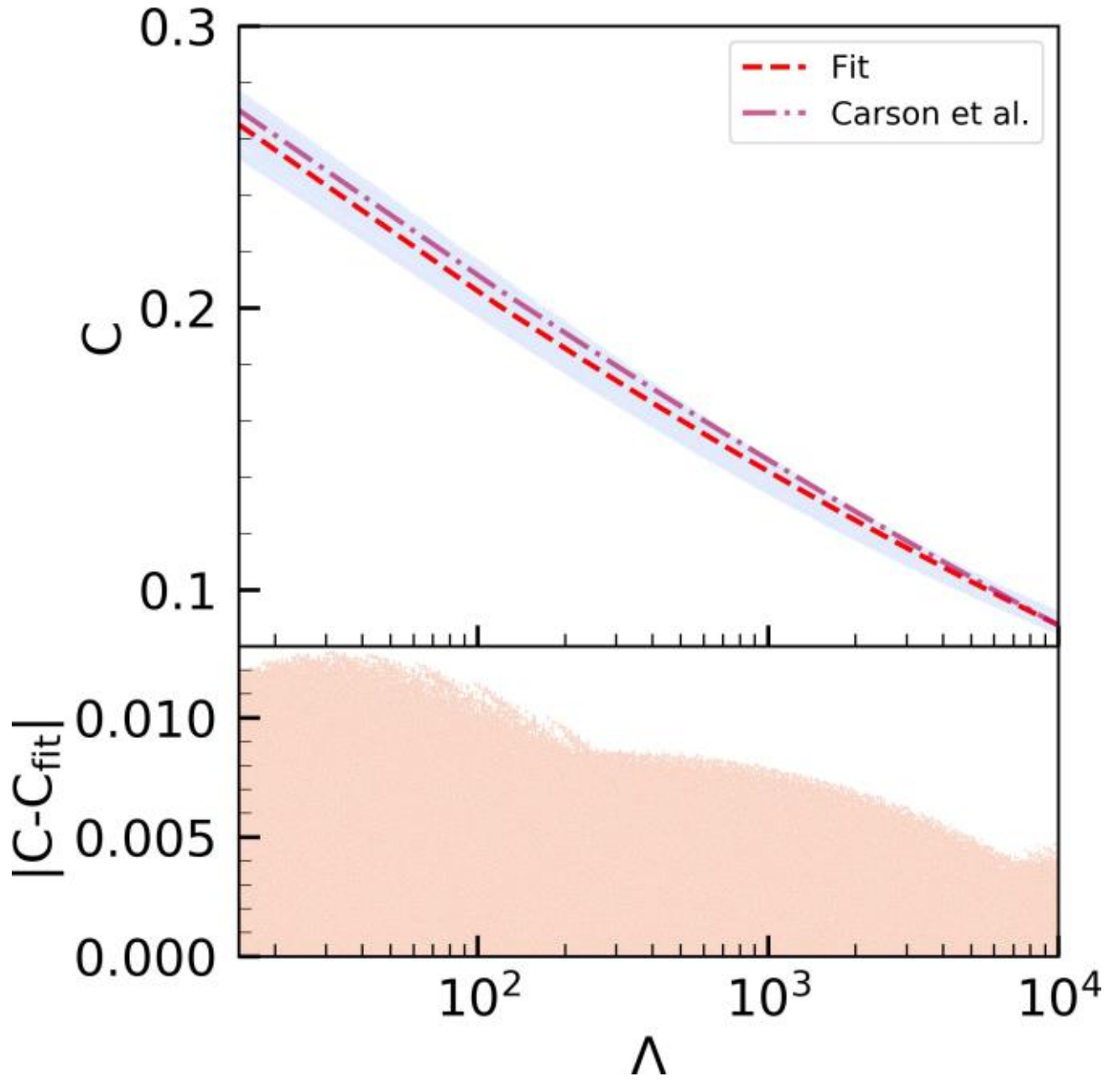

Figure 3.12: C-Love universal relation for all EOS. The red line is fitted with Eq. 3.1 In the lower panel the residuals for the fitting are calculated. Furthermore, we conduct a comparison with the findings presented in Ref. Carson *et al.* [2019] for the scenario of single fluid TOV without dark matter (highlighted in pink-red).

In a previous study conducted by Malik et al., Malik *et al.* [2022a], it was argued that the Urca mass exhibits a robust correlation with nuclear symmetry energy. The presence of dark matter may, however, affect our perception of the central baryonic density, resulting in a wrong estimation of the proton density, in particular, a larger proton fraction, and, therefore larger nuclear symmetry energy. To gain a comprehensive understanding of these phenomena, further investigations are required in the future.

There has been a large interest in the finding of universal relations that involve several NS properties and are independent of the NS mass, see for instance Carson *et al.* [2019], Maselli *et al.* [2013], Yagi & Yunes [2013, 2017]. Although universal relations for neutron stars are inherently insensitive to the EOS and, therefore, cannot be utilized to differentiate between different EOS models, they hold the potential to serve as powerful tools for inferring the properties of neutron stars in connection with other measurements. Besides, if a particular NS composition breaks the universal relation, this feature may be considered a smoking gun to identify that special composition. We may, therefore, question whether DM will break some of the known universal relations. In the following, we will analyze the universal relation $C - \Lambda$ proposed in Maselli *et al.* [2013], and discussed in Carson *et al.* [2019]. The C-Love relationships are depicted in Figure 3.12. We begin by fitting the data related to EOS to a simple curve represented by the equation:

$$C = \sum_{k=0}^{2} a_k (\ln \Lambda)^k. \tag{3.1}$$

By performing this fitting process, we obtain the following values for the coefficients: $a_0 = 0.36054566$, $a_1 = -0.0375908$, and $a_2 = 0.00086283$. In the lower panel of Fig. 3.12, the absolute difference from the fits is displayed. The absolute difference is approximately equal to 1%, and furthermore, it decreases to about 0.5% for higher values of $\Lambda$. We have also compared our results with Figure 4 in Ref. Carson *et al.* [2019], where they used $a_0 = 0.3617$, $a_1 = -0.03548$, and $a_2 = 0.0006194$ for the constrained EOS. This constrained EOS shows an absolute difference of less than 1%, while the unconstrained EOS is around 1%. We conclude that DM does not break the universal relation $C - \Lambda$.

## 3.7 Conclusion

In conclusion, this study has illuminated the connection between dark matter and neutron star properties, accounting for the uncertainties in the equation-of-state within the baryonic sector. We considered dark matter as being composed of fermionic particles with masses of a few GeVs, interacting with a dark vector meson. Using a two-fluid scenario and sampling 50,000 dark matter EOSs, we analyzed the structure of dark matter-admixed neutron stars, assuming dark matter is confined within the visible radius of neutron stars, i.e., only no-halo configurations were studied. We imposed that the dark-matter admixed star should have a maximum mass above 1.9 $M_\odot$, a value within $3\sigma$ of the PSR J0348+0432 mass, $2.01 \pm 0.04\ M_\odot$. The results revealed interesting correlations between dark matter parameters and various neutron star properties, consistent with literature findings: the larger the fraction of dark matter, the smaller the maximum mass, radius, and tidal deformability of the neutron star. In particular, the dark matter mass fraction within a neutron star was found to have a strong negative correlation with its maximum gravitational mass if a single nuclear model was considered. However, this correlation disappears when accounting for the uncertainties associated with the nuclear matter EOS. The maximum mass constraints of dark-matter admixed neutron stars depend on the softness or stiffness of the nuclear matter EOS employed, with some EOSs being able to sustain a significant fraction of dark matter and still describe approximately two-solar-mass stars.

The inclusion of dark matter led to a decrease in the radius and tidal deformability for all masses, indicating its influence on the structural characteristics of neutron stars. Through the analysis of various observational constraints and data, we demonstrated the potential of dark matter to affect the compression and central energy density of baryonic matter inside neutron stars. The presence of dark matter induces nuclear matter compression, resulting in a larger central baryonic density, smaller radius, and thinner crust. The increased central baryonic density has important consequences for neutron star properties: it favors the onset of non-nucleonic degrees of freedom in less massive stars and may affect the onset of direct Urca processes, influencing neutron star cooling. Our study has particularly highlighted the impact of dark matter on the cooling process and nuclear symmetry energy of baryonic matter. The detection of stars with similar masses but different surface temperatures could indicate that the cooler ones are dark-matter admixed stars.

Table 3.4: Summary of Neutron Star Properties and their Corresponding Values obtained from Filtered Two-Fluid Solutions with EOS1, EOS2, EOS3, and EOS4. The table presents the median values, along with the lower and upper bounds of the 90% confidence intervals (CI), for various neutron star properties. These properties include the maximum mass of neutron stars ($M_{max}$), the total radius ($R_{t,x}$) for values of $x$ in the range of [1.2, 1.4, 1.6, 1.8, 2.0], the dimensionless tidal deformability ($\Lambda_x$) for $x$ in the range of [1.4, 1.6, 1.8], the fraction of dark matter energy density over nuclear matter ($f_{d,x}$), and the nuclear matter baryon density ($\rho_{B,x}$) for $x$ in the range of [1.4, 1.6, 2.0]. The values are based on comprehensive analyses using EOS1, EOS2, EOS3, and EOS4 as the equation-of-state models for neutron stars.

| NS | Units | EOS1 | | | EOS 2 | | | EOS 3 | | | EOS 4 | | |
|---|---|---|---|---|---|---|---|---|---|---|---|---|---|
| | | median | 90 % CI | | median | 90 % CI | | median | 90 % CI | | median | 90 % CI | |
| | | | min | max | | min | max | | min | max | | min | max |
| $M_{max}$ | $M_{\odot}$ | 2.454 | 2.067 | 2.710 | 2.107 | 2.067 | 2.710 | 2.051 | 2.067 | 2.710 | 2.351 | 2.067 | 2.710 |
| $R_{t,2.0}$ | km | 13.16 | 11.77 | 13.87 | 12.68 | 11.77 | 13.87 | 11.50 | 11.77 | 13.87 | 12.50 | 11.77 | 13.87 |
| $R_{t,1.8}$ | km | 13.16 | 12.00 | 13.81 | 13.00 | 12.00 | 13.81 | 12.04 | 12.00 | 13.81 | 12.56 | 12.00 | 13.81 |
| $R_{t,1.6}$ | km | 13.11 | 12.04 | 13.73 | 13.13 | 12.04 | 13.73 | 12.27 | 12.04 | 13.73 | 12.54 | 12.04 | 13.73 |
| $R_{t,1.4}$ | km | 13.02 | 12.01 | 13.63 | 13.17 | 12.01 | 13.63 | 12.40 | 12.01 | 13.63 | 12.48 | 12.01 | 13.63 |
| $R_{t,1.2}$ | km | 12.92 | 11.94 | 13.53 | 13.18 | 11.94 | 13.53 | 12.48 | 11.94 | 13.53 | 12.40 | 11.94 | 13.53 |
| $\Lambda_{1.8}$ | ... | 18 | 7 | 29 | 16 | 7 | 29 | 9 | 7 | 29 | 13 | 7 | 29 |
| $\Lambda_{1.6}$ | ... | 36 | 17 | 57 | 36 | 17 | 57 | 23 | 17 | 57 | 28 | 17 | 57 |
| $\Lambda_{1.4}$ | ... | 76 | 39 | 114 | 82 | 39 | 114 | 58 | 39 | 114 | 62 | 39 | 114 |
| $f_{d,2.0}$ | ... | 0.47 | 0.14 | 0.68 | 0.27 | 0.14 | 0.68 | 0.20 | 0.14 | 0.68 | 0.40 | 0.14 | 0.68 |
| $f_{d,1.6}$ | ... | 0.43 | 0.12 | 0.65 | 0.26 | 0.12 | 0.65 | 0.19 | 0.12 | 0.65 | 0.37 | 0.12 | 0.65 |
| $f_{d,1.4}$ | ... | 0.41 | 0.11 | 0.63 | 0.24 | 0.11 | 0.63 | 0.18 | 0.11 | 0.63 | 0.34 | 0.11 | 0.63 |
| $\rho_{B,2.0}$ | $fm^{-3}$ | 0.408 | 0.350 | 0.576 | 0.589 | 0.350 | 0.576 | 0.788 | 0.350 | 0.576 | 0.489 | 0.350 | 0.576 |
| $\rho_{B,1.6}$ | $fm^{-3}$ | 0.339 | 0.303 | 0.394 | 0.399 | 0.303 | 0.394 | 0.500 | 0.303 | 0.394 | 0.395 | 0.303 | 0.394 |
| $\rho_{B,1.4}$ | $fm^{-3}$ | 0.314 | 0.283 | 0.354 | 0.352 | 0.283 | 0.354 | 0.435 | 0.283 | 0.354 | 0.361 | 0.283 | 0.354 |

Additionally, the study explored universal relations, known as the C-Love relationships, which provide insights into neutron star properties that are not easily measurable. We verified that within 1%, the C-Love universal relation was not broken by the presence of dark matter, confirming previous findings.

Overall, this research contributes to our understanding of the complex interplay between dark matter and neutron star properties. By uncovering these connections, we are moving closer to unraveling the mysteries concealed within neutron stars.

# Chapter 4

# Aspects of Dark Matter in Neutron stars: Machine Learning Appproach

In this chapter, we investigate the connection between dark matter and neutron star properties by utilizing advanced machine learning techniques. Random Forest classifiers are applied to analyze neutron star characteristics, aiming to determine if these stars exhibit features indicative of dark matter admixture. The dataset comprises 32,000 simulated neutron star sequences, each with mass, radius, and tidal deformability inferred from recent observations and theoretical models. The study employs a two-fluid model for the neutron star, incorporating separate equations of state for nucleonic and dark matter, with the latter assuming a fermionic dark matter scenario. The classifiers are trained and validated using various feature sets, including tidal deformability for different masses. The performance of the classifiers is thoroughly assessed using confusion matrices, which show that neutron stars with admixed dark matter can be distinguished from nuclear matter neutron stars with a misclassification rate of approximately 17%. This chapter also explores the potential of certain neutron star properties as indicators of dark matter presence. Among these properties, radius measurements, particularly at extreme mass values, display considerable promise. The findings from our study are essential for guiding future observational strategies and improving dark matter detection capabilities in neutron stars. This research is the first to demonstrate that the radii of neutron stars at 1.4 and 2.07 solar masses, determined using NICER data from the pulsars PSR J0030+0451 and PSR J0740+6620, strongly indicate that the presence of dark matter in a neutron star is more likely than a purely hadronic composition.

## 4.1 Random forest

Random Forest Fawagreh *et al.* [2014], Pal [2005], Rigatti [2017] is a widely-used machine learning algorithm that falls under the category of supervised learning. It is versatile, suitable for both regression (predicting numeric outcomes) and classification (predicting categorical outcomes) tasks. Known for its flexibility and ease of use, a Random Forest consists of multiple decision trees that operate on different subsets of the provided dataset. By averaging the results of these trees, it enhances the predictive accuracy for the dataset.

### 4.1.1 Assumptions of Random Forest

To effectively leverage Random Forest, it's essential to grasp the key principles underlying the algorithm:

- **Independence of Trees**: Each decision tree in the forest should operate independently. This is ensured through the use of bootstrap sampling and the introduction of randomness in feature selection.
- **Large Data Requirements**: Random Forest needs a substantial amount of data to construct diverse trees and attain peak performance.
- **Adequately Deep Trees**: The algorithm presumes that each tree is grown deep enough to uncover the true patterns within the data.
- **Noise Management**: While Random Forest is capable of handling noisy data, it operates under the assumption that the noise is random rather than systematic.

### 4.1.2 How Random Forest Classification works?

Imagine you're trying to make an important decision about the best route to take for a road trip, and you ask several friends for their suggestions. Each friend has traveled different routes and has unique experiences and insights. After hearing all their advice, you decide to follow the route most frequently recommended by your friends.

In a random forest classification, this collaborative approach is mirrored by creating multiple decision trees using various random subsets of the data and features. Each decision tree, akin to one of your friends, provides its recommendation on how to classify the data. The final prediction is determined by aggregating the recommendations from all the decision trees and selecting the most frequently suggested classification. For regression tasks, instead of voting, the average of the predictions from all the trees is taken. Here’s a breakdown of its working process:

- **Building Multiple Decision Trees**: Random Forest creates multiple decision trees, where each tree operates independently and examines different parts of the data. This independence helps ensure that the model isn’t overly swayed by any single tree’s peculiarities.

- **Selecting Random Features**: To make each tree unique, Random Forest employs random feature selection. This means that during the training process, each tree only considers a random subset of the available features. This method ensures that each tree provides a different perspective on the data, contributing to the overall diversity of the model.

- **Bagging (Bootstrap Aggregating)**: A fundamental part of Random Forest is the use of bagging. This involves creating various subsets of the original dataset by sampling with replacement, meaning some data points might be chosen multiple times while others not at all. Each tree is trained on a different subset, which introduces variability and increases the model's robustness.

- **Prediction and Voting**: When making a prediction, each tree in the Random Forest independently casts a vote. In classification tasks, the final prediction is based on the majority vote among all the trees. For regression tasks, the final prediction is the average of all the tree outputs. This collective voting system helps ensure that the final decision is balanced and well-considered.

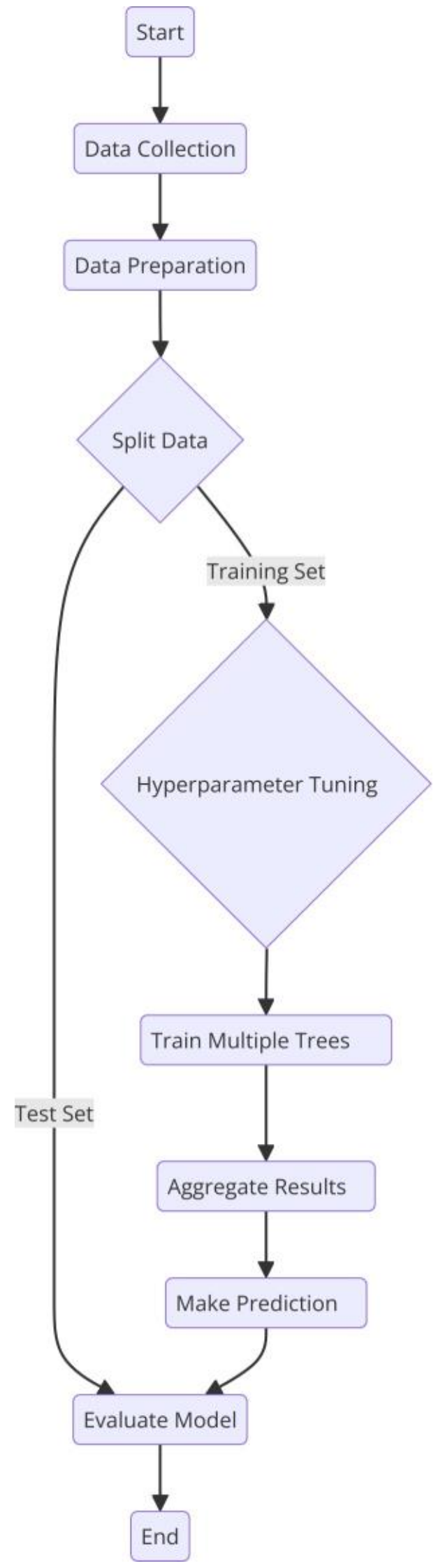

Figure 4.1: Flowchart of how Rnadom Forest Classification works

Through these methods, Random Forest effectively combines the strengths of multiple decision trees, resulting in a model that is more accurate and less prone to overfitting compared to individual decision trees. In fig. 4.1 we have shown a small flowchart representation, how random forest classification works.

### 4.1.3 Hyperparameter tuning in random forests

Hyperparameter optimization, or hyperparameter tuning, involves identifying the optimal set of hyperparameters for a machine learning model to enhance its performance on a specific task. These hyperparameters, which are configured before the training process starts, dictate various aspects of the learning algorithm's functioning and are not derived from the training data itself. If hyperparameter tuning does not occur, the model will produce errors and inaccurate results as the loss function is not minimized.

- **Grid Search:** Grid search is the simplest method for hyperparameter tuning. It involves constructing a model for every possible combination of the hyperparameter values provided, then evaluating each model to find the one that delivers the best performance. For instance, we might define a range of values to test for both `n_estimators` and `max_depth`, and grid search would create a model for each possible pairing of these values.

- **Random Search:** Unlike grid search, random search does not require a set list of values to test for each hyperparameter. Instead, it uses statistical distributions to randomly sample values for each hyperparameter. This approach allows for more varied and potentially more effective combinations to be tested.

Hyperparameter tuning is primarily driven by empirical testing rather than theoretical approaches. To find the optimal settings, it involves trying out various combinations and assessing the performance of each resulting model. Hyperparameters can be likened to the adjustable settings of an algorithm, similar to how one might adjust the temperature and time settings on an oven to bake a perfect cake. However, if models are evaluated only on the training set, it can lead to a critical issue in machine learning known as overfitting. An overfit model might perform exceptionally well on the training set but will be ineffective in real-world applications. Hence, the standard practice for hyperparameter optimization includes cross-validation to address overfitting.

Below is the list of some of the important random forest parameters:

- **max_depth**
  This parameter controls the maximum height that the trees within the forest can reach. It's crucial for enhancing the model's accuracy. As the tree depth increases, the model's accuracy also improves up to a point, after which it begins to decline due to overfitting. Therefore, setting an appropriate value for `max_depth` is essential to prevent overfitting.

- **min_sample_split**
  This parameter specifies the minimum number of observations required in a node for it to be split. The default value for `min_sample_split` is 2. This means that if a terminal node has more than two observations and is not pure, it can be further split into subnodes. However, having the default value set to 2 can lead to a problem where the tree continues to split until the nodes are entirely pure. This results in the tree growing excessively large and overfitting the data. Fig. 4.2 clearly shows how `min_sample_ split` works

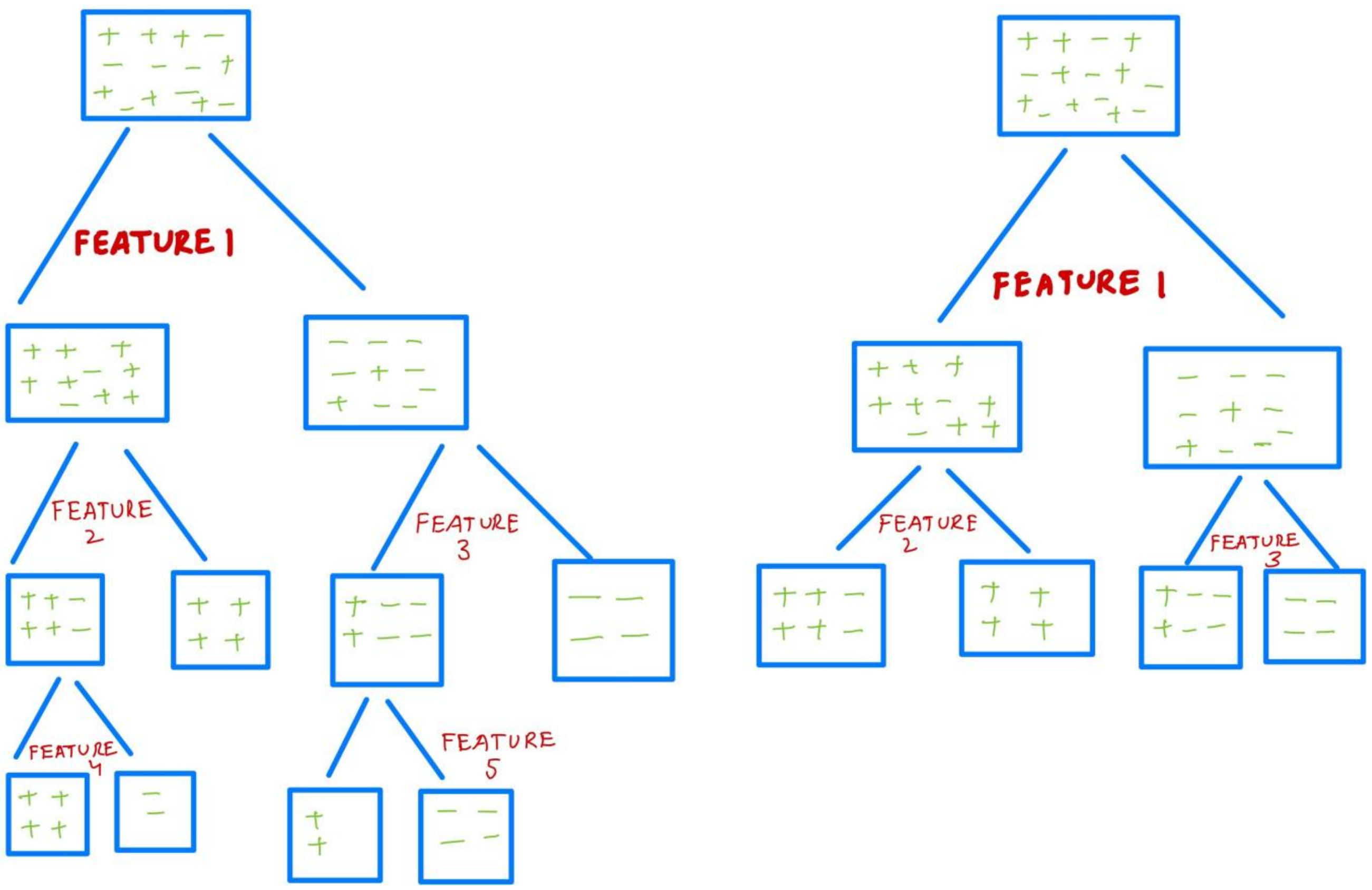

Figure 4.2: Example of how "min_sample_split" hyperparameter works. In this example, if we increase the min_sample_split from 2 to 6, the tree on the left would then look like the tree on the right.

- **max_terminal_nodes:** This hyperparameter controls the tree's node splitting, limiting the tree's expansion. If the number of terminal nodes exceeds the specified limit after a split, further splitting will cease, preventing additional tree growth.

- **min_samples_leaf:** The min_samples_leaf hyperparameter in Random Forest determines the smallest number of samples that must be present in a leaf node after a split. This prevents the creation of leaf nodes that are too small.

- **n_estimators:** The n_estimators hyperparameter in a Random Forest controls the number of decision trees used in the ensemble. While increasing the number of trees can improve generalization, it also raises the computational complexity. The default value in scikit-learn is 100.

- **max_samples:** The max_samples hyperparameter specifies the maximum number of samples to be drawn from the training dataset to train each individual tree in the Random Forest.

## 4.2 Prepration of datasets

To calculate the neutron star properties with admixed dark matter configurations, we employ a two-fluid approach, as discussed in sections 2.4.2.2 and 2.4.2.3. In this methodology, distinct equations of state (EOSs) are required for two separate fluids that interact solely through gravity, as described by the two-fluid equations. For the nuclear matter and dark matter components, we use the EOSs detailed in sections 2.1 and 2.4.2.1, respectively.

Our study encompasses a comprehensive dataset of 32,000 neutron star (NS) properties, including mass-radius and mass-tidal deformability curves. Of these, 16,000 are derived from models incorporating dark matter, while the remaining 16,000 are from models without dark matter. To facilitate the machine learning classification task, we structured our datasets accordingly. We create two features, labeled $X1$ and $X2$. Feature $X1$ includes mass and radius values, while $X2$ incorporates an additional element—tidal deformability. These values are generated using random uniform sampling, with mass and radius ranging from 1 to their respective maximum values. $X1$ consists of 10 elements, each containing 5 values of mass and radius. In contrast, $X2$ includes 5 tidal deformability measurements, resulting in a total of 15 elements. It is important to note that the target vector $Y$ contains a single element, set to zero for $X$ with only nuclear matter and 1 for admixed dark matter. In our study, we use the Random Forest classifier Breiman [2001], recognized as one of the most effective classification algorithms, to analyze our data. We begin by combining two datasets, each containing 16,000 entries, and thoroughly shuffle them to ensure uniformity in data distribution. The merged dataset is then divided into three parts: 60% for training, 20% for validation, and 20% for testing. The training set, comprising 60% of the data, is used to teach the Random Forest model the patterns and relationships inherent in the data. The validation set, 20% of the data, is crucial for fine-tuning the model. During this phase, we adjust the hyperparameters of the Random Forest classifier—configuration settings that structure the learning process and significantly impact the model's performance. By tweaking these parameters while monitoring the model's performance on the validation set, we aim to find the optimal configuration that yields the best results. This hyperparameter tuning is essential to prevent overfitting, where the model performs well on the training data but poorly on unseen data. The final 20% of the data, reserved as a test dataset, is used to evaluate the fully trained model's performance. This dataset is crucial as it represents new, unseen data for the model, providing a realistic assessment of its performance in real-world scenarios. These test data are not involved in the training or validation process, ensuring that our evaluation of the model's effectiveness is unbiased and reflective of its true predictive capabilities.

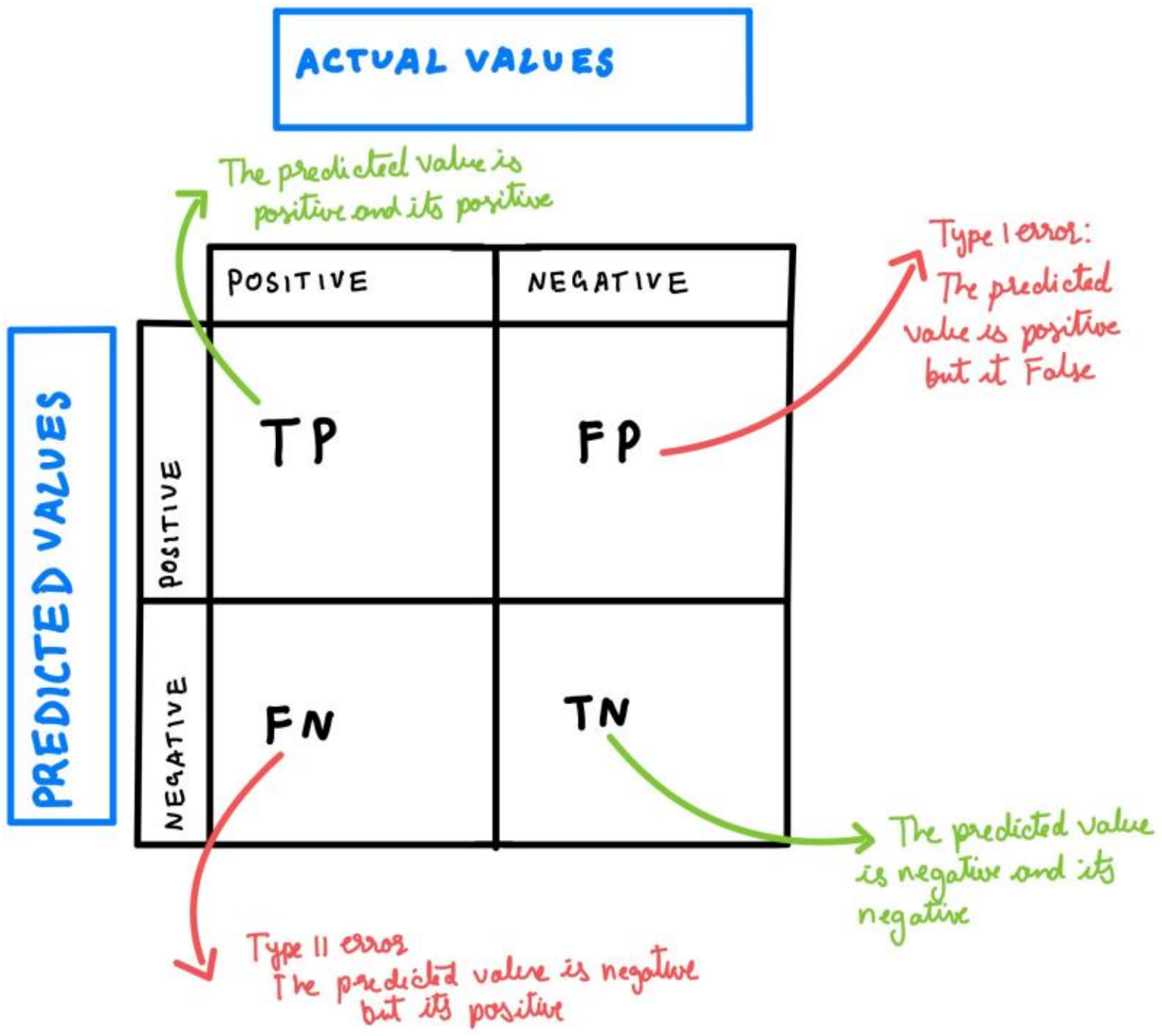

Figure 4.3: Visualization of Confusion Matrix

## 4.3 Confusion Matrices

A confusion matrix is a table that summarizes the performance of a machine learning model on a set of test data. It shows the number of correct and incorrect predictions made by the model. Confusion matrices are usually generated using test data rather than training data. This approach helps assess how well the model generalizes to new, unseen data, providing a more accurate measure of its real-world performance. A visual representation of the confusion matrix is depicted in Fig. 4.3, providing insight into the classification results.

### 4.3.1 Components of a Confusion Matrix:

The matrix shows the number of instances generated by the model on the test data.

- **True Positive (TP)**: The model correctly predicted a positive outcome (the actual outcome was positive).
- **True Negative (TN)**: The model correctly predicted a negative outcome (the actual outcome was negative).
- **False Positive (FP)**: The model incorrectly predicted a positive outcome (the actual outcome was negative). This is also called a Type I error.
- **False Negative (FN)**: The model incorrectly predicted a negative outcome (the actual outcome was positive). This is also called a Type II error.

**Why do we need a Confusion Matrix?**

- **Granular Insights**: Unlike a single metric such as accuracy, a confusion matrix provides detailed insights into how the model is performing across all classes. It breaks down the performance into true positives, true negatives, false positives, and false negatives.
- **Error Analysis**: By examining the types of errors (false positives and false negatives) the model is making, you can understand the specific areas where the model is struggling. This helps in diagnosing issues with the model and guiding further improvements.

### 4.3.2 Metrics based on Confusion Matrix Data

- **Accuracy**: The overall correctness of the model, calculated as:

$$\text{Accuracy} = \frac{TP + TN}{TP + TN + FP + FN}$$

- **Precision (Positive Predictive Value)**: The proportion of positive predictions that are actually correct, calculated as:

$$\text{Precision} = \frac{TP}{TP + FP}$$

- **Recall (Sensitivity or True Positive Rate)**: The proportion of actual positives that are correctly identified, calculated as:

$$\text{Recall} = \frac{TP}{TP + FN}$$

- **F1 Score**: The harmonic mean of precision and recall, providing a balance between the two, calculated as:

$$\text{F1 Score} = 2 \times \frac{\text{Precision} \times \text{Recall}}{\text{Precision} + \text{Recall}}$$

- **Specificity (True Negative Rate)**: The proportion of actual negatives that are correctly identified, calculated as:

$$\text{Specificity} = \frac{\text{TN}}{\text{TN} + \text{FP}}$$

## 4.4 Feature Importance

Feature importance in machine learning, specifically in random forests, refers to a technique that assigns a score to each feature based on how useful they are at predicting the target variable. This score indicates the relative importance of each feature in making predictions. Here's how feature importance is typically determined in random forests:

- **Mean Decrease in Impurity (Gini Importance)**
  This is the most common method. It measures the importance of a feature by the total decrease in node impurity (e.g., Gini impurity or entropy) brought by that feature, averaged over all the trees in the forest. Features that result in larger decreases in impurity when used in splits are considered more important.

- **Mean Decrease in Accuracy (Permutation Importance)**
  This method involves randomly shuffling the values of a feature and measuring how much the permutation decreases the model's accuracy. If shuffling a feature's values leads to a significant drop in accuracy, the feature is considered important. This approach helps in understanding the importance of a feature in the context of all other features.

Random forests leverage the aggregation of multiple decision trees, which helps in providing a more robust estimate of feature importance compared to single decision trees. Here's why feature importance is valuable:

- **Feature Selection:** Identifying the most important features can help in selecting a subset of features, potentially improving model performance and reducing overfitting.

- **Model Interpretation:** Understanding which features are most influential aids in interpreting the model and the underlying relationships in the data.

- **Data Insights:** It provides insights into the data, helping to uncover patterns and relationships that might not be obvious.

In practice, libraries such as Scikit-learn provide built-in functions to compute feature importance for random forest models.

## 4.5 Results

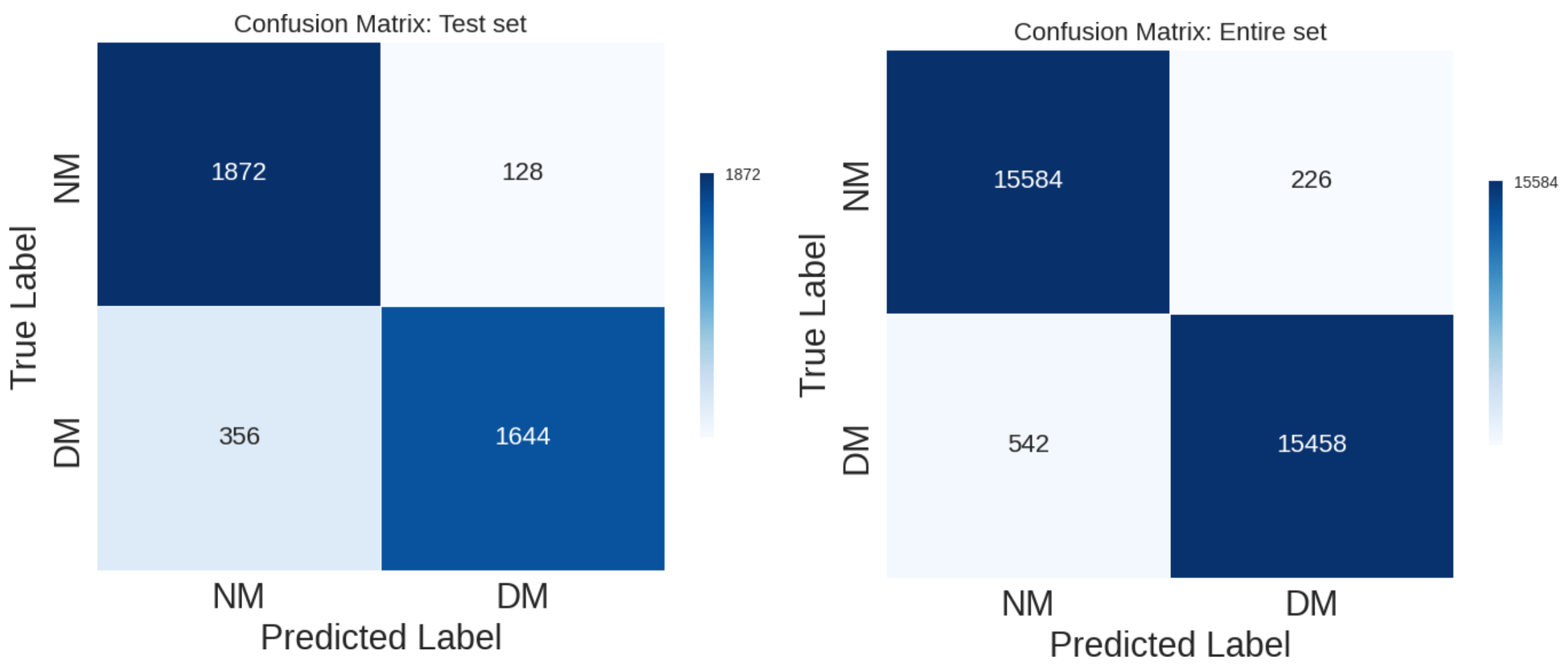

Figure 4.4: The confusion matrices displayed are for the test set and the entire set, respectively (model trained with $X1$ feature, i.e., NS mass and radius).

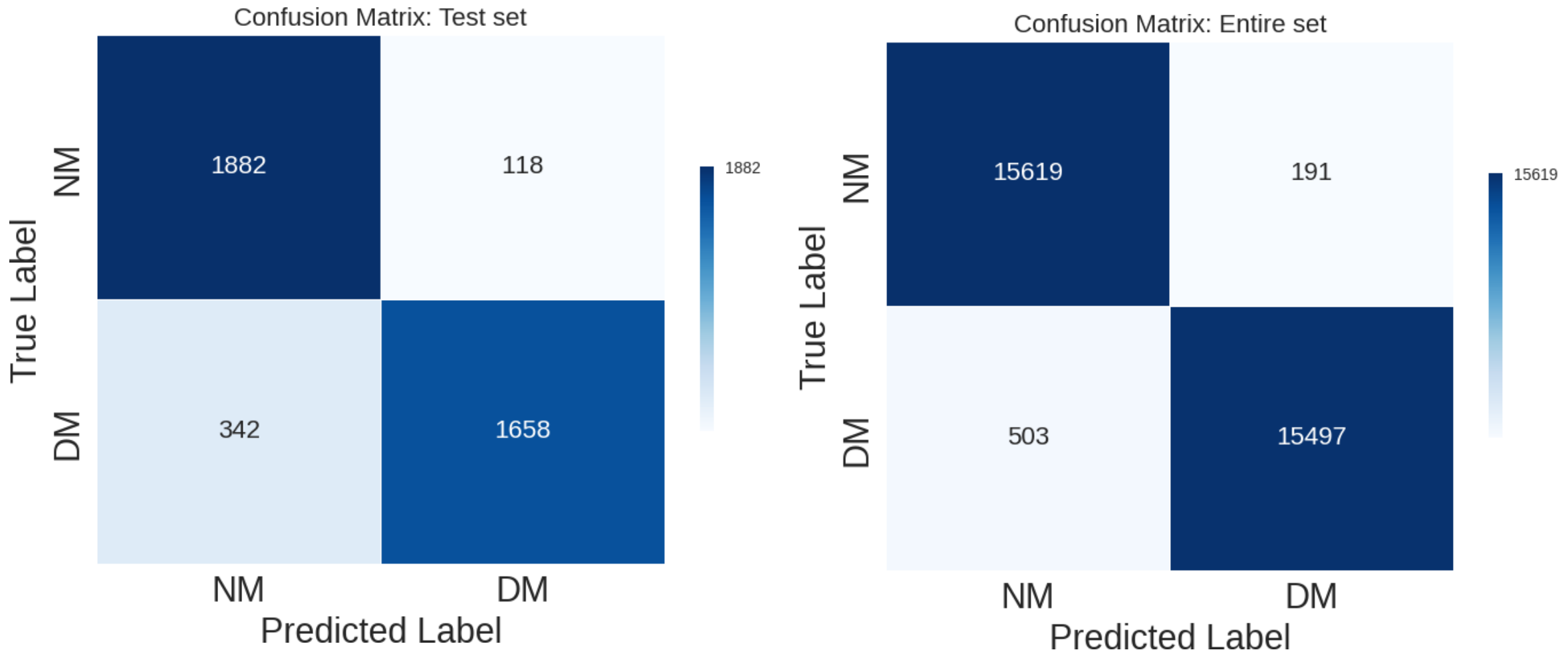

Figure 4.5: Same as shown in Figure 4.4, but with the trained model on the $X2$ feature; it has five additional features compared to the previous, which are the tidal deformability for five different masses.

In Figure 4.4, two panels represent confusion matrices for a Random Forest classifier applied to train model with only the $X1$ feature. The left panel shows the prediction of the trained model on the test set and the right panel shows the entire dataset. A confusion matrix is a table that is often used to describe the performance of a classification model on a set of data for which the true values are known. It cross-tabulates the actual class labels with the predicted class labels, providing insight into the accuracy and types of errors made by the classifier. In the left panel, the confusion matrix for the test set reveals that the classifier accurately predicted NS with only nuclear matter (NM) 1872 times and with admixed dark matter (DM) 1644 times. However, there were instances of misclassification, indicated by off-diagonal numbers: 128 instances of NM were incorrectly classified as DM and 356 instances of DM were incorrectly classified as NM. These errors highlight the instances where the classifier was challenged to distinguish between the two classes. Moving to the right panel, the confusion matrix for the entire dataset shows that the classifier correctly identified NM 15,584 times and DM 15,458 times. Misclassifications are also present here, with NM being mistaken for DM 226 times and vice versa 542 times. The percentage of false positive for NS with only nuclear matter (NM) is 6.4% (1.98%) and NS with admixed dark matter (DM) is 17.8% (4.98%) for the set of tests (entire). It should be noted that the test set contains data separated from those not involved in training, but the entire set contains data involved in training. Figure 4.5 shows a confusion matrix for the model trained in the $X2$ feature set, which includes five additional features: tidal deformability for five different mass categories. The purpose of introducing these additional features was to determine whether the inclusion of tidal deformability constraints could improve the accuracy of the predictions. The results show that the rate of false positives, incorrectly predicting NM as dark matter DM, is 5.9% for the test set and 1.7% for the entire dataset. On the other hand, the rate of false positives for neutron stars with admixed dark matter is 17.1% for the test set and 4.69% for the entire set. These figures suggest that the inclusion of tidal deformability data does not significantly enhance the predictive power of the model.

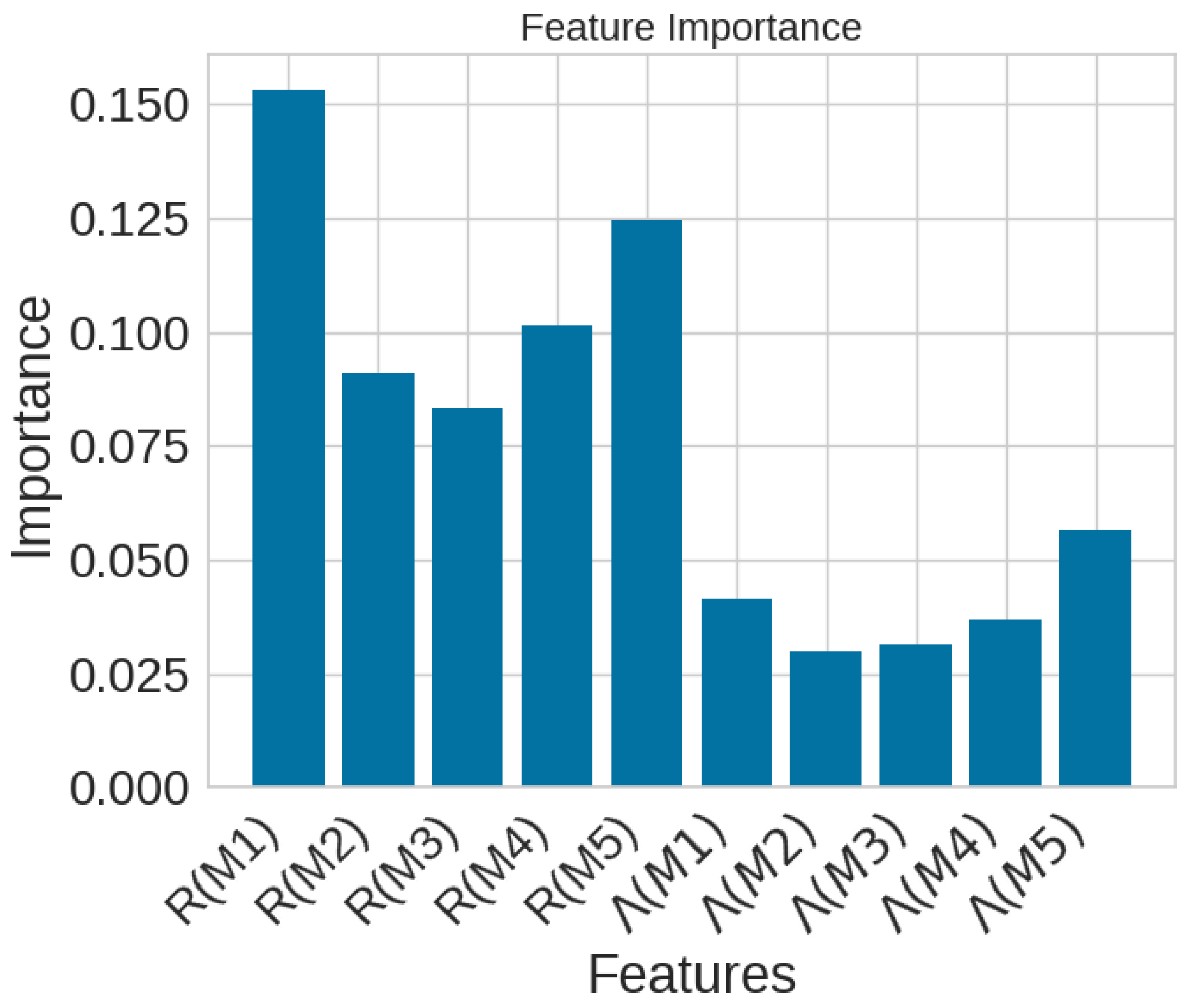

Figure 4.6: Feature importance for Random Forest classification from $X2$.

As Random Forest is a tree-based model, one can also extract the importance of different features (or the mean decrease in impurity) within a trained model in predicting the target variable. This is accomplished by monitoring the reduction in model precision when each feature is taken out of the model. To put it more simply, it shows the impact of each feature on the model's decision-making process, with higher numbers indicating a greater effect on the classification result. In Figure 4.6, we plot the importance of the feature that delineates the varying significance of the features employed by the Random Forest model in the context of classifying the existence of dark matter within neutron stars (NS). Radius measurements at various mass points ('$R(M1)$' through '$R(M5)$') dominate the feature importance, with the plot revealing that '$R(M1)$', the radius at a lower mass point, holds the greatest predictive power, followed by '$R(M4)$', a higher mass point. This pattern suggests that radius measurements, especially at the extremes of the mass range considered, are critical for the model to discern the presence of dark matter in the NS. Tidal deformability features ('$\Lambda(M1)$' through '$\Lambda(M5)$') are also included, although they exhibit less influence on the model's decisions. This insight underscores the necessity of precise radius measurements over a range of masses as a more determinant factor in revealing or classifying dark matter within NS. The comparative lower importance of the tidal deformability suggests that, while useful, they do not contribute as significantly to the model's classification ability as the radius measurements.

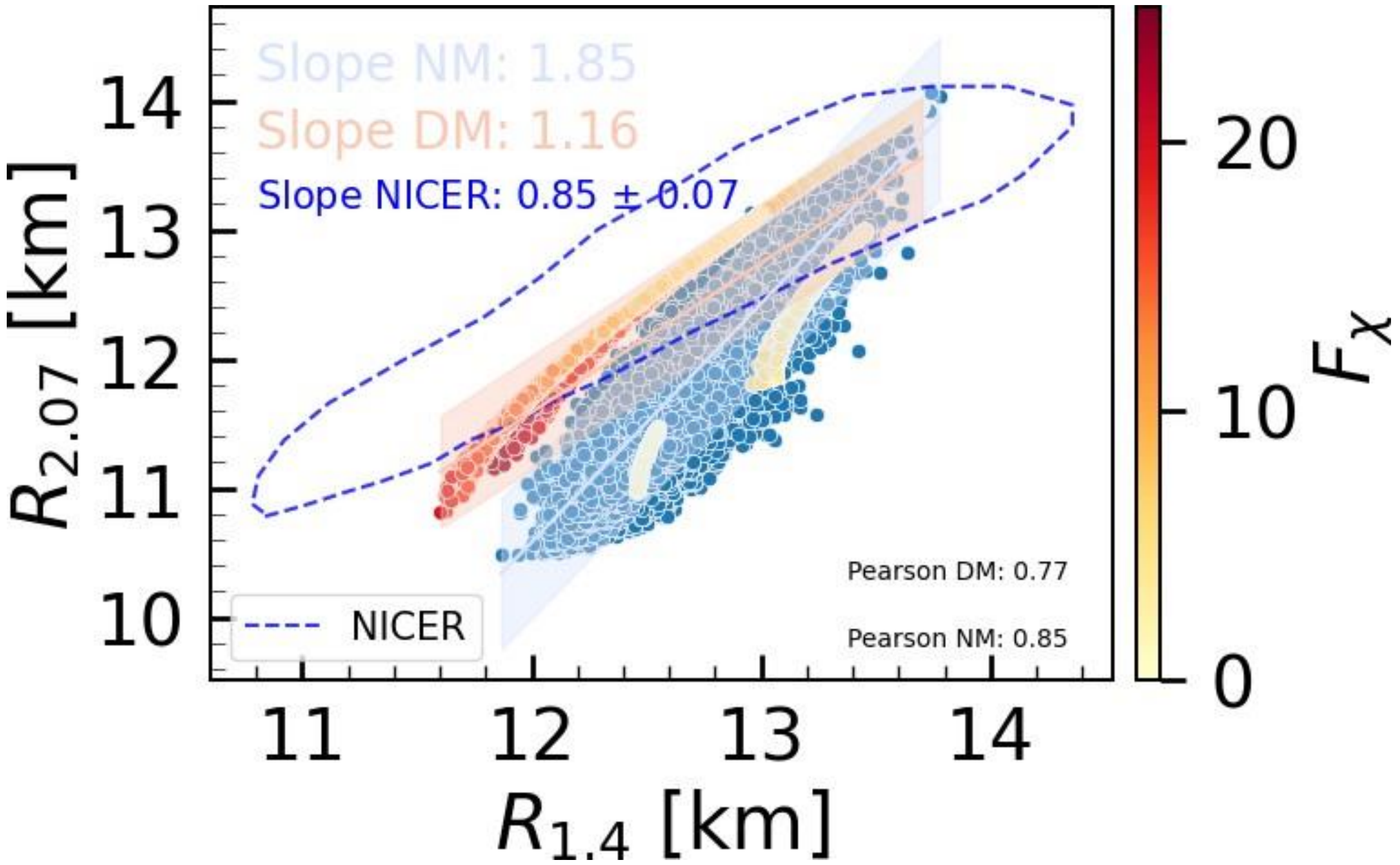

Figure 4.7: The figure displays the relationship between the radii of neutron stars with masses of 1.4 and 2.07 solar masses ($M_\odot$). The comparison is made between two different scenarios: one with dark matter and one without. The radius in the scenario that includes dark matter is the total radius of the neutron star, including the dark matter component.

In our analysis, depicted in Figure 4.7, we explore the correlation between the radii of neutron stars with masses of 1.4 and 2.07 solar masses ($M_\odot$), comparing scenarios with and without dark matter. Inspired by a recent study Lin & Steiner [2023], which presents a method to detect phase transitions in neutron stars by examining the radius correlations, including NICER's observations of PSR J0740+6620 and PSR 0030+0451, we extend this approach to investigate the presence of dark matter. We plotted the relationship between the radii of neutron stars with masses of 1.4 and 2.07 solar masses ($M_\odot$), comparing scenarios with and without dark matter. In the dark matter scenario, the total radius of the neutron star includes the dark matter component. We found that the Pearson correlation coefficient is 0.77 for the dark matter set and 0.85 for the nuclear matter set, indicating a slight weakening of the correlation with dark matter. Furthermore, we compared the constraints derived from NICER in blue, which were obtained by marginalizing the $1\sigma$ posteriors of the NICER measurements of these two pulsars over the NS mass. Then, the radius data of these two observations yielded this blue domain. This analysis considered only the data analyzed by the Riley et al. group. Significantly, we observed that the slope of the relationship between $R_{2.07}$ and $R_{1.4}$ undergoes a considerable change in the presence of dark matter. The slope is 1.85 in the absence of dark matter and 1.16 with admixed dark matter. The line fitted to each scenario is plotted along with the uncertainty band of $1\sigma$. The presence of dark matter appears to reduce this slope. In particular, the slope for the NICER constraints is $0.85 \pm 0.07$, showing substantial overlap with the dark matter admixed set. This overlap suggests that the NICER measurements of the radii of PSR J0740+6620 and PSR 0030+0451 might indicate a higher likelihood of the presence of dark matter compared to solely hadronic matter.

The research conducted thus far has strongly encouraged us to explore the potential for dark matter to be revealed in the observational properties of neutron stars, such as mass, radius, and dimensionless tidal deformability, through the use of machine learning. The machine learning tools are becoming increasingly important in neutron star physics Carvalho *et al.* [2023], Ferreira & Providência [2021], Murarka *et al.* [2022], Soma *et al.* [2024], Vidaña [2023]. We use the robust dataset of neutron star properties of both nuclear matter and admixed dark matter configurations that were previously discussed. Our goal is to create a classification system that can distinguish between the presence and absence of dark matter in neutron stars.

## 4.6 Conclusion

In this chapter, we explored the combination of dark matter and neutron stars through a two-fluid model, concentrating solely on a fermionic dark matter equation of state. To replicate this situation, we carefully employed EOS for nucleonic matter and dark matter within a relativistic mean-field (RMF) approach, which included a comprehensive collection of 16,000 different EOS. The nuclear EOS was the result of Bayesian inference with minimal constraints on nuclear saturation properties, an NS maximum mass of more than 2 solar masses, and a low-density pure neutron matter constraint. For the dark matter component, a straightforward model was adopted, featuring a singular fermionic entity interacting with a dark vector meson. The resulting dark matter EOS was contingent on the interaction strength with the dark vector meson and the intrinsic mass of the fermionic dark matter. The interplay between the dark matter EOS and the proportion of dark matter present critically determined the resultant properties of the neutron stars.

A comprehensive study was carried out on the combination of neutron stars and dark matter, using four different nucleonic equations of state and a range of dark matter equations of state. This research looked at 32,000 neutron stars properties curve, examining their mass, radius, and tidal deformability. The study also delved into the integrity of universal relations in the presence of dark matter, affirming the stability of the compactness–tidal deformability (C–Λ) relation. Furthermore, we investigate the correlation between the radii $R_{1.4}$ and $R_{2.07}$ for neutron stars with masses 1.4 and 2.07 $M_{\odot}$, respectively. The two radii are closely related, but the inclusion of dark matter in neutron stars can alter the gradient. For the first time, our analysis has taken advantage of the most recent observational data from NICER, which strongly suggests that dark matter must be included in neutron star models. As our knowledge advances and our observational powers increase, the intricate part that dark matter plays in the universe will become more distinct, giving us new perspectives on the essential character of our cosmos.

We employed machine learning techniques, particularly Random Forest classifiers, to classify neutron stars (NS) in the presence of dark matter, based on their inherent properties. We evaluated the predictive accuracy of these classifiers, trained on various feature sets, using confusion matrices. Our findings suggest that the properties of NS with a mixed dark matter configuration can be identified with approximately 17% probability of misclassification as nuclear matter NS. Adding constraints related to tidal deformability does not significantly reduce this uncertainty. This figure reflects both the inherent uncertainty of the classifier model and the variability in NS properties. Future research should include a more focused analysis to distinguish between these sources of uncertainty, as well as a broader examination using a variety of machine learning algorithms. Additionally, our analysis sought to identify critical features that would be useful for future observations to detect dark matter within NS. The results suggest that measurements of radius at both lower and higher masses appear to be very promising indicators for the presence of dark matter.

# Chapter 5

# Role of Admixed Dark Matter in Neutron Stars: Observational Evidences

The equation of state (EOS) of neutron stars is pivotal in determining their internal structure and observable characteristics, such as mass and radius. Various microscopic many-body approaches and phenomenological models are employed to construct the EOS, taking into account the interactions of nucleons, hyperons, and other exotic particles Curceanu *et al.* [2019], Gal *et al.* [2016], Tolos & Fabbietti [2020]. The EOS models are validated against both terrestrial laboratory data and astrophysical observations Burgio *et al.* [2021], Dutra *et al.* [2014], Malik & Providência [2022b], Oertel *et al.* [2017]. Furthermore, the detection of gravitational waves from neutron star mergers provides additional constraints, with postmerger emissions offering insights into the EOS by correlating specific spectral features with the star's compactness Takami *et al.* [2014]. Recent studies have shown that the onset of phase transition from hadronic matter to quark matter can significantly affect the dynamics and gravitational wave signals of binary neutron star mergers Haque *et al.* [2023], Weih *et al.* [2020]. One area of research that has recently attracted considerable interest is altered gravity theories, with the study of neutron stars within the framework of modified gravity theories being a lively field of research, offering an opportunity to compare the predictions of these theories with observational data Alam *et al.* [2024], Cooney *et al.* [2010], Nobleson *et al.* [2023], Yazadjiev *et al.* [2018]. The combined efforts from both theoretical and observational studies are essential for enhancing our understanding of the extreme conditions within neutron stars. As discussed in previous sections, Recent studies have extended this understanding by incorporating the effects of dark matter (DM) within neutron stars Barbat *et al.* [2024], Cronin *et al.* [2023], Das *et al.* [2019, 2022], Diedrichs *et al.* [2023], Ellis *et al.* [2018], Flores *et al.* [2024], Giangrandi *et al.* [2022], Karkevandi *et al.* [2022], Lenzi *et al.* [2023],

Nelson *et al.* [2019], Routaray *et al.* [2023], Rutherford *et al.* [2023], Sagun *et al.* [2023], Scordino & Bombaci [2024], Shakeri & Karkevandi [2022], Singh *et al.* [2023], Thakur *et al.* [2024a]. Dark matter, interacting with neutrons through mechanisms such as Higgs boson exchange, significantly alters the EOS, potentially leading to observable changes in neutron star characteristics. For instance, DM can affects the mass-to-radius relation Das *et al.* [2020b, 2021a, b], Guha & Sen [2021], Lourenço *et al.* [2022], Panotopoulos & Lopes [2017], Sen & Guha [2021, 2022], Shirke *et al.* [2023, 2024], and the cooling rates of neutron stars Giangrandi *et al.* [2024], Kouvaris [2008]. Studies show that neutron stars containing a mix of baryonic and dark matter can exhibit different mass-radius relations compared to those composed solely of neutron matter, influencing gravitational redshift and potentially explaining observations inconsistent with normal neutron stars Rezaei [2017].

Numerous candidates for dark matter particles, including bosonic dark matter, axions, sterile neutrinos, and various WIMPs, are discussed in Bauer & Plehn [2019], Calmet & Kuipers [2021], Silk *et al.* [2010]. The nature of dark matter and its properties, such as self-interaction and coupling with standard model particles, have been explored for their impact on neutron star dynamics Diedrichs *et al.* [2023], Ellis *et al.* [2018], Leung *et al.* [2022], Narain *et al.* [2006], Panotopoulos & Lopes [2017]. Two commonly researched methods for examining neutron stars mixed with dark matter are found in existing literature. One approach considers non-gravitational interactions, using mechanisms like the Higgs portal Das *et al.* [2019, 2020b], Dutra *et al.* [2022], Flores *et al.* [2024], Hong & Ren [2024], Lenzi *et al.* [2023], Sen & Guha [2021] and self-interacting dark matter models Shirke *et al.* [2023]. This can be treated as a single-fluid system. The other approach considers only gravitational interactions, treated as a two-fluid system Buras-Stubbs & Lopes [2024], Collier *et al.* [2022], Emma *et al.* [2022], Hong & Ren [2024], Ivanytskyi *et al.* [2020], Karkevandi *et al.* [2022], Liu *et al.* [2023], Miao *et al.* [2022], Rü ter *et al.* [2023], Rutherford *et al.* [2023]. Another method involves a single fluid model in which neutron decay into dark matter particles accounts for the neutron decay anomaly Bastero-Gil *et al.* [2024], Baym *et al.* [2018], Husain *et al.* [2022], Motta *et al.* [2018], Shirke *et al.* [2023, 2024]. Discrepancies in neutron decay lifetimes measured via bottle and beam experiments suggest more decayed neutrons than produced protons, possibly due to decay into nearly degenerate dark fermions.

To understand the mixed dark matter scenario in neutron stars (NS), knowledge of the equation of state (EOS) for both baryonic and dark matter is crucial. The dense matter EOS can be described by relativistic and nonrelativistic models. Nonrelativistic models describe nucleons within finite nuclei well but fail with infinite dense nuclear matter. Relativistic mean-field (RMF) models, suitable for describing both finite nuclei and high-density matter in NS, incorporate many-body interactions via mesons ($\sigma$, $\omega$, $\rho$). Two main RMF approaches describe nuclear properties: nonlinear meson terms in the Lagrangian density Boguta & Bodmer [1977], Mueller & Serot [1996a], Steiner *et al.* [2005b], Todd-Rutel & Piekarewicz [2005] and density-dependent coupling parameters Lalazissis *et al.* [2005], Typel & Wolter [1999], Typel *et al.* [2010]. Other metamodels are constrained by low and high-density theoretical calculations Drischler *et al.* [2016], Hebeler *et al.* [2013], Kurkela *et al.* [2010]. Approaches to accommodate all possible EOSs include piecewise polytropic interpolation, speed of sound interpolation, spectral interpolation, and Taylor expansion Annala *et al.* [2020, 2021], Kurkela *et al.* [2014], Lindblom & Indik [2012], Lope Oter *et al.* [2019], Most *et al.* [2018].

## 5.1 Motivation

Our study explores the Non-Linear model within the RMF framework. The $\sigma$-cut potential approach Maslov *et al.* [2015], proposing sharp increases in mean field self-interaction potential around the nuclear saturation density ($\rho_0$), effectively stiffens the EOS without affecting nuclear matter properties near $\rho_0$ Ma *et al.* [2022]. Few studies on the $\sigma$-cut scheme focus on EOS stiffness implications, including kaon condensation, hyperons in neutron stars Ma *et al.* [2022], stellar properties, nuclear matter constraints Dutra *et al.* [2016], strangeness neutron stars within RMF Zhang *et al.* [2018], NICER data analysis Kolomeitsev & Voskresensky [2024], and effects on pure nucleonic and hyperonic-rich NS matter Thakur *et al.* [2024b]. The stiffening obtained with the introduction of the $\sigma$ cut potential may be seen as the inclusion of an exclusion volume effect Typel [2016], or even as mimicking the onset of another baryonic phase such as the quarkyonic phase McLerran & Reddy [2019]. As upcoming observational data becomes increasingly refined, we are motivated to conduct a systematic study of the interior structure of neutron stars under three distinct scenarios. First, we consider neutron stars comprising only nucleonic degrees of freedom within the framework of the non-linear (NL) model. Second, we modify the NL model to include a $\sigma$-cut potential. Third, we investigate neutron stars that contain an admixture of dark matter, specifically focusing on fermionic dark matter for simplicity. Previous studies have indicated that the presence of dark matter tends to reduce both the mass and radius of neutron stars. Conversely, incorporating a $\sigma$-cut potential has been shown to increase these parameters. Given these contrasting effects, our goal is to calculate the Bayesian evidence for each of these three scenarios, thereby determining which model is most consistent with the latest observational data. By systematically evaluating these models, we aim to rank them based on their alignment with recent observations, providing a clearer understanding of the interior structure of neutron stars and the potential influence of dark matter and modified nucleonic interaction.

We have performed a detailed statistical analysis using the current astrophysical observational data on NS properties within the Bayesian inference framework to explore the potential existence of dark matter in NS. Both baryonic (visible) matter and dark matter are considered within the RMF framework. The EOSs are developed using empirical constraints based on experimental data regarding the properties of finite nuclei and observations from astrophysics. The nuclear matter properties taken into account include the pressure of symmetric nuclear matter ($P_{\rm SNM}$), the pressure resulting from symmetry energy ($P_{\rm sym}$), and the symmetry energy itself ($e_{\rm sym}$). These properties are empirically constrained across various densities using experimental data on the bulk characteristics of finite nuclei, such as nuclear masses, neutron skin thickness in $^{208}$Pb, dipole polarizability, isobaric analog states, and heavy ion collision (HIC) data covering the density range from 0.03 to 0.32 fm$^{-3}$. Additionally, astrophysical data utilized include the mass-radius posterior distributions for PSR J0030+0451 Miller *et al.* [2019], Riley *et al.* [2019] and PSR J0740+6620 Miller *et al.* [2021], Riley *et al.* [2021] as well as the posterior distribution for dimensionless tidal deformability for components of binary neutron stars from the GW170817 event. The datasets are presented in Table 5.1. We examined three different scenarios: i) solely nucleonic degrees of freedom within RMF, referred to as NL, ii) an EOS stiffened by a modified $\sigma$-cut potential, referred to as NL-$\sigma$ cut, and iii) impact of dark matter on neutron stars, specifically, how the presence of fermionic dark matter in admixed NS matter through neutron decay affects the nuclear equation of state, referred as NL-DM. We investigated which of these three scenarios aligns better with recent observational data on NS. Additionally, we analyzed the impact of PREX-II experimental data with or without inclusion in all cases, which often contradicts various other nuclear data reported in the literature. Finally, we examined the nonradial oscillations of neutron stars (NS), focusing on $f$ and $p$ mode oscillations using the posterior obtained for each case. This study provides a definitive answer to the probabilistic measure of the possibility of dark matter existence in NS.

## 5.2 Methodolgy:

In this chapter, we utilize the methods detailed in Sections 2.1 and 2.2 of Chapter 2 for calculating the Non-Linear and Non-Linear $\sigma$-cut models, respectively. The details regarding Bayesian inference, f and p mode calculations, and the single-fluid TOV equations used for determining neutron star properties have already been discussed in Sections 2.5, 2.3, and 2.4.1.1.

Table 5.1: The empirical values of symmetry energy ($e_{sym}$), symmetry energy pressure ($P_{sym}$), and symmetric nuclear matter pressure ($P_{SNM}$) from experimental data on the bulk properties of finite nuclei and HIC. The astrophysical observational constraints on the radii and tidal deformability of neutron stars. See Ref[Tsang *et al.* 2024] for details.

| Symmetric matter | | | |
|---|---|---|---|
| Constraints | n ($fm^{-3}$) | $P_{SNM}$ ($MeV/fm^3$) | Ref. |
| HIC(DLL) | 0.32 | 10.1 ± 3.0 | [Danielewicz *et al.* 2002] |
| HIC(FOPI) | 0.32 | 10.3 ± 2.8 | [Le Fe`vre *et al.* 2016] |

| Asymmetric matter | | | | |
|---|---|---|---|---|
| Constraints | n ($fm^{-3}$) | S(n) (MeV) | $P_{sym}$ ($MeV/fm^3$) | Ref. |
| Nuclear structure | | | | |
| $\alpha_D$ | 0.05 | 15.9 ± 1.0 | | [Zhang & Chen 2015] |
| PREX-II | 0.11 | | 2.38 ± 0.75 | [Adhikari *et al.* 2021, Lynch & Tsang 2022, Reed *et al.* 2021] |
| Nuclear masses | | | | |
| Mass(Skyrme) | 0.101 | 24.7 ± 0.8 | | [Brown & Schwenk 2014, Lynch & Tsang 2022] |
| Mass(DFT) | 0.115 | 25.4 ± 1.1 | | [Kortelainen *et al.* 2012, Lynch & Tsang 2022] |
| IAS | 0.106 | 25.5 ± 1.1 | | [Danielewicz *et al.* 2017, Lynch & Tsang 2022] |
| Heavy-ion collisions | | | | |
| HIC(Isodiff) | 0.035 | 10.3 ± 1.0 | | [Lynch & Tsang 2022, Tsang *et al.* 2009] |
| HIC(n/p ratio) | 0.069 | 16.8 ± 1.2 | | [Lynch & Tsang 2022, Morfouace *et al.* 2019] |
| HIC(π) | 0.232 | 52 ± 13 | 10.9 ± 8.7 | [Estee *et al.* 2021, Lynch & Tsang 2022] |
| HIC(n/p flow) | 0.240 | | 12.1 ± 8.4 | [Cozma 2018, Lynch & Tsang 2022, Russotto & *et. al.* 2016, Russotto *et al.* 2011] |

| Astrophysical | | | | |
|---|---|---|---|---|
| Constraints | $M_\odot$ | R (km) | $\Lambda_{1.36}$ | Ref. |
| LIGO [1] | 1.36 | | $300^{+420}_{-230}$ | [Abbott *et al.* 2019b] |
| *Riley PSR J0030+0451 [2] | 1.34 | $12.71^{+1.14}_{-1.19}$ | | [Riley *et al.* 2019] |
| *Miller PSR J0030+0451 [3] | 1.44 | $13.02^{+1.24}_{-1.06}$ | | [Miller *et al.* 2019] |
| *Riley PSR J0740+6620 [4] | 2.07 | $12.39^{+1.30}_{-0.98}$ | | [Riley *et al.* 2021] |
| *Miller PSR J0740+6620 [5] | 2.08 | $13.7^{+2.6}_{-1.5}$ | | [Miller *et al.* 2021] |
| *Choudhury PSR J0437-4715 [6] | 1.418 | $11.36^{+0.95}_{-0.63}$ | | [Choudhury *et al.* 2024] |

## 5.3 Speed Of Sound

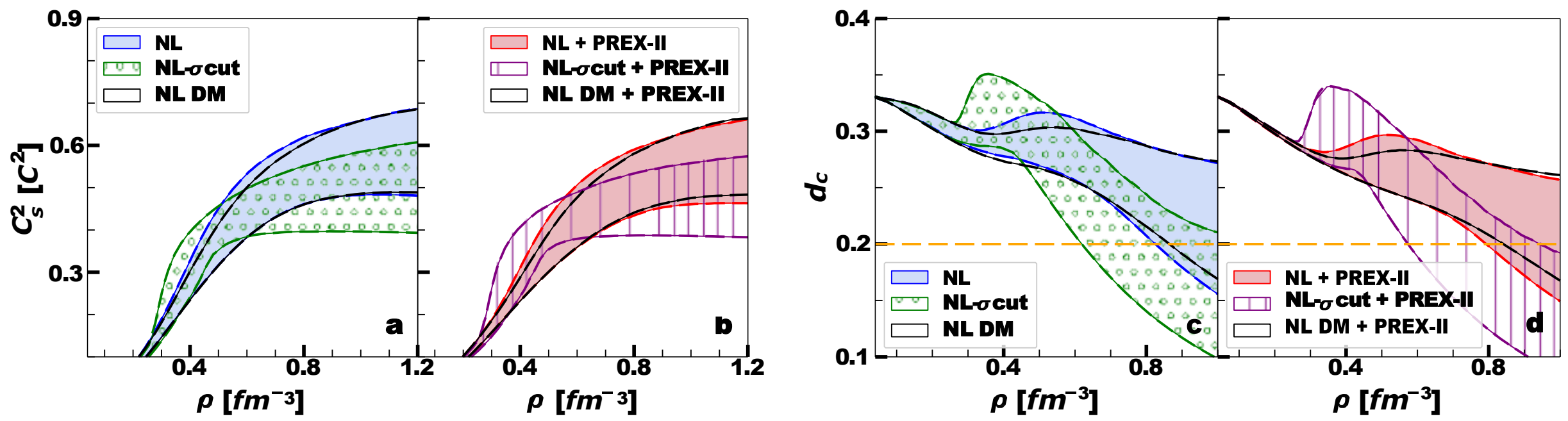

Figure 5.1: (left) 90% confidence intervals for $c_s^2$ vs. baryon density for NL, NL-$\sigma$ cut and NL-DM models. (right) 90% confidence intervals for $d_c$ vs. baryon density for NL, NL-$\sigma$ cut and NL-DM models. $d_c = \sqrt{\Delta^2 + \Delta'^2}$, where $\Delta' = c_s^2\left(\frac{1}{\gamma} - 1\right)$, and $\Delta = \frac{1}{3} - \frac{P}{\epsilon}$. The left (right) panel subfigure "a" excludes PREX-II data; the subfigure "b" includes it.

Fig. 5.1 depicts the 90% confidence intervals (CI) for the squared speed of sound ($c_s^2$) versus baryon density ($\rho$) for different models, shown in the left panel (subfigures a and b). Subfigure *a* excludes PREX-II data, while subfigure *b* includes it. The PREX-II data have an insignificant effect on $c_s^2$ since this data primarily affects the density-dependent symmetry energy, and the squared speed of sound is largely affected by the symmetric nuclear matter part of the EOS. Among the scenarios: NL, NL-$\sigma$ cut, and NL-DM, the presence of dark matter in the NL-DM model slightly affects the NL model at intermediate densities (0.4-0.8 fm$^{-3}$), reducing $c_s^2$ and thus softening the EOS. For the NL-$\sigma$ cut model, modifications to the $\sigma$ potential result in a stiffer $c_s^2$ at low densities below 0.5 fm$^{-3}$, becoming softer at higher densities compared to both the NL and the NL-DM models. This occurs because, in the NL-$\sigma$ cut model, the $\sigma$ field saturates due to a very stiff $\sigma$ potential, and the effective mass stays at a constant value that can be above 0.5 $m_N$. The repulsive $\omega$ potential becomes dominant, and at high densities, the effect of the $\omega^4$ term softens the EOS. Indeed, as can be seen from the Table 5.6, the median value of the $\xi$ coupling more than doubles, increasing from 0.005 to 0.011. As discussed in Mueller & Serot [1996b] (see also Malik *et al.* [2023]) this term causes the speed of sound squared to tend towards 1/3 at high densities, and the larger the coupling the faster this limit is reached. The parameter $d_c$, introduced in Ref. Annala *et al.* [2023], is defined as $d_c = \sqrt{\Delta^2 + (\Delta')^2}$, $\Delta = \frac{1}{3} - \frac{P}{\epsilon}$ is the renormalized trace anomaly introduced in Fujimoto *et al.* [2022], $\Delta' = c_s^2\left(\frac{1}{\gamma} - 1\right)$ denotes the logarithmic derivative of $\Delta$, and $\gamma = \frac{d \ln P}{d \ln \epsilon}$. In the conformal limit, $c_s^2$ and $\gamma$ reach 1/3 and 1 respectively. It has been proposed that a value $d_c \lesssim 0.2$ suggests proximity to the conformal limit, as both $\Delta$ and its derivative need to be small for this to hold true. Since quark matter is expected to show approximate conformal symmetry, a small $d_c$ could be indicative of the presence of quark matter. In the right panel of Fig. 5.1, we illustrate the posterior distribution of the quantity $d_c$ as a function of density for all the cases under consideration, both including and excluding PREX-II data. In the right subfigure ($c$), excluding PREX-II data, we display the confidence intervals for $d_c$ derived from the NL, NL-$\sigma$ cut and NL-DM models. The blue-shaded region represents the 90% confidence interval for the NL model, with dashed blue lines indicating its boundaries. As density increases, $d_c$ initially decreases, followed by an increase, reaching a peak around 0.5 fm$^{-3}$, and subsequently exhibits a clear downward trend. For the NL-$\sigma$ cut model, depicted by the green hatched region with circle patterns and bounded by dashed green lines, a similar decreasing trend is observed, but it peaks earlier, around 0.4 fm$^{-3}$, and it is due to the stiffening of the $\sigma$ potential it. For the NL-DM model, the confidence region is narrower than those of the NL and NL-$\sigma$ cut models, and the peak is lower compared to the other models. The value of $d_c$ falls below 0.2 even though in the present approach, only nuclear matter exists at higher densities, specifically beyond 0.7 fm$^{-3}$. In contrast, the NL-$\sigma$ cut model shows a value below 0.2 at much lower densities, specifically beyond 0.6 fm$^{-3}$, even with only nuclear degrees of freedom. This behavior is due to the dominance of the repulsive $\omega$ potential, which is modified by the $\sigma$ potential. In the right subfigure ($d$), which incorporates PREX-II data, there is no discernible impact attributable to the inclusion of PREX-II data. In case of NL-$\sigma$ cut there is hump around 0.4 $fm^{-3}$ because the sigma-cut EOS modifies the equation of state at intermediate densities, making it stiffer due to the additional sigma function. This stiffening reduces the need for the omega meson's contribution to achieve high-mass neutron stars. As a result, the coupling constant $g_\omega$ is smaller in this case, while the $\omega^4$ term coupling is larger, leading to an overall softer EOS at very high densities.

This effect is visible in the speed of sound squared ($c_s^2$), which rises rapidly at intermediate densities due to the sigma-cut contribution but then stabilizes or even decreases above $\rho$ = 0.4 fm$^{-3}$. The impact of this behavior on $d_c$ can be understood through

$$\Delta = \frac{1}{3} - \frac{P}{E}$$

and its derivative $\Delta$. Initially, as pressure increases rapidly, $P/E$ also increases, but when the speed of sound stabilizes, $P/E$ saturates at a lower value compared to other EOS scenarios. This leads to a larger $\Delta$, and since the transition from stiff to soft EOS is more pronounced in the sigma-cut case, $\Delta$ also exhibits a significant change, producing the observed bump in $d_c$. After this transition, the effect of $\Delta$ diminishes, and $d_c$ stabilizes.

Additionally, in the conformal limit at very high densities, both $\Delta$ and $\Delta$ should ideally approach zero. The presence of the $\omega^4$ term in our case accelerates the approach of $c_s^2$ toward $1/3$, similar to what is observed in agnostic EOS studies Annala *et al.* [2023]. In those cases, a bump in $d_c$ emerges because the EOS must be stiff at low densities to support $2M_\odot$ stars but softer at high densities to comply with perturbative QCD constraints. Although our study does not impose pQCD constraints, the sigma-cut EOS naturally leads to a similar effect due to the role of the $\omega^4$ term.

## 5.4 Neutron Star Properties

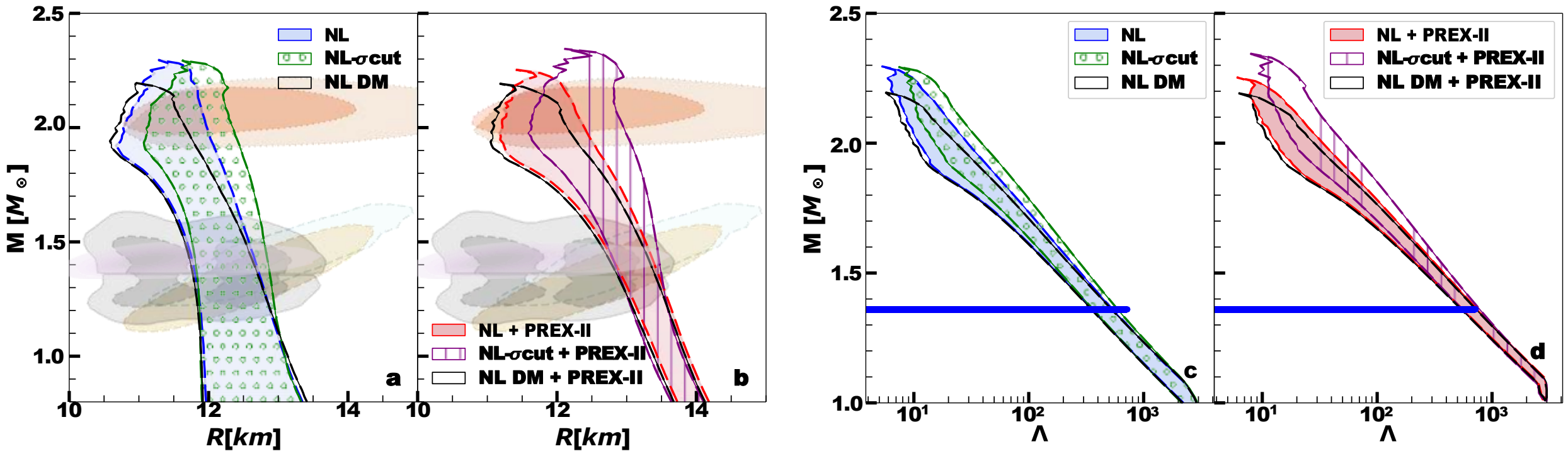

Figure 5.2: (left) The 90% credible interval (CI) region for the neutron star (NS) mass-radius posterior $P(R|M)$ is plotted for the NL, NL-$\sigma$ cut, and NL-DM models. The gray area indicates the constraints obtained from the binary components of GW170817, with their respective 90% and 50% credible intervals. Additionally, the plot includes the 1 $\sigma$ (68%) CI for the 2D mass-radius posterior distributions of the millisecond pulsars PSR J0030 + 0451 (in cyan and yellow color) Miller *et al.* [2019], Riley *et al.* [2019] and PSR J0740 + 6620 (in orange and peru color)Miller *et al.* [2021], Riley *et al.* [2021], based on NICER X-ray observations. Furthermore, we display the latest NICER measurements for the mass and radius of PSR J0437-4715 Choudhury *et al.* [2024] (lilac color). (right) The 90% CI region for the mass-tidal deformability posterior $P(\Lambda|M)$ for the same models is presented. The blue bars represent the tidal deformability constraints at 1.36 $M_\odot$ Abbott *et al.* [2018a]. In the left (right) subfigure, panels a(c) and b(d) correspond to the data without and with PREX-II data, respectively.

Fig. 5.2 depicts the mass-radius (M-R) and mass-tidal deformability (M-$\Lambda$) relationships of neutron stars based on the different scenarios considered in this study, compared with different astrophysical observational data. In panel (a), results for the NL (blue), NL-$\sigma$ cut (green with dots), and NL-DM (black dashed) models are plotted without including data from the PREX-II experiment. Panel (b) shows the same information, including the PREX-II data, in the likelihood, represented by the colors red, pink, and black, respectively. In panels (c) and (d), we plot the 90% CI of mass-tidal deformability for all the models with and without PREX-II, respectively. All distributions correspond to 90% confidence intervals. We compare our results with several observational constraints. In panels (a) and (b), the grey region shows constraints from the binary components of the gravitational wave event GW170817, including their 90% and 50% credible intervals (CI). Constraints from the NICER x-ray data for the millisecond pulsar PSRJ0030+0451 are depicted in cyan and yellow color, while those for the pulsar PSRJ0740+6620 are shown in orange and peru color, both representing the $1\sigma$ (68%) CI for the 2D posterior distribution in the M-R domain. The new pulsar data PSR J0437-4715 is highlighted in a lilac shade. In panels (c) and (d), the constraints obtained from GW170817 are included for the $1.36 M_{\odot}$ tidal deformability Abbott *et al.* [2018a].

In Fig. 5.2 panels (a) and (b), the posterior distributions of the three different cases diverge from each other starting around an NS mass of 1.4 $M_{\odot}$. The NL-$\sigma$ cut tends to shift the M-R posterior to the right, thereby increasing the radius, while the dark matter in NL-DM tends to shift it to the left. Therefore, the effects of dark matter and the $\sigma$ cut potential are opposite. We aim to test which case is favored by the current astrophysical and nuclear constraints, and therefore, we calculate the Bayes factor for each inference model. The results are given in Tables 5.2 and 5.3, where, the Bayes factor for each model and the Bayes factor ratios are given, respectively. Interestingly, our evidence calculations suggest that: i) the models that do not include the PREX-II constraints are favored, having systematically a higher Bayes factor; ii) considering the models with no PREX-II constraint, the model with the $\sigma$ cut seems to be preferred with respect to both the NL and the NL-DM models, indicating a preference for a stiffening of the EOS at high densities. These results should be reflected on the slope of the M-R curves, giving preference for a smaller negative slope or even a positive slope.

Table 5.2: Log evidence ln(Z) Values for the different Models. The *Best Model* is NL-$\sigma$ cut (without PREX-II) with the highest log evidence of $-62.18$.

| **Model** | ln(Z) | ln(Z) (With PSR J0437-471) |
|---|---|---|
| NL | $-64.14 \pm 0.16$ | $-65.25 \pm 0.15$ |
| NL + PREX-II | $-68.53 \pm 0.17$ | ... |
| NL-$\sigma$ cut | *$-62.18 \pm 0.15$* | *$-63.36 \pm 0.15$* |
| NL-$\sigma$ cut + PREX-II | $-66.15 \pm 0.17$ | ... |
| NL DM | $-64.53 \pm 0.15$ | $-65.57 \pm 0.15$ |
| NL DM + PREX-II | $-69.12 \pm 0.17$ | ... |

Table 5.3: Log Evidence Differences and Interpretations (P2 indicates with PREX-II, and NL-$\sigma$c indicates NL-$\sigma$ cut).

| **Model1/Model2** | $\Delta$ ln(Z) | **Interpretation** |
|---|---|---|
| NL-$\sigma$c P2/NL-$\sigma$c | $-3.96$ | Decisive for NL-$\sigma$c |
| NL-$\sigma$c P2/NL P2 | 2.38 | Substantial for NL-$\sigma$c P2 |
| NL-$\sigma$c P2/NL | $-2.01$ | Substantial for NL |
| NL-$\sigma$c/NL P2 | 6.35 | Decisive for NL-$\sigma$c |
| NL-$\sigma$c/NL | 1.96 | Substantial for NL-$\sigma$c |
| NL P2/NL | $-4.39$ | Decisive for NL |

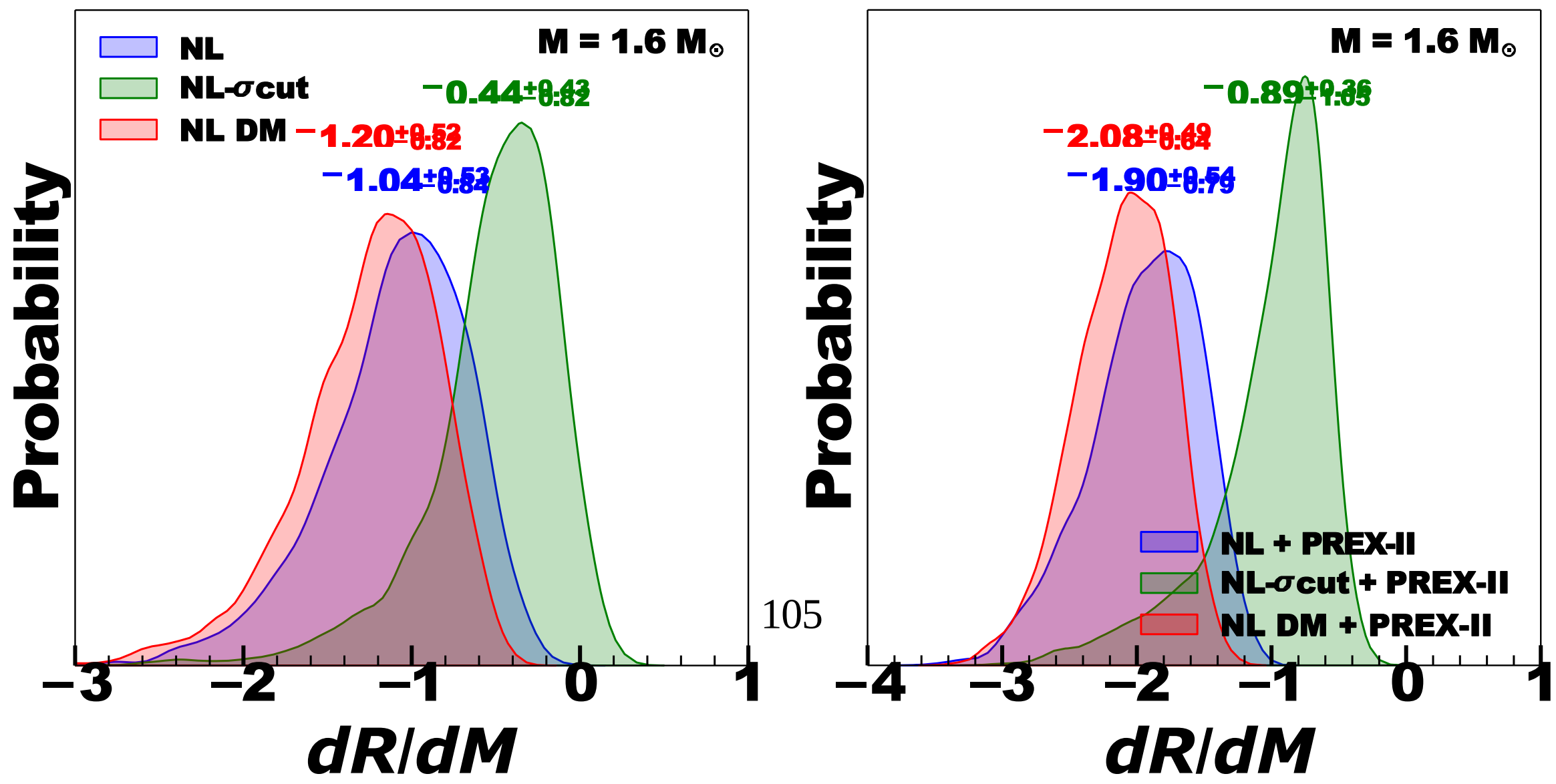

Fig. 5.3 depicts the dR/dM distribution for neutron stars with a mass of 1.6 $M_{\odot}$. The left panel shows the distribution across all three models without PREX-II data, while the right panel incorporates the PREX-II data. Among the models, the NL$-\sigma$cut is the most preferred, displaying a dR/dM slope of $-0.44^{+0.43}_{-0.82}$ for a neutron star mass of 1.6 $M_{\odot}$. The NL model has a slightly more negative slope of $-1.04^{+0.53}_{-0.84}$, and the NL-DM model yields the most negative one with a value $-1.20^{+0.52}_{-0.82}$. When additional PREX-II data is considered, all slopes become more negative, and Bayesian evidence indicates lesser support for these models. The effect of PREX-II data makes the symmetry energy stiffer, increasing the radius for lower masses below 1.6 $M_{\odot}$. It can be seen from Fig. 5.2(b) that the PREX-II data makes the M-R distribution narrower in the lower part of the MR curve. This will be explained in greater detail at the end of the section.

In Table 5.4, we summarize some properties of nuclear matter properties (NMP) and neutron star properties predicted by the NL, NL-$\sigma$ cut, and NL-DM inference models. Some conclusions can be drawn: i) independently of the inclusion or not of PREX-II constraints the symmetric nuclear matter properties are only slightly affected by the inclusion of the $\sigma$ cut potential and the presence of DM, in particular, in the last case making the EOS slightly harder to compensate the effect of DM on the neutron star properties; ii) the symmetry energy and its slope at saturation do not depend on the model but the inclusion of PREX-II constraint rises the symmetry energy and its slope at saturation from 32 MeV and 54 MeV to 38 MeV and $\sim$ 105 MeV, respectively. In Fig. 5.2(a) (no PREX-II data), the NL model's confidence interval (shaded blue) shows a broader range of radii for neutron stars, predicting a median value of 12.18 km for a 1.4$M_{\odot}$ star, $R_{1.4}$, see Table 5.4. For the NL-$\sigma$ cut model, $R_{1.4}$ slightly increases to 12.29 km, and for NL-DM it slightly decreases to 12.09 km. Including PREX-II data, the radius shifts to larger values, in particular, to 13.15 km, 13.20 km, and 13.10 km for, respectively, NL, NL-$\sigma$ cut, and NL-DM models. Including $\sigma$ cut increases the maximum mass from 2.055 $M_{\odot}$ to 2.087 $M_{\odot}$ with PREX-II. Conversely, adding DM decreases the maximum mass to 1.981 $M_{\odot}$ (or 1.975 $M_{\odot}$ with PREX-II).

In Fig., 5.2(c) and (d) showcase the 90% confidence intervals for tidal deformability $\Lambda$, respectively without and with PREX-II constraint, as it relates to neutron star mass across all considered cases. Including PREX-II results in an increase in tidal deformability for all cases due to an increase in radius. The effect of the $\sigma$ cut parameter and DM matter on the tidal deformability is clearly seen comparing $\Lambda_{1.36}$ = 478 for the NL-$\sigma$ cut model and $\Lambda_{1.36}$ = 419 for NL-DM with $\Lambda_{1.36}$ = 440 for the NL model, (see Table 5.4). PREX-II data shifts $\Lambda_{1.36}$ predictions to higher values: $\Lambda_{1.36}$ = 653 for NL-PREX-II (red), $\Lambda_{1.36}$ = 675 for NL-$\sigma$ cut-PREX-II (purple), and $\Lambda_{1.36}$ = 639 for NL-DM-PREX-II (black boundary). Both panels (c) and (d) show that all models fit within or near GW170817 constraints, even when PREX-II data are included.

Table 5.4: The median values and related 90% confidence intervals (CI) for certain nuclear matter parameters (NMPs), equation of state (EOS), and neutron star (NS) properties for the following models: NL, NL-$\sigma$ cut, and NL-DM, both with and without the inclusion of PREX-II data. In this table, saturation density ($\rho_0$) is expressed in units of fm$^{-3}$. NMPs such as $\epsilon_0$ - $Z_{\rm sym,0}$ are given in MeV. The properties of neutron stars, such as $M_{\rm max}$, are expressed in units of $M_\odot$. The radii corresponding to masses $M_{\rm i} \in [1.4, 1.6, 1.8, 2.07]$ $M_\odot$ are given in kilometers. The parameter $\Lambda_{1.36}$, representing the tidal deformability, is unitless. The square of the speed of sound, $c_s^2$, is in units of $c^2$. The frequencies for the $f$ mode and the $p$ mode for different NS masses are measured in kHz.

| | Without PREX-II | | | | | | With PREX-II | | | | | |
|---|---|---|---|---|---|---|---|---|---|---|---|---|
| **Quantity** | NL | | NL-σ cut | | NL DM | | NL | | NL-σ cut | | NL DM | |
| | **Med.** | **CI** | **Med.** | **CI** | **Med.** | **CI** | **Med.** | **CI** | **Med.** | **CI** | **Med.** | **CI** |
| $m^*$ | 0.73 | [0.69, 0.78] | 0.77 | [0.73, 0.79] | 0.74 | [0.71, 0.78] | 0.73 | [0.69, 0.78] | 0.76 | [0.72, 0.78] | 0.75 | [0.72, 0.78] |
| $\rho_0$ | 0.160 | [0.155, 0.165] | 0.160 | [0.155, 0.165] | 0.161 | [0.155, 0.165] | 0.160 | [0.155, 0.165] | 0.160 | [0.155, 0.165] | 0.161 | [0.156, 0.165] |
| $\epsilon_0$ | -15.99 | [-16.03, -15.96] | -15.99 | [-16.03, -15.96] | -16.00 | [-16.03, -15.97] | -15.99 | [-16.03, -15.96] | -16.00 | [-16.03, -15.96] | -16.00 | [-16.03, -15.97] |
| $K_0$ | 239 | [215, 263] | 240 | [222, 259] | 243 | [227, 265] | 238 | [212, 262] | 236 | [216, 257] | 244 | [230, 266] |
| $Q_0$ | -472 | [-541, -401] | -522 | [-592, -466] | -467 | [-521, -404] | -483 | [-563, -415] | -537 | [-612, -468] | -475 | [-524, -414] |
| $Z_0$ | 2465 | [1406, 3718] | 2060 | [1400, 2648] | 2221 | [1436, 2973] | 2395 | [1389, 3742] | 2181 | [1445, 2977] | 2119 | [1306, 2788] |
| $J_{\rm sym,0}$ | 32 | [29, 36] | 32 | [29, 36] | 32 | [30, 36] | 38 | [36, 39] | 38 | [36, 39] | 38 | [37, 40] |
| $L_{\rm sym,0}$ | 54 | [38, 86] | 54 | [38, 87] | 54 | [37, 90] | 106 | [97, 112] | 104 | [96, 110] | 105 | [97, 111] |
| $K_{\rm sym,0}$ | -133 | [-175, -69] | -158 | [-183, -87] | -144 | [-176, -76] | -11 | [-57, 18] | -20 | [-70, 2] | -17 | [-64, 8] |
| $Q_{\rm sym,0}$ | 1004 | [50, 1294] | 866 | [1, 1270] | 965 | [2, 1297] | 17 | [-48, 65] | 8 | [-47, 49] | 7 | [-54, 54] |
| $Z_{\rm sym,0}$ | -2926 | [-10616, 1519] | -1227 | [-8971, 1723] | -2272 | [-10276, 1649] | -600 | [-895, -22] | -515 | [-704, 176] | -567 | [-759, -2] |
| $M_{\rm max}$ | 2.014 | [1.893, 2.152] | 2.055 | [1.931, 2.163] | 1.981 | [1.876, 2.095] | 2.003 | [1.881, 2.131] | 2.087 | [1.960, 2.201] | 1.975 | [1.868, 2.079] |
| $R_{\rm max}$ | 10.65 | [10.19, 11.02] | 11.12 | [10.48, 11.47] | 10.52 | [10.16, 10.82] | 11.00 | [10.65, 11.31] | 11.62 | [10.93, 11.96] | 10.89 | [10.59, 11.15] |
| $R_{1.4}$ | 12.18 | [11.77, 12.68] | 12.29 | [11.81, 12.85] | 12.09 | [11.74, 12.65] | 13.15 | [12.84, 13.40] | 13.20 | [12.83, 13.47] | 13.10 | [12.79, 13.35] |
| $R_{1.6}$ | 12.03 | [11.57, 12.47] | 12.25 | [11.68, 12.74] | 11.91 | [11.53, 12.38] | 12.82 | [12.43, 13.13] | 13.04 | [12.54, 13.37] | 12.74 | [12.37, 13.04] |
| $R_{1.8}$ | 11.75 | [11.11, 12.24] | 12.11 | [11.38, 12.58] | 11.59 | [11.07, 12.04] | 12.35 | [11.73, 12.79] | 12.81 | [12.09, 13.22] | 12.22 | [11.65, 12.64] |
| $R_{2.07}$ | 11.32 | [10.71, 11.88] | 11.66 | [11.01, 12.25] | 11.25 | [10.96, 11.57] | 11.65 | [11.05, 12.25] | 12.29 | [11.51, 12.83] | 11.66 | [11.29, 12.10] |
| $\Lambda_{1.36}$ | 440 | [356, 536] | 478 | [363, 594] | 419 | [351, 526] | 653 | [554, 745] | 675 | [548, 781] | 639 | [541, 726] |
| $c^2_{s,\rm max}$ | 0.56 | [0.48, 0.67] | 0.47 | [0.39, 0.58] | 0.58 | [0.49, 0.68] | 0.53 | [0.46, 0.64] | 0.44 | [0.38, 0.55] | 0.56 | [0.48, 0.65] |
| $f_{1.4}$ | 2.30 | [2.19, 2.39] | 2.27 | [2.15, 2.38] | 2.33 | [2.20, 2.41] | 2.10 | [2.04, 2.16] | 2.08 | [2.03, 2.17] | 2.11 | [2.06, 2.18] |
| $f_{1.6}$ | 2.38 | [2.29, 2.50] | 2.33 | [2.23, 2.47] | 2.42 | [2.32, 2.51] | 2.22 | [2.15, 2.31] | 2.17 | [2.10, 2.28] | 2.25 | [2.18, 2.33] |
| $f_{1.8}$ | 2.48 | [2.37, 2.64] | 2.39 | [2.29, 2.57] | 2.54 | [2.44, 2.66] | 2.36 | [2.26, 2.51] | 2.26 | [2.17, 2.42] | 2.41 | [2.32, 2.54] |
| $f_{2.0}$ | 2.56 | [2.44, 2.68] | 2.48 | [2.36, 2.69] | 2.65 | [2.55, 2.76] | 2.50 | [2.37, 2.69] | 2.36 | [2.25, 2.63] | 2.57 | [2.47, 2.71] |
| $p_{1.4}$ | 6.38 | [5.69, 6.94] | 6.37 | [5.66, 6.97] | 6.43 | [5.69, 6.99] | 5.44 | [5.33, 5.57] | 5.43 | [5.33, 5.57] | 5.48 | [5.38, 5.61] |
| $p_{1.6}$ | 6.66 | [5.98, 7.11] | 6.68 | [5.93, 7.22] | 6.71 | [5.99, 7.16] | 5.71 | [5.59, 5.87] | 5.68 | [5.57, 5.84] | 5.76 | [5.65, 5.91] |

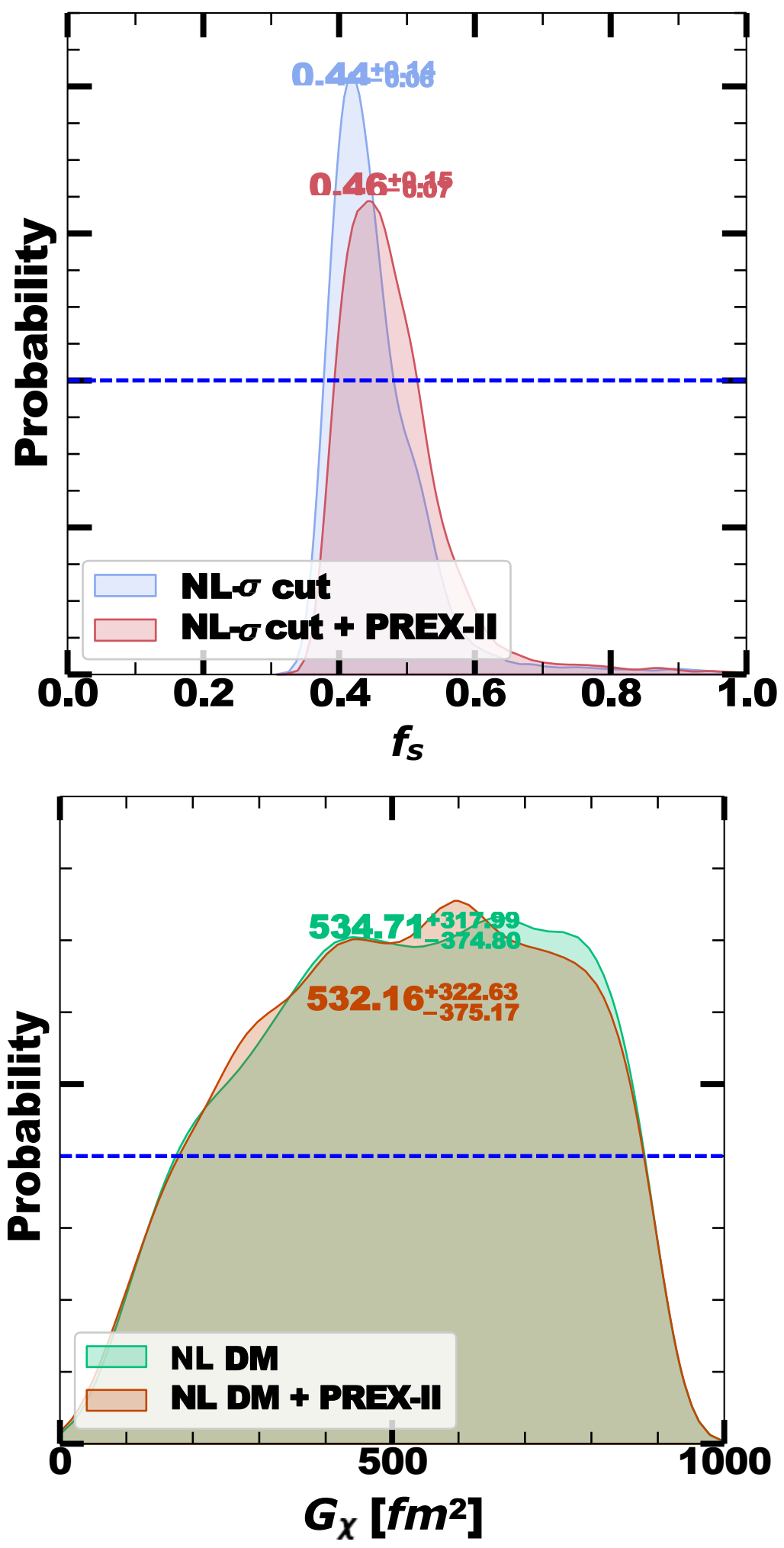

Figure 5.4: The probability distribution of the free parameter $f_s$ for the $\sigma$cut model (top) and $G_\chi$ for the DM model (bottom) is shown. Both panels include the distribution with and without PREX-II. The blue line indicates uniform prior.

Given that the inference analysis has been completed, we can now explore the constraints on the unknown parameter $f_s$ for the NL-$\sigma$ cut and the $G_\chi$ parameter for the dark matter model, under the assumption of a uniform prior for both cases. The values considered for $f_s$ ranged from 0 to 1.

The uniform prior range for $G_\chi$ (dark matter self-interaction strength) is based on Ref. Shirke *et al.* [2023]. The lower bound on $G_\chi$ is determined from the tidal deformability constraint, specifically $\Lambda_{1.4M_\odot} = 190^{+390}_{-120}$ Abbott *et al.* [2018b]. The minimum value of $G_\chi$ that ensures $\Lambda_{1.4M_\odot} > 70$ is found to be $G > 1.6\,\text{fm}^2$. However, the authors note that an upper bound on $G_\chi$ cannot be established from the upper limit of $\Lambda_{1.4M_\odot}$, as $\Lambda$ remains consistent for arbitrarily large values of $G_\chi$.

To resolve the core-cusp problem on galactic scales, the dark matter self-interaction cross-section must satisfy:

$$\frac{\sigma_\chi}{m_\chi} \leq 100\,\mathrm{cm}^2/\mathrm{g}.$$

For a dark matter particle with a mass of 938 MeV, this constraint sets an upper bound of $G_\chi \leq 943\,\mathrm{fm}^2$.

Considering these factors, we choose a prior range for $G_\chi$ from 0 to 1000 $\mathrm{fm}^2$.

Figure 5.4 shows the probability distribution of $f_s$ for the $\sigma$cut model (top panel) and of $G_\chi$ for the DM model from the Bayesian inference. In both panels, results including and without PREX-II data are shown. Including PREX-II data shifts the median value of $f_s$ slightly from 0.438 to 0.464 and broadens the distribution. Figure 5.4 bottom panel shows that $G_\chi$ is only marginally lowered by PREX-II data, suggesting a minor shift in the parameter estimate. However, the spread in $G_\chi$ remains relatively unchanged. We conclude that the wide range of constraints from the nuclear and astrophysical data considered does not significantly narrow down this parameter.

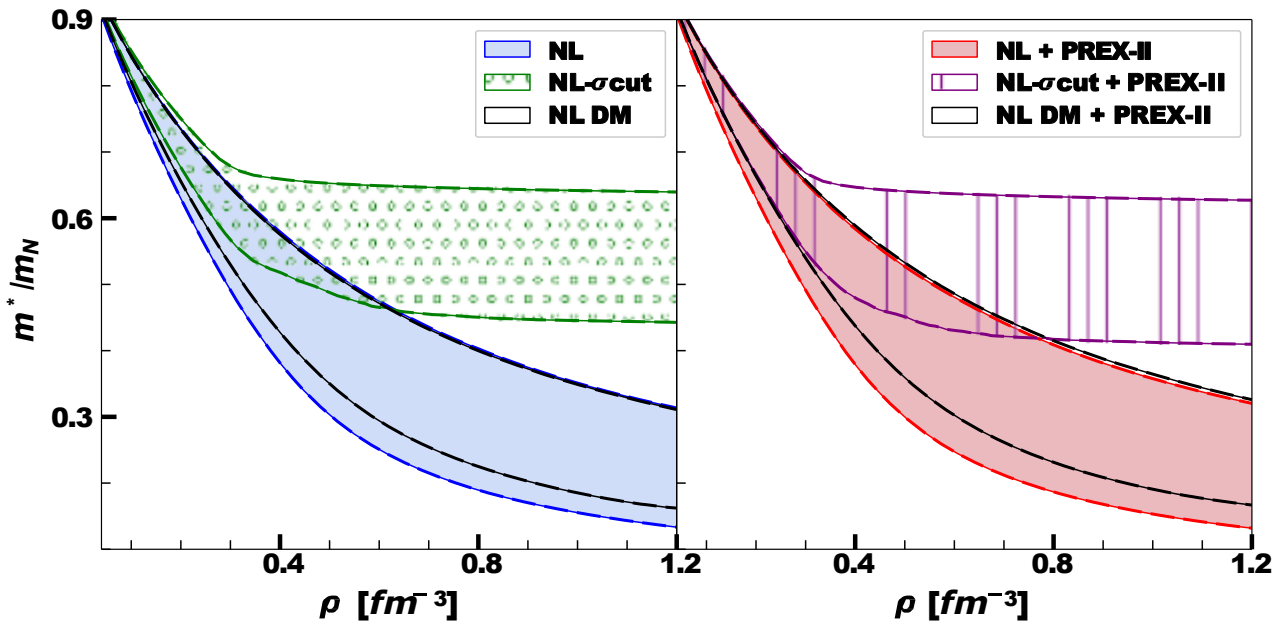

Figure 5.5: The 90% CI of effective mass ($m^*$) as a function of baryon density ($\rho$) for the NL, NL-$\sigma$-cut and NL-DM cases. The left panel with all the constraints in Table 5.1 except for the PREX-II data, while the right panel includes the PREX-II data.

Fig. 5.5 depicts the 90% CI region for the effective mass $m^*$ against baryon density $\rho$, without (left panel) and with (right panel) PREX-II constraints. The NL scenario demonstrates a decreasing trend with density, and PREX-II data exerts only a minimal effect on $m^\star$. When the $\sigma$ cut potential is applied, the effective mass stabilizes above a 0.3 $\mathrm{fm}^{-3}$ density due to the $\sigma$ potential, thereby stiffening the EOS. For the NL-DM model that includes admixed dark matter, there is a slight upward pull on the $m^\star$ posterior, but it remains encompassed by the NL model in both panels.

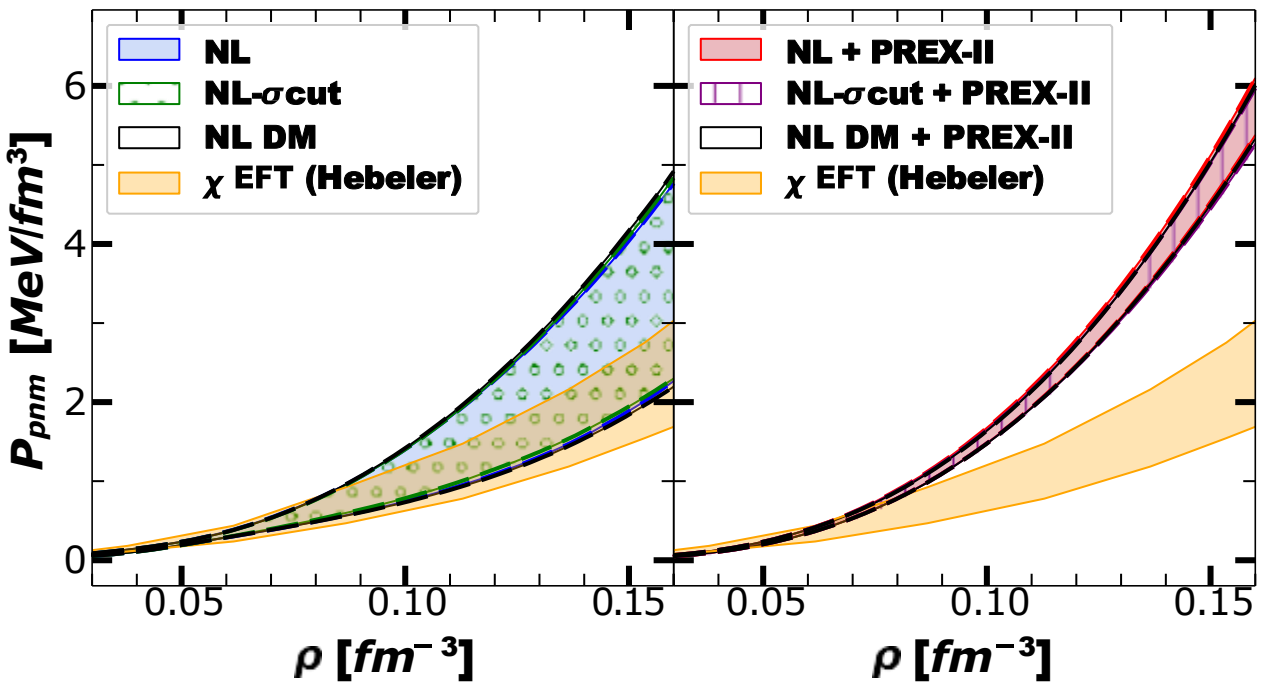

Figure 5.6: Displayed are the posterior distributions for the NL, NL-$\sigma$ cut, and NL-DM models within the 90% confidence interval (CI) regarding the pressure (P) as a function of baryon density in pure neutron matter (PNM). The plot on the left does not include PREX-II, while the plot on the right does.

In Fig. 5.6, we evaluate how well the computed posteriors match the *ab-initio* chiral effective field theoretical ($\chi$EFT) calculations for pure neutron matter (PNM) constraints across all scenarios. The figure displays the posterior distributions for the NL and NL-$\sigma$ cut models within the 90% confidence interval (CI) for the PNM pressure as a function of baryon density. At low densities near saturation density, all three cases, both with and without PREX-II, are indistinguishable. At larger densities, the inclusion of additional PREX-II data makes the pure neutron matter pressure significantly stiffer, pulling the posterior outside the PNM chEFT constraints. In the absence of PREX-II constraint, the PNM pressure posterior for all three cases shows a good overlap with $\chi$EFT data. It should be noted that the $\chi$EFT PNM constraints were not part of the constraints applied to the likelihood considered in this study. This data indicates the PREX-II data are in tension with $\chi$EFT constraints, contrary to all the other constraints included in our inference analysis, independently of the model considered.

## 5.5 f & p Mode Oscillations

In the era of multimessenger astronomy, NS asteroseismology has become crucial for insights into dense matter EOS. Gravitational wave events like GW170817 Abbott *et al.* [2017d, e] and GW190425, along with future detectors (LIGO-Virgo-KAGRA, Einstein Telescope Punturo *et al.* [2010], Cosmic Explorer), are expected to improve EOS determination. When an NS is perturbed, it oscillates in radial or non-radial modes. Radial modes involve simple expansion and contraction, maintaining a spherical shape. Non-radial oscillations (fundamental mode (f-mode), pressure mode (p-mode), and gravity mode (g-mode)) deviate from the spherical shape, driven by pressure and buoyancy Kokkotas & Schmidt [1999]. f-modes are significant for their gravitational radiation, detectable by current detectors.

The 90% credible interval for the f-mode frequency in GW170817 ranges from 1.43 kHz to 2.90 kHz for the more massive NS and from 1.48 kHz to 3.18 kHz for the less massive one Kunjipurayil *et al.* [2022]. Recent studies have examined f-mode oscillations in NS using the Cowling approximation Dimmelmeier *et al.* [2006], Pradhan & Chatterjee [2021], Ranea-Sandoval *et al.* [2018], which neglects background spacetime perturbations. These results were refined to include full general relativistic (GR) effects Kunjipurayil *et al.* [2022], Pradhan *et al.* [2022]. Studies also explored f-modes in dark matter-admixed NS. Ref. Das *et al.* [2021a] examined the Higgs interaction mode of dark matter using the Cowling approximation, while Ref. Flores *et al.* [2024] studied it in full GR. Ref. Shirke *et al.* [2024] investigated the neutron decay anomaly model of dark matter on f-mode oscillations within a full GR framework. This study focuses on non-radial oscillations of NS, such as $f$ and $p$ modes. Mode frequencies are calculated using the relativistic Cowling approximation Chirenti *et al.* [2015]. This approximation is often used in the literature for a first-step calculation. The reason is the Cowling approximation can greatly simplify the pulsating equations in full GR simulations. The Cowling approximation might overestimate the frequency by 10-30% depending on the compactness, in comparison to those frequencies obtained from linearized GR treatments Yoshida & Kojima [1997]. The overestimation in the case of Cowling compared to the linearized GR approach decreases with increasing stellar compactness. An explanation for this trend was suggested by Yoshida & Eriguchi [1997], indicating that as compactness increases, the influence of metric perturbations on the f-mode eigenfunction diminishes, as the eigenfunction is concentrated near the surface Chirenti *et al.* [2015].

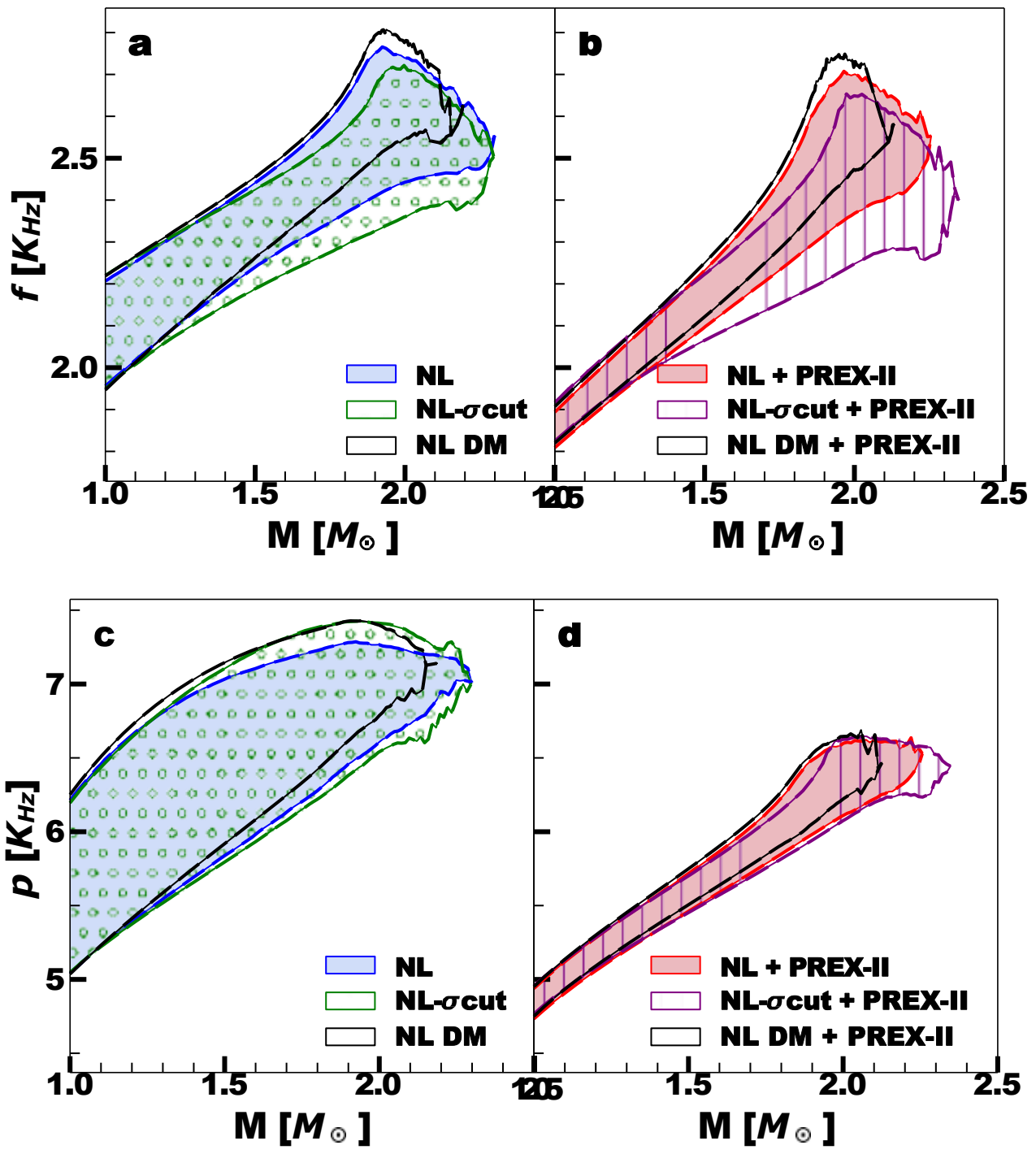

Figure 5.7: The graphs illustrate the relationship between neutron star (NS) mass ($M$ in units of solar mass $M_{\odot}$) and the frequencies of non-radial oscillation modes: $f$ (fundamental mode, upper plots) and $p$ (pressure mode, lower plots). The left panels show results for the NL, NL $\sigma$-cut, and NL-DM with dark matter, while the right panels compare the same but with additional PREX-II data. The domain represents the 90% CI region.

We have computed the non-radial oscillation modes of neutron stars, specifically the frequencies of the $f$ and $p$ modes, within the Cowling approximation using the posterior results of the three cases: Nl, NL-$\sigma$ cut, and NL-DM. Figure 5.7 illustrates the $f$ -mode (fundamental) and $p$-mode (pressure) frequencies of neutron stars as functions of their mass (in solar masses). The right column displays corresponding data incorporating PREX-II results. The $f$-mode frequency ranges approximately from 2.0 kHz to 2.8 kHz as the neutron star mass increases from 1.0 to 2.3 solar masses. The $p$-mode frequency spans from 5 kHz to 7.5 kHz over the same mass range. For PREX-II, the distribution is narrower at lower neutron star masses, an effect similar to the one obtained in the mass-radius plot. The feature *dM/df* varies among the three cases for the $f$ -mode frequency. In contrast, the $p$-mode frequency is only marginally distinguishable across all three cases and columns. The *dM/dp* slope for the $p$-mode and mass domain is similar for all three scenarios. Compared to the $p$ mode, the $f$ -mode is more sensitive to the symmetry energy and conveys analogous slope information as the mass-radius relationship. This correlation is evident as Ref Roy *et al.* [2024] established a strong relationship between NS radius and $f$-mode frequency.

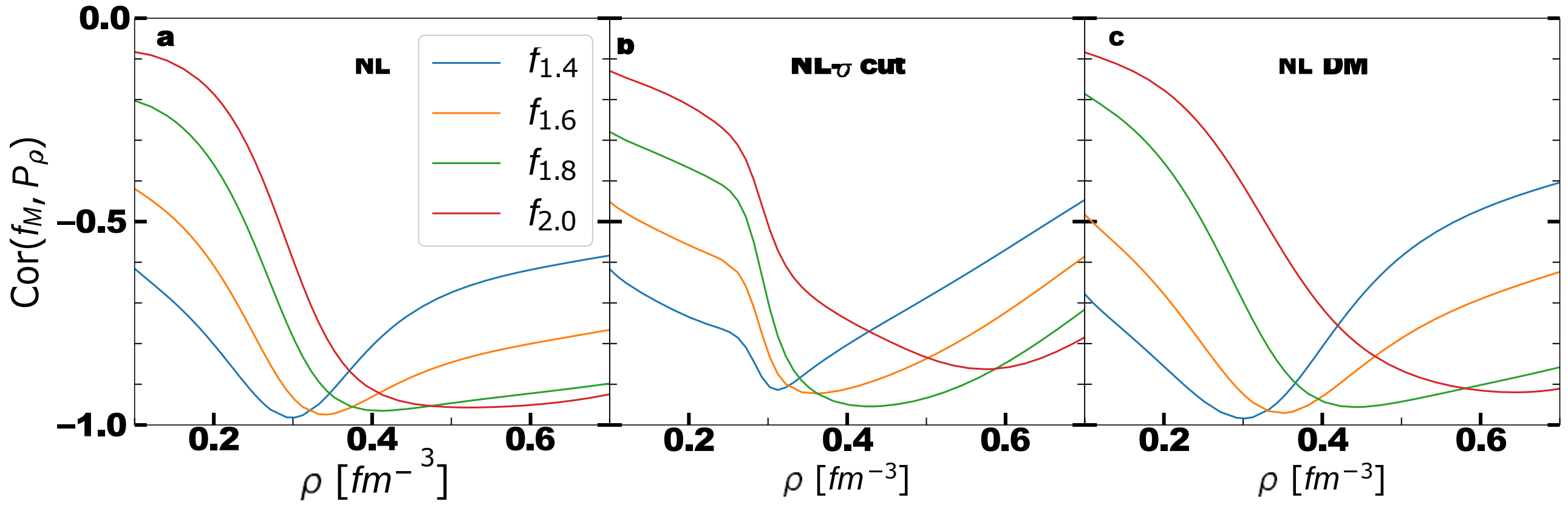

Figure 5.8: Pearson's correlation coefficients between the $f$ mode oscillation frequency $f_{\mathrm{M}}$ for neutron star masses $M \in [1.4, 1.6, 1.8, 2.0]$ $M_\odot$ and the neutron star matter pressure $P(\rho)$ across different baryon densities $\rho$, for the models without PREX-II.

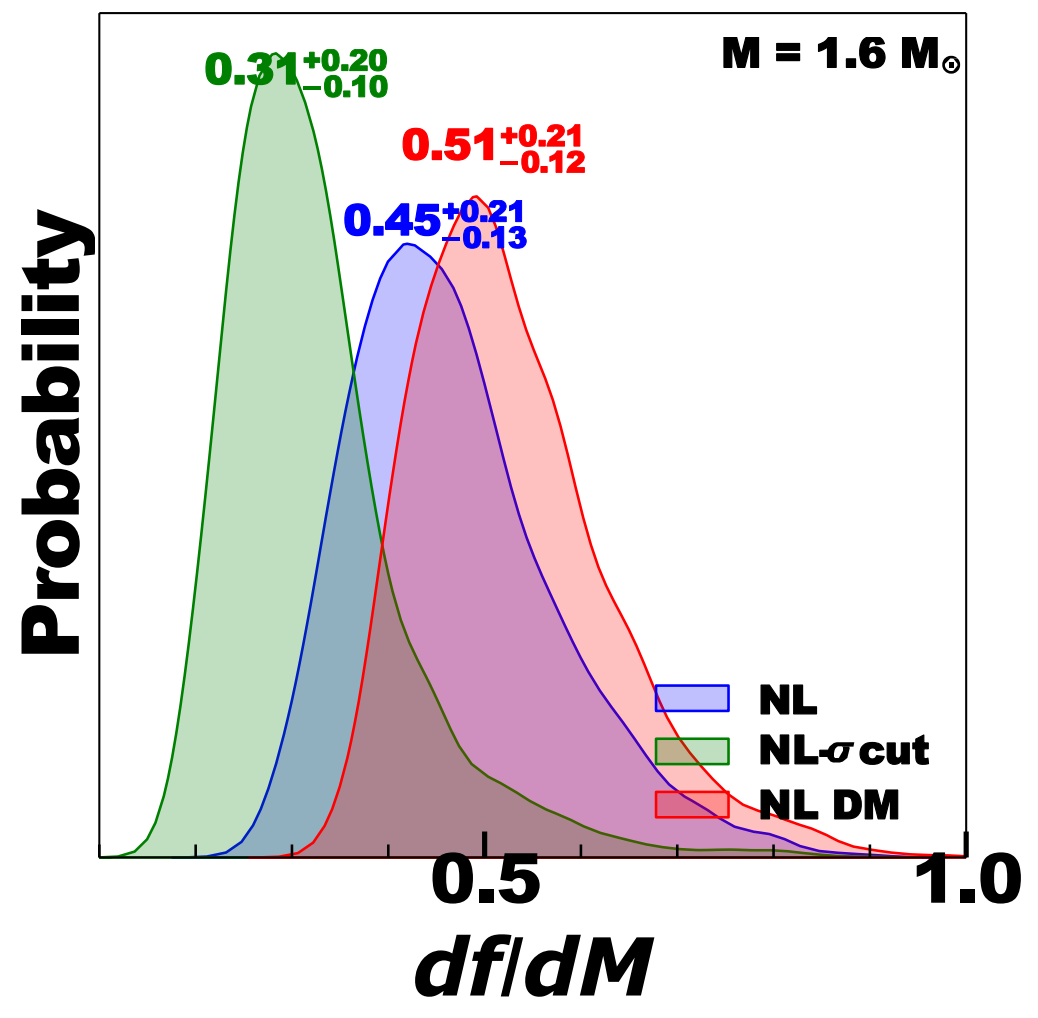

Figure 5.9: The $df/dM$ distribution at a neutron star mass of 1.6 $M_\odot$ for three scenarios: NL, NL$\sigma$cut, and NL-DM, without PREX-II data.

Fig. 5.8 displays the Pearson correlation coefficients between pressure and $f$ -mode frequencies for neutron stars of different masses. These coefficients are plotted against the baryon number density ($\rho_{\mathrm{B}}$). The figure comprises three panels corresponding to the models NL (left), NL-$\sigma$ cut (middle), and NL-DM (right, all not including the PREX-II constraint. The different colored lines correspond to neutron stars of varying masses, specifically 1.4$M_\odot$ (blue), 1.6$M_\odot$ (orange), 1.8$M_\odot$ (green), and 2.0$M_\odot$ (red). The correlation peaks at different densities for the $f$ mode, depending on the mass. As the mass increases from 1.4 to 2 $M_\odot$, the correlation peak shifts to a higher density, which is consistent with the expectation that central density for a higher mass star resides at a higher density. The correlations are stronger in the NL model, although they are all comparable: for the NL-$\sigma$ cut model, the onset of the $\sigma$cut potential seems to reduce the correlation associated with the lower mass stars. This potential also affects the correlation in the most massive stars; for the NL-DM model, the correlation is more affected for the massive stars when the effect of DM is stronger. For higher NS masses, the correlation is weaker compared to others, in particular, for NL-DM and NL-$\sigma$ cut models.

In Fig. 5.9, we plot the $df/dM$ distribution at a neutron star mass of 1.6 $M_{\odot}$ for three scenarios: NL, NL$\sigma$cut, and NL-DM, without PREX-II data. The slope $df/dM$ is positive in all these cases, in contrast to the slope $dR/dM$ (see Fig. 5.3). Moreover, it becomes apparent that there exists an inverse correlation between the neutron star (NS) radius and the $f$-mode oscillation frequency Kumar *et al.* [2023], Roy *et al.* [2024]. The slope $df/dM$ for NL, NL-$\sigma$ cut, and NL-DM are $0.45^{+0.21}_{-0.13}$, $0.31^{+0.20}_{-0.10}$, and $0.51^{+0.21}_{-0.12}$, respectively. Including dark matter in the NL-DM model results in the steepest slope in the mass-frequency plane for $f$ mode oscillation. Given that the NL-$\sigma$ cut case exhibits the strongest Bayesian evidence compared to others (see Table 5.2), we can conclude that the collective data from nuclear to astrophysical sources favors a smaller slope $df/dM$, indicative of a stiff EOS.

## 5.6 Impact of PSR J0437-4715

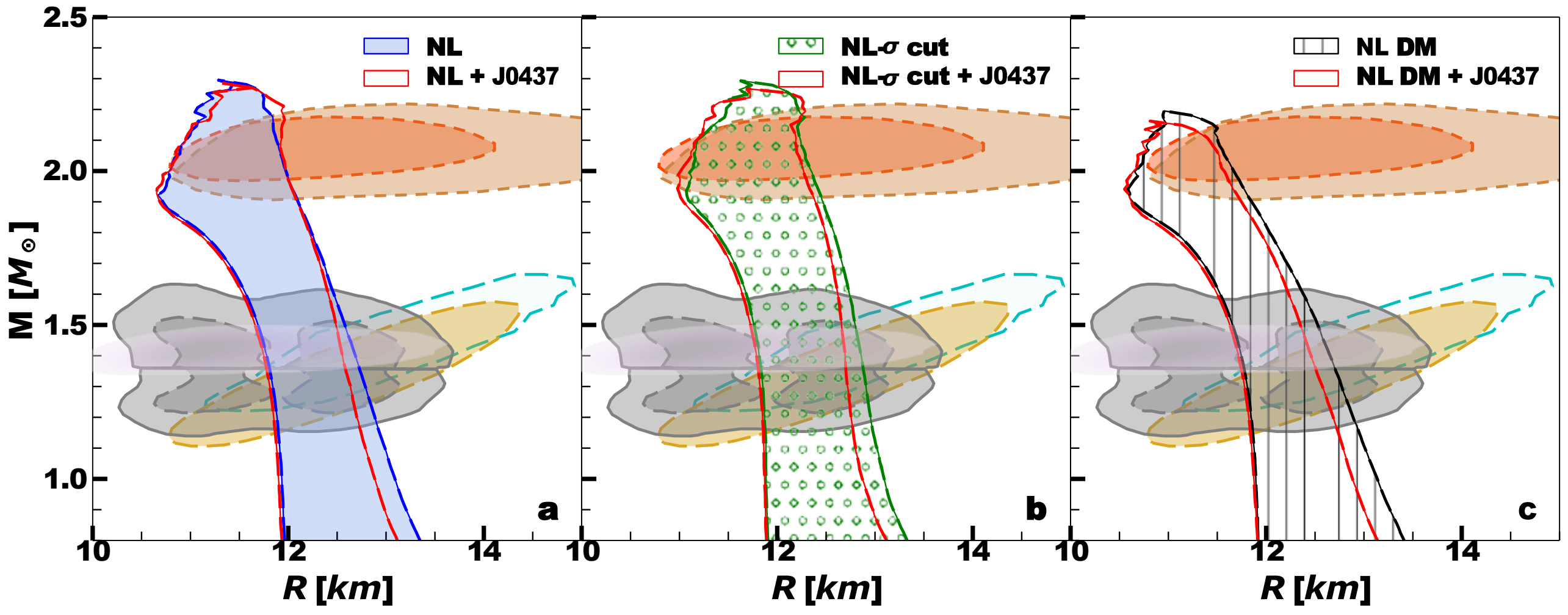

Figure 5.10: The posterior distribution of the neutron star mass-radius $P(R|M)$ for the models (a) NL, (b) NL-$\sigma$cut, and (c) NL-DM is compared with the distribution that includes the new PSR J0437-4715 NICER mass-radius measurements. The legends NL+J0437, NL-$\sigma$cut+J0437, and NL-DM+J0437 indicate the integration of PSR J0437-4715 NICER data with the previous NICER measurements. The additional constraints being compared are identical to those in Fig. 5.2. Refer to its caption for more information.

NICER's nearest and brightest target is the 174 Hz millisecond pulsar PSR J0437-4715. Using NICER data from July 2017 to July 2021, and incorporating NICER background estimates, the latest mass-radius measurements of PSR J0437-4715 are reported Choudhury *et al.* [2024]. We have investigated the effect of these measurements, together with the two old NICER measurements for pulsars PSR J0030+0451 and PSR J0740+6620, on NL, NL-$\sigma$ cut, and NL-DM. All were compared without PREX-II. Fig. 5.10 shows the posterior distribution of neutron star mass-radius relations for the NL, NL-$\sigma$ cut, and NL-DM models, incorporating the new data from PSR J0437-4715 by NICER. Panels (a), (b), and (c) represent NL, NL-$\sigma$ cut, and NL-DM, respectively, and illustrate the effect of the new NICER measurements on the mass-radius posterior. The inclusion of data from PSR J0437-4715 particularly affects the estimated radius for neutron stars in the 1 to 1.5 $M_\odot$ mass range. This new data reduces the upper limit of the 90% confidence interval by about 200 m and the lower limit by less than $\sim$ 30 m, with a consistent effect across all models, see Table 5.5 where the NS radius for masses of 1.2, 1.4, and 1.6 $M_\odot$ for the three cases, including the new PSR constraints are given. We have computed the Bayes evidence for each model that incorporates PSR J0437-4715 data but excludes PREX-II and observed a decrease of $\sim 1$ in the logarithm of the Bayes evidence in all instances, see Table 5.2. This suggests that the new NICER data conflicts with the old data or that the current EOS model lacks the flexibility to simultaneously accommodate all NICER data.

Table 5.5: The median and 90% confidence intervals (CI) of the radius $R_M$ (km) for neutron star masses $M \in [1.2, 1.4, 1.6]$ $M_\odot$, comparing results for obtained posterior with older NICER data and the additional inclusion of new PSR J0437-4715 NICER mass-radius measurements. The models include NL, NL-$\sigma$ cut, and NL-DM. The terms NL+J0437, NL-$\sigma$cut+J0437, and NL-DM+J0437 represent the inclusion of PSR J0437-4715 NICER measurements along with the older NICER data.

| Quantity | NL | | NL +J0437 | | NL-σ cut | | NL-σ cut + J0437 | | NL DM | | NL DM + J0437 | |
|---|---|---|---|---|---|---|---|---|---|---|---|---|
| | **Med.** | **90% CI** | **Med.** | **90% CI** | **Med.** | **90% CI** | **Med.** | **90% CI** | **Med.** | **90% CI** | **Med.** | **90% CI** |
| $R_{1.2}$ | 12.27 | [11.89, 12.89] | 12.23 | [11.85, 12.70] | 12.29 | [11.87, 12.95] | 12.21 | [11.85, 12.76] | 12.20 | [11.85, 12.89] | 12.14 | [11.82, 12.69] |
| $R_{1.4}$ | 12.18 | [11.78, 12.69] | 12.14 | [11.75, 12.55] | 12.29 | [11.81, 12.86] | 12.21 | [11.79, 12.69] | 12.09 | [11.75, 12.65] | 12.03 | [11.72, 12.46] |
| $R_{1.6}$ | 12.03 | [11.57, 12.47] | 11.98 | [11.54, 12.41] | 12.25 | [11.68, 12.74] | 12.17 | [11.64, 12.62] | 11.91 | [11.53, 12.38] | 11.85 | [11.50, 12.23] |

## 5.7 Conclusion

In this study, we have examined the equation of state (EOS) for neutron stars under three distinct scenarios: a purely nucleonic composition, a nucleonic composition with a $\sigma$-cut potential, and a nucleonic composition with an admixture of dark matter. The effect of the $\sigma$-cut potential is to stiffen the EOS above saturation density, having a net effect similar to the presence of a quarkyonic phase, see McLerran & Reddy [2019], or an exclusion volume Typel [2016]. By employing Bayesian inference and incorporating the latest constraints from nuclear physics and astrophysical observations, we have been able to evaluate the plausibility and impact of each scenario. Our analysis reveals that the inclusion of dark matter and modified potentials in the EOS significantly affects the macroscopic properties of neutron stars, such as their mass-radius relationships and non-radial oscillation modes.

Our results indicate that the inclusion of PREX-II constraints has a strong effect on several NS properties, such as mass-radius curves or $f$ -modes, and in particular, on the slope of these curves. PREX-II data shifts the radius of low-mass stars to very large radii of the order of 13.5 - 14 km. The models including PREX-II data entirely unsuccessful to reproduce $\chi$EFT PNM pressure. Also, the calculation of the Bayes factor has shown decisive or substantial evidence against these models when compared with the models, not including PREX-II data.

Table 5.6: The median and 90% confidence interval (CI) values for the derived parameters across various posteriors. Specifically, $B$ and $C$ are $b \times 10^3$ and $c \times 10^3$, respectively. The parameter $f_s$ in the NL - $\sigma$ cut model is dimensionless, while the parameter $G_\chi$ in the NL DM model is measured in units of fm$^2$.

| Quantity | Without PREX-II | | | | | | With PREX-II | | | | | |
|---|---|---|---|---|---|---|---|---|---|---|---|---|
| | NL | | NL-σ cut | | NL DM | | NL | | NL-σ cut | | NL DM | |
| | **Med.** | **CI** | **Med.** | **CI** | **Med.** | **CI** | **Med.** | **CI** | **Med.** | **CI** | **Med.** | **CI** |
| $g_\sigma$ | 8.438 | [7.827, 8.915] | 7.965 | [7.681, 8.470] | 8.257 | [7.842, 8.653] | 8.369 | [7.775, 8.907] | 8.073 | [7.714, 8.576] | 8.144 | [7.753, 8.552] |
| $g_\omega$ | 9.903 | [8.689, 10.715] | 8.901 | [8.381, 9.927] | 9.540 | [8.719, 10.242] | 9.767 | [8.612, 10.687] | 9.120 | [8.475, 10.121] | 9.336 | [8.557, 10.082] |
| $g_\rho$ | 10.108 | [9.282, 10.950] | 10.059 | [9.311, 10.840] | 10.097 | [9.274, 10.927] | 9.288 | [9.048, 9.562] | 9.364 | [9.142, 9.625] | 9.342 | [9.118, 9.586] |
| B | 5.199 | [4.167, 7.897] | 7.253 | [4.961, 8.731] | 5.746 | [4.578, 7.880] | 5.346 | [4.172, 8.125] | 6.722 | [4.678, 8.612] | 6.066 | [4.767, 8.257] |
| C | -4.106 | [-4.916, 0.346] | -2.124 | [-4.692, 3.478] | -3.838 | [-4.901, -0.032] | -3.969 | [-4.897, 1.540] | -2.613 | [-4.758, 2.709] | -3.441 | [-4.861, 1.784] |
| ξ | 0.005 | [0.000, 0.012] | 0.011 | [0.001, 0.028] | 0.004 | [0.000, 0.012] | 0.006 | [0.000, 0.015] | 0.015 | [0.002, 0.032] | 0.005 | [0.000, 0.013] |
| $\Lambda_\omega$ | 0.047 | [0.013, 0.092] | 0.060 | [0.014, 0.104] | 0.051 | [0.012, 0.096] | 0.001 | [0.000, 0.006] | 0.002 | [0.000, 0.008] | 0.002 | [0.000, 0.007] |
| $f_S$ | - | - | 0.44 | [0.38, 0.58] | - | - | - | - | 0.46 | [0.39, 0.61] | - | - |
| $G_\chi$ | - | - | - | - | 534.71 | [159.91, 852.7] | - | - | - | - | 532.16 | [156.99, 854.79] |

The analysis of the effect of the $\sigma$ cut potential has shown that the constraints imposed in our Bayesian inference calculation favor this model, giving larger Bayes factors. The NL-$\sigma$ cut model gives rise to a stiffening of the EOS at large densities and, therefore, predicts massive stars with larger radii and smaller $f$ mode frequencies. It also presents a very distinctive effect on the speed of sound, giving rise to a steep increase above 0.2 fm$^{-3}$ and a leveling out above 0.4 fm$^{-3}$. Also, the trace anomaly-related quantity was affected, showing a clear peak for $\rho \sim 0.3$ fm$^{-3}$ followed by a steep decrease attaining values below 0.2 at 0.6 fm$^{-3}$ while for the other models, there is no distinctive peak and the values 0.2 is only reached above 0.8 fm$^{-3}$.

Moreover, our investigation also examined the non-radial oscillations, specifically the $f$ and $p$ modes. We identified a large sensitivity of the $f$ oscillations to changes in the neutron star's composition and EOS. Although working within the Cowling approximation, which in future work should be generalized to incorporate full general relativistic effects, it was shown the existence of a strong correlation between the $f$ -modes and the NS pressure at different densities, with the correlation peak shifting to a higher density as the mass increases. We also analyzed the slope of the $f$ mode curve with respect to the star mass. The smallest one was associated with the NL-$\sigma$ cut model, the model that presented the largest values of $dR/dM$, possibly even positive, due to the presence of a stiff EOS at high densities. This is the most favored model.

We have also investigated the impact of the new PSR J0437-4715 measurements on the neutron star mass-radius posterior distribution, observing a consistent reduction of approximately 0.2 km in the upper boundary of the 90% confidence interval across all models. This refinement enhances the model fit, as evidenced by the notable decrease in the logarithmic Bayes evidence ($\sim 1$), suggesting either a conflict with previous measurements or a need for more flexible theoretical models to accommodate the updated data.

# Chapter 6

# Summary and Conclusion

This thesis presents an extensive study on the impact of dark matter on neutron star properties, utilizing both theoretical and statistical approaches to explore their astrophysical implications. By integrating both single fluid and two fluid formalisms, Bayesian inference, and machine learning techniques, we systematically examine how dark matter influences neutron star structure, mass-radius relationships, and oscillation modes, while also incorporating the latest observational constraints.

Our study reveals that the presence of dark matter within neutron stars significantly alters their macroscopic properties, including mass, radius, and tidal deformability. The two-fluid formalism indicates that dark matter admixture leads to higher central densities, which may enable exotic processes such as direct Urca cooling. Furthermore, we confirm that the semi-universal C-Love relation remains largely intact, suggesting that in the present scenario it is difficult to distinguish between with and without dark matter admixed neutron stars.

To overcome challenges in identifying dark matter signatures, we employed machine learning techniques, training Random Forest classifiers on large neutron star datasets. While our results indicate that neutron stars with dark matter can be statistically distinguished from purely hadronic stars, a 17% misclassification rate highlights the inherent complexity of this problem. The inclusion of tidal deformability data did not significantly improve classification accuracy, suggesting that neutron star radii—especially at extreme mass points—may serve as more reliable indicators of dark matter presence.

We employed a Bayesian framework to assess the feasibility of dark matter admixed neutron stars by considering recent observational constraints. Using the Relativistic Mean Field (RMF) model with various equations of state (EOS), we explored three distinct scenarios: neutron stars composed solely of nucleonic matter, neutron stars with a modified $\sigma$-cut potential to stiffen the EOS, and neutron stars admixed with fermionic dark matter based on neutron decay anomaly. Our analysis shows that the $\sigma$-cut potential model, which stiffens the EOS at high densities, provides the best fit to observational constraints. However, our results also highlight tension with new PREX-II data as it fails to match $\chi$EFT PNM pressure predictions. Also, the calculation of the Bayes factor has shown decisive or substantial evidence against these models when compared with the models, not including PREX-II data. Additionally, we find that non-radial oscillations, particularly f-modes, serve as promising probes of neutron star composition, with their frequencies being highly sensitive to EOS variations. We have also investigated the impact of the new PSR J0437- 4715 measurements on the neutron star mass-radius posterior distribution, observing a consistent reduction of approximately $\simeq 0.2$ km in the upper boundary of the 90% confidence interval across all models, suggesting either a conflict with prior measurements or a need for more adaptable theoretical models to account for the updated data.

### 6.0.1 Future Scope:

Future advancements in multi-messenger astronomy will greatly improve our understanding of dark matter in neutron stars. The combination of third-generation gravitational wave detectors (Einstein Telescope, Cosmic Explorer) and high-precision X-ray observatories (NICER, ATHENA) will provide more stringent constraints on neutron star mass, radius, and tidal deformability. A Bayesian inference framework, integrating diverse nuclear and dark matter models with new astrophysical data, will enhance our ability to differentiate between purely hadronic and dark matter-admixed neutron stars. Additionally, progress in machine learning techniques, such as deep neural networks and Gaussian process regression, will refine our capacity to infer the equation of state directly from observational data, providing a novel method for studying dense matter physics.

Further research on neutron star oscillations, modified gravity theories, and highly magnetized stars will open new frontiers in compact object astrophysics. Investigating both radial and non-radial oscillations of dark matter-admixed neutron stars will shed light on their structural and dynamical properties. Additionally, studies on hybrid stars with hadron-quark phase transitions, the impact of strong magnetic fields on the equation of state, and anisotropic stellar configurations in alternative gravity models will offer deeper insights into the behavior of matter under extreme conditions. With increasing observational precision, these theoretical advancements will be essential for interpreting neutron star data, bringing us closer to uncovering the nature of dark matter and dense matter physics.

# Mr. Prashant Thakur orcid link-

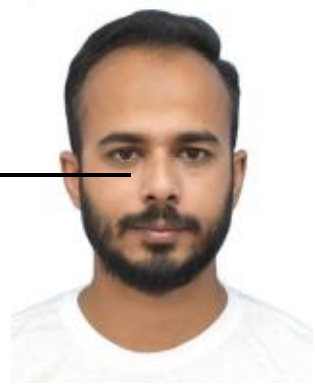

**CONTACT INFORMATION**

**Research Scholar**
Department of Physics, BITS Pilani, K. K. Birla Goa Campus
Room No- PhD Hall
State: Goa, PIN:403726, Country: India.

mobile: +91 8580404641
e-mail: p20190072@goa.bits-pilani.ac.in

**PERMANENT ADDRESS**

Singh's House, Near boys hostel
Vivekanand Vihar
City: Solan ; State: Himachal Pradesh,
PIN: 173212; Country: India

Home: + 91 8580404641
e-mail: prashantthakur1921@gmail.com

**EDUCATION**

- **Ph.D.** (2020-2024) **(Submitted on 03/09/2024)**
  -**Birla Institute of Science & Technology** – Pilani, Goa, India
  -Supervisor: **Prof. Tarun Kumar Jha**
  -Advisor: **Dr. Tuhin Malik**

- **Masters of Science (M.Sc) Physics:**
  -**Shoolini University, Solan, India**,
  -**Year of Completion :** 2018
  -**Degree Grade Point Average :** 7.77/10.00

- **Bachelor of Science (B.Sc) Physics:**
  -**Centre of Excellence, Sanjauli Degree College**, Shimla, Himachal Pradesh
  -**Year of Completion :**2015
  -**Pass Subjects :** Physics(Hons.) Mathematics, Chemistry, English, Hindi

- **Higher Secondary Exam:**
  -**Chapslee School**, Shimla, Himachal Pradesh
  -**Year of Completion:** 2011

- **Secondary Exam:**
  -**Chapslee School**, Shimla, Himachal Pradesh
  -**Year of Completion :** 2009

**RESEARCH INTERESTS**

*My research focuses on neutron stars and their equations of state, particularly their properties, interactions with dark matter, and how these interactions can constrain from astrophysical observations. I aim to understand the internal structure, dynamics, and role of exotic particles in neutron stars. During my PhD, I have had the opportunity to gain expertise in various advanced methodologies and tools that are pivotal for state-of-the-art research in high energy nuclear physics and astrophysics. Specifically, I have developed a strong proficiency in Bayesian Inference, which has been instrumental in my data analysis and modeling efforts. My work with Machine Learning has enabled me to leverage complex algorithms and models to uncover patterns and insights from large datasets. Furthermore, my experience with Big Data Analysis has equipped me with the skills to handle and interpret vast amounts of data efficiently. Additionally, I have become adept at using Mathematica for symbolic computations and analytical derivations, which has greatly enhanced my problem-solving capabilities. My proficiency with the RNS (Rotating Neutron Star) code has allowed me to perform detailed simulations and analyses of neutron star properties, contributing significantly to my research on dense matter physics and dark matter interactions. Currently, and in the foreseeable future, my research aims to explore the internal structure of neutron stars (NS) and investigate the presence of dark matter within them through a phenomenological methodology, in which I am well-versed.*

**RESEARCH VISIT**

- Departamento de Física, University of Coimbra, Coimbra, Portugal, from 1st May 2023 - 30th July 2023.
  .
- Inter-University Centre for Astronomy and Astrophysics, India, from 13th September 2024 - 30th September 2024.
  Visited Dr. Apratim Ganguly.

**PUBLICATIONS**

**JOURNALS**

## In Communication

### 1. Feasibility of Dark Matter Admixed Neutron Star Based on Recent Observational Constraints

**Authors:** *Prashant Thakur, Tuhin Malik, Arpan Das, B.K. Sharma, T.K. Jha, Constança Providência*

**arXiv:** arXiv:2408.03780v1

**In Communication to:** Astronomy & Astrophysics

### 2. Non-Radial Oscillation Modes in Hybrid Stars with Hyperons and Delta Baryons: Full General Relativity Formalism vs. Cowling Approximation

**Authors:** *Ishfaq Ahmad Rather, Kau D. Marquez, Prashant Thakur, Odilon Lourenço*

**e-Print:** e-Print: 2412.12002 [astro-ph.HE]

**In Communication to:** Physical Review D

**Date:** Dec 16, 2024

### 4. Supernova Remnants with Mirror Dark Matter and Hyperons

**Authors:** *Adamu Issifu (Espirito Santo U.), Prashant Thakur (Birla Inst. Tech. Sci.), Franciele M. da Silva (Londrina U.), Kau D. Marquez (Espirito Santo U.), Débora P. Menezes (Londrina U.) et al.*

**e-Print:** e-Print: 2412.17946 [hep-ph]

**In Communication to:** Physical Review D

**Date:** Dec 23, 2024

### 3. Impact of $\sigma$-cut Potential on Antikaon Condensation in Neutron Stars within the Relativistic Mean Field Model

**Authors:** *Prashant Thakur, B. K. Sharma, Lakshana Sudarsan, Krishna Kunnampully, T. K. Jha*

**In Communication to:** Physical Review C

## Published

### 1. Hyperon Bulk Viscosity and r-Modes of Neutron Stars

**Authors:** *O P Jyothilakshmi, P E Sravan Krishnan, Prashant Thakur, V Sreekanth, T.K. Jha*

**DOI:** 10.1093/mnras/stac2360

**Journal:** Monthly Notices of the Royal Astronomical Society, 516 (2022) 3, 3381-3388

### 2. Exploring Robust Correlations Between Fermionic Dark Matter Model Parameters and Neutron Star Properties: A Two-Fluid Perspective

**Authors:** *Prashant Thakur, Tuhin Malik, Arpan Das, T.K. Jha, Constança Providência*

**DOI:** 10.1103/PhysRevD.109.043030

**Journal:** Physical Review D, 109 (2024) 4, 043030

### 3. Towards Uncovering Dark Matter Effects on Neutron Star Properties: A Machine Learning Approach

**Authors:** *Prashant Thakur, Tuhin Malik, T.K. Jha*

**DOI:** 10.3390/particles7010005

**Journal:** Particles, 7 (2024) 1, 80-95

### 4. Influence of the Symmetry Energy and $\sigma$-cut Potential on the Properties of Pure Nucleonic and Hyperon-Rich Neutron Star Matter

**Authors:** *Prashant Thakur, B. K. Sharma, A. Ashika, S. Srivishnu, T.K. Jha*

**DOI:** 10.1103/PhysRevC.109.025805

**Journal:** Physical Review C, 109 (2024) 2, 025805

*

## Conference Proceedings

### 1. Neutron Stars with Fermionic Dark Matter: A Two-Fluid Approach

**Authors:** *Prashant Thakur, T.K. Jha*

**Proceedings:** DAE Symp.Nucl.Phys., 66 (2023) 776-777

### 2. Antikaon Condensates with Dark Vector Meson in Neutron Stars

**Authors:** *Prashant Thakur, T.K. Jha*

**Proceedings:** DAE Symp.Nucl.Phys., 66 (2023) 804-805

### 3. HESSJ1731-347 Supernova Remnant as Possible Dark Matter Admixtured Candidate

**Authors:** *Prashant Thakur, T.K. Jha, B.K. Sharma*

**Proceedings:** DAE Symp.Nucl.Phys., 67 (2024) 817-818

### 4. Neutron Stars Anisotropic Nature: A Study of Exotic States of Matter and Cosmic Observations

**Authors:** *Premachand Mahapatra, Prashant Thakur*

**Proceedings:** DAE Symp.Nucl.Phys., 67 (2024) 819-820

### 5. On the Possibility of a 2.6$M_\odot$ Neutron Star

**Authors:** *Tamanna Iqbal, R. Chandra, B.K. Sharma, Prashant Thakur, T.K. Jha*

**Proceedings:** DAE Symp.Nucl.Phys., 66 (2023) 772-773

**Conferences & Workshops Attended**

- **Gravitational-Wave Astronomy Summer School (Online)**
  Organized by ICTS-TIFR, Bengaluru, India — July 5-16, 2021
- **ICTS Summer School on Gravitational-Wave Astronomy 2022**
  Hosted offline at ICTS-TIFR, Bengaluru, India — May 30 - June 10, 2022
- **Workshop on Lunar Gravitational-Wave Detection**
  ICTS-TIFR, Bengaluru, India — April 17-20, 2023

- **DAE Symposium on Nuclear Physics 2022**
  Cotton University, Guwahati, Assam, India — December 1-5, 2022
  *Presented Poster*

- **Dark Matter and Stars: Multi-Messenger Probes of Dark Matter and Modified Gravity**
  Centro de Congressos, CENTRA, IST, University of Lisbon, Portugal — May 3-5, 2023
  *Presented Poster*

- **DAE Symposium on Nuclear Physics 2023**
  IIT Indore, Madhya Pradesh, India — December 9-13, 2023
  *Presented Poster*

- **NEOSGrav2024: International Conference on Neutron Star Equation of State and Gravitational Waves**
  Kenilworth Hotel, Goa, India — October 1-4, 2024
  *Invited Talk*

- **3rd International Conference on Neutrinos and Dark Matter**
  Cairo, Egypt — Dec 11-14, 2024
  *Invited talk*

**TEACHING**

· Teaching Assistance (TA) at BITS-Pilani Goa (Mechanics Lab, Electrodynamics and Optics Lab)

**PROGRAMMING LANGUAGES/SKILLS**

**Coding Skills:**

- Python, FORTRAN 90, Linux Shell scripting

· **Software:**

- Wolfram Mathematica, RNS, LORENE, NMMA, LaTeX

· **Gravitational Wave Analysis:**

- BILBY

· **Neutron Star Related Codes:**

- Equation of States (Relativistic Mean Field Theory)
- Dark Matter Modeling (Fermionic and Bosonic)
- Tolman-Oppenheimer-Volkoff (TOV) Equation Solver
- Two-Fluid TOV Solver
- Non-Radial Oscillations of Neutron Stars (f, p, and g modes) using both Cowling Approximation and Full GR Framework
- Modified theory of Gravity f(R,T)
- Anisotropic Neutron Stars

**KNOWN LANGUAGES**

· English, Hindi, Punjabi, Pahadi

**PERSONAL DETAILS**

· Name of Father : Mr. HarKrishan Singh (Retired Government officer).
· Name of Mother : Mrs. Sangeeta (House Wife)
· Date of Birth : 19th March, 1992
· Nationality : Indian
· Marital Status : Married
· Spouse name: Prachie Sharma

**REFEREES**

**Prof. Tarun Kumar Jha**
Associate Professor
Physics Department
Birla Institute of Technology & Science, Pilani
K K Birla Goa Campus
India
e-mail: *tkjha@goa.bits-pilani.ac.in*

**Prof. Bharat Kishore Sharma**
Assistant Professor
Department of Sciences
Amrita School of Physical Sciences, Amrita Vishwa Vidyapeetham, Coimbatore 641112
India
e-mail: *bk_sharma@cb.amrita.edu*

**Prof. Constança Providência**
Professor
CFisUC
Department of Physics, University of Coimbra,
P-3004 - 516 Coimbra, Portugal
e-mail: *cp@uc.pt*

**Dr. Tuhin Malik**
Researcher
CFisUC
University of Coimbra
Portugal
e-mail: *tuhin.malik@uc.pt*

**Dr. Arpan Das**
Assistant Professor
Physics Department
Birla Institute of Technology & Science, Pilani
Pilani Campus, India
e-mail: *arpan.das@pilani.bits-pilani.ac.in*

# Tarun Kumar Jha, Ph.D.

**Associate Professor, Department of Physics**
**BITS Pilani K K Birla Goa Campus, Goa - 403726; INDIA**

tkjha@goa.bits-pilani.ac.in
https://www.bits-pilani.ac.in/goa/tarun-kumar-jha/
**ORCID** 0000-0002-9334-240X
**Google Scholar**
https://scholar.google.co.in/citations?user=gmFaVeIAAAAJ&hl=en

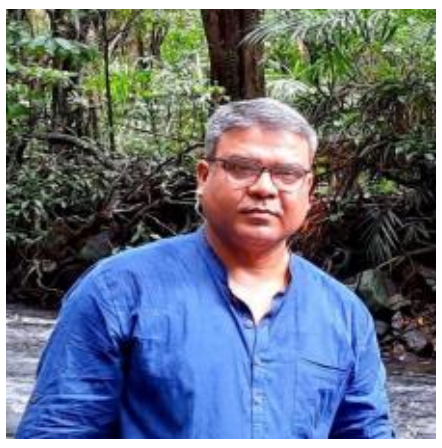

## Employment History

2021 – · · · · · **Associate Professor** Department of Physics
BITS Pilani K K Birla Goa Campus; Goa - 403 726, INDIA.

2009 – 2020 **Assistant Professor** Department of Physics
BITS Pilani K K Birla Goa Campus; Goa - 403 726, INDIA.

2007 – 2009 **Post Doctoral Fellow** Theoretical Physics Division
Physical Research Laboratory, Ahmedabad - 380 009; INDIA.

## Education

2002 – 2007 **Ph.D., Indian Institute of Technology - Kharagpur** Physics,
West Bengal - 721302, INDIA
Thesis title: *Relativistic Nuclear Equation of State in an Effective Model.*

1997 – 1999 **M.Sc. Nuclear Physics**
Sambalpur University, Sambalpur - 768 - 019; Odisa, INDIA.

1993 – 1996 **B.Sc. Physics (H)** Gangadhar Meher University,
Sambalpur - 768 001; Odisa, INDIA.

1991 – 1993 **Class XII (CBSE, Science (PCMB))** Atomic Energy Central School, Jaduguda,
Jharkhand - 832102, INDIA.

1991 **Class X (CBSE)** Atomic Energy Central School, Jaduguda,
Jharkhand - 832102, INDIA.

## Ph.D. Thesis Supervision

### Ph.D. Awarded

16th August 2019 **Debashree Sen (2013 PHXF 0414G)**
Department of Physics, BITS PIlani K K Birla Goa Campus, Goa - 403726
**Thesis Title: Cold Dense Matter Phases and Neutron Star Structure in the Light of Recent Observation**

11th March 2020 **Tuhin Malik (2014 PHXF 0402G)**
Department of Physics, BITS PIlani K K Birla Goa Campus, Goa - 403726
**Thesis Title: Equation of State of Dense Matter From Finite Nuclei to Neutron Star Merger**

24th April 2024 **Naresh Kumar Patra (2019 PHXF 0033G)**
Department of Physics, BITS PIlani K K Birla Goa Campus, Goa - 403726
**Thesis Title: Bayesian and Principal Component Analyses of Neutron Star Properties**

## Ph.D. Thesis Supervision (continued)

### Ph.D. Ongoing

- **Prashant Thakur (2019 PHXF 0072G)**
  Department of Physics, BITS PIlani K K Birla Goa Campus, Goa - 403726
  **Thesis Title:**

- **Harsh Chandrakar (2022 PHXF 0023G)**
  Department of Physics, BITS PIlani K K Birla Goa Campus, Goa - 403726
  **Thesis Title:**

## Research Publications

### Journal Articles

1. P. Thakur, T. Malik, A. Das, **T. K. Jha**, and C. ç. Providência, "Exploring robust correlations between fermionic dark matter model parameters and neutron star properties: A two-fluid perspective," *Phys. Rev. D*, vol. 109, p. 043 030, 4 Feb. 2024. DOI: 10.1103/PhysRevD.109.043030.
2. P. Thakur, T. Malik, A. Das, **T. K. Jha**, B. K. Sharma, and C. Providência, "Feasibility of dark matter admixed neutron star based on recent observational constraints," 2024. arXiv: 2408.03780 [nucl-th]. Url: https://arxiv.org/abs/2408.03780.
3. P. Thakur, T. Malik, and **T. K. Jha**, "Towards uncovering dark matter effects on neutron star properties: A machine learning approach," *Particles*, vol. 7, no. 1, pp. 80–95, 2024, ISSN: 2571-712X. DOI: 10.3390/particles7010005.
4. P. Thakur, B. K. Sharma, A. Ashika, S. Srivishnu, and **T. K. Jha**, "Influence of the symmetry energy and $\sigma$-cut potential on the properties of pure nucleonic and hyperon-rich neutron star matter," *Phys. Rev. C*, vol. 109, p. 025 805, 2 Feb. 2024. DOI: 10.1103/PhysRevC.109.025805.
5. N. K. Patra, P. Saxena, B. K. Agrawal, and **T. K. Jha**, "Establishing connection between neutron star properties and nuclear matter parameters through a comprehensive multivariate analysis," *Phys. Rev. D*, vol. 108, p. 123 015, 12 Dec. 2023. DOI: 10.1103/PhysRevD.108.123015.
6. P. Thakur, N. K. Patra, **T. K. Jha**, and T. Malik, "High-density behavior of symmetry energy and speed of sound in the dense matter within an effective chiral model," 2023. arXiv: 2301.12690 [nucl-th]. Url: https://arxiv.org/abs/2301.12690.
7. O. P. Jyothilakshmi, P. E. S. Krishnan, P. Thakur, V. Sreekanth, and **T. K. Jha**, "Hyperon bulk viscosity and r-modes of neutron stars," *Monthly Notices of the Royal Astronomical Society*, vol. 516, no. 3, pp. 3381–3388, Aug. 2022, ISSN: 0035-8711. DOI: 10.1093/mnras/stac2360. eprint: https://academic.oup.com/mnras/article-pdf/516/3/3381/45882806/stac2360.pdf.
8. N. K. Patra, B. K. Sharma, A. Reghunath, A. K. H. Das, and **T. K. Jha**, "Effect of the $\sigma$-cut potential on the properties of neutron stars with or without a hyperonic core," *Phys. Rev. C*, vol. 106, p. 055 806, 5 Nov. 2022. DOI: 10.1103/PhysRevC.106.055806.
9. N. K. Patra, T. Malik, D. Sen, **T. K. Jha**, and H. Mishra, "An equation of state for magnetized neutron star matter and tidal deformation in neutron star mergers," *The Astrophysical Journal*, vol. 900, no. 1, p. 49, Aug. 2020. DOI: 10.3847/1538-4357/aba8fc.
10. B. K. Sharma, S. Sathees, M. K. Meghaa, and **T. K. Jha**, "Effect of v coupling on liquid gas phase transition in warm asymmetric nuclear matter," *Nuclear Physics A*, vol. 1002, p. 121 974, 2020, ISSN: 0375-9474. DOI: https://doi.org/10.1016/j.nuclphysa.2020.121974.

## Conference Proceedings

## Research Projects

### Projects Completed: As PI

2013 - 2016 **DAE BRNS (18 Lacs)**
**Neutron Stars as potential continuous gravitational wave emitter.**

2012 - 2013 **BITS Research Initiation Grant (2 Lacs)**
**Neutron Stars as potential continuous gravitational wave emitter.**

### Projects Completed: As Co-PI

2015 - 2018 **DAE BRNS (20 Lacs)**
**Aspects of some astrophysical systems like supernovae and neutron stars in the light of new physics.**

## Skills

Operating System — Linux, Dos, Windows.

Languages — Strong reading, writing and speaking competencies for English.

Coding — Python, Fortran & C, Mathematica, LaTeX, Xmgrace, Gnuplot, Origin

Misc. — Academic research, teaching, training, consultation, LaTeX typesetting and publishing.

## Miscellaneous Experience

### Other Informations

1999 **M.Sc. Project**, in "Linear Electron Acceleration" at Inter University Consortium for Department of Atomic Energy Facilities, Kolkata Center under the supervision of Dr. J V M Krishna.

2002 - 2004 **Research**, Institute of Physics, Bhubaneshwar, under the supervision of Prof. S. K. Patra.

2003 - 2004 **Category A speaker, TPSC**, in Nuclear Physics.

2004 - 2005 **Lab Assignments**, for M.Sc. Ist year and Nuclear Physics Experiments at IIT Kharagpur.

2006 - 2007 **Category A speaker, TPSC**, in Nuclear Physics.

### Talks at important places

- Neutron stars as potential continuous gravitational wave emitter, talk at Institute of Physics, Bhubaneswar, Odisa, India (26 May - 5 June 2013).

## Miscellaneous Experience (continued)

- Neutron Stars: New window to the Universe, an invited talk in 'National Conference in Nuclear Physics (NCNP-2013)' at Sambalpur university, Odisa, India (1 - 3 March 2013).
- Applicability of Relativistic Mean Field Theory to Nuclear Reactions, an invited talk in 'SURROGATE - 2013' at M S university, Baroda, India (24 - 25 January 2013).
- Equation of State and Neutron Star Structure, an invited talk at the workshop on 'Gravitational Wave Data Analysis' at BITS Pilani K K Birla Goa Campus, Goa, India (17 - 21 December 2012).
- Recent trends in Neutron star Physics, an invited talk delivered at the International NUSTAR WEEK 2012 at VECC, Kolkata, India (8 - 12 October 2012).
- Nuclear Astrophysics: Focusing telescopes on nuclear structure, an invited lecture delivered at the UGC Sponsored state level workshop Recent Advances in Theoretical Physics at Navyug Science College Surat, India (January 2012).
- Gravitaional waves from rapidly rotating neutron stars, a talk delivered at the International Conference on Theoretical and Applied Physics, at Indian Institute of Technology, Kharagpur, India (Dec 2011).
- Compact Stars, a Lecture delivered at the UGC Sponsored state level workshop Recent trends in Astronomy and Astrophysics at Navyug Science College Surat, India ( January 2011).
- Application of Relativistic Mean Field Theory (RMF) in Nuclear Physics, Invited Talk, delivered at the DAE-BRNS symposium in Nuclear physics, BITS Pilani, Pilani, India (December 2010).
- An Aesthetic journey to the Neutron Stars, a colloquium, delivered at Indian Institute of Astrophysics, Bangalore, India (April 2009).
- Aspects of Nuclear Matter to Neutron Stars a seminar delivered at Indian Institute of Science and Raman Research Institute, Bangalore, India (April 2009).
- From Nuclear Matter to Neutron Stars a review talk in DAE-BRNS Symposium on High Energy Physics held at Banaras Hindu University (BHU), Varanasi, India (December 2008).
- Attributes of a Rotating Neutron Star in DAE-BRNS Symposium on Nuclear Physics held at Sambalpur University, Odisha, India (December 2007).
- Nuclear Equation of State and Compact Stars in DAE-BRNS Symposium on High Energy Physics held at Indian Institute of Technology- Kharagpur, India (December 2006).
- Effect of Hyperons on Nuclear Equation of state and neutron star structure in an Effective Model in Workshop on Physics and Astrophysics of Hadrons and Hadronic Matter held at Visva Bharati University, Shantiniketan, India (November 2006).

## Teaching

Compulsory Courses

- PHY C341 Nuclear Physics (IC)
  PHY F343 Nuclear and Particle Physics (IC)
  PHY F111 Mechanics, Oscillations and Waves (IC)
  PHY F110 Physics Laboratory - I (IC)
  PHY C352 Atomic and Molecular Spectroscopy (IC)
  Physics Laboratory - II/ Advanced physics Laboratory (IC)
  Modern Physics Laboratory (IC)
  BITS F111 Thermodynamics (I)
  PHY F212 Electromagnetic Theory - I (IC)

## Teaching (continued)

Electives
- PHY F215 Introduction to Astronomy & Astrophysics (IC)
  PHY F317 Introduction to Radio Astronomy (IC)
  Nuclear Astrophysics (Reading Course) (IC)

Project Courses
- Study Oriented Projects
  Lab Oriented Projects
  Design Oriented Projects